\documentclass[a4paper,12pt,nofootinbib]{article}
\usepackage[english]{babel} 
\usepackage[T1]{fontenc}
\usepackage{float}
\usepackage{setspace}
\usepackage{appendix}
\usepackage{parcolumns}
\usepackage{cite}
\usepackage{times}
\usepackage[OT1]{fontenc}
\usepackage{type1cm}
\usepackage{url}
\usepackage{subfigure}
\usepackage{braket}
\usepackage{graphicx}
\usepackage{sidecap}
\usepackage{xspace}
\usepackage{tikz}
\usepackage{blkarray}
\usepackage{indentfirst}
\usepackage[mathscr]{eucal}
\usepackage{geometry}
\usepackage{booktabs}
\usepackage{indentfirst}
\usepackage{bm}
\usepackage{multirow}
\usepackage{dcolumn}
\usepackage{graphicx}
\usepackage{mathrsfs}  
\usepackage{enumerate}
\usepackage[utf8]{inputenc}
\usepackage{amsmath}
\usepackage{amssymb}
\usepackage{amsfonts}
\usepackage{amsthm}
\usepackage{mathrsfs}
\usepackage{graphicx}
\usepackage{multirow}
\usepackage{ulem}	
\usepackage{amsmath,mathtools}
\usepackage{comment}
\usepackage{amsbsy}
\usepackage{hyperref}
\usepackage{bookmark}

\usepackage{nccmath}   

\makeatletter
\@addtoreset{equation}{section}

\makeatother

\newcommand{\be}{\begin{equation}}
\newcommand{\ee}{\end{equation}}
\newcommand{\bea}{\begin{eqnarray}}
\newcommand{\eea}{\end{eqnarray}}
\newcommand{\vs}[1]{\vspace{#1 mm}}

\newcommand{\nn}{\nonumber\\}
\newcommand{\lc}{{\mit\Gamma}}
\newcommand{\con}{K}
\newcommand{\dis}{L}

\def\cL{{\cal L}}
\def\cM{{\cal M}}

\def\cP{{\cal P}}

\def\cO{{\cal O}}

\newcommand{\trp}[2]{\mathrm{tr}_{(#1#2)}\phi}
\newcommand{\divp}[1]{\mathrm{div}_{(#1)}\phi}
\newcommand{\divdivp}[2]{\mathrm{div}_{(#1#2)}\phi}
\newcommand{\divtrp}[2]{\mathrm{div}\,\mathrm{tr}_{(#1#2)}\phi}
\newcommand{\trdivp}[1]{\mathrm{tr}\mathrm{div}_{(#1)}\phi}
\newcommand{\Divp}[1]{\mathrm{Div}_{(#1)}\phi}

\newcommand{\trT}[2]{\mathrm{tr}_{(#1#2)}T}
\newcommand{\divT}[1]{\mathrm{div}_{(#1)}T}

\newcommand{\trdivT}[1]{\mathrm{tr}\mathrm{div}_{(#1)}T}
\newcommand{\DivT}[1]{\mathrm{Div}_{(#1)}T}

\newcommand{\trDivT}[1]{\mathrm{tr}\mathrm{Div}_{(#1)}T}
\newcommand{\trQ}[2]{\mathrm{tr}_{(#1#2)}Q}
\newcommand{\divQ}[1]{\mathrm{div}_{(#1)}Q}
\newcommand{\divdivQ}[2]{\mathrm{div}_{(#1#2)}Q}

\newcommand{\trdivQ}[1]{\mathrm{tr}\mathrm{div}_{(#1)}Q}
\newcommand{\DivQ}[1]{\mathrm{Div}_{(#1)}Q}

\newcommand{\trDivQ}[1]{\mathrm{tr}\mathrm{Div}_{(#1)}Q}
\newcommand{\deveq}{\nonumber\end{aligned}\end{equation*}\begin{equation*}\begin{aligned}}

\begin{document}

\vs{3}
\begin{center}
{\large\bf Metric-Affine Gravity\\
as an Effective Field Theory}
\vs{8}

{\large
A. Baldazzi${}^*$\footnote{e-mail address: abaldazz@sissa.it}
\quad
O. Melichev${}^*$\footnote{e-mail address: omeliche@sissa.it}
\quad
R. Percacci${}^*$\footnote{e-mail address: percacci@sissa.it}
}
\medskip

${}^*${International School for Advanced Studies, via Bonomea 265, I-34136 Trieste, Italy}
{and INFN, Sezione di Trieste, Italy}
\end{center}
\bigskip

{\narrower
{\bf Abstract.}
We discuss theories of gravity with independent metric (or frame field)
and connection, from the point of view of effective field theory.
We count the parity-even Lagrangian terms of dimension up to four
and give explicit bases for the independent terms that 
contribute to the two-point function.
We then give the decomposition of the linearized action 
on a complete basis of spin projectors and 
consider various subclasses of MAGs.
We show that teleparallel theories can be dynamically equivalent to
any metric theory of gravity and give the particle content
of those whose Lagrangian contains only dimension-two terms.
We point out the existence of a class of MAGs whose EOMs do not
admit propagating degrees of freedoms.
Finally, we construct simple MAGs that contain only 
a massless graviton and a state of spin/parity $2^-$ or $3^-$.
As a side result, we write the relativistic wave equation
for a spin/parity $2^-$ state.
}

\section{Introduction}

General Relativity (GR) can be regarded as an Effective Field Theory (EFT) 
with a range of validity that goes from macroscopic scales to the Planck scale.
Near this upper limit, correction terms are expected to
become significant.
These corrections are arranged systematically in powers
of derivatives (or better, in a covariant formalism, 
in powers of curvatures and covariant derivatives of curvatures)
and the first corrections are quadratic in curvature.
The theory of gravity with Lagrangian quadratic in curvature
will be referred to as four-derivative gravity (4DG)
(a.k.a. quadratic gravity) and has been studied,
independently of the EFT framework, for a long time.
It is known to be renormalizable \cite{stelle}
asymptotically free \cite{avrabar} (for the right signs of the couplings)
and to contain ghosts, but the ghosts are massive and
do not appear at low energy.
Even though the general EFT of gravity is not renormalizable,
quantum corrections affecting low-energy observables
can be unambiguously calculated \cite{Donoghue:1995cz}.

The theories mentioned so far are metric theories of gravity,
meaning that the metric is the only dynamical field
(apart of matter fields of course, but we will be mostly
concerned with pure gravity).
In metric theories of gravity the connection is constrained
to be the Levi-Civita (LC) connection, that is the unique
connection that is torsion-free and metric-compatible.
Metric-Affine Gravity (MAG) is a very large class of theories of gravity where the connection is treated as an independent variable.
Thus, they can have nonvanishing torsion, or non-metricity, or both.
Probably the best motivation for studying MAGs is that,
in many ways and more than Einstein's General Relativity,
they resemble the theories of the other fundamental interactions.
This has generated a large literature on ``gauge theories of gravity'',
where one tries to apply ideas and tools of Yang-Mills theories
to gravity. We refer to \cite{Blagojevic:2013xpa}
for a useful collection of references,
covering also the history of the subject.
Further, one may hope that this enhanced similarity
is a step towards a possible unification.
We will not discuss this further in this paper
and only refer to \cite{Krasnov:2017epi} for a review of such attempts.

In this paper we view MAGs as EFTs.
The main advantage of the EFT point of view is that it gives
a systematic criterion to write Lagrangians,
based on their impact on low-energy physics.
The general idea is to expand in the small parameter 
$E/\Lambda$, where $E$ is the typical energy of the processes
that one wants to study, and $\Lambda$ is ``the cutoff'' of the theory,
which is not to be regarded as an artificial regularization device, 
but rather as the energy where the given
field theory is expected to break down.
Since MAGs are theories of gravity, one may expect that
the field theoretic description breaks down at the Planck scale.
One would then typically identify the cutoff of the EFT with the Planck mass.
It is possible that the field theory still makes sense above
this cutoff, by some non-perturbative effect \cite{perbook2,Reuter:2019byg},
but in this paper we will mostly restrict our attention to
sub-Planckian physics.

In many cases $E$ can be traded for momentum,
and hence for derivatives of the fields.
In these cases the EFT expansion can be viewed as a derivative expansion.
In a gravitational context, a more covariant definition
would be based on powers of curvature, or covariant derivatives.
We will discuss the difficulties that 
a derivative expansion would entail for MAG, 
so in this paper we will consider an energy expansion,
where the terms in the Lagrangian are classified
according to their canonical dimension.
According to this criterion, the dominant term in the infrared 
would be the cosmological term.
It corresponds to the unit (dimensionless) operator in the Lagrangian.
However, being interested mostly in an expansion around flat space,
we implicitly assume that the cosmological term is negligible,
for reasons that we do not need to know.
Then, the leading terms in the EFT are the 
terms of dimension-two,
and the next-to-leading ones are the terms of dimension four.
The latter are utterly negligible at all reachable energy scales,
but are expected to become significant when we
come close to the Planck scale.

Then, there are two possible scenarios.
The most natural one is that all the masses that arise in
the theory are comparable to the Planck mass.
In this case the only physical particle in the MAG 
would be the graviton, and the EFT would be very similar to the
metric EFT of gravity already discussed in the literature.
All the massive states would already be ``integrated out''
and would only contribute tiny effects through quantum loops.
We emphasize that even in this apparently very dull case,
MAGs have a greater explanatory power
than metric EFTs of gravity,
because the vanishing of torsion and nonmetricity
can be shown to be generic consequences of the dynamics at low energy,
whereas in the metric theories it has to be postulated.

The more interesting scenario would occur if some of the massive
states are much lighter than the Planck scale,
so there would be an energy interval where these massive states
could exist as physical particles.
There is no difficulty in arranging this at the level of the
Lagrangian parameters, but this scenario would give rise 
to various issues.
The first is that maintaining the mass hierarchy in the 
presence of loop corrections would likely entail
some degree of fine tuning.
The second and more important issue is related to the fact 
that tree-level unitarity could be violated already at energies 
much below the Planck scale.
This has been discussed for higher spin fields 
in \cite{Cucchieri:1994tx}, and MAGs are generally 
higher-spin theories, because the connection is a three-index field
and generally contains a spin-3 degree of freedom.
Third, it is in general difficult to find
Lagrangians for MAG that do not contain pathological features such
as ghosts or tachyons
\cite{neville,Sezgin:1979zf,sezgin2,Lin:2018awc,Percacci:2020ddy,Marzo:2021esg,Marzo:2021iok}.

From this point of view, MAGs are similar to 4DG.
There has been recently a revived interest in possible
mechanisms to avoid these issues in 4DG
\cite{Mannheim:2006rd,Salvio:2014soa,Anselmi:2018ibi,Anselmi:2018tmf,Donoghue:2019fcb,Holdom:2021hlo,Holdom:2021oii},
and there have been some first steps to carry them over
also to MAGs \cite{Anselmi:2020opi,Piva:2021nyj}.
If these ideas are successful, one potential consequence
is that the spectrum of MAG may be very different from
what a naive tree-level analysis would indicate.
\footnote{It is even possible that no bosonic field propagates
above the Planck mass, a statement that has sometimes
been made in the context of noncommutative geometry
\cite{Kurkov:2013kfa}.}
In this paper we shall not venture so far,
but it is important to keep in mind that all our statements
may be subject to important changes when quantum corrections
are taken into account.

As a starting point for more detailed explorations
of these issues in MAG, the first goal of this paper will be
to discuss the most general MAG Lagrangian involving
parity-even terms of mass dimension two and four,
and to give a complete basis of such invariants for the two-point functions.
This is a task that in ordinary field theories is typically quite
straightforward, but in MAG, due to the tensorial nature of the fields,
already presents a certain degree of algebraic complexity.
In this connection we note that much of the 
older literature on MAGs was based
on the idea of gauging the Poincar\'e group.
Since the tetrad can be viewed as translational gauge field
and torsion as its curvature,
it was natural to restrict attention to Lagrangians
quadratic in curvature and torsion.
The most general Lagrangian of this type has 12 terms of dimension 2
and 16 terms of dimension 4, for a total of 28 free parameters.
This is also the class of Lagrangians that was considered in 
\cite{Percacci:2020ddy}.
However, such a restriction is unnatural from the point of view of EFT, 
where many more terms are expected to arise at dimension four.
To the best of our knowledge, there has been no systematic
exploration of this huge class of theories.

In general, at a fixed order in the EFT expansion,
MAGs contain more degrees of freedom (d.o.f.'s) than purely metric
theories of gravity.
In an expansion around flat space, these can be classified
according to their spin (from 0 to 3) and parity ($+$=even or $-$=odd).
In metric theories of gravity the only field is a symmetric tensor,
that can be decomposed in one spin $2^+$ component,
one spin $1^-$ component and two spin $0^+$ component.
\footnote{The spin  $1^-$ and one $0^+$ correspond to the action
of an infinitesimal diffeomorphism and therefore are pure gauge.
Here we just count states at a kinematical level.}
In MAG there is an additional three-index tensor,
which in general can carry one $3^-$ state,
three $2^+$ states, two $2^-$ states,
three $1^+$ states, six $1^-$ states,
four $0^+$ states, one $0^-$ state.
Depending on the Lagrangian, several of these could correspond to
propagating particles.
Furthermore, the physical graviton could be a mixture of the metric fluctuation with some of the $2^+$ states of the three-index field.

It may be useful to observe that in particle physics,
every particle state is carried by a different field.
Even though Lorentz vector and tensor fields carry several
representations of the three-dimensional rotation group,
as a rule only one of them is physical,
the others playing the role of auxiliary variables.
Some of the previously studied examples of 
ghost- and tachyon-free MAGs represent
interesting exceptions to this rule:
they contain more particle states than fields.
By ``simple MAG'' we mean a MAG that conforms to the rule:
they contain a massless graviton
and only one other particle carried by the three-index field.
We will briefly review the spin projector technique here and apply it to
describe general teleparallel theories
and two special simple MAGs
with spin/parity $2^-$ or $3^-$.
These examples suggest a systematic way of constructing MAGs with
a predetermined particle content.

The plan of the paper is as follows.
In Section 2 we introduce the basic objects,
without delving too much in their geometric meaning.
Then we note that every MAG has two equivalent descriptions
that differ by a field redefinition.
We name them after Cartan and Einstein.
The first is perhaps more appealing to a geometrical mind,
while the second comes more natural to particle physicists.
\footnote{In this connection, see e.g. the exchange in
\cite{weinberg-hehl}.}
They are both equally valid, but the latter is often
easier to work with.
We then give a rough classification of MAGs
based on the presence or absence of curvature, torsion and nonmetricity.
We give a proof of the fact that any metric theory of gravity
has a teleparallel equivalent.
In Section 3 we count all terms of dimension two and four,
and explicilty enumerate those that
are quadratic in curvature and/or covariant derivatives of
torsion and nonmetricity.
We do this both in the Einstein and in the Cartan forms,
and show the correspondence between them.
In Section 4 we review the spin-projector technique for MAG
and discuss the effect of gauge invariances.
Section 5 is devoted to theories with dimension-two Lagrangians,
and in particular to teleparallel theories.
In Section 6 we discuss a special class of MAGs that,
viewed in the Cartan form, would look perfectly normal,
but have no propagating degrees of freedom.
In Section 8 we construct two ``simple'' MAGs,
by which we mean MAGs with only the graviton
and a single other propagating particle.
In particular, we consider the cases of spin $2^-$ or $3^-$,
which require a three-index tensor for a Lorentz-covariant description.

Appendices \ref{sec:app.lin} and \ref{sec:app.projective} 
contain sprawling formulae, that are ubiquitous in this subject
and would make the main text hard to follow.
In Appendix \ref{sec:app.propagators} we collect 
the standard forms of the propagators for any particle of spin up to $3^-$ 
(spin $3^+$ would need a four-index tensor).
As a side result, we use the spin projector formalism to
derive the relativistic wave equation for a $2^-$ particle.
Appendix \ref{sec:app.coefficients} contains the kinetic coefficient
for the general MAG, which is one of the main results of this paper.

\subsection{Notation and conventions}
\label{sec:not}

We use standard GR notation for the Levi-Civita connection
and standard Yang-Mills notation for the dynamically independent
connection, as in the following table
\medskip

\begin{center}
\begin{tabular}{|c|c|c|c|}
\hline
& coefficients 
& covariant derivative
& curvature 
\\
\hline
LC connection 
&  $\lc_\mu{}^\rho{}_\sigma$
& $\nabla_\mu$ 
& $R_{\mu\nu}{}^\rho{}_\sigma$
\\
\hline
Independent connection 
&  $A_\mu{}^\rho{}_\sigma$
& $D_\mu$ 
& $F_{\mu\nu}{}^\rho{}_\sigma$
\\
\hline
\end{tabular}
\end{center}
\medskip

We will use same symbol for a given geometrical object in any frame,
thus for example $A_\mu{}^\rho{}_\sigma$ 
are the connection coefficients in a coordinate frame
and $A_\mu{}^a{}_b$ are the connection coefficients
in a frame (\ref{frames}).

In order to identify more easily expressions
involving the same tensors with indices contracted
in different ways, it proves convenient to use the following notation.
Given a tensor $\phi_{abc}$, we define
\bea
\trp12_c&\equiv&\phi^{(12)}_c=\phi_a{}^a{}_c\ ,\quad
\trp13^b\equiv\phi^{(13)b}=\phi_a{}^{ba}\ ,\ \mathrm{etc.}
\nonumber\\
\divp1^b{}_c&=&\nabla_a\phi^{ab}{}_{c}\ ,\quad
\divp2_{ac}=\nabla_b\phi_a{}^b{}_c\ ,\  \mathrm{etc.}
\nonumber\\
\divdivp23_c&=&\nabla_a\nabla_b\phi^{ab}{}_c\ ,\ \mathrm{etc.}
\nonumber\\
\trdivp1&=&\divp1^a{}_a\ ,\quad
\divtrp12=\nabla_a\trp12^a\ ,\ \mathrm{etc.}
\nonumber
\eea
Note that with the LC connection
$\divtrp12=\trdivp3$, etc.

When the divergence is calculated with the
independent dynamical connection $A$, it will be written as ``Div''.
In this case one has to be more
careful about raising and lowering indices,
because the covariant derivative of the metric may not be zero.
Then one has to make conventions, for example,
$\Divp1^b{}_c=D_a(g^{ad}\phi_d{}^b{}_{c})$
or
$\Divp1^b{}_c=g^{ad}D_a\phi_d{}^b{}_{c}$.
We will not need to commit ourselves to such
choices in this paper.

\section{General connections}

\subsection{Torsion, nonmetricity and curvature in various bases}
\label{sec:TQF}

In this section we use arbitrary bases $\{e_a\}$
in the tangent spaces
and $\{e^a\}$ in the cotangent spaces.
We use interchangeably ``basis'', ``frame'' and ``gauge''.
Given a coordinate system $x^\mu$,
they are related to the coordinate bases by
\be
e_a=\theta_a{}^\mu\partial_\mu\ ,\qquad
e^a=\theta^a{}_\mu dx^\mu\ .
\label{frames}
\ee
The transformation matrices $\theta_a{}^\mu$
and $\theta^a{}_\mu$ are called the frame field
and coframe field (a.k.a. soldering form).
They can also be given a global geometrical
meaning as isomorphisms between two bundles.
We do not need that here.

The components of the metric in the general frames
are $g_{ab}$ and the components of a general
linear connection are $A_a{}^b{}_c$.
They are related to the components in the coordinate bases by
\begin{eqnarray}
\label{pullbackmetric}
g_{\mu\nu}&=&\theta^a{}_\mu\, \theta^b{}_\nu\, g_{ab} \ ,\\
\label{pullbackconnection}
A_\lambda{}^\mu{}_\nu&=&\theta_a{}^\mu A_\lambda{}^a{}_b \theta^b{}_\nu
+\theta_a{}^\mu \partial_\lambda \theta^a{}_\nu \ .
\end{eqnarray}
When one works with generic frames,
the dynamical variables of gravity are
the metric $g$, the frame field $\theta$ and
(in a MAG) the connection $A$.
In this formalism, the theory is invariant
under local changes of frame i.e. local $GL(4)$ transformations
\cite{Floreanini:1989hq,percacci2,Siegel:1993xq,Kirsch:2005st,Leclerc:2005qc}.

The nonmetricity tensor $Q$ is (minus) the covariant derivative 
of the metric and torsion is the
exterior covariant derivative of the frame field:
\begin{eqnarray}
\label{nonmetricity}
Q_{\lambda ab}&=&
-\partial_\lambda g_{ab}
+A_\lambda{}^c{}_a\, g_{cb}
+A_\lambda{}^c{}_b\, g_{ac} \ ,\\
\label{torsion}
T_\mu{}^a{}_\nu&=&
\partial_\mu \theta^a{}_\nu-\partial_\nu \theta^a{}_\mu+
A_\mu{}^a{}_b\, \theta^b{}_\nu-A_\nu{}^a{}_b\, 
\theta^b{}_\mu \ .
\end{eqnarray}
These are called the nonmetricity and torsion, respectively.

Given a metric $g_{ab}$ and frame field $\theta^a{}_\mu$, 
there is a unique connection, 
called the Levi Civita connection, such that 
$T=0$ and $Q=0$.
Its components are
\be
\lc_{abc}=\frac12\left(
E_{acb}+E_{cab}-E_{bac}\right)
-\frac12\left(f_{abc}+f_{cab}-f_{bca}\right) \ ,
\ee
where 
\bea
E_{cab}&=&{\theta}_c{}^\lambda\, \partial_\lambda g_{ab} \ ,
\\
f_{bc}{}^a&=&
\left(\theta_b{}^\mu\,
\partial_\mu\theta_c{}^\lambda-
\theta_c{}^\mu\, \partial_\mu\theta_b{}^\lambda
\right)\theta^a{}_\lambda \ .
\eea
Note that $E$ and $f$ are not tensors
($f$ are the structure functions of the frame fields).

It is cumbersome to work with generic bases.
There are two types of more convenient bases:
coordinate (a.k.a. natural) bases and orthonormal bases.

In a coordinate basis the frame field has components
$$
\theta_a{}^\mu=\delta_a^\mu\ .
$$
The structure functions $f$ vanish
(since $[\partial_\mu,\partial_\nu]=0$)
and in the formula for the LC connection
only the first term remains.
In this gauge one recognizes that
$\lc_\lambda{}^\mu{}_\nu$ are the Christoffel symbols.
Then, torsion is a purely algebraic object:
\be
T_\mu{}^\rho{}_\nu=
A_\mu{}^\rho{}_\nu-A_\nu{}^\rho{}_\mu
\label{TA}
\ee
whereas non-metricity always involves a derivative of $g$.

In an orthonormal basis the metric has components
$$
g_{ab}=\eta_{ab}\ .
$$
Then (\ref{pullbackmetric}) becomes the defining relation
for the tetrad.
$E=0$ and in the formula for the LC connection
only the second term remains.
One recognizes the resulting formula as
the ``spin connection''.
\footnote{The spin connection is often called $\omega_{\mu ab}$.
Here we stick to the convention that the components of
the same geometrical object in different bases
should not be given different names.}
In this gauge the nonmetricity is a purely algebraic object:
\be
Q_{cab}=A_{cab}+A_{cba}
\label{QA}
\ee
whereas torsion still involves a derivative of $\theta$.

The curvature of the independent connection is,
in a coordinate basis,
\be
F_{\rho\sigma}{}^\mu{}_\nu=
\partial_\rho A_\sigma{}^\mu{}_\nu
-\partial_\sigma A_\rho{}^\mu{}_\nu
+A_\rho{}^\mu{}_\lambda A_\sigma{}^\lambda{}_\nu
-A_\sigma{}^\mu{}_\lambda A_\rho{}^\lambda{}_\nu
\ee
where, in line with standard Yang-Mills conventions,
the last two indices can be viewed as a single index
for the Lie algebra of $GL(4)$.
We use the same convention for the curvature of the
LC connection, which is the Riemann tensor:
\be
R_{\rho\sigma}{}^\mu{}_\nu=
\partial_\rho\lc_\sigma{}^\mu{}_\nu
-\partial_\sigma\lc_\rho{}^\mu{}_\nu
+\lc_\rho{}^\mu{}_\lambda\lc_\sigma{}^\lambda{}_\nu
-\lc_\sigma{}^\mu{}_\lambda\lc_\rho{}^\lambda{}_\nu\ .
\ee

An important role will be played by the Bianchi identities.
For the independent dynamical connection, they read
\bea
F_{[\alpha\beta}{}^\gamma{}_{\delta]}
-D_{[\alpha} T_\beta{}^\gamma{}_{\delta]}
-T_{[\alpha}{}^\epsilon{}_{\beta|} T_\epsilon{}^\gamma{}_{|\delta]}&=&0 \ ,
\label{bianchi1}
\\
D_{[\alpha}F_{\beta\gamma]}{}^\delta{}_\epsilon+T_{[\alpha}{}^\eta{}_{\beta|} F_{\eta|\gamma]}{}^\delta{}_\epsilon&=&0 \ .
\label{bianchi2}
\eea
The Bianchi identities of the LC connections are the same,
except that the torsion terms are missing.

In order to minimize the number of fields,
in this paper we shall work mostly with coordinate bases.
One has to bear in mind that this is already a partial gauge choice
(we have eliminated the freedom of choosing a frame independently
of the coordinate system)
and that it hides some general features of the theory.

\subsection{The distortion}
\label{sec:phi}

We will denote $A_\mu{}^\rho{}_\nu$ a generic connection in the tangent bundle.
Given a metric $g_{\mu\nu}$, it can be uniquely decomposed into
\be
\label{phi}
A_{\alpha\beta\gamma}=\lc_{\alpha\beta\gamma}+\phi_{\alpha\beta\gamma} \ ,
\ee
where $\lc_{\alpha\beta\gamma}$ is the LC connection of $g_{\mu\nu}$
and $\phi_{\alpha\beta\gamma}$ is a tensor that,
following \cite{Hehl:1994ue}, we will call ``distorsion''.
In general, it has no symmetry properties.
Indices are raised and lowered with $g_{\mu\nu}$.
From (\ref{torsion}) and (\ref{nonmetricity}) one finds
\be
\label{TQphi}
T_{\alpha\beta\gamma} = \phi_{\alpha\beta\gamma}-\phi_{\gamma\beta\alpha}\ ,\qquad
Q_{\alpha\beta\gamma} = \phi_{\alpha\beta\gamma}+\phi_{\alpha\gamma\beta}\ .
\ee
These relations can be inverted, to give the distortion
as a function of torsion and nonmetricity.
In fact we can write
\be
\label{discon}
\phi_{\alpha\beta\gamma}= \dis_{\alpha\beta\gamma}+\con_{\alpha\beta\gamma}\ ,
\ee
where 
\begin{eqnarray}
\dis_{\alpha\beta\gamma} & = & \frac{1}{2}\left(Q_{\alpha\beta\gamma}+Q_{\gamma\beta\alpha}-Q_{\beta\alpha\gamma}\right)\ ,
\nonumber\\
\con_{\alpha\beta\gamma} & = & \frac{1}{2}\left(T_{\alpha\beta\gamma}+T_{\beta\alpha\gamma}-T_{\alpha\gamma\beta}\right)\ .
\label{emanuele}
\end{eqnarray}
Note that the tensor $\con_{\alpha\beta\gamma}$, called the contortion, 
is antisymmetric in the second and third index
(whereas $T$ is antisymmetric in the first and third).
The tensor $\dis_{\alpha\beta\gamma}$, that does not seem to have
a commonly accepted name,
is symmetric in the first and third index (whereas $Q$ is symmetric in the second and third index).

Notice that (\ref{TQphi}) can then also be written as
\be
\label{TQalphabeta}
T_{\alpha\beta\gamma} =  \con_{\alpha\beta\gamma}-\con_{\gamma\beta\alpha}\ ,
\qquad
Q_{\alpha\beta\gamma} = \dis_{\alpha\beta\gamma}+\dis_{\alpha\gamma\beta}\ ,
\ee
so $\dis$ contains all the nonmetricity and $\con$ contains all the torsion.
Another way of saying this is that $\lc+\con$ is torsion-free
and $\lc+\dis$ is metric.
We shall actually not use the tensors $\con$ and $\dis$ in the following
and prefer to express everything either in terms of $\phi$
or of $T$ and $Q$.
 
We denote $F_{\mu\nu}{}^\rho{}_\sigma$ the curvature tensor of $A_\mu{}^\rho{}_\sigma$,
and $R_{\mu\nu}{}^\rho{}_\sigma$ the curvature tensor of $\lc_\mu{}^\rho{}_\sigma$.
They are related as follows:
\begin{eqnarray}
F_{\mu\nu}{}^\alpha{}_\beta & = & 
R_{\mu\nu}{}^\alpha{}_\beta
+\nabla_{\mu}\phi_{\nu \,\,\, \beta}^{\,\,\, \alpha}-\nabla_{\nu}\phi_{\mu\,\,\, \beta}^{\,\,\, \alpha}
+\phi_{\mu\,\,\, \gamma}^{\,\,\, \alpha}\phi_{\nu \,\,\, \beta}^{\,\,\, \gamma}-\phi_{\nu \,\,\, \gamma}^{\,\,\, \alpha}\phi_{\mu \,\,\, \beta}^{\,\,\, \gamma} \ .
\label{FtoR}
\end{eqnarray}
In general, $F$ is only antisymmetric in the first two indices.
It has three independent contractions:
the Ricci-like tensors
$$
F^{(13)}_{\mu\nu}=F_{\lambda\mu}{}^\lambda{}_\nu\ ,\qquad
F^{(14)}_{\mu\nu}=g^{\alpha\beta}F_{\alpha\mu\nu\beta}
$$
that do not have symmetry properties in general,
and the antisymmetric tensor 
$$
F^{(34)}_{\mu\nu}=F_{\mu\nu}{}^\lambda{}_\lambda\ .
$$

The analog of the Ricci scalar for the connection $A_\mu{}^\alpha{}_\beta$
is the unique contraction $F_{\mu\nu}{}^{\mu\nu}$,
which, up to total derivatives, can be written as
\begin{eqnarray}
F_{\mu\nu}{}^{\mu\nu} & = & R
+\phi_{\mu \,\,\,\, \gamma}^{\,\,\,\, \mu}\phi_{\nu}^{\,\,\,\, \gamma\nu}-\phi_{\nu\mu\gamma}\phi^{\mu\gamma\nu}\ .
\end{eqnarray}
This can be reexpressed in terms of non-metricity and
torsion as
\begin{eqnarray}
F_{\mu\nu}{}^{\mu\nu} & = & R
+\frac{1}{4}T_{\alpha\beta\gamma}T^{\alpha\beta\gamma}
 +\frac{1}{2}T_{\alpha\beta\gamma}T^{\alpha\gamma\beta}
 -\trT12_{\alpha}\trT12^{\alpha}
\nonumber\\
 &  & 
+\frac{1}{4}Q_{\alpha\beta\gamma}Q^{\alpha\beta\gamma}-\frac{1}{2}Q_{\alpha\beta\gamma}Q^{\beta\alpha\gamma}
-\frac{1}{4}\trQ23_{\alpha}\trQ23^{\alpha}
+\frac{1}{2} \trQ12_{\alpha}\trQ23^{\alpha}\nonumber \\
 &  & 
 -Q_{\alpha\beta\gamma}T^{\alpha\beta\gamma}
  -\trQ23_{\alpha}\trT12^{\alpha}
  +\trQ12_{\alpha}\trT12^{\alpha}\ .
\label{palatiniTQ}
\end{eqnarray}

\subsection{The Higgs phenomenon}
\label{sec:Higgs}

While not strictly necessary for the rest of the paper,
this section is useful to understand in what sense MAG 
is closer to other gauge theories of physics than GR,
and its limitations as an EFT.
We give here a minimal account, and refer to \cite{percacci2,cuiaba}
for a more detailed discussion.

We start by noting that the frame field is subject to the (nonlinear)
constraint of non-degeneracy, such that locally it can be viewed as
a having values in the linear group: $\theta^a{}_\mu\in GL(4)$.
The metric is also subject to nonlinear constraints
(the eigenvalues must have definite signs)
and locally can be seen as having values in the coset $GL(4)/O(1,3)$.
Thus, metric and frame carry nonlinear realizations of the linear group
\cite{{Isham:1971dv}}.
Local linear transformations of the frame
are represented by left multiplication on $\theta$
and similarity transformations on $g$.
By writing all tensorial formulas in arbitrary frames,
they become covariant under local $GL(4)$ transformations
and in this way one can see gravity as a gauge theory of the
linear group, with $A_\lambda{}^\mu{}_\nu$ a connection
for the linear group.
In this theory a choice of gauge is equivalent to a choice of frame.
In an explicitly $GL(4)$--invariant formulation,
the fields $\theta^a{}_\mu(x)$ and $g_{ab}(x)$
must be simultaneously present,
but either one of them can be gauged away,
by choosing either coordinate frames ($\theta^a{}_\mu=\delta^a_\mu$)
or orthonormal frames ($g_{ab}=\eta_{ab}$), respectively.
Thus these fields can be seen as ``gauged Goldstone bosons'',
akin to $SU(N)$ scalar fields $U$ coupled to an $SU(N)$ gauge field.
Unlike in the particle physics examples, however,
in MAG there are two Goldstone bosons 
and there is not enough gauge freedom to gauge both away.

The gauged sigma model is the low-energy description of
the Higgs phenomenon, whose essence is the following.
The kinetic term of the Goldstone bosons is
$f^2(DU)^2$, where $D_\mu U=\partial_\mu U-A_\mu U$
and $f$ is a coupling with dimension of mass.
One assumes that in the ground state of the theory  
$DU=0$ and $F=0$ ($F$ is the curvature of $A$).
One can choose a gauge such that this state is represented by $A=0$, $U=1$
and expanding the fields around this VEV, the leading term
is a mass for $A$.
More generally, and independently of the ground state,
one can always choose the unitary gauge $U=1$, 
and the kinetic term of the Goldstone bosons
becomes $f^2 A_\mu A^\mu$, a mass term for the connection.

One sees that something very similar happens in MAG.
From (\ref{pullbackmetric}) and (\ref{pullbackconnection})
it is clear that in a formulation of MAG that is explicitly
invariant under local $GL(4)$ transformations,
the kinetic terms of $\theta^a{}_\mu$ and $g_{ab}$
are of the form $a T^2$ and $a Q^2$,
with indices on $T$ and $Q$ contracted in various possible ways
and $a$ a coupling with dimension of mass squared.
Let us then assume that the ground state of the theory
is given by $F=0$, $T=0$ and $Q=0$, i.e. flat spacetime.
One can choose a gauge such that this state is represented
by $A_\lambda{}^\mu{}_\nu=0,$
$\theta^a{}_\mu=\delta^a_\mu$ and $g_{ab}=\eta_{ab}$.
Expanding the fields around their VEVs,
the leading term of $aT^2$ and $aQ^2$ is just $aA^2$,
a mass term for the connection.
Alternatively, in the gauge $\theta^a{}_\mu=\delta^a_\mu$,
Equation (\ref{TA}) shows that
$aT^2$ is just a mass term for certain antisymmetric components
of the connection,
and in the gauge $g_{ab}=\eta_{ab}$, Equation (\ref{QA})
shows that $aQ^2$ is a mass term
for certain symmetric components of the connection.
Thus, these two gauge choices are analogs of the unitary gauge.
Even more covariantly, and independently of the state,
Equation (\ref{TQphi}) shows that
$aT^2$ and $aQ^2$ are mass terms for the deviation of
the connection from the LC connection.
As in the ordinary Higgs phenomenon,
the kinetic term of the ``Goldstone bosons'' is just a gauge-invariant
way of writing a mass term for the gauge field.

The analogy with the particle physics models
only falls short in one respect:
whereas in the particle physics example any fluctuation of
the Goldstone boson can be absorbed by the choice of unitary gauge,
(i.e. after letting $U$ fluctuate we can readjust the gauge
so that $U=1$ again),
the $GL(4)$ gauge freedom is not sufficient to absorb
the fluctuation of both $\theta$ and $g$.
One of them contains physical degrees of freedom, 
and gives rise to the graviton.
What is true in both cases is that the connection degree of freedom
becomes massive and therefore cannot be excited at sufficiently low energy as an independent degree of freedom.

This discussion makes it clear that at a formal level MAGs are 
very close in spirit to gauged nonlinear sigma models or, for a phenomenologically important example, to the so-called 
``electroweak chiral perturbation theory'',
which is obtained as the limit of electroweak theory
when the Higgs particle becomes infinitely massive
\cite{Appelquist:1980vg,Longhitano:1980iz,Donoghue:1989qw,Herrero:1993nc}
As mentioned in the introduction, whether 
there exists a window in which MAG can act as an EFT
with propagating graviton and connection
(or equivalently graviton and distortion)
depends on the mass matrix for the latter.
Furthermore, the EFT has nothing to say on the theory it comes from
as a low energy approximation.
Nevertheless, the model suggest the possibility that
it could be a theory where $GL(4)$ is linearly realized.
This would correspond to an ``unbroken'' phase where the dynamics
is independent of the metric, i.e. a topological field theory.

\subsection{Equivalent forms}
\label{sec:equiv}

It appears from the discussion in the previous sections
that any MAG can be described in two equivalent ways,
depending on what connection is used to write covariant derivatives.
\begin{itemize}
\item 
if the connection $A_\mu{}^\alpha{}_\beta$ is used to write
the covariant derivatives,
the Lagrangian will be a combination of curvature tensors $F_{\alpha\beta\gamma\delta}$,
their covariant derivatives, the tensors $T$, $Q$
and their covariant derivatives $D_\mu T_{\alpha\beta\gamma}$,
$D_\mu Q_{\alpha\beta\gamma}$.
In this form, the theory is very similar to a Yang-Mills theory.
We will call this {\bf ``the Cartan form''} of MAG. 
\item 
if the LC connection $\lc_\mu{}^\alpha{}_\beta$
is used to write the covariant derivatives,
the Lagrangian will be a combination of the Riemann tensor
$R_{\alpha\beta\gamma\delta}$ and
its covariant derivatives, the distortion $\phi_{\alpha\beta\gamma}$
and its covariant derivatives $\nabla_\mu\phi_{\alpha\beta\gamma}$,
(or equivalently $T$, $Q$ and their covariant derivatives).
In this form, MAG looks like ordinary metric gravity
coupled to a peculiar matter field.
We will call this {\bf ``the Einstein form''} of MAG.
\end{itemize}

Using equation (\ref{phi}), any action for a MAG in Cartan form
can be rewritten in Einstein form
\be
S_C(g,A)=S_c(g,\lc+\phi)=S_E(g,\phi)\ .
\label{C2E}
\ee
We see that the transformation from Cartan to Einstein form
is just a change of field variables.
\footnote{A choice of variables in field theory is sometimes called a ``frame''.
Thus we could also speak of ``Cartan frame'' and ``Einstein frame''.
We prefer not to do so, in order to avoid confusion with the
Einstein frame of conformal geometry,
and more importantly because we are already using the term ``frame''
in its more standard meaning of linear basis in the tangent space.}
The two forms of the theory are physically equivalent.

Because of this choice, and of the possibility of using different
frames (either general or natural or orthonormal), the same MAG can be presented
in several ways, that may not be immediately recognizable.
It is thus important to distinguish physical statements
that do not depend on the gauge (i.e. the choice of frame)
or on the choice of field variables,
from statements that depend upon these choices
and have no physical content.

One such aspect is the number of derivatives,
which in the EFT approach is often used
to assess the relative importance of different terms
in the Lagrangian.
In the Einstein form of MAG, the independent fields are
the metric $g_{\mu\nu}$ and the distortion $\phi_\rho{}^\mu{}_\nu$.
The torsion and non-metricity tensors
are algebraic linear combinations of the distortion
and can themselves be taken as independent dynamical variables.
Thus for example, a term like $T^2$ has no derivatives
and counts as a mass term, while a term like $(\nabla T)^2$
has two derivatives and counts as an ordinary kinetic term.

In the Cartan form of MAG, the status of torsion and non-metricity
depends on the choice of basis, i.e. on the gauge.
In a general linear basis, they are the covariant derivatives
of the fundamental dynamical variables $\theta$ and $g$.
Thus terms like $T^2$ or $Q^2$ have two derivatives, 
while $(\nabla T)^2$ or $(\nabla Q)^2$ have four derivatives.
Things will look different if we use special frames.
In coordinate frames, a term like $T^2$ has no derivatives
and $(DT)^2$ has two derivatives
but $Q^2$ has two derivatives and $(DQ)^2$ has four derivatives.
Conversely, in an orthonormal frame  $Q^2$ has no derivatives
and $(DQ)^2$ has two derivatives
but $T^2$ has two derivatives and $(DT)^2$ has four derivatives.
Obviously the physics cannot change.
In particular, the physical propagating degrees of freedom
must be the same in all these different versions of the theory.
We see that the number of derivatives depends on the choice
of field variables, and on the choice of gauge.
This highlights that the derivative expansion
is not a useful approach in MAG.
When we regard MAG as an EFT,
we shall therefore classify the terms in the Lagrangian 
according to their canonical dimension.
\footnote{
It is worth emphasizing that similar, though somewhat simpler,
considerations apply also to EFT's containing Yang-Mills fields.}

\subsection{Basic classification}
\label{sec:class}

Even in its simplest form (using coordinate bases), 
a general MAG contains 74 component functions and, 
as we shall discuss later, 
its Lagrangian has hundreds of free parameters.
There are two ways in which one can reduce this complexity.
One is to impose additional gauge invariances, on top of diffeomorphisms.
These gauge invariances have two effects:
they make some field components unphysical,
and they constrain the form of the Lagrangian,
reducing the number of free parameters.
We shall discuss in Section \ref{sec:gaugeinv} some examples of gauge invariances.
It is important that such symmetries should be present at the full
nonlinear level, because in this case one could hope that they
persist when quantum corrections are taken into account.
Accidental symmetries that may be present at linearized level but not in the full theory, will generally be broken
by quantum effects.

The other way is to impose kinematical constraints on the fields.
There are very many ways of doing this,
but here we shall discuss only the most basic possibilities,
which are suggested by the discussion in the previous sections:
we will say that
a MAG is {\bf symmetric} if $\phi_{abc}$ is symmetric in $a$, $c$,
{\bf antisymmetric} if $\phi_{abc}$ is antisymmetric in $b$, $c$,
or {\bf general} if $\phi_{abc}$ has no symmetry property.
\footnote{A three-index tensor that is simultaneously symmetric in one
pair of indices and antisymmetric in another is zero.
Thus a MAG that is simultaneously symmetric and antisymmetric
is not a MAG - it does not have an independent connection.} 
Then, from (\ref{TQphi}) we see the following:
\begin{itemize}
\item 
{\bf ``Antisymmetric MAG''}. In this case $Q=0$,
so the connection is metric-compatible.
These may also be called ``metric MAGs'',
but we will refrain from doing so in order not to
confuse them with metric theories of gravity
(where the only variable is the metric).
\item 
{\bf ``Symmetric MAG''}. 
In this case $T=0$, so this type of theory can be equivalently characterized as being torsion-free. 
\item 
{\bf ``General MAG''}. 
In this case both $T$ and $Q$ are generally nonzero.
\end{itemize}
\medskip

More restrictive kinematical constraints could consist
in assuming that torsion or nonmetricity are of a special form,
for example 
$T_{\alpha\beta\gamma}=v^\delta\epsilon_{\alpha\beta\gamma\delta}$
(this example arises in supergravity) or
$Q_{\lambda\mu\nu}=b_\lambda g_{\mu\nu}$ (as in Weyl's theory).
Another interesting class of MAGs are the teleparallel theories,
where one imposes $F_{\alpha\beta\gamma\delta}=0$.
We emphasize that at this stage these are just kinematical
restrictions on the theory, without implications for the dynamics.

\begin{figure}
\begin{center}
\begin{tikzpicture}
[>=stealth,scale=0.55,baseline=5mm]
\fill[fill=yellow!60,draw,thick]
(0,0) node[circle,fill=red!20,align=center,draw,scale=0.8]{$F=Q=0$\\antisymmetric\\teleparallel}
-- (6,0) node[fill=red!20,align=center,draw,rounded corners,scale=0.8]{$F=0$\\general\\teleparallel}
-- (12,0) node[circle,fill=red!20,align=center,draw,scale=0.8]{$F=T=0$\\symmetric\\teleparallel}
-- (9,5.2) node[fill=red!20,align=center,rotate=-60,draw,rounded corners,scale=0.8]{$T=0$\\symmetric MAG}
-- (6,10.4) node[circle,fill=red!20,align=center,draw,scale=0.8]{$T=Q=0$\\metric theories}
-- (3,5.2) node[fill=red!20,align=center,rotate=60,draw,rounded corners,scale=0.8]{$Q=0$\\antisymmetric MAG}
-- (0,0);
\draw (6,3.45) node{ General MAG};
\end{tikzpicture}
\end{center}
\caption{The interior of the triangle represents general MAGs,
the sides MAGs with one kinematical constraint,
the vertices MAGs with two kinematical constraints.
This figure had been used in \cite{Jarv:2018bgs,BeltranJimenez:2019tjy} 
as a representation of the relation between GR and its teleparallel equivalents, but it can be used in a broader context.}
\label{fig:triangle}
\end{figure}
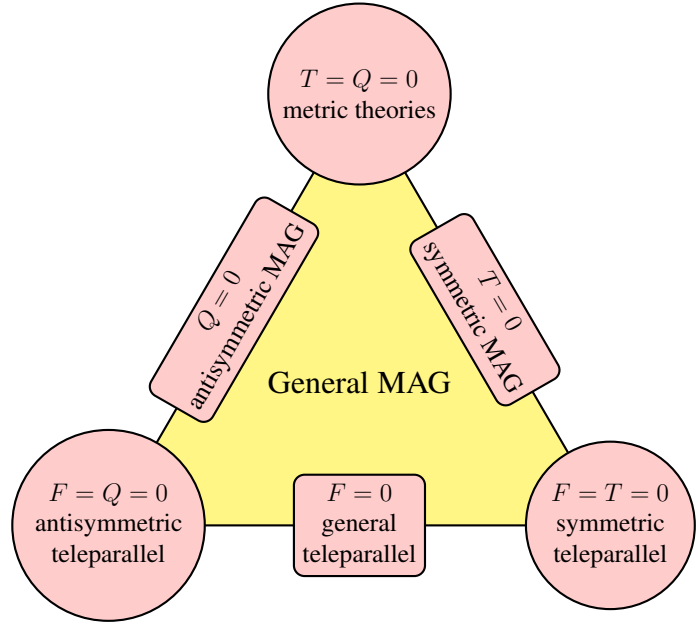
\medskip

According to the presence or absence of kinematical constraints, 
MAGs can be arranged in a triangle, as in Fig.\ref{fig:triangle}.
The theories in the top vertex are formulated in terms
of the metric (and possibly a frame field, but this is just
a different gauge choice) and the connection is the LC one.
The geometry they use is Riemannian geometry.
These are the metric theories of gravity.
GR is the metric theory of gravity whose Lagrangian
contains at most two derivatives of the metric,
but there are infinitely many more complicated ones,
containing higher powers of curvature.

The base of the triangle contains the teleparallel theories.
Historically the first and still best known example is
the Weitzenb\"ock theory, or antisymmetric teleparallel theory,
that contains only torsion and resides in the bottom left corner.
Slightly less well-known are teleparallel theories constructed 
only with nonmetricity, that occupy the right corner
\cite{Vitagliano:2013rna,BeltranJimenez:2017tkd,Jarv:2018bgs,Conroy:2017yln}
General teleparallel theories,
filling the base of the triangle, have only been discussed more recently
\cite{BeltranJimenez:2019tjy,BeltranJimenez:2019odq}.
We shall discuss them further in the next section.

For many purposes it is enough to consider theories that
contain only torsion or only nonmetricity.
These simplified models correspond to the sides of the triangle.
They have fewer fields (34 and 50, respectively,
when one uses coordinate frames) and correspondingly
fewer terms in the action.
In the following we will sometimes describe these cases separately.

\subsection{Universality of teleparallelism}

At the dynamical level it is known how to formulate actions
for any teleparallel geometry that yield equations that
are equivalent to Einstein's equations
(``teleparallel equivalents of GR'').
Their Lagrangian is
\be
\mathbb{T}=\frac{1}{4}T_{\alpha\beta\gamma}T^{\alpha\beta\gamma}
 +\frac{1}{2}T_{\alpha\beta\gamma}T^{\alpha\gamma\beta}
 -\trT12_{\alpha}\trT12^{\alpha}
 \label{ATEGR}
\ee
for the antisymmetric case,
\be
\mathbb{Q}=\frac{1}{4}Q_{\alpha\beta\gamma}Q^{\alpha\beta\gamma}-\frac{1}{2}Q_{\alpha\beta\gamma}Q^{\beta\alpha\gamma}
-\frac{1}{4}\trQ23_\alpha \trQ23^{\alpha}
+\frac{1}{2}\trQ23_{\alpha}\trQ12^{\alpha}
\label{STEGR}
\ee
for the symmetric case and
\be
\mathbb{G}=\mathbb{T}+\mathbb{Q}
 -Q_{\alpha\beta\gamma}T^{\alpha\beta\gamma}
 -\trQ23_{\alpha}\trT12^{\alpha}
 +\trQ12_\alpha \trT12^{\alpha}\ .
 \label{GTEGR}
\ee
for the general case.
These combinations differ from the Hilbert term only by
a total derivative, as is seen from (\ref{palatiniTQ}).
More general teleparallel theories with actions
of the form $f(\mathbb{T})$ or $f(\mathbb{Q})$ have also been studied in some detail.
They are in some sense analogous to the Lagrangians 
for metric theories of the form $f(R)$,
but not equivalent to them.

It is an interesting question,
whether any metric theory of gravity has a teleparallel equivalent.
We can answer this question in the affirmative.
To begin with, let us consider a general 
action for a metric theory of gravity
that contains only powers of undifferentiated
curvature tensors:
\be
S_M(g)=\int d^4x\sqrt{-g}\,\cL(g_{\mu\nu},
R_{\mu\nu}{}^\rho{}_\sigma)\ .
\ee
While ultimately everything only depends on the metric,
we have separated the dependence of the Lagrangian
on the Riemann tensor 
and on the metric, which is used to contract all indices.

The EOM is obtained from the variation
\be
\delta S_M=\int d^4x\sqrt{-g}\left[
\frac{1}{2}\cL\, g^{\alpha\beta}\delta g_{\alpha\beta}
-\frac{\partial \cL}{\partial g^{\mu \nu}} g^{\mu\alpha}g^{\nu\beta}
\delta g_{\alpha\beta}
+Z^{\mu\nu}{}_\rho{}^\sigma 
\delta R_{\mu\nu}{}^\rho{}_\sigma\right]
\ee
where $Z^{\mu\nu}{}_\rho{}^\sigma
=\frac{\partial\cL}{\partial R_{\mu\nu}{}^\rho{}_\sigma}$.
Thus the EOM is
\be
\frac{1}{2}\cL\, g^{\alpha\beta} 
-\frac{\partial \cL}{\partial g^{\mu\nu}} 
g^{\mu\alpha}g^{\nu\beta}
+
\left(\delta^{(\alpha}_\sigma\delta^{\beta)}_{[\mu}\nabla^\rho\nabla_{\nu]}
-g^{\rho(\alpha}\delta^{\beta)}_\sigma\nabla_{[\mu}\nabla_{\nu]}  
-g^{\rho(\alpha}\nabla_\sigma\delta^{\beta)}_{[\mu}\nabla_{\nu]}\right) 
Z^{\mu\nu}{}_\rho{}^\sigma
=0\ .
\ee

For a teleparallel theory, $F_{\mu\nu}{}^\rho{}_\sigma=0$
so equation (\ref{FtoR}) implies that
\be
R_{\mu\nu}{}^\alpha{}_\beta=-P_{\mu\nu}{}^\alpha{}_\beta
\label{ennio}
\ee
where
$$
P_{\mu\nu}{}^\alpha{}_\beta=
\nabla_{\mu}\phi_{\nu \,\,\, \beta}^{\,\,\, \alpha}-\nabla_{\nu}\phi_{\mu\,\,\, \beta}^{\,\,\, \alpha}
+\phi_{\mu\,\,\, \gamma}^{\,\,\, \alpha}\phi_{\nu \,\,\, \beta}^{\,\,\, \gamma}-\phi_{\nu \,\,\, \gamma}^{\,\,\, \alpha}\phi_{\mu \,\,\, \beta}^{\,\,\, \gamma}\ .
$$
Now consider the following action for a teleparallel
theory in Einstein form:
\be
S_T(g,\phi)=\int d^4x\sqrt{-g}\,\cL(g_{\mu\nu},
-P_{\mu\nu}{}^\rho{}_\sigma)\ .
\label{genTaction}
\ee
where $\cL$ is the same as in $S_M$.

The constraint $F_{\mu\nu}{}^\rho{}_\sigma=0$
also implies that
\be
A_\nu{}^\rho{}_\sigma = \left(\Lambda^{-1}\right)^{\rho}{}_\alpha \partial_\nu \Lambda^\alpha{}_\sigma\ ,
\label{puregauge}
\ee
which in turn implies
\be
\phi_\nu{}^\rho{}_\sigma = \left(\Lambda^{-1}\right)^{\rho}{}_\alpha \partial_\nu \Lambda^\alpha{}_\sigma
- \lc_\nu{}^\rho{}_\sigma \,.
\label{constraintsol}
\ee
Inserting in (\ref{genTaction}) we obtain a new unconstrained action
$S_T'(g_{\mu\nu},\Lambda^\alpha{}_\beta)$.
Now $\Lambda$ is a pure gauge degree of freedom
and its EOM is empty,
as follows from the observation that due to (\ref{ennio}),
$P$ does not depend on $\Lambda$.
The only nontrivial equation follows from the variation of the metric:
\be
\delta S_T=\int d^4x\sqrt{-g}\left[
\frac{1}{2}\cL\, g^{\alpha\beta}\delta g_{\alpha\beta}
-\frac{\partial \cL}{\partial g^{\mu \nu}} g^{\mu\alpha}g^{\nu\beta}
\delta g_{\alpha\beta}
+W^{\mu\nu}{}_\rho{}^\sigma 
\delta P_{\mu\nu}{}^\rho{}_\sigma\right]
\ee
where $W^{\mu\nu}{}_\rho{}^\sigma
=\frac{\partial\cL}{\partial P_{\mu\nu}{}^\rho{}_\sigma}$.
But
$$
W^{\mu\nu}{}_\rho{}^\sigma
=-Z^{\mu\nu}{}_\rho{}^\sigma\Big|_{R\to-P}
$$
and
$$
\delta P_{\mu\nu}{}^\rho{}_\sigma
=-\delta R_{\mu\nu}{}^\rho{}_\sigma\ ,
$$
so the EOM of this teleparallel theory is the same
as the one of the original metric theory.

Let us now come to the more general case when the action contains also up to $n$-times differentiated Riemann tensors:
\be
S_M(g)=\int d^4x\sqrt{-g}\,\cL(g_{\mu\nu},
R_{\mu\nu}{}^\rho{}_\sigma,
\nabla_\alpha R_{\mu\nu}{}^\rho{}_\sigma,
\ldots,
\nabla_{\alpha_1}\cdots\nabla_{\alpha_n}R_{\mu\nu}{}^\rho{}_\sigma
)\ .
\ee
In this case the variation will contain $n$
additional terms:
$$
\sum_{i=1}^n Z_i^{\alpha_1\ldots\alpha_i\mu\nu}{}_\rho{}^\sigma
\delta (\nabla_{\alpha_1}\cdots\nabla_{\alpha_i}R_{\mu\nu}{}^\rho{}_\sigma)\ ,
$$
where $Z_i^{\alpha_1\ldots\alpha_i\mu\nu}{}_\rho{}^\sigma
=\frac{\partial\cL}{\partial(\nabla_{\alpha_1}\cdots\nabla_{\alpha_i}R_{\mu\nu}{}^\rho{}_\sigma)}$.
The teleparallel equivalent action is
\be
S_T(g,\phi)=\int d^4x\sqrt{-g}\,\cL(g_{\mu\nu},
-P_{\mu\nu}{}^\rho{}_\sigma,
-\nabla_\alpha P_{\mu\nu}{}^\rho{}_\sigma,
\ldots,
-\nabla_{\alpha_1}\cdots\nabla_{\alpha_n}P_{\mu\nu}{}^\rho{}_\sigma
)\ .
\ee
Following the same argument as above,
based on the constraint (\ref{ennio}),
the EOMs of this theory
are the same as those of the original metric theory.

\section{Lagrangians}
\label{sec:lags}

\subsection{General structure}

As discussed in Section \ref{sec:equiv},
it is not meaningful to organize the terms of the Lagrangian
according to the number of derivatives.
We shall instead order them based on canonical dimension,
with the understanding that terms of lower dimension
are generally more important at low energy.
We shall now discuss the possible Lagrangians for MAG
containing terms of dimension two and four.
In a first overview we will entirely omit all indices
and only consider the structures that can appear.
This is useful to understand the relation between the 
Cartan and Einstein forms of the Lagrangian,
in an uncluttered environment.
In the rest of this section we shall count, 
and in part enumerate, all the structures.

We start from the Cartan form of the theory.
The covariant field strengths are the curvature $F$,
of mass dimension two, the torsion $T$ and non-metricity $Q$,
both of mass dimension one.
The scalars of dimension two that can be formed with these ingredients
are either linear in $F$ or quadratic in $T$ and $Q$.
These terms will appear in the action with coefficients of dimension two.
The scalars of dimension four are of the forms $F^2$
or $FDT/FDQ$ or quadratic in $DT/DQ$,
or cubic in $T/Q$ with one derivative, or quartic in $T/Q$.
All these terms appear in the action with dimensionless coefficients.

In order not to introduce too many different symbols,
we shall use a slightly cumbersome but helpful notation,
where all the dimension-two couplings are called $a$
and all dimensionless ones are called $c$,
and the type of term they multiply is indicated by
a superscript in brackets.
Once indices are reinstated, different couplings 
of the same type will be distinguished by a subscript.
Thus, ignoring all numerical factors and signs,
we write the Lagrangian in the schematic form
\bea
\cL_C&=&a^F F+a^{TT}TT+a^{TQ}TQ+a^{QQ}QQ
\nonumber\\
&&+c^{FF}FF+c^{FT}FDT+c^{FQ}FDQ
+c^{TT}(DT)^2+c^{TQ}DTDQ+c^{QQ}(DQ)^2
\nonumber\\
&&+c^{FTT}FTT+c^{FTQ}FTQ+c^{FQQ}FQQ
\nonumber\\
&&+c^{TTT}TTDT+\ldots+c^{QQQ}QQDQ
\nonumber\\
&&
+c^{TTTT}TTTT+\ldots+c^{QQQQ}QQQQ\ ,
\label{lagC}
\eea
where the ellipses stand for cubic and quartic terms
involving different powers of $T$ and $Q$.

The action in EInstin form is related to the action in Cartan form
by (\ref{C2E}).
In practice the transformations achieved by using $D=\nabla+\phi$ 
and equations (\ref{FtoR}) and (\ref{TQphi}), 
that we can write schematically as
$$
F\sim R+\nabla\phi+\phi\phi\ ,\quad
T\sim\phi\ ,\quad
Q\sim\phi\ .
$$
One then obtains the Lagrangian in Einstein form
\bea
\cL_E&=&m^R R+m^{\phi\phi}\phi\phi
\nonumber\\
&&+b^{RR}RR+b^{R\phi}R\nabla\phi
+b^{\phi\phi}(\nabla\phi)^2
\nonumber\\
&&+b^{R\phi\phi}R\phi\phi
+b^{\phi\phi\phi}\phi\phi\nabla\phi
+b^{\phi\phi\phi\phi}\phi\phi\phi\phi\ ,
\label{lagE}
\eea
where the dimension-two couplings are now called $m$
and the dimensionless ones are called $b$.
This is the most general Lagrangian for the Einstein form of MAG,
involving terms of dimension two and four.

At this point one can use (\ref{discon}) and (\ref{emanuele})
to reexpress $\phi$ in terms of $T$ and $Q$.
The Lagrangian then looks again more similar to (\ref{lagC}),
but there is a difference:
in (\ref{lagC}), $T$ and $Q$ have to be thought of as
depending on $A$ and $g$, whereas here they have to
be treated as independent variables.
To distinguish the two Lagrangians,
in $\cL_E$ the coefficients will be called 
$b^{RT}$, $b^{RQ}$, $b^{TT}$ etc.

In this paper we will be interested mainly in the
linearization of the theory around flat space.
We observe that in this approximation only the first two
lines of (\ref{lagC}) and (\ref{lagE}) contribute
to the propagator, while all the other terms are interactions.
Also we note that whereas the dependence on the metric
is nonpolynomial, as usual, the dependence on distortion
is at most quartic.

\subsection{Dimension-two terms}

Let us look more carefully at the dimension-two part of the Lagrangian.
In the Cartan form, it is
\be
\cL^{(2)}_C=-\frac12\left[-a^F F+\sum_{i=1}^{3}a^{TT}_i M^{TT}_i
+\sum_{i=1}^{3}a^{TQ}_i M^{TQ}_i
+\sum_{i=1}^{5}a^{QQ}_i M^{QQ}_i\right] \ ,
\label{genPalC}
\ee
where $F=F_{\mu\nu}{}^{\mu\nu}$ is the unique scalar
that can be constructed from the curvature
and $a^F=m_P^2$, where $m_P$ is the Planck mass.
This will be referred to as the Palatini term.
The other scalars are
\bea
M^{TT}_1&=&T^{\mu\rho\nu}T_{\mu\rho\nu}\ ,\quad
M^{TT}_2= T^{\mu\rho\nu}T_{\mu\nu\rho}\ ,\quad
M^{TT}_3= \trT12^{\mu} \trT12_\mu\ ,
\nonumber\\
M^{QQ}_1&=&Q^{\rho\mu\nu}Q_{\rho\mu\nu}\ ,\quad 
M^{QQ}_2=Q^{\rho\mu\nu}Q_{\nu\mu\rho}\ ,
\nonumber
\\
M^{QQ}_3&=&\trQ23^{\mu} \trQ23_\mu \ ,\quad
M^{QQ}_4=\trQ12^{\mu}\trQ12_\mu\ ,\quad
M^{QQ}_5=\trQ23^{\mu}\trQ12_\mu\ ,
\nonumber
\\
M^{TQ}_1&=&T^{\mu\rho\nu} Q_{\mu\rho\nu}\ ,\quad  
M^{TQ}_2=\trT12^{\mu}\trQ23_\mu \ ,\quad
M^{TQ}_3=\trT12^{\mu}\trQ12_\mu\ .
\label{ttqq}
\eea
Going from the Cartan to the Einstein form,
as discussed in the previous subsection,
yields
\be
\cL^{(2)}_E=-\frac12\left[-m^R R
+\sum_{i=1}^{11}m^{\phi\phi}_i M^{\phi\phi}_i\right] \ ,
\label{genPalE}
\ee
where
\bea
M^{\phi\phi}_1&=&\phi_{\mu\nu\rho} \phi^{\mu\nu\rho}\ ,\
M^{\phi\phi}_2= \phi_{\mu\nu\rho} \phi^{\mu\rho\nu}\ ,\
M^{\phi\phi}_3= \phi_{\mu\nu\rho} \phi^{\rho\nu\mu}\ ,\
M^{\phi\phi}_4= \phi_{\mu\nu\rho} \phi^{\nu\mu\rho}\ ,\
M^{\phi\phi}_5= \phi_{\mu\nu\rho} \phi^{\nu\rho\mu}\ ,
\nonumber
\\
M^{\phi\phi}_6&=& \trp12_\mu\trp12^\mu\ ,\quad
M^{\phi\phi}_7= \trp13_\mu\trp13^\mu\ ,\quad
M^{\phi\phi}_8= \trp23_\mu\trp23^\mu\ ,\quad
\nonumber
\\
M^{\phi\phi}_9&=& \trp12_\mu\trp13^\mu\ ,\quad
M^{\phi\phi}_{10}= \trp12_\mu\trp23^\mu\ ,\quad
M^{\phi\phi}_{11}= \trp13_\mu\trp23^\mu\ .\quad
\eea
The first term is now the Hilbert Lagrangian
and the rest are mass terms for $\phi$.
The correspondence between the parameters $m_i$ and $a_i$ is
\begin{align}
& 
m^R = a^F,\quad
m^{\phi\phi}_1 =  2 a^{TT}_1+2 a^{QQ}_1+a^{TQ}_1,\quad 
m^{\phi\phi}_2 =  a^{TT}_2+2a^{QQ}_1+a^{TQ}_1,\quad \nn
&
m^{\phi\phi}_3 =  -2 a^{TT}_1+a^{QQ}_2-a^{TQ}_1, \quad 
m^{\phi\phi}_4 =  a^{TT}_2+a^{QQ}_2,\quad 
m^{\phi\phi}_5 = a^F-2 a^{TT}_2+2 a^{QQ}_2-a^{TQ}_1,\quad\nn
& 
m^{\phi\phi}_6 =  a^{TT}_3+a^{QQ}_4+a^{TQ}_3,\quad 
m^{\phi\phi}_7 =  a^{QQ}_4, \quad 
m^{\phi\phi}_8 =  a^{TT}_3+4 a^{QQ}_3-2 a^{TQ}_2,\nn 
&
m^{\phi\phi}_9 =  -a^F+2 a^{QQ}_4+a^{TQ}_3,\quad 
m^{\phi\phi}_{10} =  -2 a^{TT}_3+2 a^{QQ}_5+2 a^{TQ}_2-a^{TQ}_3,\quad 
m^{\phi\phi}_{11} = 2 a^{QQ}_5- a^{TQ}_3.
\label{mTQtophi}
\end{align}
The inverse map is given in Appendix \ref{sec:app.lin.1},
Equation (\ref{mphitoTQ}).
Reexpressing $\phi$ in terms of $T$ and $Q$ we obtain
\be
\cL^{(2)}_E=-\frac12\left[-m^R R+\sum_{i=1}^{3}m^{TT}_i M^{TT}_i
+\sum_{i=1}^{3}m^{TQ}_i M^{TQ}_i
+\sum_{i=1}^{5}m^{QQ}_i M^{QQ}_i\right] \ ,
\label{genPalETQ}
\ee
where $T$, $Q$ are now independent variables and
\begin{align}
& 
m^{TT}_1 = a^{TT}_1-\frac14 a^F,\quad 
m^{TT}_2 = a^{TT}_2-\frac12 a^F,\quad 
m^{TT}_3 = a^{TT}_3+a^F, \nn
&
m^{QQ}_1 = a^{QQ}_1-\frac14 a^F ,\quad 
m^{QQ}_2 = a^{QQ}_2+\frac12 a^F,\quad
m^{QQ}_3 = a^{QQ}_3+\frac14 a^F,\nn
&
m^{QQ}_4 = a^{QQ}_4, \quad 
m^{QQ}_5 = a^{QQ}_5-\frac12 a^F ,\nn 
&
m^{TQ}_1 = a^{TQ}_1+a^F ,\quad 
m^{TQ}_2 = a^{TQ}_2+a^F ,\quad 
m^{TQ}_3 = a^{TQ}_3-a^F .
\end{align}
These formulae can be specialized to antisymmetric and symmetric MAG,
simply setting $Q=0$ and $T=0$, respectively.

In the rest of the section we will perform a similar
analysis for the dimension-four terms.

\subsection{Dimension-four terms in Einstein form}

The counting of independent terms turns out to be far easier 
in the Einstein point of view, where we use the variables $(g,\phi)$.
We therefore start from this case.
We will loosely refer to scalar monomials in the fields 
which appear in the Lagrangian as ``invariants''.
In the Einstein form of the theory, they will be denoted
$$
H^{X,Y}_i\, \quad\mathrm{where}\quad X,Y\in\{R,T,Q\}
$$
and $i$ is an index labelling different monomials.
We shall discuss first the antisymmetric MAG, which is simplest,
then the symmetric MAG and finally the general case.

\subsubsection{Antisymmetric MAG \'a la Einstein}
\label{sec:AMAGE}

We start from the subclass of antisymmetric MAGs,
taking $g$ and $T$ as basic variables.
The numbers of independent terms or each type turns out to be

\begin{center}
\begin{tabular}{|c|c|c|c|c|c|c|}
\hline
$R^2$    & $(\nabla T)^2$ & $R\,\nabla T$ &  $R\,T^2$& $T^2\nabla T$ & $T^4$ 
 & Total\\
\hline
3   & 9  &  2 & 14 & 31  & 33   &  92 \\
\hline
\end{tabular}
\end{center}
Let us list explicitly the terms of the first three columns,
that are relevant for the flat space propagators.
We have three $RR$ terms
\be
H^{RR}_1=R_{\mu\nu\rho\sigma}R^{\mu\nu\rho\sigma}\ ,\quad
H^{RR}_2=R_{\mu\nu}R^{\mu\nu}\ ,\quad
H^{RR}_3=R^2\ ,\quad
\label{RR}
\ee
nine $(\nabla T)^2$ terms
\be
\begin{tabular}{lll}
$H^{TT}_{1} = \nabla^\alpha T^{\beta\gamma\delta} \nabla_\alpha T_{\beta\gamma\delta}$ \ ,
& $H^{TT}_{2} = \nabla^\alpha T^{\beta\gamma\delta} \nabla_\alpha T_{\beta\delta\gamma}$ \ ,
\\
$H^{TT}_{3} = \nabla^\alpha\trT12^\beta \nabla_\alpha\trT12_\beta$ \ ,
\\
$H^{TT}_{4} = \divT1^{\alpha\beta}\divT1_{\alpha\beta}$ \ ,
& $H^{TT}_{5} = \divT1^{\alpha\beta}\divT1_{\beta\alpha}$ \ ,
\\
$H^{TT}_{6} = \divT2^{\alpha\beta}\divT2_{\alpha\beta}$ \ ,
& $H^{TT}_{7} = \divT1^{\alpha\beta}\divT2_{\alpha\beta}$ \ ,
\\
$H^{TT}_{8} = 
\divT2^{\alpha\beta}\nabla_\alpha\trT12_\beta$ \ ,
& $H^{TT}_{9} = (\trdivT1)^2$ \ ,
\label{nablaTnablaT}
\end{tabular}
\ee
and just cosidering the independent contractions
one has five $R\nabla T$-type terms
\be
\begin{tabular}{lll}
$H^{RT}_{1} = R^{\alpha\beta\gamma\delta} 
\nabla_\alpha T_{\beta\gamma\delta}$ \ ,
& $H^{RT}_{2} = R^{\alpha\gamma\beta\delta} 
\nabla_\alpha T_{\beta\gamma\delta}$ \ ,
&\\
$H^{RT}_{3} = R^{\beta\gamma} \divT1_{\beta\gamma}$ \ ,
& $H^{RT}_{4} = R^{\alpha\beta}\nabla_\alpha\trT12_{\beta}$ \ ,
& $H^{RT}_{5} = R\,\trdivT1$ \ .
\label{RnablaT}
\end{tabular}
\ee
However, these invariants are not all independent.
Indeed we note that contracting the first (algebraic) 
Bianchi identity with $\nabla T$ we obtain the relation
\be
H^{RT}_2=2 H^{RT}_1 \ ,
\label{einantisymrel1}
\ee
while using the second Bianchi identity, contracted with $T$,
and integrating by parts we obtain the relations
\bea
H^{RT}_3&=&H^{RT}_1\ ,
\nonumber\\
H^{RT}_5&=&-2 H^{RT}_4\ .
\label{einantisymrel2}
\eea
A possible choice consists of keeping
\be
\label{einantisymbasis}
\{H^{RT}_3 \ , \ H^{RT}_5 \}
\ee
as independent invariants of type $R\nabla T$.
Thus, there are $3+9+2=14$ independent terms quadratic in the fields.

In the table we also give the number of interaction terms.
We have determined these numbers using the function
{\tt AllContractions} of the {\it xTras} package
for {\it Mathematica}.
\footnote{While this counting may still be possible by hand
in this case, it becomes practically impossible for general MAG.}
For the $RTT$ terms, this gives 18 different contractions,
but the first Bianchi identity, contracted with $TT$,
gives 4 relations between these terms, leading to 14.
For $TT\nabla T$, {\tt AllContractions} gives 46 terms,
but there are 15 total derivative terms of this type,
so the number of independent ones is 31.
\footnote{The number of total derivative terms
can be determined applying {\tt AllContractions} to $qTTT$,
where $q^\mu$ is any vector (it can be thought of as the momentum).}

\subsubsection{Symmetric MAG \'a la Einstein}
\label{sec:SMAGE}

For symmetric (torsionfree) theories,
one can take $g$ and $Q$ as fundamental variables.
Then, the counting of dimension-four terms is as follows:

\bigskip
\begin{center}
\begin{tabular}{|c|c|c|c|c|c|c|}
\hline
$R^2$   &  $(\nabla Q)^2$ & $R\,\nabla Q$ &  $R\,Q^2$
& $Q^2\nabla Q$ & $Q^4$  & Total\\
\hline
3   & 16 &  4 & 22 & 59  & 69   & 173 \\
\hline
\end{tabular}
\end{center}
The quadratic invariants are the three $R^2$ terms
already listed in (\ref{RR}), plus the following $(\nabla Q)^2$ terms

\be
\begin{tabular}{lll}
$H^{QQ}_{1} = \nabla^\alpha Q^{\beta\gamma\delta}\, 
\nabla_\alpha Q_{\beta\gamma\delta}$ \ ,
& $H^{QQ}_{2} = \nabla^\alpha Q^{\beta\gamma\delta}\, 
\nabla_\alpha Q_{\gamma\beta\delta}$ \ ,
\\
$H^{QQ}_{3} = \nabla^\alpha \trQ12^\beta\, \nabla_\alpha\trQ12_\beta$ \ ,
& $H^{QQ}_{4} = \nabla^\alpha \trQ23^\beta\, \nabla_\alpha\trQ23_\beta$ \ ,
\\
$H^{QQ}_{5} = \nabla^\alpha \trQ12^\beta\, \nabla_\alpha\trQ23_\beta$ \ ,
\\
$H^{QQ}_{6} = \divQ1^{\alpha\beta}\,\divQ1_{\alpha\beta}$ \ ,
& $H^{QQ}_{7} = \divQ2^{\alpha\beta}\,\divQ2_{\alpha\beta}$ \ ,
\\
$H^{QQ}_{8} = \divQ2^{\alpha\beta}\,\divQ2_{\beta\alpha}$ \ ,
& $H^{QQ}_{9} = \divQ1^{\alpha\beta}\,\divQ2_{\alpha\beta}$ \ ,
\\
$H^{QQ}_{10} = 
\divQ2^{\alpha\beta}\nabla_\alpha\trQ12_\beta$ \ ,
& $H^{QQ}_{11} =  
\divQ2^{\alpha\beta}\nabla_\alpha\trQ23_\beta$ \ ,
\\
$H^{QQ}_{12} = 
\divQ2^{\alpha\beta}\nabla_\beta\trQ12_\alpha$ \ ,
& $H^{QQ}_{13} = 
\divQ2^{\alpha\beta}\nabla_\beta\trQ23_\alpha$ \ ,
\\
$H^{QQ}_{14} = (\trdivQ1)^2$ \ ,
& $H^{QQ}_{15} = (\trdivQ2)^2$ \ ,
\\
$H^{QQ}_{16} =  \trdivQ1\,\trdivQ2$ \ ,
\end{tabular}
\label{nablaQnablaQ}
\ee
and the $R\nabla Q$ terms
\be
\begin{tabular}{lll}
$H^{RQ}_1=R^{\alpha\gamma\beta\delta}\nabla_\alpha Q_{\beta\gamma\delta}$ \ ,
& $H^{RQ}_2=R^{\alpha\beta} \nabla_\alpha\trQ12_{\beta}$ \ ,
& $H^{RQ}_3=R^{\alpha\beta} \nabla_\beta\trQ23_{\alpha}$ \ ,
\\
$H^{RQ}_4=R^{\alpha\beta}\, \divQ1_{\alpha\beta}$ \ ,
& $H^{RQ}_5=R^{\alpha\beta}\, \divQ2_{\alpha\beta}$ \ ,
\\
$H^{RQ}_6=R\, \trdivQ1$ \ ,
& $H^{RQ}_7=R\, \trdivQ2$ \ .
\end{tabular}
\label{RnablaQ}
\ee

Once again, not all these invariants are independent.
We note that using the second Bianchi identity
contracted with $Q$, and allowing integrations by parts,
we obtain three relations
\bea
H^{RQ}_1&=&H^{RQ}_4-H^{RQ}_5 \ ,
\nonumber\\
2H^{RQ}_2&=&H^{RQ}_7 \ ,
\nonumber\\
2H^{RQ}_3&=&H^{RQ}_6 \ .
\label{einsymrel1}
\eea
For example, we can solve for $H^{RQ}_1$, $H^{RQ}_2$, $H^{RQ}_3$
and keep 
\be\label{einsymbasis}
\{H^{RQ}_4 \ , \ H^{RQ}_5 \ , \ H^{RQ}_6 \ , \ H^{RQ}_7  \}
\ee
as independent invariants.
There are therefore $3+16+4=23$ independent invariants
quadratic in the fields.

The numbers of cubic and quartic interaction terms
are determined as in the previous subsection.
{\tt AllContractions} gives 23 terms of the type $RQQ$,
and the first Bianchi identity contracted with $QQ$
gives one relation between them, bringing the number of independent
terms of this type to 22.
For $QQ\nabla Q$ terms, {\tt AllContractions} gives 95 terms,
but 36 of them are total derivatives,
so the number of independent ones is 59.

\subsubsection{General MAG \'a la Einstein}
\label{sec:genMAGE}

In the general case the counting is simpler if we use $\phi$
as a variable, rather than $T$ and $Q$.
Then we have

\begin{center}
\begin{tabular}{|c|c|c|c|c|c|c|}
\hline
$R^2$    & $(\nabla\phi)^2$ & 
$R\,\nabla\phi$ &  $R\,\phi^2$& $\phi^2\nabla\phi$ & $\phi^4$  & Total\\
\hline
3   & 38  &  6 & 56 & 315  & 504   & 922 \\
\hline
\end{tabular}
\end{center}
The list of the $(\nabla\phi)^2$ terms is
\footnote{note that up to terms of the form $R\nabla\phi$,
$\divp1^{\alpha\beta}\nabla_\beta\trp12_\alpha
= \divp3^{\alpha\beta}\nabla_\alpha\trp12_\beta$ etc.}

\be
\begin{tabular}{lll}
$H^{\phi\phi}_{1} = \nabla^\alpha\phi^{\beta\gamma\delta} \nabla_\alpha\phi_{\beta\gamma\delta}$ \ ,
& $H^{\phi\phi}_{2} = \nabla^\alpha\phi^{\beta\gamma\delta} \nabla_\alpha\phi_{\beta\delta\gamma}$ \ ,
& $H^{\phi\phi}_{3} = \nabla^\alpha\phi^{\beta\gamma\delta} \nabla_\alpha\phi_{\delta\gamma\beta}$ \ ,
\\
$H^{\phi\phi}_{4} = \nabla^\alpha\phi^{\beta\gamma\delta} \nabla_\alpha\phi_{\gamma\beta\delta}$ \ ,
& $H^{\phi\phi}_{5} = \nabla^\alpha\phi^{\beta\gamma\delta} \nabla_\alpha\phi_{\delta\beta\gamma}$ \ ,
\\
$H^{\phi\phi}_{6} = \nabla^\alpha\trp12^\beta\nabla_\alpha\trp12_\beta$ \ ,
& $H^{\phi\phi}_{7} = \nabla^\alpha\trp13^\beta\nabla_\alpha\trp13_\beta$ \ ,
& $H^{\phi\phi}_{8} = \nabla^\alpha\trp23^\beta\nabla_\alpha\trp23_\beta$ \ ,
\\
$H^{\phi\phi}_{9} = \nabla^\alpha\trp12^\beta\nabla_\alpha\trp13_\beta$ \ ,
& $H^{\phi\phi}_{10} = \nabla^\alpha\trp12^\beta\nabla_\alpha\trp23_\beta$ \ ,
& $H^{\phi\phi}_{11} = \nabla^\alpha\trp13^\beta\nabla_\alpha\trp23_\beta$ \ ,
\\
$H^{\phi\phi}_{12} = \divp1^{\alpha\beta}\divp1_{\alpha\beta}$ \ ,
& $H^{\phi\phi}_{13} = \divp1^{\alpha\beta}\divp1_{\beta\alpha}$ \ ,
\\
$H^{\phi\phi}_{14} = \divp2^{\alpha\beta}\divp2_{\alpha\beta}$ \ ,
& $H^{\phi\phi}_{15} = \divp2^{\alpha\beta}\divp2_{\beta\alpha}$ \ ,
\\
$H^{\phi\phi}_{16} = \divp3^{\alpha\beta}\divp3_{\alpha\beta}$ \ ,
& $H^{\phi\phi}_{17} = \divp3^{\alpha\beta}\divp3_{\beta\alpha}$ \ ,
\\
$H^{\phi\phi}_{18} = \divp1^{\alpha\beta}\divp2_{\alpha\beta}$ \ ,
& $H^{\phi\phi}_{19} = \divp1^{\alpha\beta}\divp2_{\beta\alpha}$ \ ,
\\
$H^{\phi\phi}_{20} = \divp1^{\alpha\beta}\divp3_{\alpha\beta}$ \ ,
& $H^{\phi\phi}_{21} = \divp1^{\alpha\beta}\divp3_{\beta\alpha}$ \ ,
\\
$H^{\phi\phi}_{22} = \divp2^{\alpha\beta}\divp3_{\alpha\beta}$ \ ,
& $H^{\phi\phi}_{23} = \divp2^{\alpha\beta}\divp3_{\beta\alpha}$ \ ,
\\
$H^{\phi\phi}_{24} = \divp1^{\alpha\beta}\nabla_\alpha\trp12_\beta$ \ ,
& $H^{\phi\phi}_{25} = \divp1^{\alpha\beta}\nabla_\alpha\trp13_\beta$ \ ,
& $H^{\phi\phi}_{26} = \divp1^{\alpha\beta}\nabla_\alpha\trp23_\beta$ \ ,
\\
$H^{\phi\phi}_{27} = \divp3^{\alpha\beta}\nabla_\alpha\trp12_\beta$ \ ,
& $H^{\phi\phi}_{28} = \divp3^{\alpha\beta}\nabla_\alpha\trp13_\beta$ \ ,
& $H^{\phi\phi}_{29} = \divp3^{\alpha\beta}\nabla_\alpha\trp23_\beta$ \ ,
\\
$H^{\phi\phi}_{30} = \divp2^{\alpha\beta}\nabla_\beta\trp12_\alpha$ \ ,
& $H^{\phi\phi}_{31} = \divp2^{\alpha\beta}\nabla_\beta\trp13_\alpha$ \ ,
& $H^{\phi\phi}_{32} = \divp2^{\alpha\beta}\nabla_\beta\trp23_\alpha$ \ ,
\\
$H^{\phi\phi}_{33} = (\trdivp1)^2$ \ ,
& $H^{\phi\phi}_{34} = (\trdivp2)^2$ \ ,
& $H^{\phi\phi}_{35} = (\trdivp3)^2$ \ ,
\\
$H^{\phi\phi}_{36} = \trdivp1\,\trdivp2$ \ , 
& $H^{\phi\phi}_{37} = \trdivp1\,\trdivp3$ \ ,
& $H^{\phi\phi}_{38} = \trdivp2\,\trdivp3$ \ .
\end{tabular}
\ee
Note that the contraction of indices in the terms 
$H^{\phi\phi}_{30}$ - $H^{\phi\phi}_{32}$
is different from the order in the preceding six terms.
This is necessary to make them independent.
In fact another way of writing those nine terms is
\be
\begin{tabular}{lll}
$H^{\phi\phi}_{24} = - \divdivp12^\alpha\trp12_\alpha$ \ ,
& $H^{\phi\phi}_{25} = - \divdivp12^\alpha\trp13_\alpha$ \ ,
& $H^{\phi\phi}_{26} = - \divdivp12^\alpha\trp23_\alpha$ \ ,
\\
$H^{\phi\phi}_{27} = - \divdivp13^\alpha\trp12_\alpha$ \ ,
& $H^{\phi\phi}_{28} = - \divdivp13^\alpha\trp13_\alpha$ \ ,
& $H^{\phi\phi}_{29} = - \divdivp13^\alpha\trp23_\alpha$ \ ,
\\
$H^{\phi\phi}_{30} = - \divdivp23^\alpha\trp12_\alpha $ \ ,
& $H^{\phi\phi}_{31} = - \divdivp23^\alpha\trp13_\alpha$ \ ,
& $H^{\phi\phi}_{32} = - \divdivp23^\alpha\trp23_\alpha$ \ .
\end{tabular}
\ee
The $R\nabla\phi$ terms are
\be
\begin{tabular}{lll}
$H^{R\phi}_{1} = R^{\alpha\beta\gamma\delta}\nabla_\alpha\phi_{\beta\gamma\delta}$ \ ,
& $H^{R\phi}_{2} = R^{\alpha\beta\gamma\delta}\nabla_\delta\phi_{\alpha\gamma\beta}$ \ ,
& $H^{R\phi}_{3} = R^{\alpha\beta\gamma\delta}\nabla_\delta\phi_{\alpha\beta\gamma}$ \ ,
\\
$H^{R\phi}_{4} = R^{\alpha\beta}\nabla_\alpha\trp12_\beta$ \ ,
& $H^{R\phi}_{5} = R^{\alpha\beta}\nabla_\alpha\trp13_\beta$ \ ,
& $H^{R\phi}_{6} = R^{\alpha\beta}\nabla_\alpha\trp23_\beta$ \ ,
\\
$H^{R\phi}_{7} = R^{\alpha\beta}\divp1_{\alpha\beta}$ \ ,
& $H^{R\phi}_{8} = R^{\alpha\beta}\divp2_{\alpha\beta}$ \ ,
& $H^{R\phi}_{9} = R^{\alpha\beta}\divp3_{\alpha\beta}$ \ ,
\\
$H^{R\phi}_{10} =R\,\trdivp1$ \ ,
& $H^{R\phi}_{11} = R\,\trdivp2$ \ ,
& $H^{R\phi}_{12} = R\,\trdivp3$ \ .
\end{tabular}
\label{Rnabphi}
\ee

\bigskip

Using the first Bianchi identity for $\nabla$ and contracting with
$\nabla\phi$ we obtain the relation
\be
H^{R\phi}_1+H^{R\phi}_2-H^{R\phi}_3=0 \ ,
\label{anguilla}
\ee
Contracting the second Bianchi identity
with $\phi$ and using integrations by parts,
one finds:
\bea
H^{R\phi}_{1}-H^{R\phi}_{8}+H^{R\phi}_{9}&=&0 \ ,
\nonumber\\
H^{R\phi}_{2}+H^{R\phi}_{7}-H^{R\phi}_{9}&=&0 \ ,
\nonumber\\
H^{R\phi}_{3}+H^{R\phi}_{7}-H^{R\phi}_{8}&=&0 \ ,
\nonumber\\
2H^{R\phi}_{4}-H^{R\phi}_{12}&=&0 \ ,
\nonumber\\
2H^{R\phi}_{5}-H^{R\phi}_{11}&=&0 \ ,
\nonumber\\
2H^{R\phi}_{6}-H^{R\phi}_{10}&=&0 \ .
\eea
A linear combination of the first three is equivalent
to (\ref{anguilla}), so there are six independent relations.
Using these we can eliminate six invariants,
bringing the number of $R\nabla\phi$ terms
from 12 to 6, as indicated in the table.
For example, we can solve for 
$H^{R\phi}_1$, $H^{R\phi}_2$, $H^{R\phi}_3$, $H^{R\phi}_4$, $H^{R\phi}_5$, $H^{R\phi}_6$
and keep 
\be
\label{eingenbasis}
\{H^{R\phi}_7 \ , \ H^{R\phi}_8 \ , \ H^{R\phi}_9 \ , \ H^{R\phi}_{10} \ , \ H^{R\phi}_{11} \ , \ H^{R\phi}_{12}  \}
\ee
as independent invariants.
There are therefore $3+38+6=47$ independent invariants
quadratic in the fields.

In the following we will mostly use $T$ and $Q$ as independent fields
instead of $\phi$.
Then the kinetic terms for these fields would be given by
(\ref{nablaTnablaT},\ref{nablaQnablaQ}) and by the
following $\nabla T\nabla Q$ terms:
\be
\begin{tabular}{lll}
$H^{TQ}_{1} = \nabla^\alpha T^{\beta\gamma\delta}\, 
\nabla_\alpha Q_{\beta\gamma\delta}$ \ ,
& $H^{TQ}_{2} = \nabla^\alpha \trT12^\beta\, \nabla_\alpha\trQ12_\beta$ \ ,
\\
$H^{TQ}_{3} = \nabla^\alpha \trT12^\beta\, \nabla_\alpha\trQ23_\beta$ \ ,
\\
$H^{TQ}_{4} = \divT1^{\alpha\beta}\,\divQ1_{\alpha\beta}$ \ ,
& $H^{TQ}_{5} = \divT2^{\alpha\beta}\,\divQ2_{\alpha\beta}$ \ ,
\\
$H^{TQ}_{6} = \divT1^{\alpha\beta}\,\divQ2_{\alpha\beta}$ \ ,
& $H^{TQ}_{7} = \divT1^{\alpha\beta}\,\divQ2_{\beta\alpha}$ \ ,
\\
$H^{TQ}_{8} = 
\divT2^{\alpha\beta} \nabla_\alpha\trQ12_\beta$ \ ,
& $H^{TQ}_{9} =  
\divT2^{\alpha\beta} \nabla_\alpha\trQ23_\beta$ \ ,
\\
$H^{TQ}_{10} = 
\divQ2^{\alpha\beta} \nabla_\alpha\trT12_\beta$ \ ,
& $H^{TQ}_{11} = 
\divQ2^{\alpha\beta} \nabla_\beta\trT12_\alpha$ \ ,
\\
$H^{TQ}_{12} = 
\trdivT1\,\trdivQ1$ \ ,
& $H^{TQ}_{13} = 
\trdivT1\,\trdivQ2$ \ .
\end{tabular}
\label{nablaTnablaQ}
\ee
We count 9 $(\nabla T)^2$ terms, 16 $(\nabla Q)^2$ terms
and 13 $\nabla T\nabla Q$ terms.
In total they amount to 38 terms, that can be used interchangeably with
the 38 $(\nabla \phi)^2$ terms listed above.
A basis for the quadratic terms is given by these 38 terms,
plus the three $R^2$ terms, 
plus 
\be
\label{eingenTQbasis}
\{H^{RT}_3 \ , \ H^{RT}_5 \ , \ H^{RQ}_4 \ , \ H^{RQ}_5 \ , \ H^{RQ}_6 \ , \ H^{RQ}_7  \} \ ,
\ee
(which is the union of \eqref{einantisymbasis} and \eqref{einsymbasis}),
for a total 47 terms.

For the cubic interactions,
{\tt AllContractions} gives 65 terms of the type $R\phi\phi$,
but the first Bianchi identity, contracted with $\phi\phi$,
yields 9 relations between them, so that the number
of independent ones is 56.
\footnote{The nine relations can be most easily counted
in terms of $R\phi\phi$, but they are equivalent to
the 4 relations that we have already mentioned for the $RTT$ terms,
one relation already mentioned for the $RQQ$ terms
and four additional ones for the $RTQ$ terms.}
{\tt AllContractions} also gives 483 terms of the form
$\phi\phi\nabla\phi$, out of which 168 are total derivatives,
so the number of independent ones is 315.
The numbers are obviously the same if one uses $T$ and $Q$ as variables.

\subsection{Dimension-four terms in Cartan form}

The count of the possible terms in the Lagrangian
in the Cartan form of the theory is more tricky.
The invariants that can appear in the Lagrangian in Cartan form
are denoted
$$
L^{X,Y}_i\ ,\quad\mathrm{where}\quad X,Y\in\{F,T,Q\}\ ,
$$
to distinguish them from the $H^{XY}_i$ of the Einstein form of the theory.
We shall begin by listing all the terms that can appear
in the first three terms of (\ref{lagC}).

$FF$ terms:
\bea
L^{FF}_1\!\!&=\!\!&F^{\mu\nu\rho\sigma}F_{\mu\nu\rho\sigma}\ , \quad
L^{FF}_2=F^{\mu\nu\rho\sigma}F_{\mu\nu\sigma\rho}\ , \quad
L^{FF}_3=F^{\mu\nu\rho\sigma}F_{\rho\sigma\mu\nu} \ ,
\nonumber
\\
L^{FF}_4&\!\!=\!\!&F^{\mu\nu\rho\sigma}F_{\mu\rho\nu\sigma}\ ,\quad
L^{FF}_5=F^{\mu\nu\rho\sigma}F_{\mu\sigma\nu\rho}\ ,\quad
L^{FF}_6=F^{\mu\nu\rho\sigma}F_{\mu\sigma\rho\nu} \ ,
\nonumber
\\
L^{FF}_7&=&F^{(13)\mu\nu}F^{(13)}_{\mu\nu}\ ,\quad
L^{FF}_8=F^{(13)\mu\nu}F^{(13)}_{\nu\mu}\ ,
\nonumber
\\
L^{FF}_9&=&F^{(14)\mu\nu}F^{(14)}_{\mu\nu}\ ,\quad 
L^{FF}_{10}=F^{(14)\mu\nu}F^{(14)}_{\nu\mu}\ , 
\nonumber
\\ 
L^{FF}_{11}&=&F^{(13)\mu\nu}F^{(14)}_{\mu\nu}\ ,\quad
L^{FF}_{12}=F^{(13)\mu\nu}F^{(14)}_{\nu\mu}\ ,
\nonumber
\\
L^{FF}_{13}&=&F^{(34)\mu\nu}F^{(34)}_{\mu\nu}\ ,\quad
L^{FF}_{14}=F^{(34)\mu\nu}F^{(13)}_{\mu\nu}\ ,\quad
L^{FF}_{15}=F^{(34)\mu\nu}F^{(14)}_{\mu\nu}\ ,
\nonumber
\\
L^{FF}_{16}&=&F^2\ .
\label{HFF}
\eea

$(DT)^2$ terms:
\be
\begin{tabular}{lll}
$L^{TT}_{1} = D^\alpha T^{\beta\gamma\delta} D_\alpha T_{\beta\gamma\delta}$ \ ,
& $L^{TT}_{2} = D^\alpha T^{\beta\gamma\delta} D_\alpha T_{\beta\delta\gamma}$ \ ,
\\
$L^{TT}_{3} = D^\alpha \trT12^\beta D_\alpha\trT12_\beta$ \ ,
\\
$L^{TT}_{4} = \DivT1^{\alpha\beta}\DivT1_{\alpha\beta}$ \ ,
& $L^{TT}_{5} = \DivT1^{\alpha\beta}\DivT1_{\beta\alpha}$ \ ,
\\
$L^{TT}_{6} = \DivT2^{\alpha\beta}\DivT2_{\alpha\beta}$ \ ,
& $L^{TT}_{7} = \DivT1^{\alpha\beta}\DivT2_{\alpha\beta}$ \ ,
\\
$L^{TT}_{8} = \DivT2^{\alpha\beta}D_\alpha\trT12_\beta$ \ ,
& $L^{TT}_{9} = (\trDivT1)^2$ \ .
\end{tabular}
\label{HTT}
\ee

$(DQ)^2$ terms:
\be
\begin{tabular}{lll}
$L^{QQ}_{1} = D^\alpha Q^{\beta\gamma\delta}\,
D_\alpha Q_{\beta\gamma\delta}$ \ ,
& $L^{QQ}_{2} = D^\alpha Q^{\beta\gamma\delta}\, 
D_\alpha Q_{\gamma\beta\delta}$ \ ,
\\
$L^{QQ}_{3} = D^\alpha \trQ12^\beta\, D_\alpha\trQ12_\beta$ \ ,
& $L^{QQ}_{4} = D^\alpha \trQ23^\beta\, D_\alpha\trQ23_\beta$ \ ,
\\
 $L^{QQ}_{5} = D^\alpha \trQ12^\beta\, D_\alpha\trQ23_\beta$ \ ,
\\
$L^{QQ}_{6} = \DivQ1^{\alpha\beta}\,\DivQ1_{\alpha\beta}$ \ ,
& $L^{QQ}_{7} = \DivQ2^{\alpha\beta}\,\DivQ2_{\alpha\beta}$ \ ,
\\
$L^{QQ}_{8} = \DivQ2^{\alpha\beta}\,\DivQ2_{\beta\alpha}$ \ ,
& $L^{QQ}_{9} = \DivQ1^{\alpha\beta}\,\DivQ2_{\alpha\beta}$ \ ,
\\
$L^{QQ}_{10} = 
\DivQ2^{\alpha\beta}\,D_\alpha\trQ12_\beta$ \ ,
& $L^{QQ}_{11} =  
\DivQ2^{\alpha\beta}\,D_\alpha\trQ23_\beta$ \ ,
\\
$L^{QQ}_{12} = 
\DivQ2^{\alpha\beta}\,D_\beta\trQ12_\alpha$ \ ,
& $L^{QQ}_{13} = 
\DivQ3^{\alpha\beta}\,D_\beta\trQ23_\alpha$ \ ,
\\
$L^{QQ}_{14} = (\trDivQ1)^2$ \ ,
& $L^{QQ}_{15} = (\trDivQ2)^2$ \ ,
\\
$L^{QQ}_{16} =  \trDivQ1\,\trDivQ2$  \ .
\end{tabular}
\label{HQQ}
\ee

$DTDQ$ terms
\be
\begin{tabular}{lll}
$L^{TQ}_{1} = D^\alpha T^{\beta\gamma\delta}\, 
D_\alpha Q_{\beta\gamma\delta}$ \ ,
& $L^{TQ}_{2} = D^\alpha \trT12^\beta\, D_\alpha\trQ12_\beta$ \ ,
\\
$L^{TQ}_{3} = D^\alpha \trT12^\beta\, D_\alpha\trQ23_\beta$ \ ,
\\
$L^{TQ}_{4} = \DivT1^{\alpha\beta}\,\DivQ1_{\alpha\beta}$ \ ,
& $L^{TQ}_{5} = \DivT2^{\alpha\beta}\,\DivQ2_{\alpha\beta}$ \ ,
\\
$L^{TQ}_{6} = \DivT1^{\alpha\beta}\,\DivQ2_{\alpha\beta}$ \ ,
& $L^{TQ}_{7} = \DivT1^{\alpha\beta}\,\DivQ2_{\beta\alpha}$ \ ,
\\
$L^{TQ}_{8} = 
\DivT2^{\alpha\beta} D_\alpha\trQ12_\beta$ \ ,
& $L^{TQ}_{9} =  
\DivT2^{\alpha\beta} D_\alpha\trQ23_\beta$ \ ,
\\
$L^{TQ}_{10} = 
\DivQ2^{\alpha\beta} D_\alpha\trT12_\beta$ \ ,
& $L^{TQ}_{11} = 
\DivQ2^{\alpha\beta} D_\beta\trT12_\alpha$ \ ,
\\
$L^{TQ}_{12} = 
\trDivT1\,\trDivQ1$ \ ,
& $L^{TQ}_{13} = 
\trDivT1\,\trDivQ2$ \ .
\end{tabular}
\label{HTQ}
\ee

$FDT$ terms:
\be
\begin{tabular}{lll}
$L^{FT}_1=F^{\mu\nu\rho\sigma}D_\mu T_{\nu\rho\sigma}$ \ ,
& $L^{FT}_2=F^{\mu\nu\rho\sigma}D_\mu T_{\nu\sigma\rho}$ \ ,
& $L^{FT}_3=F^{\mu\nu\rho\sigma}D_\mu T_{\rho\nu\sigma}$ \ ,
\\
$L^{FT}_4=F^{\mu\nu\rho\sigma}D_\rho T_{\mu\nu\sigma}$ \ ,
& $L^{FT}_5=F^{\mu\nu\rho\sigma}D_\rho T_{\mu\sigma\nu}$ \ ,
& $L^{FT}_6=F^{\mu\nu\rho\sigma}D_\sigma T_{\mu\nu\rho}$ \ ,
\\
$L^{FT}_7=F^{\mu\nu\rho\sigma}D_\sigma T_{\mu\rho\nu}$ \ ,
\\
$L^{FT}_8=F^{(13)\mu\nu} D_\mu\trT12_{\nu}$ \ ,
& $L^{FT}_9=F^{(13)\mu\nu} D_\nu\trT12_{\mu}$ \ ,
\\
$L^{FT}_{10}=F^{(14)\mu\nu} D_\mu\trT12_{\nu}$ \ ,
& $L^{FT}_{11}=F^{(14)\mu\nu} D_\nu\trT12_{\mu}$ \ ,
& $L^{FT}_{12}=F^{(34)\mu\nu} D_\mu\trT12_{\nu}$ \ ,
\\
$L^{FT}_{13}=F^{(13)\mu\nu}\, \DivT1_{\mu\nu}$ \ ,
& $L^{FT}_{14}=F^{(13)\mu\nu}\, \DivT1_{\nu\mu}$ \ ,
\\
$L^{FT}_{15}=F^{(14)\mu\nu}\, \DivT1_{\mu\nu}$ \ ,
& $L^{FT}_{16}=F^{(14)\mu\nu}\, \DivT1_{\nu\mu}$ \ ,
\\
$L^{FT}_{17}=F^{(13)\mu\nu}\, \DivT2_{\mu\nu}$ \ ,
& $L^{FT}_{18}=F^{(14)\mu\nu}\, \DivT2_{\mu\nu}$ \ ,
\\
$L^{FT}_{19}=F^{(34)\mu\nu}\, \DivT1_{\mu\nu}$ \ ,
& $L^{FT}_{20}=F^{(34)\mu\nu}\, \DivT2_{\mu\nu}$ \ ,
& $L^{FT}_{21}=F\, \trDivT1$ \ .
\end{tabular}
\label{HFT}
\ee

$FDQ$ terms:
\be
\begin{tabular}{lll}
$L^{FQ}_1=F^{\mu\nu\rho\sigma}D_\mu Q_{\nu\rho\sigma}$ \ ,
& $L^{FQ}_2=F^{\mu\nu\rho\sigma}D_\nu Q_{\rho\sigma\mu}$ \ ,
& $L^{FQ}_3=F^{\mu\nu\rho\sigma}D_\nu Q_{\sigma\rho\mu}$ \ ,
\\
$L^{FQ}_4=F^{\mu\nu\rho\sigma}D_\rho Q_{\mu\nu\sigma}$ \ ,
& $L^{FQ}_5=F^{\mu\nu\rho\sigma}D_\sigma Q_{\mu\nu\rho}$ \ ,
\\
$L^{FQ}_6=F^{(13)\mu\nu} D_\mu\trQ12_{\nu}$ \ ,
& $L^{FQ}_7=F^{(13)\mu\nu} D_\nu\trQ12_{\mu}$ \ ,
\\
$L^{FQ}_8=F^{(13)\mu\nu} D_\mu\trQ23_{\nu}$ \ ,
& $L^{FQ}_9=F^{(13)\mu\nu} D_\nu\trQ23_{\mu}$ \ ,
\\
$L^{FQ}_{10}=F^{(14)\mu\nu} D_\mu\trQ12_{\nu}$ \ ,
& $L^{FQ}_{11}=F^{(14)\mu\nu} D_\nu\trQ12_{\mu}$ \ ,
\\
$L^{FQ}_{12}=F^{(14)\mu\nu} D_\mu\trQ23_{\nu}$ \ ,
& $L^{FQ}_{13}=F^{(14)\mu\nu} D_\nu\trQ23_{\mu}$ \ ,
\\
$L^{FQ}_{14}=F^{(34)\mu\nu} D_\mu\trQ12_{\nu}$ \ ,
& $L^{FQ}_{15}=F^{(34)\mu\nu} D_\mu\trQ23_{\nu}$ \ ,
\\
$L^{FQ}_{16}=F^{(13)\mu\nu}\, \DivQ1_{\mu\nu}$ \ ,
& $L^{FQ}_{17}=F^{(14)\mu\nu}\, \DivQ1_{\mu\nu}$ \ ,
\\
$L^{FQ}_{18}=F^{(13)\mu\nu}\, \DivQ2_{\mu\nu}$ \ ,
& $L^{FQ}_{19}=F^{(13)\mu\nu}\, \DivQ2_{\nu\mu}$ \ ,
\\
$L^{FQ}_{20}=F^{(14)\mu\nu}\, \DivQ2_{\mu\nu}$ \ ,
& $L^{FQ}_{21}=F^{(14)\mu\nu}\, \DivQ2_{\nu\mu}$ \ ,
& $L^{FQ}_{22}=F^{(34)\mu\nu}\, \DivQ2_{\mu\nu}$ \ ,
\\
$L^{FQ}_{23}=F\, \trDivQ1$ \ ,
& $L^{FQ}_{24}=F\, \trDivQ2$ \ .
\end{tabular}
\label{HFQ}
\ee

\noindent
We observe that whereas the 38 terms 
$L^{TT}$, $L^{QQ}$, $L^{TQ}$
in (\ref{HTT},\ref{HQQ},\ref{HTQ})
are in one-to-one correspondence with the terms 
$H^{TT}$, $H^{QQ}$, $H^{TQ}$ in
(\ref{nablaTnablaT},\ref{nablaQnablaQ},\ref{nablaTnablaQ}),
there are many more terms of type $FF$, $FDT$, $FDQ$
than $RR$, $R\nabla T$, $R\nabla Q$.
This is due to the fact that the curvature tensor $F$
has less symmetries than the Riemann tensor.
This also means that there will also be many more relations.
Our goal now will be to uncover these relations,
exhibit a basis of invariants
and construct the map between the couplings
in the Cartan basis and those in the 
previously established Einstein basis.

Concerning the cubic and quartic interaction terms,
we shall not attempt to count them here,
as this would be overly complicated.
However we know that ultimately they will be in
one-to-one correspondence with those of the Einstein formulation,
that have been counted in the previous sections.

\subsubsection{Antisymmetric MAG \'a la Cartan}
\label{AMAGC}

Let us list all the quadratic invariants.
Since in antisymmetric MAG $F$ is antisymmetric 
in both pairs of indices,
there are fewer independent terms than in general MAG.
We keep $L^{FF}_i$ with $i=1,3,4,7,8,16$, while
\bea
&&
L^{FF}_2=-L^{FF}_1\ ,\quad
L^{FF}_5=-L^{FF}_4\ ,\quad
L^{FF}_6=L^{FF}_4\ ,\quad
\nonumber\\
&&
L^{FF}_9=L^{FF}_7\ ,\quad
L^{FF}_{10}=L^{FF}_8\ ,\quad
L^{FF}_{11}=-L^{FF}_7\ ,\quad
L^{FF}_{12}=-L^{FF}_8\ ,\quad
\nonumber\\
&&
L^{FF}_{13}=L^{FF}_{14}=L^{FF}_{15}=0\ .
\eea
We keep all the terms $L^{TT}$.
They are the same as the invariants of type $(\nabla T)^2$,
except for the replacement of $\nabla$ by $D$.
We keep $L^{FT}_i$ with $i=1,3,4,5,8,9,13,14,17,21$, while
\bea
&&
L^{FT}_2=-L^{FT}_1\ ,\quad
L^{FT}_6=-L^{FT}_4\ ,\quad
L^{FT}_7=-L^{FT}_5\ ,\quad
\nonumber\\
&&
L^{FT}_{10}=-L^{FT}_8\ ,\quad
L^{FT}_{11}=-L^{FT}_9\ ,\quad
L^{FT}_{15}=-L^{FT}_{13}\ ,\quad
L^{FT}_{16}=-L^{FT}_{14}\ ,\quad
\nonumber\\
&&
L^{FT}_{18}=-L^{FT}_{17}\ ,\quad
L^{FT}_{12}=L^{FT}_{19}=L^{FT}_{20}=0\ .
\eea
We now have 25 quadratic terms, compared to the 14
quadratic terms in the Einstein form of antisymmetric MAG.

There are several additional relations.
Multiplying (\ref{bianchi1}) by $F$, we obtain,
up to interaction term of the form $FTT$,
\bea
L^{FT}_8-L^{FT}_9-L^{FT}_{17}&=&-L^{FF}_7+L^{FF}_8 \ ,
\nonumber\\
-2L^{FT}_1+L^{FT}_5&=&-L^{FF}_1+2L^{FF}_4 \ ,
\nonumber\\
L^{FT}_3-2L^{FT}_4&=&-L^{FF}_3+2L^{FF}_4 \ ,
\eea
and multiplying (\ref{bianchi2}) by $T$ 
(and using integrations by parts) gives,
again up to terms of the form $FTT$,
\bea
L^{FT}_5-2L^{FT}_{14}&=&0 \ ,
\nonumber\\
-L^{FT}_4+L^{FT}_{13}-L^{FT}_{17}&=&0 \ ,
\nonumber\\
2L^{FT}_9+L^{FT}_{21}&=&0 \ .
\label{relsA}
\eea
Furthermore, multiplying (\ref{bianchi1}) by $\nabla T$
gives, up to terms cubic in $T$,
\bea
L^{FT}_{17}&=&1/2 L^{TT}_6-L^{TT}_8  \ ,
\nonumber\\
L^{FT}_{13}-L^{FT}_{14}&=&L^{TT}_7-L^{TT}_8 \ ,
\nonumber\\
L^{FT}_8-L^{FT}_9&=& -L^{TT}_3+L^{TT}_8+L^{TT}_9 \ ,
\nonumber\\
2L^{FT}_4-L^{FT}_5&=&- L^{TT}_6+2L^{TT}_7 \ ,
\nonumber\\
L^{FT}_1-L^{FT}_3+L^{FT}_4&=&- L^{TT}_2+L^{TT}_5+L^{TT}_7 \ ,
\nonumber\\
2L^{FT}_1-L^{FT}_5&=&L^{TT}_1 -2L^{TT}_4 \ .
\label{relsB}
\eea
Altogether we have obtained 12 relations,
of which 11 turn out to be linearly independent.
Therefore we can eliminate 22 out the 36 invariants
listed in (\ref{HFF},\ref{HTT},\ref{HFT}),
and we remain with 14 independent quadratic invariants,
exactly as in the counting in the Einstein form.

There are many ways of solving these relations,
but we shall consider here only two.
The first is to retain all the nine $L^{TT}$ terms, 
plus 
\be\label{carantisymbasisonetone}
\{L^{FF}_1 \ , \ L^{FF}_7 \ , \ L^{FF}_{16} \} \quad
\mathrm{and}
\quad \{ L^{FT}_{13} \ , \ L^{FT}_{21}  \} \ ,
\ee
which is in one-to-one correspondence with \eqref{RR} and \eqref{einantisymbasis}.
Thus, the elements of this basis are in one-to-one correspondence with the elements
of the basis in the Einstein form,
from which they are obtained just by replacing
$R\to F$ and $\nabla\to D$.
The remaining invariants are given in Equation (\ref{AMAGCbasis1})
in Appendix \ref{sec:app.lin.3}.

Due to the geometrical meaning of the curvature, when we use
the independent connection $A$, it seems desirable to keep
all terms that contain $F$, and instead remove others.
We can choose as a basis the six $L^{FF}$ invariants $\{ L^{FF}_1 \ , \ L^{FF}_3 \ , \ L^{FF}_4 \ , \ L^{FF}_7 \ , \ L^{FF}_8 \ , L^{FF}_{16} \}$ ,
plus 
\be\label{carantisymbasisFF}
\{L^{TT}_1 \ , \ L^{TT}_2 \ , \ L^{TT}_3\ , \ L^{TT}_5 \} \quad
\mathrm{and}
\quad \{ L^{FT}_1 \ , \ L^{FT}_{8} \ , \ L^{FT}_9 \ , \ L^{FT}_{13}   \} \ .
\ee
The remaining invariants are given in Equation (\ref{AMAGCbasis2})
in Appendix \ref{sec:app.lin.3}.

\subsubsection{Symmetric MAG \'a la Cartan}
\label{SMAGcartan}

In symmetric (torsionfree) MAG, the curvature tensor
is only antisymmetric in the first pair of indices,
but the first Bianchi identity (\ref{bianchi1})
leads to six independent relations
\bea 
L^{FF}_1-2L^{FF}_6&=&0 \ ,
\nonumber\\
L^{FF}_2-2L^{FF}_5&=&0 \ ,
\nonumber\\
L^{FF}_3-L^{FF}_4+L^{FF}_5&=&0 \ ,
\nonumber\\
L^{FF}_{13}+2L^{FF}_{14}&=&0 \ ,
\nonumber\\
L^{FF}_7-L^{FF}_8+L^{FF}_{14}&=&0 \ ,
\nonumber\\
L^{FF}_{11}-L^{FF}_{12}+L^{FF}_{15}&=&0 \ .
\eea
This reduces the number of independent 
curvature squared terms to 10.
We keep the invariants $L^{FF}_i$ with
$i=1,2,3,7,8,9,10,11,12,16$ and solve for the others:
\bea 
L^{FF}_4\!\!&=\!\!&1/2 L^{FF}_2+L^{FF}_3  \ ,\qquad
L^{FF}_5=1/2 L^{FF}_2 \ ,
\nonumber\\
L^{FF}_6&=&1/2 L^{FF}_1 \ , \qquad\quad\qquad
L^{FF}_{13}=2(L^{FF}_7-L^{FF}_8) \ ,
\nonumber\\
L^{FF}_{14}&=&-L^{FF}_7+L^{FF}_8 \ ,\qquad\ \
L^{FF}_{15}=-L^{FF}_{11}+L^{FF}_{12} \ .
\eea
Multiplying (\ref{bianchi1})
by $DQ$ we obtain, up to interaction terms, the relations
\bea
L^{FQ}_{18}-L^{FQ}_{19}+L^{FQ}_{22}&=&0\ ,
\nonumber\\
L^{FQ}_{6}-L^{FQ}_{7}+L^{FQ}_{14}&=&0\ ,
\nonumber\\
L^{FQ}_{8}-L^{FQ}_{9}+L^{FQ}_{15}&=&0\ ,
\nonumber\\
L^{FQ}_{1}+L^{FQ}_{3}+L^{FQ}_{5}&=&0\ ,
\label{relB1Q}
\eea
and multiplying (\ref{bianchi2}) by $Q$ we obtain,
again up to interaction terms, the relations
\bea
L^{FQ}_{7}-L^{FQ}_{11}-L^{FQ}_{24}&=&0\ ,
\nonumber\\
L^{FQ}_{9}-L^{FQ}_{13}-L^{FQ}_{23}&=&0\ ,
\nonumber\\
L^{FQ}_{5}-L^{FQ}_{17}+L^{FQ}_{20}&=&0\ ,
\nonumber\\
L^{FQ}_{4}-L^{FQ}_{16}+L^{FQ}_{18}&=&0\ .
\label{relB2Q}
\eea
In this case the Bianchi identities are not enough to
uncover all the relations and we have to resort to another method.
We can use (\ref{FtoR}) in the $FF$ terms;
this will give among other things
$RR$ terms, $R\nabla T$ and $R\nabla Q$.
We can look for linear combinations of the $FF$ terms
such that these terms involving $R$ in the r.h.s. cancel out.
In this way, up to cubic and quartic terms, 
we will relate $FF$ terms to $(DT)^2$ terms etc.
From the $FF$ terms we obtain
\bea
L^{FF}_{1}+L^{FF}_2&=&L^{QQ}_1-L^{QQ}_6\ ,
\nonumber\\
2(L^{FF}_1-L^{FF}_3)&=&3 L^{QQ}_1-2 L^{QQ}_2-3 L^{QQ}_6
-2 L^{QQ}_7+4 L^{QQ}_9 \ ,
\nonumber\\
4(L^{FF}_{7}-L^{FF}_8)&=&L^{QQ}_4-L^{QQ}_{14}\ ,
\nonumber\\
4(L^{FF}_{9}-L^{FF}_{10})&=&4L^{QQ}_{3}+L^{QQ}_{4}
-4L^{QQ}_{5}+4L^{QQ}_{7}-4L^{QQ}_{8}-8L^{QQ}_{10}
\nonumber\\
&&
\!\!\!\!\!+4L^{QQ}_{11}+8L^{QQ}_{12}-4L^{QQ}_{13}-L^{QQ}_{14}
-4L^{QQ}_{15}+4L^{QQ}_{16}\ ,
\nonumber\\
4(L^{FF}_{11}-L^{FF}_{12})&=&-L^{QQ}_{4}+2L^{QQ}_{5}
-2L^{QQ}_{11}+2L^{QQ}_{13}+L^{QQ}_{14}-2L^{QQ}_{16}\ ,
\nonumber\\
L^{FF}_{7}\!+\!L^{FF}_{8}\!+\!L^{FF}_{9}+
\nonumber\\\qquad\quad\!L^{FF}_{10}
\!+\!2L^{FF}_{11}\!+\!2L^{FF}_{12}&=&
L^{QQ}_{3}\!+\!L^{QQ}_{7}\!+\!L^{QQ}_{8}\!-\!2L^{QQ}_{10}\!-\!2L^{QQ}_{12}\!+\!L^{QQ}_{15}\ .
\label{Saddrel1}
\eea

Operating in a similar way on the $FDQ$ terms we obtain
\bea
2L^{FQ}_{1}&=&L^{QQ}_1-L^{QQ}_6\ ,
\nonumber\\
L^{FQ}_{2}+L^{FQ}_3&=&-L^{QQ}_2+L^{QQ}_9\ ,
\nonumber\\
2(L^{FQ}_{2}+L^{FQ}_4)&=&L^{QQ}_1-2L^{QQ}_2-L^{QQ}_6
-2L^{QQ}_7+4L^{QQ}_9\ ,
\nonumber\\
2(L^{FQ}_{2}-L^{FQ}_5)&=&L^{QQ}_1-2L^{QQ}_2-L^{QQ}_6
+2L^{QQ}_9\ ,
\nonumber\\
2(L^{FQ}_{6}-L^{FQ}_7)&=&-L^{QQ}_5+L^{QQ}_{16}\ ,
\nonumber\\
L^{FQ}_{6}+L^{FQ}_{10}&=&-L^{QQ}_3+L^{QQ}_{10}\ ,
\nonumber\\
2(L^{FQ}_{10}-L^{FQ}_{11})&=&-2L^{QQ}_3+L^{QQ}_5
+2L^{QQ}_{10}-2L^{QQ}_{12}+2L^{QQ}_{15}-L^{QQ}_{16}\ ,
\nonumber\\
2(L^{FQ}_{8}-L^{FQ}_{9})&=&-L^{QQ}_4+L^{QQ}_{14}\ ,
\nonumber\\
L^{FQ}_{8}+L^{FQ}_{12}&=&-L^{QQ}_5+L^{QQ}_{11}\ ,
\nonumber\\
2(L^{FQ}_{12}-L^{FQ}_{13})&=&L^{QQ}_4-2L^{QQ}_5
+2L^{QQ}_{11}-2L^{QQ}_{13}-L^{QQ}_{14}+2L^{QQ}_{16}\ ,
\nonumber\\
2L^{FQ}_{14}&=&L^{QQ}_5-L^{QQ}_{16}\ ,
\nonumber\\
2L^{FQ}_{15}&=&L^{QQ}_4-L^{QQ}_{14}\ ,
\nonumber\\
L^{FQ}_{16}+L^{FQ}_{17}&=&L^{QQ}_9-L^{QQ}_{10}\ ,
\nonumber\\
2(L^{FQ}_{18}-L^{FQ}_{19})&=&-L^{QQ}_{11}+L^{QQ}_{13}\ ,
\nonumber\\
L^{FQ}_{18}+L^{FQ}_{20}&=&L^{QQ}_7-L^{QQ}_{10}\ ,
\nonumber\\
2(L^{FQ}_{20}-L^{FQ}_{21})&=&2L^{QQ}_7-2L^{QQ}_{8}
-2L^{QQ}_{10}+L^{QQ}_{11}+2L^{QQ}_{12}\ ,
\nonumber\\
2L^{FQ}_{22}&=&L^{QQ}_{11}-L^{QQ}_{13}\ .
\label{Saddrel2}
\eea
We need one additional relation involving both $FF$ and $FDQ$:
\be
2(L^{FF}_{10}+L^{FF}_{12}-L^{FQ}_6+L^{FQ}_{18})
=L^{QQ}_5+2L^{QQ}_8
-L^{QQ}_{11}-4L^{QQ}_{12}+L^{QQ}_{13}
+2L^{QQ}_{15}-L^{QQ}_{16}\ .
\label{Saddrel3}
\ee
These 24 relations are all linearly independent,
but they are not independent when one takes them together
with the 8 relations (\ref{relB1Q},\ref{relB2Q}) coming from
the Bianchi identities.
In fact the system of all 32 relations has rank 27.
This means that we have 50-27=23 independent invariants,
in agreement with the counting in section \ref{sec:SMAGE}.

We can choose as an independent set the relations 
(\ref{Saddrel1},\ref{Saddrel2},\ref{Saddrel3}),
plus the first three relations in (\ref{relB2Q}).
There are many ways of solving these relations,
but we shall consider here only two.
The first is to retain all the 16 $L^{QQ}$ terms, 
plus 
\be
\label{carsymbasisonetone}
\{L^{FF}_1 \ , \ L^{FF}_7 \ , \ L^{FF}_{16} \} \quad
\mathrm{and}
\quad \{ L^{FQ}_{16} \ , \ L^{FQ}_{18} \ , \ L^{FQ}_{23} \ , \ L^{FQ}_{24}  \} \ ,
\ee
which is in one-to-one correspondence with the sum of 
\eqref{RR} and \eqref{einsymbasis}.
The remaining invariants are given in Equation (\ref{SMAGCbasis1})
in Appendix \ref{sec:app.lin.4}.

As in the antisymmetric case,
we can also keep in the basis the ten $L^{FF}$ invariants 
$$\{ L^{FF}_1 \ , \ L^{FF}_2 \ , \ L^{FF}_3 \ , \ L^{FF}_7 \ , \ L^{FF}_8 \ , \ L^{FF}_9 \ , \ L^{FF}_{10} \ , \ L^{FF}_{11} \ , \ L^{FF}_{12} \ , L^{FF}_{16} \}$$ 
plus 
\be
\label{carsymbasisFF}
\{L^{QQ}_1 \ , \ L^{QQ}_{10} \ , \ L^{QQ}_{11}\ , \ L^{QQ}_{12} \ , \ L^{QQ}_{14} \} \quad\mathrm{and}\quad
\{ L^{FQ}_{10} \ , \ L^{FQ}_{11} \ , \ L^{FQ}_{12} \ , \ L^{FQ}_{14} \ , \ L^{FQ}_{16} \ , \ L^{FQ}_{17} \ , \ L^{FQ}_{18} \ , \ L^{FQ}_{23}  \} \ .
\ee
The remaining invariants are given in Equation (\ref{SMAGCbasis2})
in Appendix \ref{sec:app.lin.4}.

\subsubsection{General MAG \'a la Cartan}
\label{GMAGcartan}

We have listed in (\ref{HFF},\ref{HTT},\ref{HQQ},\ref{HTQ},\ref{HFT},\ref{HFQ}) 16 terms of type $FF$,
38 terms of type $D(T/Q)^2$ and 45 terms of type $FD(T/Q)$.
We thus have 99 quadratic terms, compared to the 47 ones
of the Einstein form of the theory.
We now look for linear relations between these terms.
As in the previous sections, these relations hold up
to terms cubic and quartic in $F$, $T$, $Q$.

Multiplying the first Bianchi identity by $F$ we get
\bea
L^{FF}_1-2 L^{FF}_6&=&2 L^{FT}_1 + L^{FT}_7\ ,\nonumber\\
L^{FF}_2-2 L^{FF}_5&=&2 L^{FT}_2 + L^{FT}_5\ ,\nonumber\\
L^{FF}_3-L^{FF}_4+L^{FF}_5&=&-L^{FT}_3+L^{FT}_4-L^{FT}_6\ ,\nonumber\\
L^{FF}_7-L^{FF}_8+L^{FF}_{14}&=&-L^{FT}_8+L^{FT}_9+L^{FT}_{17}\ ,\nonumber\\
L^{FF}_{11}-L^{FF}_{12}+L^{FF}_{15}&=&-L^{FT}_{10}
+L^{FT}_{11}+L^{FT}_{18}\ ,\nonumber\\
L^{FF}_{13}+2 L^{FF}_{14}&=&-2 L^{FT}_{12} + L^{FT}_{20}\ ,
\label{B1F}
\eea
while multiplying it by $DT$ we get
\bea
2 L^{FT}_1+L^{FT}_7&=& L^{TT}_1-2 L^{TT}_4\ ,\nonumber\\
L^{FT}_2+L^{FT}_3+L^{FT}_6&=& L^{TT}_2-L^{TT}_5-L^{TT}_7\ ,\nonumber\\
2 L^{FT}_4-L^{FT}_5&=& - L^{TT}_6 + 2 L^{TT}_7\ ,\nonumber\\
L^{FT}_8-L^{FT}_9+L^{FT}_{12}&=&-L^{TT}_3+L^{TT}_8+L^{TT}_9\ ,\nonumber\\
L^{FT}_{13}-L^{FT}_{14}+L^{FT}_{19}&=& L^{TT}_7-L^{TT}_8\ ,\nonumber\\
2 L^{FT}_{17}+L^{FT}_{20}&=& L^{TT}_6-2 L^{TT}_8\ ,
\label{B1T}
\eea
and multiplying it by $DQ$ we get
\bea
L^{FQ}_1 + L^{FQ}_3+L^{FQ}_5&=&
L^{TQ}_1-L^{TQ}_4+L^{TQ}_7\ ,\nonumber\\
L^{FQ}_6 - L^{FQ}_7+L^{FQ}_{14} &=&
-L^{TQ}_2+L^{TQ}_8-L^{TQ}_{13}\ ,\nonumber\\
L^{FQ}_8 - L^{FQ}_9+L^{FQ}_{15}&=&
-L^{TQ}_3+L^{TQ}_9-L^{TQ}_{12}\ ,\nonumber\\
L^{FQ}_{18}-L^{FQ}_{19}+L^{FQ}_{22}&=&
L^{TQ}_5-L^{TQ}_{10}+L^{TQ}_{11}\ .
\label{B1Q}
\eea
Multiplying the second Bianchi identity by $T$ we get
\bea
L^{FT}_5-2L^{FT}_{14}&=&0\ ,\nonumber\\
L^{FT}_7-2L^{FT}_{16}&=&0\ ,\nonumber\\
L^{FT}_4 - L^{FT}_{13} + L^{FT}_{17}&=&0\ ,\nonumber\\
L^{FT}_6 - L^{FT}_{15} + L^{FT}_{18}&=&0\ ,\nonumber\\
L^{FT}_9-L^{FT}_{11} + L^{FT}_{21}&=&0\ ,\nonumber\\
2L^{FT}_{19}-L^{FT}_{20}&=&0\ ,
\label{B2T}
\eea
and multiplying it by $Q$ we get
\bea
L^{FQ}_4-L^{FQ}_{16} + L^{FQ}_{18}&=&0\ ,\nonumber\\
L^{FQ}_5-L^{FQ}_{17} + L^{FQ}_{20}&=&0\ ,\nonumber\\
L^{FQ}_7-L^{FQ}_{11} - L^{FQ}_{24}&=&0\ ,\nonumber\\
L^{FQ}_9-L^{FQ}_{13} - L^{FQ}_{23}&=&0\ .
\label{B2Q}
\eea
In total these are 26 relations, of which 25 are independent.
\footnote{
Twice the first of (\ref{B1F}), minus the second minus the fourth,
minus the second of (\ref{B2T}), minus twice the third plus the fifth,
is identically zero.}
One can obtain an independent set by eliminating
for example the fifth relation in (\ref{B2T}).

As in the case of symmetric MAG, the Bianchi identities
do not exhaust the set of linear relations between the invariants.
The additional ones can be obtained by the same method that
we used for symmetric MAGs,
namely using (\ref{FtoR}) and eliminating terms of the form
$R^2$, $R\nabla T$ and $R\nabla Q$ from the right hand side.
This gives many additional relations that are listed in
Appendix \ref{sec:app.lin.2}.
Considering also these,
we have altogether a system of 70 relations of which 52 are independent.
Since the initial number of invariants is 99,
we remain with 47 independent invariants,
in agreement with the counting in the Einstein form of the theory.

We can now exhibit two bases.
The first consists of all the 38 $L^{TT}$, $L^{QQ}$ and $L^{TQ}$ terms, 
plus 
\be
\label{cargenbasisonetone}
\{L^{FF}_1 \ , \ L^{FF}_7 \ , \ L^{FF}_{16} \} 
\quad\mathrm{and}\quad
\{ L^{FT}_{13} \ , \ L^{FT}_{21} \}
\quad\mathrm{and}\quad
\{ L^{FQ}_{16} \ , \ L^{FQ}_{18} \ , \ L^{FQ}_{23} \ , \ L^{FQ}_{24}  \} \ ,
\ee
which is the sum of \eqref{carantisymbasisonetone} and \eqref{carsymbasisonetone}, and thus
is in one-to-one correspondence with the sum of \eqref{RR} and \eqref{eingenTQbasis}.
The remaining invariants are given in Equations (\ref{GMAGCbasis1}-\ref{GMAGCbasis1bis}-\ref{GMAGCbasis1bisbis})
of Appendix \ref{sec:app.lin.5}.

As before,
we can also choose as a basis all the 16 $L^{FF}$ invariants in \eqref{HFF}
plus 
\bea
\label{cargenbasisFF}
&&\{ L^{TT}_1\ , \ L^{TT}_2\ , \ L^{TT}_3\ , \ L^{TT}_5 \}\nn
&&\{L^{QQ}_1 \  , \ L^{QQ}_{10} \ , \ L^{QQ}_{11}\ , \ L^{QQ}_{12} \ , \ L^{QQ}_{14} \} \nn
&& \{ L^{TQ}_1,L^{TQ}_{10},L^{TQ}_{11},L^{TQ}_{12} \}\nn
&&\{L^{FT}_1\ , \ L^{FT}_8\ , \ L^{FT}_9 , \ L^{FT}_{12}\ , \ L^{FT}_{13}\ , \ L^{FT}_{14}\ ,  \ L^{FT}_{15}\ , \ L^{FT}_{18}\ , \ L^{FT}_{21}\}\nn
&&\{ L^{FQ}_{10} \ , \ L^{FQ}_{11} \ , \ L^{FQ}_{12} \ , \ L^{FQ}_{14} \ , \ L^{FQ}_{16} \ , \ L^{FQ}_{17} \ , \ L^{FQ}_{18} \ , \ L^{FQ}_{19}\ , \ L^{FQ}_{23}  \} \ .
\eea
The remaining invariants can be expressed as linear combination of these.
Explicit formulas are given in Equations (\ref{GMAGCbasis2}-\ref{GMAGCbasis2bis}-\ref{GMAGCbasis2bisbis})
in Appendix \ref{sec:app.lin.5}.

We observe that the bases given for antisymmetric and symmetric MAGs
can be obtained from these by dropping the terms that contain $Q$ and $T$
respectively.
In the case of the first basis this is enough.
In the case of the second basis, one has to further eliminate
certain terms of type $FF$, $FDT$ or $FDQ$.

\subsection{The map}

For certain purposes it is useful to have the map
between the coefficients of the Lagrangian in the
Cartan form and in the Einstein form.
This has already been discussed in the case of the
terms of dimension two.
For the terms of dimension four, we shall limit ourselves
to the transformation of the 47 quadratic terms.
The procedure has already been described in sect.3.1.
Inserting (\ref{FtoR}) in (\ref{lagC}),
a straightforward calculation leads to a Lagrangian
of the form (\ref{lagE}), whose $b$ coefficients
are functions of the original $c$ coefficients.
These linear relations are given in Appendix \ref{sec:app.lin.6}.

\subsection{Redundant couplings}

In the preceding sections we have given
bases for the quadratic part of the Lagrangian.
We have also listed other terms that could be written as linear combinations of the basis elements, either by integrations by parts or
by use of Bianchi identities or of additional relations.
There is an additional way to reduce the number of
independent terms, namely by using field redefinitions.
In a quantum field theory with a field $\varphi$, consider
a general local field redefinition
\be
\varphi'=\varphi+F(\varphi)
\ee
where $F$ depends on $\varphi$ and its derivatives.
Further suppose that there is a coupling $\zeta$ such that
\be
\frac{\partial S}{\partial\zeta}
=\int\frac{\delta S}{\delta\varphi}F(\varphi)\ .
\ee 
Then one says that $\zeta$ is ``inessential'' or ``redundant''.
Couplings that do not have this property are ``essential''.
The importance of this notion is that field redefinitions
do not affect the $S$-matrix and therefore
redundant couplings do not enter in the expressions
of quantum field theoretic observables.

The standard example of a redundant coupling is the
wave function renormalization $Z$, that multiplies
the kinetic term. 
For example, for a scalar field, we have the kinetic term
$\frac12 Z(\partial\varphi)^2$.
It is customary (but by no means necessary!) to set $Z=1$,
with the redefinition $F=(\sqrt{Z}-1)\varphi$.
Quantum corrections generally change the value of $Z$,
but one can always go back to the canonical value
by redefining the field.
In particular, in the Wilsonian renormalization group,
an infinitesimal renormalization group transformation is always
followed by an infinitesimal field redefinition,
in such a way that $Z$ remains constant and equal to one
along the flow.

In an EFT there is a criterion to establish what 
types of transformations $F(\varphi)$ are admissible
\cite{Manohar:2018aog}.
Since the transformation is local, it must involve powers of
the field and its derivatives, with suitable coefficients.
The simplest term is of the form $\delta\varphi=\alpha\varphi$ with
a dimensionless coefficient $\alpha$.
All the other terms will have coefficients with negative
mass dimension, and therefore have to be written as dimensionless
numbers times inverse powers of the cutoff $\Lambda$.
For example, one may consider
$\delta\varphi=(\beta/\Lambda^2)\partial^2\varphi$.
When acting on the mass term $m^2\varphi^2$,
this changes the kinetic term by $\delta Z=\beta (m/\Lambda)^2$.
If the mass is very light compared to the cutoff,
this change is negligible.
Terms with more derivatives will be even more suppressed.
Similarly, terms with higher powers of $\varphi$,
that would affect interactions, are similarly suppressed.
If the theory is renormalizable, one can take $\Lambda\to\infty$,
the only allowed transformation is linear in $\varphi$
and the only redundant coupling is $Z$.

We will now discuss redundancies within the quadratic terms
of our MAG Lagrangians.
The couplings $c_i$ (in the Cartan form)
or $b_i$ (in the Einstein form) multiply
the kinetic terms and are akin to wave function renormalizations,
while the couplings $a_i$ (in the Cartan form) or $m_i$
(in the Einstein form) are mass terms.
The analogy with the scalar case would suggest 
that $c_i$ and $b_i$ are redundant,
but we have to take into account the complications due to the indices.

We will work in the Einstein form of MAG, where things are easier.
First we consider redefinitions of the metric of the form
\be
\delta g_{\mu\nu}=\alpha g_{\mu\nu}
+(\beta_1/\Lambda^2) R_{\mu\nu}
+(\beta_2/\Lambda^2) R g_{\mu\nu}\ .
\ee
It is clear that the redefinition with parameter $\alpha$
preserves the number of derivatives of the terms it acts on.
It amounts just to a rescaling of the metric and
one can use it to fix the cosmological constant.
The redefinitions with parameters $\beta$ raise 
the number of derivatives by two.
When applied to the Hilbert term, they change the coefficients
$b^{RR}_2$, $b^{RR}_3$ by an amount of order $\beta(m_P/\Lambda)^2$.
Unlike the case of the scalar field discussed above,
$m_P$ and $\Lambda$ are of the same order.
Thus, this is an allowed transformation and
$b^{RR}_2$, $b^{RR}_3$ are redundant.
\footnote{It was already well known in the perturbative quantization
of GR that with such redefinitions
one can eliminate the one-loop divergences
\cite{Kallosh:1978wt}.}

Next we consider redefinitions of the distortion.
It is enough to consider redefinitions that 
are linear in the distortion and either
ultralocal (i.e. do not contain derivatives)
or contain two derivatives.
The former map mass terms to mass terms and kinetic terms
to kinetic terms; the latter map mass terms to kinetic terms.
More complicated redefinitions will only affect the interaction terms.
The linear ultralocal redefinitions of $\phi$ are
\bea
\delta\phi_{\alpha\beta\gamma}&=&
\alpha_1 \phi_{\alpha\beta\gamma}
+\alpha_2 \phi_{\beta\gamma\alpha}
+\alpha_3 \phi_{\gamma\alpha\beta}
+\alpha_4 \phi_{\alpha\gamma\beta}
+\alpha_5 \phi_{\gamma\beta\alpha}
+\alpha_6 \phi_{\beta\alpha\gamma}
\nonumber\\
&&
+g_{\alpha\beta}\left(\alpha_7\trp12_\gamma
+\alpha_8\trp13_\gamma
+\alpha_9\trp23_\gamma\right)
\nonumber\\
&&
+g_{\alpha\gamma}\left(\alpha_{10}\trp12_\beta
+\alpha_{11}\trp13_\beta
+\alpha_{12}\trp23_\beta\right)
\nonumber\\
&&
+g_{\beta\gamma}\left(\alpha_{13}\trp12_\alpha
+\alpha_{14}\trp13_\alpha
+\alpha_{15}\trp23_\alpha\right)\ .
\label{redef}
\eea

Let us consider the case of scale-invariant MAGs.
This precludes the existence of dimension-two terms
in the Lagrangian.
Since the $38\times 15$ matrix $\frac{\delta b^{\phi\phi}_i}{\delta\alpha_j}$ has rank 15,
we can use the 15 parameters $\alpha_j$ to fix 15 
among the $b^{\phi\phi}_i$.
This means that only 23 among the $b^{\phi\phi}_i$
are essential.

In the general case, dimension-two terms will be present
and in the case when all the masses
are of the order of the Planck mass,
additional redefinitions have to be considered.
There are 6 redefinitions of the form
$\delta\phi\approx (\beta/\Lambda^2)\nabla R$
(where in the r.h.s. $R$ stands for the Riemann tensor and
two indices are contracted).
Applied to a mass term, they produce terms of the form 
$\beta m/\Lambda^2 R\nabla\phi$.
Since we are assuming $m\approx \Lambda^2$,
(remember that the parameters $m_i$ have dimension of mass squared) this is unsuppressed.
Therefore these transformations
can be used to fix the parameters $b^{R\phi}_i$.
Similarly there are 60 transformations
of the form $\delta\phi=(\beta/\Lambda^2)\nabla\nabla\phi$
(where in the r.h.s. two indices are contracted),
that applied to the mass terms produce terms of the form 
$\beta m/\Lambda^2(\nabla\phi)^2$.
These transformations can be used to fix the values
of all the 38 parameters $b^{\phi\phi}_i$.
Finally, the transformations (\ref{redef})
can be used to fix all the 11 mass terms of $\phi$.
Thus in this trivial case where the only propagating state
is the graviton, all the couplings except $m^R$ are inessential.

In cases when the masses are much lower than the cutoff,
some of these transformations will be suppressed,
and there will remain some essential parameters.
This will have to be discussed on a case by case basis.

\section{Flat space propagators}
\label{sec:lin}

\subsection{Linearized action}
\label{sec:linac}

We consider the linearization of the action around Minkowski space
\be
g_{\mu\nu}=\eta_{\mu\nu}\ ,\quad
A_\rho{}^\mu{}_\nu=0\ ,\quad
\phi_\rho{}^\mu{}_\nu=0\ .
\ee
The terms in the Lagrangian that contribute at quadratic
order in the fluctuation fields
are those that are quadratic in $F$, $T$ and $Q$, 
including also covariant derivatives of $T$ and $Q$.
In the Cartan form, these are
\bea
\label{qLagC}
&&\cL_C=
-\frac12\Big[-a^F F
+\sum_i a^{TT}_i M^{TT}_i
+\sum_i a^{TQ}_i M^{TQ}_i
+\sum_i a^{QQ}_i M^{QQ}_i
\\
&&
\!+\!\sum_i c^{FF}_i L^{FF}_i
\!+\!\sum_i c^{FT}_i L^{FT}_i
\!+\!\sum_i c^{FQ}_i L^{FQ}_i
\!+\!\sum_i c^{TT}_i L^{TT}_i
\!+\!\sum_i c^{TQ}_i L^{TQ}_i
\!+\!\sum_i c^{QQ}_i L^{QQ}_i
\Big]\,,
\nonumber
\eea
where the first line contains all the dimension-two terms
and the second contains the dimension-four terms.
We do not specify the ranges of the sums, because
they depend upon the choice of basis.
In the Einstein form, the terms contributing to the two-point function are
\bea
\label{qLagE}
&&\cL_E=
-\frac12\Big[-m^R R
+\sum_i m^{TT}_i M^{TT}_i
+\sum_i m^{TQ}_i M^{TQ}_i
+\sum_i m^{QQ}_i M^{QQ}_i
\\
&&
\!+\!\sum_i b^{RR}_i H^{RR}_i
\!+\!\sum_i b^{RT}_i H^{RT}_i
\!+\!\sum_i b^{RQ}_i H^{RQ}_i
\!+\!\sum_i b^{TT}_i H^{TT}_i
\!+\!\sum_i b^{TQ}_i H^{TQ}_i
\!+\!\sum_i b^{QQ}_i H^{QQ}_i
\Big]\,.
\nonumber
\eea

The metric fluctuation field is $h_{\mu\nu}=g_{\mu\nu}-\eta_{\mu\nu}$.
Since the VEV of $A$ (and $\phi$) is zero,
we shall not use a different symbol for its fluctuation
and identify it with $A$.
By Poincare invariance, the quadratic Lagrangian
in the Cartan form of the theory, takes the form
\be
S^{(2)} = \frac12\int \frac{d^4q}{(2\pi)^4} \left(\, 
A^{\lambda\mu\nu}\,{\cal O}^{(AA)\ \ \tau\rho\sigma}_{(C)\ \lambda\mu\nu}\, A_{\tau\rho\sigma} 
+2A^{\lambda\mu\nu}\,{\cal O}^{(Ah)\ \ \rho\sigma}_{(C)\ \lambda\mu\nu}\, h_{\rho\sigma} 
+ h^{\mu\nu}\,{\cal O}^{(hh)\ \ \rho\sigma}_{(C)\ \mu\nu}\,h_{\rho\sigma}\, \right) \ ,
\label{linC}
\ee
where, after Fourier transforming,
$\cO$ is constructed only with the metric $\eta_{\mu\nu}$
and with momentum $q^\mu$.
Similarly, in the Einstein form of the theory one obtains
\be
S^{(2)} = \frac12\int \frac{d^4q}{(2\pi)^4} \left(\, 
\phi^{\lambda\mu\nu}\,{\cal O}^{(\phi\phi)\ \ \tau\rho\sigma}_{(E)\ \lambda\mu\nu}\, \phi_{\tau\rho\sigma} 
+2\phi^{\lambda\mu\nu}\,{\cal O}^{(\phi h)\ \ \rho\sigma}_{(E)\ \lambda\mu\nu}\, h_{\rho\sigma} 
+ h^{\mu\nu}\,{\cal O}^{(hh)\ \ \rho\sigma}_{(E)\ \mu\nu}\,h_{\rho\sigma}\, \right) \ .
\label{linE}
\ee
From (\ref{phi}) one finds that
\be
A_{\lambda\mu\nu}=\phi_{\lambda\mu\nu}+J_{\lambda\mu\nu}{}^{\rho\sigma}h_{\rho\sigma}\ ,
\ee
where 
$$
J_{\lambda\mu\nu}{}^{\rho\sigma}
=\frac{i}{2}\left(
q_\lambda\delta^\rho_\mu\delta^\sigma_\nu
+q_\nu\delta^\rho_\lambda\delta^\sigma_\mu
-q_\mu\delta^\rho_\lambda\delta^\sigma_\nu\right)\ .
$$
Then we obtain the following relations between the linearized
operators in the Cartan and Einstein formulations:
\bea
{\cal O}^{(\phi\phi)\ \ \tau\rho\sigma}_{(E)\ \lambda\mu\nu}
&=&{\cal O}^{(AA)\ \ \tau\rho\sigma}_{(C)\ \lambda\mu\nu}\ ,
\nonumber\\
{\cal O}^{(\phi h)\ \ \rho\sigma}_{(E)\ \lambda\mu\nu}
&=&{\cal O}^{(Ah)\ \ \rho\sigma}_{(C)\ \lambda\mu\nu}
+{\cal O}^{(AA)\ \ \tau\alpha\beta}_{(C)\ \lambda\mu\nu}
J_{\tau\alpha\beta}{}^{\rho\sigma}\ ,\nonumber\\
{\cal O}^{(hh)\ \ \rho\sigma}_{(E)\ \mu\nu}
&=&{\cal O}^{(hh)\ \ \rho\sigma}_{(C)\ \mu\nu}
+2J^{\lambda\gamma\delta}{}_{\mu\nu}
{\cal O}^{(Ah)\ \ \rho\sigma}_{(C)\ \lambda\gamma\delta}
+J^{\lambda\gamma\delta}{}_{\mu\nu}
{\cal O}^{(AA)\ \ \tau\alpha\beta}_{(C)\ \lambda\gamma\delta}
J_{\tau\alpha\beta}{}^{\rho\sigma}\ .
\eea

\subsection{Spin projectors}

In the analysis of the spectrum of operators acting on
multi-index fields in flat space, it is very convenient to
use spin-projection operators, which can be used to
decompose the fields in their irreducible components
under the three-dimensional rotation group
\cite{rivers,barnes,aurilia}.
This is familiar in the case of vectors and two-index tensors:
a vector can be decomposed in its transverse and longitudinal components;
a two index tensor can be decomposed into its symmetric
and antisymmetric components, and each of these can be further
decomposed in its transverse and longitudinal parts in each index.
This gives rise to representations of $O(3)$ 
labelled by spin and parity, and listed in the following table:
\smallskip

\begin{table}[ht]
\begin{center}
\begin{tabular}{|c|c|c|}
\hline
 & $s$ & $a$ \\
\hline
$TT$ & $2^+_4$, $0^+_5$ & $1^+_4$ \\
\hline
$TL$  & $1^-_7$ & $1^-_8$ \\
\hline
$LL$   &  $0^+_6$ & - \\
\hline
\end{tabular}
\end{center}
\caption{$SO(3)$ spin content of projection operators for a two-index tensor in $d=4$ ($s$=symmetric, $a$=antisymmetric).}
\label{t1}
\end{table}
\smallskip

Here the subscript distinguishes different instances of the same
representation.
These representations arise as perturbations of the tetrad.
If one works only with the metric, the antisymmetric parts can be dropped.
The analogous decomposition for a three-index tensor
is given in the following table,
that is explained in more detail in \cite{Percacci:2020ddy}:
\smallskip

\begin{table}[ht]
\begin{center}
\begin{tabular}{|c|c|c|c|c|}
\hline
   & $ts$ & $hs$ & $ha$ & $ta$ \\
\hline
$TTT$ & $3^-$, $1^-_1$ & $2^-_1$, $1^-_2$  & $2^-_2$, $1^-_3$ & $0^-$ \\
\hline
$TTL+TLT+LTT$  & $2^+_1$, $0^+_1$ & - & -  & $1^+_3$  \\
\hline
$\frac32 LTT$   & - & $2^+_2$, $0^+_2$  & $1^+_2$, & - \\
\hline
$TTL+TLT- \frac12 LTT$ & - & $1^+_1$ & $2^+_3$, $0^+_3$  & -  \\
\hline
$TLL+LTL+LLT$   & $1^-_4$ & $1^-_5$  & $1^-_6$   &  - \\
\hline
$LLL$    & $0^+_4$ & -  & -  &  - \\
\hline
\end{tabular}
\end{center}
\caption{$SO(3)$ spin content of projection operators for a
three-index tensor in $d=4$. ($ts/ta$=totally (anti)symmetric; $hs/ha$=hook (anti)symmetric}
\label{t2}
\end{table}
\smallskip

In antisymmetric or symmetric MAG, only the last two or the
first two columns appear, respectively.
For antisymmetric tensors, the spin projectors were given in 
\cite{neville,Sezgin:1979zf,sezgin2} and used to study 
ghost- and tachyon-free theories
that do not have accidental symmetries
(i.e. symmetries that are present at linearized level but not in the
full nonlinear theory).
The general case where accidental symmetries are present has
been discussed in \cite{Lin:2018awc}.
The spin projectors for general three-tensors
have been given in \cite{Percacci:2020ddy,schimidt}.

For each representation $J^\cP_i$ there is a
projector denoted $P_{ii}(J^\cP)$.
In addition, for each pair of representations with the
same spin-parity, labelled by $i$, $j$, there is an
intertwining operator $P_{ij}(J^\cP)$.
We collectively refer to all the projectors
and intertwiners as the ``spin-projectors''.

Using these spin projectors, the quadratic action
can be rewritten in the form
\be
S^{(2)}= \frac12\int \frac{d^4q}{(2\pi)^4}\sum_{J Pij}\Phi(-q)\cdot
a_{ij}(J^\cP)\,P_{ij}(J^\cP)\cdot \Phi(q)\ ,
\label{linactSP}
\ee
where $\Phi=(A,h)$ in Cartan form and $\Phi=(\phi,h)$ in Einstein form,
the dot implies contraction of all indices as appropriate
and $a_{ij}(J^P)$ are matrices of coefficients.
For example, the $A-A$ part of (\ref{linC}) is
$$
\frac12\int \frac{d^4q}{(2\pi)^4}\sum_{J Pij}
a_{ij}(J^\cP)\,A^{\lambda\mu\nu}P_{ij}(J^\cP)_{\lambda\mu\nu}{}^{\tau\rho\sigma}
A_{\tau\rho\sigma}\ ,
$$
with the sums running over all the representations listed in
the preceding table.

\subsection{Gauge invariances}
\label{sec:gaugeinv}

As mentioned in Section \ref{sec:class}, one way of reducing the
complexity of MAG is to introduce additional gauge invariances.
These will eliminate degrees of freedom
and at the same time constrain the form of the Lagrangian.
One could try to analyze systematically all possible
such transformations, for example one could classify
them as having a scalar, vector or tensor parameter.
As we shall note, such a general analysis would contain
a large number of arbitrary parameters.
Here we shall content ourselves to only mention a few 
important examples.

\subsubsection{Diffeomorphisms}
\label{sec:diffeo}

The action of MAG, when written in a coordinate basis,
is in general invariant only under diffeomorphisms
\bea
g'_{\mu\nu}(x') &=& \frac{\partial x^\alpha}{\partial x'^\mu}
\frac{\partial x^\beta}{\partial x'^\nu}
g_{\alpha\beta}(x)\ ,
\\
A'_\mu{}^\alpha{}_\beta(x')&=&
\frac{\partial x^\nu}{\partial x'^\mu}
\frac{\partial x'^\alpha}{\partial x^\gamma}
\frac{\partial x^\delta}{\partial x'^\beta}A_\nu{}^\gamma{}_\delta(x) 
+ \frac{\partial x'^\alpha}{\partial x^\gamma}
\frac{\partial^2 x^\gamma}{\partial x'^\mu\partial x'^\beta}\ .
\label{diffeo}
\eea
For an infinitesimal transformation $x^{\prime\mu}=x^\mu-\xi^\mu(x)$
the transformation is given by the Lie derivatives, plus
an inhomogeneous term for the connection:
\be
\delta g_{\mu\nu} = \cL_\xi g_{\mu\nu}\ ,\qquad 
\delta A_\rho{}^\mu{}_\nu =\cL_\xi A_\rho{}^\mu{}_\nu
+\partial_\rho \partial_\nu \xi^\mu\ ,
\label{lindiff}
\ee
where $\cL_\xi A_\rho{}^\mu{}_\nu
=\xi^\lambda\partial_\lambda A_\rho{}^\mu{}_\nu
+A_\lambda{}^\mu{}_\nu \partial_\rho\xi^\lambda
-A_\rho{}^\lambda{}_\nu \partial_\lambda\xi^\mu
+A_\rho{}^\mu{}_\lambda \partial_\rho\xi^\lambda$.
On a flat background $A=0$ and the Lie derivative term is absent.

Invariance under diffeomorphisms lowers by one the rank
of the coefficient matrices $a(1^-)$ and $a(0^+)$.
(This is because the transformation parameter $\xi_\mu$
can be decomposed as a three-scalar and a three-vector).
This is particularly clear in the Einstein form of the theory,
where diffeomorphism invariance implies
\be\label{fiamma}
a(1^-)_{i7}=a(1^-)_{7i}=0\ ,\qquad
a(0^+)_{i6}=a(0^+)_{6i}=0\ .
\ee

\subsubsection{Vector transformations of \texorpdfstring{$A$}{A}}
\label{sec:vectr}

Certain classes of MAGs are invariant under additional transformations 
of the connection, parametrized by a vector $\lambda_\mu(x)$:
\bea
\delta_1 A_\mu{}^\rho{}_\nu &=& \lambda_\mu \delta^\rho_\nu\ ,\qquad \,\,\,\delta_1 g_{\mu\nu}=0\ ,
\label{pt1}
\\
\delta_2 A_\mu{}^\rho{}_\nu &=& \lambda^\rho g_{\mu\nu}\ ,\qquad 
\delta_2 g_{\mu\nu}=0\ ,
\label{pt2}
\\
\delta_3 A_\mu{}^\rho{}_\nu &=& \delta^\rho_\mu\lambda_\nu\ ,\qquad \,\,\,\delta_3 g_{\mu\nu}=0\ .
\label{pt3}
\eea
The first of these is the projective transformation.
In order to spell out the conditions for invariance of the action,
it is easier to work in the Einstein formulation.
Since the metric (and therefore the Christoffel coefficients)
transforms trivially, the tranformations of $\phi_\mu{}^\rho{}_\nu$
are the same as those of $A_\mu{}^\rho{}_\nu$ given above.
The conditions on the kinetic coefficients for invariance of the
Lagrangian, are listed in Appendix \ref{sec:app.projective}.
(See \cite{Iosifidis:2018zwo}) for earlier related work).
We note that these transformations could also be present
in arbitrary linear combinations, each yielding different
conditions on the coefficients.

Each one of these invariances, when present, lowers by one the 
rank of the coefficient matrices $a(1^-)$ and $a(0^+)$.

\subsubsection{Weyl transformations}
\label{sec:weyl}

By definition, Weyl transformations are local rescalings of the metric:
\be
\delta g_{\mu\nu}=2\omega g_{\mu\nu}\ .
\ee
This implies that the LC connection transforms as:
\be
\delta \lc_\mu{}^\rho{}_\nu=
\partial_\mu\omega \delta^\rho_\nu
+\partial_\nu\omega \delta^\rho_\mu
-g^{\rho\tau}\partial_\tau\omega g_{\mu\nu} 
\ .
\ee
If we now consider the decomposition (\ref{phi}),
we see that there are infinitely many
ways of splitting this transformation between $A$ and $\phi$.
We consider here only
\be
\delta A_\mu{}^\rho{}_\nu =0 \qquad
\delta\phi_\mu{}^\rho{}_\nu=
-\partial_\mu\omega \delta^\rho_\nu
-\partial_\nu\omega \delta^\rho_\mu
+g^{\rho\tau}\partial_\tau\omega g_{\mu\nu} \ ,
\ee
which is the usual way in which Weyl transformations 
are realized on Yang-Mills fields.

The action (\ref{qLagE}) is invariant under this transformation if all the dimension 2 terms are absent and, additionally, the following relations hold:

\begin{fleqn}[0pt]
\begin{equation*}
\begin{aligned}
& 4 b^{RR}_{1} + 2 b^{RR}_{2} + b^{RQ}_{1} + b^{RQ}_{2} + 4 b^{RQ}_{3} + b^{RQ}_{5} = 0\ , 
\deveq
& 6 b^{RR}_{1} + 6 b^{RR}_{2} + 18 b^{RR}_{3} -  b^{RQ}_{1} + 3 b^{RQ}_{2} + 2 b^{RQ}_{4} + 3 b^{RQ}_{5} + 6 b^{RQ}_{7} - 8 b^{QQ}_{1} + 2 b^{QQ}_{2} + 2 b^{QQ}_{3} \\ & \quad\quad- 32 b^{QQ}_{4} - 8 b^{QQ}_{6} + 2 b^{QQ}_{7} + 2 b^{QQ}_{8} + 2 b^{QQ}_{9} + 2 b^{QQ}_{10} + 2 b^{QQ}_{12} - 32 b^{QQ}_{14} + 2 b^{QQ}_{15} = 0\ , \\ 
& b^{RR}_{2} + 6 b^{RR}_{3} + b^{RQ}_{4} + 4 b^{RQ}_{6} + b^{RQ}_{7} = 0\ , 
\deveq
& b^{RT}_{1} + 2 b^{RT}_{2} + b^{RT}_{3} - 3 b^{RT}_{4} + 6 b^{RT}_{5} + 2 b^{TQ}_{1} - 2 b^{TQ}_{2} - 8 b^{TQ}_{3} + 2 b^{TQ}_{4} \\ & \quad\quad- 2 b^{TQ}_{10} - 2 b^{TQ}_{11} + 8 b^{TQ}_{12} + 2 b^{TQ}_{13} = 0\ , 
\deveq
& b^{RQ}_{1} - 3 b^{RQ}_{2} -  b^{RQ}_{5} - 6 b^{RQ}_{7} - 4 b^{QQ}_{2} - 4 b^{QQ}_{3} - 8 b^{QQ}_{5} - 2 b^{QQ}_{9} - 2 b^{QQ}_{10} \\ & \quad\quad - 2 b^{QQ}_{12} - 4 b^{QQ}_{15} - 8 b^{QQ}_{16} = 0\ , 
\deveq
& b^{RQ}_{4} + b^{RQ}_{5} + 2 b^{QQ}_{7} + 2 b^{QQ}_{8} + b^{QQ}_{9} + b^{QQ}_{10} + 4 b^{QQ}_{11} + b^{QQ}_{12} + 4 b^{QQ}_{13}=0\ .
\end{aligned}
\end{equation*}
\end{fleqn}
As a check we observe that if all the coefficients of type
$b^{RT}$, $b^{RQ}$, $b^{QQ}$ and $b^{TQ}$ are zero,
the remaining relations imply that the $R^2$ terms
appear in the combination 
$$
b^{RR}_1\left( R_{\mu\nu\rho\sigma}R^{\mu\nu\rho\sigma}
-2R_{\mu\nu}R^{\mu\nu}+\frac13R^2\right)
=b^{RR}_1 C_{\mu\nu\rho\sigma}C^{\mu\nu\rho\sigma}\ ,
$$
which is the square of the Weyl tensor.

\section{MAGs with dimension-two terms only}

In this section we discuss, at linearized level, the case of theories without
dimension-4 operators in the Lagrangian.
In an EFT, the dimension-two terms will be the dominant ones
at very low energy.

Consider again Fig.1.
At the top vertex of the triangle ($Q=T=0$)
one has Riemannian geometry, and the only invariant
of dimension two is the Hilbert action.
At linearized level we get the Fierz-Pauli action
\be
S^{(2)} = \frac{m^R}{2}\int \frac{d^4q}{(2\pi)^4} \left(
-\frac{1}{4} q^2 h_{\mu \nu} h^{\mu \nu}
+ \frac{1}{2} q_\mu q_\lambda h^{\mu \nu} h^\lambda_\nu
- \frac{1}{2} q_\mu q_\nu h^{\mu \nu} h
+ \frac{1}{4} q^2 h^2
\right) \,.
\ee

In the interior of the triangle we have the generalized Palatini action
(\ref{genPalC}).
The generalization consists of the following.
In the ``standard'' Palatini approach, the action is just $a^F F$.
When varied, this is not enough to constrain the connection completely.
One can either assume $T=0$ and obtain $Q=0$ as an equation,
or assume $Q=0$ and obtain $T=0$ as an equation.
Thus, the standard Palatini action works on the left and right
sides of the triangle, but not in the interior.
This is due to the fact that the Palatini action is invariant under the
projective tranformations (\ref{pt1})
The addition of the other terms in (\ref{genPalC}),
which is only natural from the point of view of EFT,
generically breaks projective invariance and fixes this problem.
In the Einstein form,
the action becomes (\ref{genPalE}) (or (\ref{genPalETQ})),
which consists just of the Hilbert action for $g$
and a mass term for the distortion 
(or equivalently torsion and nonmetricity).
Generically, this mass term will be non-degenerate and the
EOM will imply that distortion vanishes.
Thus the theory is dynamically equivalent to GR, on shell.
We note that the addition to the standard Palatini action
of torsion-squared terms in antisymmetric MAG
or nonmetricity-squared terms in symmetric MAG,
will generically not change the EOMs.
Still, these terms are expected to be present
when we think of MAG as an EFT.

We now turn to the bottom of the triangle,
which does not follow the generic behavior of the interior.
We first look at the left and right corners,
then at the bottom edge.
The following analyses will be carried out in the
Cartan version of the theory.

\subsection{Antisymmetric teleparallel theory}

This is also known as Weitzenb\"ock theory.
We have $F=Q=0$, so the action must be quadratic in torsion
\be
S = -\frac12\int d^4x \sqrt{|g|} \left[ 
a^{TT}_1 T_{\mu \rho \nu} T^{\mu \rho \nu} 
+a^{TT}_2 T_{\mu \rho \nu} T^{\mu \nu \rho} 
+a^{TT}_3 \trT12_\mu \trT12^\mu
\right]  \,.
\ee
The condition $F=0$ implies (\ref{puregauge}).
When the theory is linearized around flat space,
this becomes
$A_\mu{}^\rho{}_\nu=\partial_\mu\lambda^\rho{}_\nu$,
where $\Lambda^\rho{}_\sigma=\delta^\rho_\sigma+\lambda^\rho{}_\sigma$.
The condition $Q=0$ implies for the metric fluctuation that 
$A_{\mu \rho \nu}+A_{\mu \nu \rho} = \partial_\mu h_{\rho \nu}$ .
Putting these conditions together we have 
\be
A_{\mu \rho \nu}= \frac{1}{2}\partial_\mu h_{\rho \nu} 
+ \partial_\mu \Omega_{\rho \nu}\ ,
\label{didone}
\ee
where $\Omega$ is the antisymmetric part of $\lambda$. 
So the action 
of the linearized theory becomes
\bea
S &=& -\frac12\int \frac{d^4q}{(2\pi)^4} \Bigg[
-\frac{(2a^{TT}_1+a^{TT}_2)}{4} q^2 h_{\mu \nu} h^{\mu \nu}
+\frac{(2a^{TT}_1+a^{TT}_2-a^{TT}_3)}{4} q_\mu q_\lambda h^{\mu \nu} h^\lambda_\nu
\nonumber\\
&&
+ \frac{a^{TT}_3}{2} q_\mu q_\nu h^{\mu \nu} h
-\frac{a^{TT}_3}{4} q^2 h^2
- (2a^{TT}_1-a^{TT}_2) q^2 \Omega^{\mu \nu} \Omega_{\mu \nu}
\nonumber\\
&&
+ (2a^{TT}_1-3a^{TT}_2-a^{TT}_3) q_\mu q_\lambda \Omega^{\mu \nu} \Omega^\lambda{}_\nu
- (2a^{TT}_1+a^{TT}_2+a^{TT}_3) q_\mu q_\lambda \Omega^{\mu \nu} h^\lambda_\nu
\Bigg] \,.
\nonumber
\eea

The linearized action can then be written in a form
analogous to (\ref{linactSP}):
\bea
S &=& \frac{1}{2} \sum_{P,i,j}\int \frac{d^4q}{(2\pi)^4} \left( \Omega(-q) \,\, h(-q) \right) \cdot 
\begin{pmatrix}
a_{ij}^{\Omega \Omega}
P_{ij}^{\Omega \Omega}
& a_{ij}^{\Omega h}
P_{ij}^{\Omega h} \\
a_{ij}^{h \Omega }
P_{ij}^{h \Omega }
& a_{ij}^{hh}
P_{ij}^{hh}
\end{pmatrix}
\cdot  
\begin{pmatrix}
\Omega(q) \\
h(q)
\end{pmatrix} 
\nn
&& \hspace{2cm}+
\int \frac{d^4q}{(2\pi)^4} \left\{ \sigma(-q)\cdot h(q)
+ \tau(-q)\cdot \Omega(q)
\right\}\,,
\eea
where
\begin{subequations}
\begin{align}
& a\left(2^+\right) = \frac{(2a^{TT}_1+a^{TT}_2)}{4} q^2  \,,
\\
& a\left( 1^+\right) = \left(2a^{TT}_1-a^{TT}_2 \right) q^2  \,,
\\
&a\left(1^-\right) = \frac{a^{TT}_4}{8} q^2 \,\,\,
\begin{pmatrix}
  4 & -2  \\
  -2 & 1 
\end{pmatrix} \,,
\\
&a\left(0^+\right) = \frac{(a^{TT}_4+2a^{TT}_3)}{4}q^2
\begin{pmatrix}
1 & 0 \\
0 & 0
\end{pmatrix}	\ ,
\end{align}
\end{subequations}
where
\be
a^{TT}_4 \equiv 2a^{TT}_1+a^{TT}_2+a^{TT}_3\ ,
\ee
and the additional projectors are defined in Appendix~\ref{sec:pro12sa}.
In the $1^-$ sector, the order of the rows and columns 
is $(\Omega,h)$.
Note that the matrices $a\left(1^-\right)$ and $a\left(0^+\right)$ have rank $1$
because of the diffeomorphism invariance.
We fix the gauge by removing the second row and column.
At the linearized level the diffeomorphism transformation reads
\begin{subequations}
\begin{align}
& h_{\mu\nu} \to h_{\mu\nu} + \partial_\mu \xi_\nu + \partial_\nu \xi_\mu \,,
\\
& \Omega_{\mu\nu} \to \Omega_{\mu\nu} 
- \frac{1}{2}\partial_\mu \xi_\nu + \frac{1}{2}\partial_\nu \xi_\mu \ ,
\end{align}
\end{subequations}
where the transformation of $\Omega$ follows from those
of $A$ and $h$ and formula (\ref{didone})
So the sources satisfy the following constraint
\be
-2 q^\mu \sigma_{\mu\nu} + q^\mu \tau_{\mu\nu} = 0 \,.
\ee
The saturated propagator is
\bea
\Pi\!\! &=&\!\! -\frac{1}{2} \int \frac{d^4q}{(2\pi)^4} \left\{
\frac{4}{(2a^{TT}_1+a^{TT}_2)q^2} \left[ \sigma_{\mu\nu}\sigma^{\mu\nu} 
- \frac{a^{TT}_3}{2a^{TT}_1+a^{TT}_2+3a^{TT}_3} \left( \sigma^\mu_\mu \right)^2\right]
\right.\\
&& \left. 
+\frac{1}{(2a^{TT}_1-a^{TT}_2)q^2} \left[ \tau_{\mu\nu}\tau^{\mu\nu} 
- \frac{4(a^{TT\,2}_2+2a^{TT}_1a^{TT}_2+2a^{TT}_1a^{TT}_3)}{(2a^{TT}_1+a^{TT}_2)(2a^{TT}_1+a^{TT}_2+a^{TT}_3)} \frac{q^\mu q^\nu}{q^2} \tau_{\mu\rho}\tau_{\nu}{}^{\rho} \right]
\right\} \,.
\nonumber
\eea
Making the following redefinitions
\begin{subequations}
\begin{align}
& \tilde \sigma_{\mu\nu} \equiv \sigma_{\mu\nu} + C \,\sigma^\rho_\rho\,\eta_{\mu\nu} \,,
\\
& \tau_{\mu\nu} \equiv -\frac{i}{q^2}\left(q_\mu \chi_\nu - q_\nu \chi_\mu \right) 
+ \tilde \tau_{\mu\nu} 
\hspace{2cm} \mbox{with} \,\,\, 
q^\mu \chi_\mu = q^\mu \tilde \tau_{\mu\nu} =0  \,,
\end{align}
\end{subequations}
and adjusting the parameter $C$ , we can reduce the saturated propagator to the following form
\bea
\Pi &=& -\frac{1}{2} \int \frac{d^4q}{(2\pi)^4} \left[
\frac{4}{(2a^{TT}_1+a^{TT}_2)q^2} 
\left(\tilde \sigma_{\mu\nu} \tilde\sigma^{\mu\nu} 
- \frac{1}{2} \left( \tilde\sigma^\mu_\mu \right)^2\right)
\right.\\
&& \left. 
+\frac{1}{(2a^{TT}_1-a^{TT}_2)q^2} \tilde \tau_{\mu\nu} \tilde \tau^{\mu\nu} 
-\frac{2(2a^{TT}_1+a^{TT}_2-a^{TT}_3)}{(2a^{TT}_1+a^{TT}_2)(2a^{TT}_1+a^{TT}_2+a^{TT}_3)q^4} \chi_{\mu}\chi^{\mu}
\right] \,.
\nonumber
\eea
In the first term we recognize the usual graviton, 
in the second one we have a massless spin $1^+$ state
and in the last a dipole ghost with spin $1^-$.

The latter is pathological and in order to eliminate it, 
we have to impose
\be
a^{TT}_4=0 \,.
\label{relation1}
\ee
With this constraint, we recover linearized
GR together with a spin $1^+$ particle.
If we impose that $2a^{TT}_1-a^{TT}_2>0$
its propagator assumes the proper form (\ref{1+satprop}).
In this theory there are two different gauge invariances:
the previously mentioned diffeomorphisms, and
\be
\Omega_{\mu\nu} \to \Omega_{\mu\nu} 
+ \partial_\mu \chi_\nu - \partial_\nu \chi_\mu \,.
\ee
The additional degree of freedom can be removed by imposing
\be
2a^{TT}_1-a^{TT}_2=0\ ,
\label{relation2}
\ee
in which case $\Omega$ disappears from the action
(it is a pure gauge degree of freedom)
and the rest reduces to the antisymmetric teleparallel
equivalent of the Hilbert action, (\ref{ATEGR}).

\subsection{Symmetric teleparallel theory}

Now we have $F=T=0$, so the action is a generic
quadratic combination of non-metricity:
\bea
S&=&-\frac12\int d^4x \sqrt{|g|} \left[ 
a^{QQ}_1 Q_{\rho \mu \nu} Q^{\rho \mu \nu} 
+ a^{QQ}_2 Q_{\rho \mu \nu} Q^{\mu \rho \nu} 
\right.\\
&&\left.\qquad\qquad
+ a^{QQ}_3 \trQ23_\mu \trQ23^\mu
+ a^{QQ}_4 \trQ12_\mu \trQ12^\mu
+ a^{QQ}_5 \trQ23_\mu \trQ12^\mu\right] \nonumber \,.
\eea
As in the antisymmetric case, in the linearized theory $F=0$
implies $A_\mu{}^\rho{}_\nu=\partial_\mu\lambda^\rho{}_\nu$.
The condition $T=0$ implies that 
$A_\mu{}^\rho{}_\nu=A_\nu{}^\rho{}_\mu$.
Putting these conditions together we have 
\be
A_\mu{}^\rho{}_\nu=\partial_\mu\partial_\nu u^\rho  \,.
\ee
Substituting 
$Q_{\rho \mu \nu} = -\partial_\rho h_{\mu \nu} +\partial_\rho \partial_\mu u_\nu +\partial_\rho\partial_\nu u_\mu$
and linearizing, the action becomes
\bea
S&=& -\frac12\int \frac{d^4q}{(2\pi)^4} \Big[
-a^{QQ}_1 q^2 h_{\mu \nu} h^{\mu \nu}
- (a^{QQ}_2+a^{QQ}_4) q_\mu q_\lambda h^{\mu \nu} h^\lambda_\nu
-a^{QQ}_5 q_\mu q_\nu h^{\mu \nu} h
-a^{QQ}_3 q^2 h^2
\nonumber\\
&&  +(2a^{QQ}_1+a^{QQ}_2+a^{QQ}_4) q^4 u_\lambda u^\lambda 
+(2a^{QQ}_1+3a^{QQ}_2+4a^{QQ}_3+3a^{QQ}_4+4a^{QQ}_5)q^2 q_\mu q_\nu u^\mu u^\nu 
\nonumber\\
&&  
-2i (2a^{QQ}_1+a^{QQ}_2+a^{QQ}_4) q^2 q_\mu u_\nu h^{\mu \nu}-2 i (a^{QQ}_2+a^{QQ}_4+a^{QQ}_5) q_\lambda q_\mu q_\nu u^\lambda h^{\mu \nu}
\nonumber\\
&&
-2 i (2a^{QQ}_3+a^{QQ}_5) q^2 q_\lambda u^\lambda h
\Big] \,.
\nonumber
\eea

For a generic choice of coefficients
the linearized action is
\bea
S &=& \frac{1}{2} \sum_{P,i,j}\int \frac{d^4q}{(2\pi)^4} \left( u(-q) \,\, h(-q) \right) \cdot 
\begin{pmatrix}
a_{ij}^{uu}
P_{ij}^{uu}
& a_{ij}^{uh}
P_{ij}^{uh} \\
a_{ij}^{hu}
P_{ij}^{hu}
& a_{ij}^{hh}
P_{ij}^{hh}
\end{pmatrix}
\cdot  
\begin{pmatrix}
u(q) \\
h(q)
\end{pmatrix}
\nn
&& \hspace{2cm}+
\int \frac{d^4q}{(2\pi)^4} \left\{ \sigma(-q)\cdot h(q)
+ \tau(-q)\cdot u(q)
\right\}\,,
\eea
where
\begin{subequations}
\begin{align}
& a\left( 2^+\right) = a^{QQ}_1 q^2 \,,
\\
& a(1^-) = \frac{1}{2}\left( 2a^{QQ}_1+ a^{QQ}_6\right) q^2 \,\,\,
\begin{pmatrix}
  -2\,q^2 & i\sqrt{2} \,|q| \\
  i\sqrt{2}\,|q| & 1  \\
\end{pmatrix} \,,
\\
& a(0^+) = q^2 \,\,
\begin{pmatrix}
-4\,a^{QQ}_7\,q^2 & i\sqrt{3}(2a^{QQ}_3+a^{QQ}_5)|q| & 2i\,a^{QQ}_7\,|q|  \\
i\sqrt{3}(2a^{QQ}_3+a^{QQ}_5)|q| &  (a^{QQ}_1+3 a^{QQ}_3) & \sqrt{3}(2a^{QQ}_3+a^{QQ}_5)/2  \\
2i\,a^{QQ}_7\,|q| & \sqrt{3}(2a^{QQ}_3+a^{QQ}_5)/2 &  a^{QQ}_7\\
\end{pmatrix}  \,,
\end{align}
\end{subequations}
where the rows/columns of $a(1^-)$ refer to $u$, $h$, in this order,
those of $a(0^+)$ to $u$, $h$, $h$.
We defined
\begin{subequations}
\begin{align}
& a^{QQ}_6 \equiv a^{QQ}_2+a^{QQ}_4 \,,
\\
& a^{QQ}_7 \equiv a^{QQ}_1+a^{QQ}_2+a^{QQ}_3+a^{QQ}_4+a^{QQ}_5 \, ,
\end{align}
\end{subequations}
and the new projectors are defined in Appendix~\ref{sec:pro12sa}.
Note that the matrix $a\left(1^-\right)$ has rank $1$ and $a\left(0^+\right)$ has rank $2$
because of the diffeomorphism invariance.

At the linearized level the diffeomorphism transformation reads
\begin{subequations}
\begin{align}
& h_{\mu\nu} \to h_{\mu\nu} + \partial_\mu \xi_\nu + \partial_\nu \xi_\mu \,,
\\
& u_\mu \to u_\mu + \xi_\mu \ ,
\end{align}
\end{subequations}
so the sources satisfy the following constraint
\be
-2 i q^\mu \sigma_{\mu\nu} + \tau_{\nu} = 0 \,.
\ee
The saturated propagator is
\bea
\Pi &=& -\frac{1}{2} \int \frac{d^4q}{(2\pi)^4} \left\{
\frac{1}{a^{QQ}_1q^2} \left[ \sigma_{\mu\nu}\sigma^{\mu\nu} 
+(\ldots)\left( \sigma^\mu_\mu \right)^2\right]
-\frac{i}{2} 
\frac{(\ldots)}{q^4} q^\mu \tau_\mu \sigma_\nu^\nu
\right.\nn
&& \left. 
+\frac{a^{QQ}_6}{2a_4(2a^{QQ}_1+a^{QQ}_6)q^4} \left( \tau_{\mu}\tau^{\mu} 
+ (\ldots)
\frac{q^\mu q^\nu}{q^2} \tau_{\mu}\tau_{\nu} \right)
\right\} \ ,
\nonumber
\eea
where the ellipses stand for complicated combinations of couplings
whose explicit form is not very relevant.
Making the redefinitions
\begin{subequations}
\begin{align}
& \tilde \sigma_{\mu\nu} \equiv \sigma_{\mu\nu} - \frac{i A}{q^2} \left( q_\mu \tau_\nu +q_\nu \tau_\mu \right) 
+ C \,\sigma^\rho_\rho\,\eta_{\mu\nu} \,,
\\
& \tau_{\mu} \equiv -\frac{i}{q^2}q_\mu j + \tilde \tau_{\mu} 
\hspace{2cm} \mbox{with} \,\,\,  q^\mu \tilde \tau_{\mu} =0 \,,
\end{align}
\end{subequations}
and adjusting the parameters $(A,C)$ , we can reduce the saturated propagator to the form
\bea
\Pi &=& -\frac{1}{2} \int \frac{d^4q}{(2\pi)^4} \left\{
\frac{1}{a^{QQ}_1q^2} \left[ \tilde \sigma_{\mu\nu} \tilde \sigma^{\mu\nu} 
- \frac{1}{2} \left( \tilde \sigma^\mu_\mu \right)^2\right]
+\frac{(\ldots)}{q^4}  \tau^{\mu} \tilde \tau^{\nu}  + \frac{(\ldots)}{q^6} j^2 
\right\} \ .
\eea
These dipole and tripole ghosts can be eliminated
imposing the conditions
\begin{subequations}
\begin{align}
& 2a^{QQ}_1+a^{QQ}_6=0 \,,
\\
& a^{QQ}_6 + a^{QQ}_5 =0 \,,
\\
& 2a^{QQ}_3+a^{QQ}_5 =0 \ ,
\end{align}
\label{relations3}
\end{subequations}
leaving us just with the standard graviton saturated propagator.
With these constraints, $u$ becomes a pure gauge and we recover
the symmetric teleparallel equivalent of GR (\ref{STEGR}).
This is in agreement with the results of 
\cite{BeltranJimenez:2017tkd,Conroy:2017yln}.

\subsection{General teleparallel theory}

We now only assume $F=0$.
The action is
\bea
S &=& -\frac12\int d^4x \sqrt{|g|} \bigg[ 
a^{TT}_1 T_{\mu \rho \nu} T^{\mu \rho \nu} 
+ a^{TT}_2 T_{\mu \rho \nu} T^{\mu \nu \rho} 
+a^{TT}_3 \trT12_\mu \trT12^\mu
\nonumber\\
&&\hspace{2cm} 
+ a^{QQ}_1 Q_{\rho \mu \nu} Q^{\rho \mu \nu} 
+ a^{QQ}_2 Q_{\rho \mu \nu} Q^{\mu \rho \nu} 
 \nonumber\\
&&\hspace{2cm} 
+ a^{QQ}_3 \trQ23_\mu \trQ23^\mu 
+ a^{QQ}_4 \trQ12_\mu \trQ12^\mu
+ a^{QQ}_5 \trQ23_\mu \trQ12^\mu
\nonumber\\
&&\hspace{2cm} 
+ a^{TQ}_1 T_{\mu \rho \nu} Q^{\mu \rho \nu} 
+ a^{TQ}_2 \trT12_\mu \trQ23^\mu
+ a^{TQ}_3 \trT12_\mu \trQ12^\mu
\bigg]  \,.
\eea
As in the previous cases, in the linearized theory $F=0$
implies $A_\mu{}^\rho{}_\nu=\partial_\mu\lambda^\rho{}_\nu$,
but now both the symmetric part $H$ 
and the antisymmetric part $\Omega$ of $\lambda$
have to be treated as dynamical fields.

So the action becomes
\bea
S &=& -\frac{1}{2}\int \frac{d^4q}{(2\pi)^4} \left(
-a^{QQ}_1 q^2 h_{\mu \nu} h^{\mu \nu}
- (a^{QQ}_2+a^{QQ}_4) q_\mu q_\lambda h^{\mu \nu} h^\lambda_\nu
-a^{QQ}_5 q_\mu q_\nu h^{\mu \nu} h
-a^{QQ}_3 q^2 h^2
\right.\nonumber\\
&& \left. -(2a^{TT}_1+a^{TT}_2+4a^{QQ}_1+2a^{TQ}_1) q^2 H_{\mu \nu} H^{\mu \nu}
\right.\nonumber\\
&& \left.
+(2a^{TT}_1+a^{TT}_2-a^{TT}_3-4a^{QQ}_2-4a^{QQ}_4+2a^{TQ}_1-2a^{TQ}_3) q_\mu q_\lambda H^{\mu \nu} H^\lambda_\nu
\right.\nonumber\\
&& \left.
+2(a^{TT}_3-2a^{QQ}_5-a^{TQ}_2+a^{TQ}_3) q_\mu q_\nu H^{\mu \nu} H
-(a^{TT}_3+4a^{QQ}_3-2a^{TQ}_2) q^2 H^2
\right.\nonumber\\
&& \left. -(2a^{TT}_1-a^{TT}_2) q^2 \Omega_{\mu \nu} \Omega^{\mu \nu}
+(2a^{TT}_1-3a^{TT}_2-a^{TT}_3) q_\mu q_\lambda \Omega^{\mu \nu} \Omega^\lambda_\nu
\right.\nonumber\\
&& \left. +(4a^{QQ}_1+a^{TQ}_1) q^2 H_{\mu \nu} h^{\mu \nu}
+ (4a^{QQ}_2+4a^{QQ}_4-a^{TQ}_1+a^{TQ}_3) q_\mu q_\lambda H^{\mu \nu} h_\nu^\lambda
\right.\nonumber\\
&& \left. +(2a^{QQ}_5+a^{TQ}_2) q_\mu q_\nu H^{\mu \nu} h
+(2a^{QQ}_5-a^{TQ}_3) q_\mu q_\nu H h^{\mu \nu}
+(4a^{QQ}_3-a^{TQ}_2) q^2 H h
\right.\nonumber\\
&& \left. + (a^{TQ}_1+a^{TQ}_3) q_\mu q_\lambda \Omega^{\mu \nu} h_\nu^\lambda
-2 (2a^{TT}_1+a^{TT}_2+a^{TT}_3+a^{TQ}_1+a^{TQ}_3) q_\mu q_\lambda \Omega^{\mu \nu} H_\nu^\lambda
\right) \,.
\nonumber
\eea

For a generic choice we write
\bea
S &=& \frac{1}{2} \sum_{P,i,j}\int \frac{d^4q}{(2\pi)^4} 
\left( \Omega\,\, H\,\, h\right) \cdot 
\begin{pmatrix}
a_{ij}^{\Omega\Omega}
P_{ij}^{\Omega\Omega}
& a_{ij}^{\Omega H}
P_{ij}^{\Omega H}
& a_{ij}^{\Omega h}
P_{ij}^{\Omega h}\\
a_{ij}^{H \Omega}
P_{ij}^{H \Omega}
& a_{ij}^{HH}
P_{ij}^{HH}
& a_{ij}^{Hh}
P_{ij}^{Hh}\\
a_{ij}^{h\Omega }
P_{ij}^{h\Omega }
& a_{ij}^{hH}
P_{ij}^{hH}
& a_{ij}^{hh}
P_{ij}^{hh}
\end{pmatrix}
\cdot  
\begin{pmatrix}
\Omega\\
H\\
h
\end{pmatrix}
\nn
&& \hspace{2cm}+
\int \frac{d^4q}{(2\pi)^4} \left\{ \sigma\cdot h
+ \Sigma\cdot H
+ \tau\cdot \Omega
\right\}\,,
\eea
where
\begin{subequations}
\begin{align}
& a\left( 2^+\right)\! = q^2
\begin{pmatrix}
  (2a^{TT}_1+a^{TT}_2+4a^{QQ}_1+2a^{TQ}_1) & -(4a^{QQ}_1+a^{TQ}_1)/2 \\
  -(4a^{QQ}_1+a^{TQ}_1)/2 & a^{QQ}_1 \\
\end{pmatrix} \,,
\\
&a\left( 1^+\right)\! = \left(2a^{TT}_1-a^{TT}_2 \right) q^2 \,,
\\
&a\left( 1^-\right)\! = q^2 \!
\begin{pmatrix}
  a^{TT}_4/2 & - (a^{TT}_4+a^{TQ}_1+a^{TQ}_3)/2 & (a^{TQ}_1+a^{TQ}_3)/4 \\
  - (a^{TT}_4+a^{TQ}_1+a^{TQ}_3)/2 & \!\left( a^{TT}_4 + a^{TQ}_5 +a^{TQ}_1 +a^{TQ}_3\right)/2\! & -a^{TQ}_5/4 \\
  (a^{TQ}_1+a^{TQ}_3)/4 & -a^{TQ}_5/4 & \left( 2a^{QQ}_1 +a^{QQ}_6\right)/2 \\
\end{pmatrix} \, ,
\\
& a\left( 0^+\right)\! = q^2\!
\begin{pmatrix}
   a^{TQ}_4 & \sqrt{3} \,a^{TQ}_6  & -a^{TQ}_7/2 & -\sqrt{3} \,a^{TQ}_6/2 \\
   \sqrt{3} \,a^{TQ}_6 & 4\, a^{QQ}_7  & \!-\sqrt{3}(2a^{QQ}_3+a^{QQ}_5)\! & -2\,a^{QQ}_7 \\
   -a^{TQ}_7/2 & \!-\sqrt{3}(2a^{QQ}_3+a^{QQ}_5)\!  &   (a^{QQ}_1+3 a^{QQ}_3) & \!\sqrt{3}(2a^{QQ}_3+a^{QQ}_5)/2\! \\
   -\sqrt{3} \,a^{TQ}_6/2 & -2\,a^{QQ}_7  &  \!\sqrt{3}(2a^{QQ}_3+a^{QQ}_5)/2\! &  a^{QQ}_7 
\end{pmatrix}\,,
\end{align}
\end{subequations}
where the rows/columns of $a(2^+)$ refer to $H$ and $h$ (in this order),
those of $a(1^-)$ to $\Omega$, $H$, $h$,
those of $a(0^+)$ to $H$, $H$, $h$, $h$,
and we defined
\begin{subequations}
\begin{align}
& a^{TQ}_4\equiv a^{TT}_4+2a^{TT}_3+4a^{QQ}_1+12a^{QQ}_3+2a^{TQ}_1-6a^{TQ}_2 \,,
\\
& a^{TQ}_5\equiv 8a^{QQ}_1+ 4a^{QQ}_6+a^{TQ}_1+a^{TQ}_3 \,,
\\
& a^{TQ}_6\equiv 4a^{QQ}_3+2a^{QQ}_5-a^{TQ}_2-a^{TQ}_3  \,,
\\
& a^{TQ}_7\equiv 4a^{QQ}_1+12a^{QQ}_3+a^{TQ}_1-3a^{TQ}_2 \,,
\end{align}
\end{subequations}
and the projectors are defined in Appendix~\ref{sec:pro12sa}.
As usual the matrix $a\left(1^-\right)$ has rank $2$ and $a\left(0^+\right)$ has rank $3$
because of diffeomorphism invariance.

At the linearized level the diffeomorphism transformation reads
\begin{subequations}
\begin{align}
& h_{\mu\nu} \to h_{\mu\nu} + \partial_\mu \xi_\nu + \partial_\nu \xi_\mu \,,
\\
& H_\mu \to H_{\mu\nu} + \frac{1}{2}\partial_\mu \xi_\nu + \frac{1}{2} \partial_\nu \xi_\mu \,,
\\
& \Omega_{\mu\nu} \to \Omega_{\mu\nu} - \frac{1}{2}\partial_\mu \xi_\nu + \frac{1}{2} \partial_\nu \xi_\mu \ ,
\end{align}
\end{subequations}
so the sources satisfy the following constraint
\be
2 q^\mu \sigma_{\mu\nu} + q^\mu \Sigma_{\mu\nu} - q^\mu \tau_{\mu\nu} = 0 \,.
\ee
The saturated propagator is
\bea
\Pi &=& -\frac{1}{2} \int \frac{d^4q}{(2\pi)^4} \left\{
\frac{(\ldots)}{q^2} \left[ \sigma_{\mu\nu}\sigma^{\mu\nu} 
+ (\ldots) \left( \sigma^\mu_\mu \right)^2\right]
+\frac{(\ldots)}{q^2} \left[ \Sigma_{\mu\nu}\Sigma^{\mu\nu} 
+ (\ldots) \left( \Sigma^\mu_\mu \right)^2 \right]
\right.
\nonumber\\
&& \left. \hspace{2cm}
+\frac{q^\mu q^\nu}{q^4} \left[ (\ldots) \Sigma_{\mu\nu} \Sigma^\rho_\rho 
+ (\ldots) \Sigma_{\mu\rho} \Sigma{}_\nu{}^\rho 
+ (\ldots) \frac{ q^\rho q^\lambda}{q^2}\Sigma_{\mu\nu} \Sigma_{\rho\lambda}
\right]
\right.\nn
&& \left. \hspace{2cm}
+\frac{(\ldots)}{q^2} \left[ \tau_{\mu\nu}\tau^{\mu\nu} 
+ (\ldots) \frac{q^\mu q^\nu}{q^2}\tau_{\mu\rho} \tau{}_\nu{}^\rho \right]
+ (\ldots) \frac{q^\mu q^\nu}{q^2}\Sigma_{\mu\rho} \tau{}_\nu{}^\rho
\right.\nn
&& \left. \hspace{2cm}
+\frac{(\ldots)}{q^2} \left[ \Sigma_{\mu\nu}\sigma^{\mu\nu} 
+ (\ldots) \frac{q^\mu q^\nu}{q^2}\Sigma_{\mu\nu} \sigma^\rho_\rho
+ (\ldots)  \Sigma^\mu_\mu\,\sigma^\nu_\nu \right]
\right\}  \,.
\eea
Making the following redefinitions
\begin{subequations}
\begin{align}
& \tilde \sigma_{\mu\nu} \equiv \sigma_{\mu\nu} + A \,\Sigma_{\mu\nu}
+ \left( C \,\sigma^\rho_\rho + D \,\Sigma^\rho_\rho \right)\,\eta_{\mu\nu} \,,
\\
& \tilde \Sigma_{\mu\nu} \equiv  \Sigma_{\mu\nu} + B \,\sigma_{\mu\nu} 
+ \left( E \,\Sigma^\rho_\rho + F \,\sigma^\rho_\rho \right)\,\eta_{\mu\nu} \,,
\end{align}
\end{subequations}
and adjusting the parameters $(A,B,C,D,E)$ , 
we can reduce the saturated propagator to the following form
\bea
\Pi &=& -\frac{1}{2} \int \frac{d^4q}{(2\pi)^4} \left\{
\frac{(\ldots)}{q^2} \left[ \tilde \sigma_{\mu\nu}\tilde \sigma^{\mu\nu} 
-\frac{1}{2} \left( \tilde\sigma^\mu_\mu \right)^2\right]
+\frac{(\ldots)}{q^2} \left[ \tilde \Sigma_{\mu\nu} \tilde\Sigma^{\mu\nu} 
+(\ldots) \left( \tilde\Sigma^\mu_\mu \right)^2
\right]
\right.\nonumber\\
&& \left. \hspace{2.5cm}
+\frac{q^\mu q^\nu}{q^4} \left[ (\ldots) \tilde \Sigma_{\mu\nu} \tilde  \Sigma^\rho_\rho 
+ (\ldots) \tilde \Sigma_{\mu\rho} \tilde \Sigma{}_\nu{}^\rho 
+ (\ldots) \frac{ q^\rho q^\lambda}{q^2} \tilde \Sigma_{\mu\nu} \tilde \Sigma_{\rho\lambda}
\right]
\right.\nn
&& \left. \hspace{2.5cm}
+\frac{(\ldots)}{q^2} \left[ \tau_{\mu\nu}\tau^{\mu\nu} 
+ (\ldots) \frac{q^\mu q^\nu}{q^2}\tau_{\mu\rho} \tau{}_\nu{}^\rho \right]
\right\} \,.
\eea
Now that we have decoupled the sources, we decompose
\begin{subequations}
\begin{align}
& \tilde \Sigma_{\mu\nu} \equiv \tilde\Sigma_{\mu\nu}^T 
- \frac{i}{q^2} \left(q_\mu \kappa_\nu +q_\nu \kappa_\mu \right) 
+\frac{1}{q^2} \left( L_{\mu\nu} j_1 + T_{\mu\nu} j_2 \right)
\quad \mbox{with} \,\,\, q^\mu \kappa_\mu = q^\mu \tilde \Sigma_{\mu\nu}^T =0  \,,
\\
& \tau_{\mu\nu} \equiv -\frac{i}{q^2}\left(q_\mu \upsilon_\nu - q_\nu \upsilon_\mu \right) 
+ \tilde \tau_{\mu\nu} 
\quad \mbox{with} \,\,\, q^\mu \upsilon_\mu = q^\mu \tilde \tau_{\mu\nu} =0  \,,
\end{align}
\end{subequations}
and adjusting the parameter $F$, the saturated propagator becomes
\bea
\Pi &=& -\frac{1}{2} \int \frac{d^4q}{(2\pi)^4} \left\{
\frac{(\ldots)}{q^2} \left[ \tilde \sigma_{\mu\nu}\tilde \sigma^{\mu\nu} 
-\frac{1}{2} \left( \tilde\sigma^\mu_\mu \right)^2\right]
+\frac{(\ldots)}{q^2} \left[ \tilde \Sigma^T{}_{\mu\nu} \tilde\Sigma^T{}^{\mu\nu} 
-\frac{1}{2} \left( \tilde\Sigma^T{}^\mu_\mu \right)^2
\right]
\right.\nonumber\\
&& \left. \hspace{2.5cm}
+\frac{(\ldots)}{q^2} \tilde \tau_{\mu\nu} \tilde \tau^{\mu\nu} 
+ \frac{(\ldots)}{q^4}\kappa_{\mu} \kappa^\mu
+ \frac{(\ldots)}{q^4}\upsilon_{\mu} \upsilon^\mu
+ \frac{(\ldots)}{q^6} J \cdot M \cdot J
\right\} \,,
\eea
where $J = (j_1,j_2)$ .
The first term gives the GR contribution, the second one another massless spin $2^+$, the third is a massless $1^+$ state,
the remaining ones are two spin $1^-$ dipole ghosts
and two $0^+$ tripole ghosts.
The last four are pathological and must be eliminated.
This can be achieved by adjusting the coefficients
so that the various terms $(\ldots)$ diverge
(this is equivalent to setting to zero some terms in the $a$-matrices).
In the process new gauge invariances appear.

The dipole ghost $v_\mu$ coming from $\Omega$, 
can be eliminated imposing (\ref{relation1}) and
\be
a^{TQ}_1 + a^{TQ}_3 =0 \,.
\ee
In this way the following gauge invariance appears
\begin{subequations}
\begin{align}
& h_{\mu\nu} \to h_{\mu\nu}  \,,
\hspace{2cm} H_{\mu\nu} \to H_{\mu\nu}  \,,
\\
& \Omega_{\mu\nu} \to \Omega_{\mu\nu} 
+ \partial_\mu \chi_\nu - \partial_\nu \chi_\mu \,.
\end{align}
\end{subequations}

Instead to eliminate $\kappa_{\mu}$ and $J$ we impose 
the constraints (\ref{relations3}), and
\be
a^{TQ}_1 - a^{TQ}_2 =0 \,.
\label{relation4}
\ee
This amounts to imposing separate ``Diff-invariance'' 
on $h$ and $H$, i.e.
\begin{subequations}
\begin{align}
& h_{\mu\nu} \to h_{\mu\nu} + \partial_\mu \xi_\nu + \partial_\nu \xi_\mu \,,
\\
& H_{\mu\nu} \to H_{\mu\nu} + \partial_\mu \Xi_\nu + \partial_\nu \Xi_\mu \,,
\\
& \Omega_{\mu\nu} \to \Omega_{\mu\nu}  \,.
\end{align}
\end{subequations}
Using these constraints, we find a well defined theory containing two massless particles with spin $2^+$ and one with spin $1^+$ with three different gauge invariances.
In such a theory the graviton is a combination of $h$ and $H$. 
Then if we want to decouple $h$ from $H$, we have to impose
\be
4a^{QQ}_1 + a^{TQ}_1 =0 \,.
\label{relation5}
\ee
At this point if we want to have a single massless graviton
we have to kill the (non-pathological) degrees of freedom 
$1^+$ and $2^+$.
From the coefficient matrices, this is achieved by
imposing (\ref{relation2}) and
\be
2a^{TT}_1+a^{TT}_2+4a^{QQ}_1+2a^{TQ}_1=0\ .
\label{relation6}
\ee
Imposing relations (\ref{relation1},\ref{relation2},
\ref{relations3},\ref{relation4},\ref{relation5},\ref{relation6}),
the unique solution is the choice
$a^{TT}_1=-\frac{1}{4}m^R$,
$a^{TT}_2= -\frac{1}{2}m^R$,
$a^{TT}_3=m^R$,
$a^{QQ}_1=-\frac{1}{4} m^R$,
$a^{QQ}_2 +a^{QQ}_4=\frac{1}{2} m^R$,
$a^{QQ}_3=\frac{1}{4}m^R$,
$a^{QQ}_5=-\frac{1}{2}m^R$,
$a^{TQ}_1=m^R$,
$a^{TQ}_2=m^R$,
$a^{TQ}_3=-m^R$,
which reproduce the general teleparallel equivalent of GR (\ref{GTEGR}).
This analysis agrees with the findings of \cite{BeltranJimenez:2019odq}.

\section{MAGs without propagation}

There are classes of MAGs that look perfectly normal when
presented in the Cartan form, but have no propagating degrees
of freedom.
\footnote{This observation came up in discussions with E. Sezgin.}
The initial step towards these theories is the observation
that known ghost- and tachyon-free MAGs,
when presented in Einstein form,
do not contain terms quadratic in curvature \cite{Percacci:2020ddy}.
This is reasonable, insofar as 4DG is known to contain ghosts
or tachyons.

However, we can now demand more:
in the notation of equation (\ref{lagE}),
suppose that $m^R=0$, $b^{RR}=0$ and $b^{R\phi}=0$.
This means that the Hilbert term is absent,
as well as the terms quadratic in curvature
and mixed terms of the form $R\nabla\phi$.
The first two lines of the Lagrangian can therefore
be written in the form
\be
\phi_{\alpha\beta\gamma}
\left(
K^{\alpha\beta\gamma|\rho\sigma|\lambda\mu\nu}
\nabla_\rho\nabla_\sigma
+\cM^{\alpha\beta\gamma|\lambda\mu\nu}\right)
\phi_{\lambda\mu\nu}\ ,
\label{nonprop}
\ee
where $K$ and $\cM$ are tensors constructed exclusively with the
metric.
The remaining terms do not contribute to the
propagator in flat space, but only to interactions.
For simplicity we shall ignore them in the subsequent
discussion, but they do not change the conclusions.
When the Lagrangian is linearized, it gives a kinetic
operator of the form (\ref{nonprop}),
where all the metrics are Minkowski metrics and
all covariant derivatives are replaced by partial derivatives.

If we were just considering this as a theory of a field $\phi$
propagating in a fixed background metric,
it would have, in general, propagating degrees of freedom
obeying the field equation
\be
\left(
K^{\alpha\beta\gamma|\rho\sigma|\lambda\mu\nu}
\nabla_\rho\nabla_\sigma
+\cM^{\alpha\beta\gamma|\lambda\mu\nu}\right)
\phi_{\lambda\mu\nu}\ .
\ee
However, in a MAG we have to satisfy also the equation for the
metric, which in the basence of matter simply says that the
energy-momentum tensor of $\phi$ has to vanish.
Since plane waves carry nonzero energy and momentum,
it is already clear that this will forbid normal propagation.
To see this more explicitly, write the Lagrangian as
\be
\phi_{\alpha\beta\gamma}
\cO^{\alpha\beta\gamma|\lambda\mu\nu}
\phi_{\lambda\mu\nu}\ .
\ee
In flat space one can Fourier transform and write
$$
\cO^{\alpha\beta\gamma|\lambda\mu\nu}=
-K^{\alpha\beta\gamma|\rho\sigma|\lambda\mu\nu}
q_\rho q_\sigma
+\cM^{\alpha\beta\gamma|\lambda\mu\nu}\ .
$$
The energy-momentum tensor is
\be
T^{\rho\sigma}
=\frac{2}{\sqrt{-g}}\phi_{\alpha\beta\gamma} 
\frac{\partial (\sqrt{-g}\, \cO^{\alpha\beta\gamma|\lambda\mu\nu})}{\partial g_{\rho\sigma}}
\, \phi_{\lambda\mu\nu} \,.
\ee
The operator $\cO$ has zero modes corresponding to infinitesimal
coordinate transformations, but generically there will
be no others.
When this is the case, demanding $T^{\rho\sigma}=0$
implies that $\phi$ can be at most a coordinate transform of zero.

Let us observe that while the absence of terms containing
the curvature $R_{\alpha\beta\gamma\delta}$ (and its contractions)
is immediately conspicuous in the Einstein form,
it is not in the Cartan form.
We can now ask, in the Cartan form of MAG, 
what choices of coefficients will produce a theory of this type.
From (\ref{themapRR}) we see that the vanishing of the
$R^2$ terms implies
\bea
c^{FF}_1 -c^{FF}_2 + c^{FF}_3 + 1/2(c^{FF}_4 - c^{FF}_5 + c^{FF}_6) &=& 0  \,,
\nn
c^{FF}_7  + c^{FF}_8 + c^{FF}_9 + c^{FF}_{10} - c^{FF}_{11} - c^{FF}_{12} &=& 0 \,,
\nn
c^{FF}_{16} &=& 0  \,,
\nonumber
\eea
and from (\ref{themapRTQ}), the vanishing of the terms $R\nabla(T/Q)$ implies
\bea
2(4 c^{FF}_1\!-4c^{FF}_2\!+4c^{FF}_3\!+2c^{FF}_4\!-2c^{FF}_5\!
+2c^{FF}_6\!+c^{FF}_7\!+c^{FF}_8\!+c^{FF}_9\!
+c^{FF}_{10}\!-c^{FF}_{11}\!-c^{FF}_{12})&=&0\ ,
\nonumber\\
- c^{FF}_7 - c^{FF}_8 - c^{FF}_9-c^{FF}_{10} + c^{FF}_{11} + c^{FF}_{12} + 4 c^{FF}_{16} &=&0\ ,
\nonumber\\
2(-2c^{FF}_1\!+2c^{FF}_2\!-2c^{FF}_3\!-c^{FF}_4\!+c^{FF}_5\!-c^{FF}_6\!
+c^{FF}_7\!+c^{FF}_8)\!-c^{FF}_{11}\! - c^{FF}_{12}&=&0\ ,
\nonumber\\
- 3 c^{FF}_7 - 3 c^{FF}_8 - c^{FF}_9
-c^{FF}_{10} + 2 c^{FF}_{11} + 2 c^{FF}_{12}&=&0\ ,
\nonumber\\
c^{FF}_9+c^{FF}_{10} - c^{FF}_{11}/2 - c^{FF}_{12}/2 - 2 c^{FF}_{16}
&=&0\ ,
\nonumber\\
1/2(- c^{FF}_7 - c^{FF}_8 - c^{FF}_9
-c^{FF}_{10} + c^{FF}_{11} + c^{FF}_{12} + 4 c^{FF}_{16} )&=&0\ .
\nonumber
\eea

Furthermore, it is also important to notice that this phenomenon
will not be apparent in the linearized form of the theory:
the energy-momentum tensor is quadratic in $\phi$ and
the linearized EOM for the metric on a flat background will just be $0=0$.
Instead, the linearized theory will contain some accidental symmetry.

Probably the simplest and most illuminating example
is the action where we retain only $c^{FF}_{13}=2$,
all the others being zero:
$$
\cL=-F^{(34)}_{\mu\nu}F^{(34)\mu\nu}\ .
$$
Using (\ref{FtoR}), 
$$
F^{(34)}_{\mu\nu}
=\frac12\left(\nabla_\mu \trQ23_\nu-\nabla_\nu \trQ23_\mu\right)\ .
$$
Thus, in spite of appearances, this is just a free Maxwell field
coupled to a metric that does not have a kinetic term.
There is an EOM stating that the
electromagnetic energy-momentum tensor is zero,
which implies that $F_{\mu\nu}=0$.
On the other hand, if we study this theory with the methods of Section \ref{sec:lin},
we find that all coefficient matrices are zero
except for $a(1^-)$, that has rank one.
All the nonzero rows/columns are proportional to $q^2$
and choosing a gauge appropriately one would conclude that
the theory contains a free massless spin one particle.

A less trivial example is obtained by setting all coefficients to zero except $c^{FF}_2=c^{FF}_1=2$:
$$
\cL=-F_{\mu\nu(\rho\sigma)}F^{\mu\nu(\rho\sigma)} \,.
$$
In this case the linearized analysis seems to indicate
several propagating (and interacting) particles, but this conclusion
is false in the full nonlinear theory.

\section{DIY MAGs}

The spin-projector formalism has been used to look
for MAGs that are free of ghosts and tachyons
\cite{neville,Sezgin:1979zf,sezgin2,Lin:2018awc,Percacci:2020ddy,Marzo:2021esg,Marzo:2021iok}.
The general procedure has been to impose conditions on
the kinetic coefficients and see what kind of particles
the theory describes.
Here we would like to use a different approach:
to decide a priori what particles we want
and then construct a MAG that has the right propagator
for those particles.
This goes as follows: we know the correct forms of the
propagators for particles of any spin/parity.
These are listed in Appendix \ref{sec:app.propagators}.
At the linearized level, one can write down a kinetic term 
that gives the correct
propagators for the desired states, and nothing else.
Then one can turn this kinetic term into a full nonlinear
Lagrangian for a MAG in Einstein form by the simple procedure
of minimal coupling.
The Lagrangian obtained in this way is highly non-unique:
the order of the covariant derivatives is arbitrary
and all the cubic and quartic terms are absent.
Nevertheless, this is a MAG that has the desired propagators.
As a subsequent step one can try to add the cubic and
quartic terms, and, if necessary, adjust the ordering of the derivatives
at the cost of adding terms of the form $R\phi\phi$.
We note that this procedure will work if we remain in the
context of the general Lagrangians of Section \ref{sec:lags}.
This is because the general linearized kinetic term 
for MAG has 47 free parameters,
corresponding to the 47 independent terms of a general Lagrangian.
It would not work in general for the Lagrangians
that only have dimension-four terms of the form $F^2$,
that depend altogether on 28 free parameters.

In this section we will give two examples of this construction.
Being a three-index tensor, distortion can carry any
of the states listed in Table \ref{t2}.
From the point of view of particle physics,
it may seem redundant to use distortion to describe
a particle of spin $0^\pm$, $1^\pm$ or $2^+$,
because all these particles can be described by tensor
fields of lower rank.
The only states that do require a three index tensor
have spin $2^-$ and $3^-$.
We will therefore analyze here these two cases
at the linearized level.
We stress that the MAGs constructed in this way can only be
said to be consistent at the linearized level.
We do not make any claim as to their consistency
when interactions are turned on.

\subsection{Simple MAG with a \texorpdfstring{$2^-$}{2-minus} state}

We start from a general MAG.
A look at Table \ref{t2} shows that there are two possible
d.o.f.'s with spin $2^-$: $2^-_1$ being hook-symmetric
and $2^-_2$ being hook-antisymmetric
(recall that in this context we refer here to 
symmetry or antisymmetry in the last two indices).
The free Lagrangians for a spin/parity $2^-$ state
carried by an antisymmetric or symmetric tensor, 
and the corresponding propagators, are given in Appendix \ref{2minus}.
Here we show how to recover those linearized Lagrangians
from MAGs.

We will use the coefficient matrices for the theory 
in the Einstein form, for which the last row and column are
identically zero as a result of diffeomorphism invariance.
In order to remove the unwanted propagating dof's we impose
various conditions on the coefficient matrices.
Demanding that the matrices for spins $3^-$, $1^+$ and $0^-$
have no terms proportional to $q^2$ leads to the constraints:
\bea
b^{TT}_2 &=& b^{TT}_1\ ,\qquad\qquad
b^{TT}_5 = b^{TT}_1+b^{TT}_4, \nonumber\\
b^{TT}_6 &=& -b^{TT}_1\ ,\qquad\quad
b^{TT}_7 = -2b^{TT}_1\ , \nonumber\\
b^{QQ}_2 &=& -b^{QQ}_1\ ,\qquad\quad
b^{QQ}_8 = 3b^{QQ}_1 + b^{QQ}_7\ ,\nonumber\\ 
b^{TQ}_5 &=& -b^{TQ}_1\ ,\qquad\quad
b^{TQ}_7 = b^{TQ}_1 + b^{TQ}_6\ .
\eea

Next, in the sectors $2^+$ and $0^+$ we demand that the mixed $a$-$h$
terms vanish and that all the other terms, except for
those corresponding to the standard graviton, have no $q^2$ terms.
This leads to
\bea
b^{TT}_4 &=& -2b^{TT}_1+2b^{TT}_2+b^{TT}_7+2b^{QQ}_1+b^{QQ}_7+b^{TQ}_1+b^{TQ}_5, \nonumber\\
b^{TT}_6 &=& -b^{TT}_1-b^{QQ}_1-1/2b^{QQ}_7-1/2b^{TQ}_1-1/2b^{TQ}_5,\qquad
b^{TT}_9 = -b^{TT}_3, \nonumber\\
b^{QQ}_6 &=& 3b^{TT}_1-3/2b^{TT}_2+1/2b^{TT}_5+1/2b^{TT}_7
-2b^{QQ}_1 - b^{QQ}_2, \nonumber\\
b^{QQ}_{12} &=& -b^{QQ}_{10}, \quad
b^{QQ}_{13} = -b^{QQ}_{11}, \quad
b^{QQ}_{14} = -b^{QQ}_{4}\ , \quad
b^{QQ}_{15} = -b^{QQ}_3 \ , \quad
b^{QQ}_{16} = -b^{QQ}_5\ , \nonumber\\
b^{TQ}_4 &=& -6b^{TT}_1+2b^{TT}_2-2b^{TT}_5-b^{TT}_7+2b^{QQ}_1+2b^{QQ}_2-b^{TQ}_1,\nonumber\\
b^{TQ}_6 &=& -4b^{TT}_1+2b^{TT}_2-b^{TT}_7+4b^{QQ}_1+2b^{QQ}_2-b^{QQ}_9, \nonumber\\
b^{TQ}_7 &=& 2b^{TT}_2+b^{TT}_7+4b^{QQ}_1+2b^{QQ}_7+2b^{TQ}_1+b^{TQ}_5,  \nonumber\\
b^{TQ}_{11} &=& -b^{TQ}_{10} \qquad\quad
b^{TQ}_{12} = b^{TQ}_3\ ,\qquad\quad
b^{TQ}_{13} = b^{TQ}_2,
\eea
and further six relations for the $R\nabla T$ and $R\nabla Q$ that,
together withe Bianchi identities,
remove all the terms of this type.

Then we impose that the spin $2^-$ and $1^-$
are properly related, as discussed in Appendix \ref{2minus}. 
This leads to
\bea
b^{TT}_2 &=& 2 b^{TT}_1 + 1/3b^{TT}_3\ ,\qquad
b^{TT}_8 = -2b^{TT}_3\ , \nonumber\\
b^{QQ}_2 &=& -2 b^{QQ}_1 - b^{QQ}_3\ , \ \ \
b^{QQ}_4 = b^{QQ}_3\ , \nonumber\\
b^{QQ}_5 &=& b^{QQ}_{10}=-b^{QQ}_{11}=-2b^{QQ}_3\ , \nonumber\\
b^{TQ}_3 &=& 
-b^{TQ}_2=b^{TQ}_8=-b^{TQ}_9=b^{TQ}_{10}=
2b^{TT}_1 + 2/3 b^{TT}_3 + 2 b^{QQ}_1+2b^{QQ}_3+b^{TQ}_1\ .
\eea
The same requirement for the mass parameters implies
\bea
m^{TT}_3&=&-2m^{TT}_1-m^{TT}_2\ ,\qquad\qquad\,\,\,
m^{QQ}_4=2m^{QQ}_1-m^{QQ}_2+4m^{QQ}_3\ ,
\nonumber\\
m^{QQ}_5&=&4m^{QQ}_1-2m^{QQ}_2+4m^{QQ}_3\ ,
\quad
m^{TQ}_2\!=\!-m^{TQ}_3=m^{TQ}_1\ .
\eea
The coefficient matrices now depend only on $b^{TT}_1$, $b^{QQ}_1$, $b^{TQ}_1$
and on the mass parameters $m^{TT}_1$, $m^{TT}_2$, $m^{QQ}_1$, $m^{QQ}_2$, 
$m^{QQ}_3$, $m^{TQ}_1$.
In particular the sectors $2^+_{44}$ and $0^+_{55}$ 
have the right form to propagate a massless graviton.
Similarly the matrix for the $2^-$ and the submatrix
$1^-_{22}$, $1^-_{23}$, $1^-_{32}$, $1^-_{33}$
describe two mixed spin $2^-$ dof's.
All the remaining components are either zero on shell
(if the mass is nonzero) or a gauge dof (if the mass is zero).
In particular the matrix $a(2^-)$ is
\bea
a_{11}(2^-) &=&\frac{1}{2}\left(
3(3b^{TT}_1+4b^{QQ}_1+2b^{TQ}_1)(-q^2)
\right.\nonumber\\
&&\left.\quad
+(-6m^{TT}_1-3m^{TT}_2-8m^{QQ}_1+4m^{QQ}_2-6m^{TQ}_1)\right) \ , \nn
a_{12}(2^-) &=&\frac{\sqrt3}{2}
((3b^{TT}_1+b^{TQ}_1)(-q^2)-(2m^{TT}_1+m^{TT}_2+m^{TQ}_1)) \ ,\nn
a_{22}(2^-) &=&\frac{1}{2}\left(
3b^{TT}_1 (-q^2)+(-2m^{TT}_1-m^{TT}_2) \right) \ ,
\eea
and the submatrix
$1^-_{22}$, $1^-_{23}$, $1^-_{32}$, $1^-_{33}$ is the same up to the sign.
Then, there is the graviton contribution inside $a(2^+)_{44}$ and $a(0^+)_{55}$, with the correct proportionality discussed in Appendix \ref{2plus}.
Finally, except for the entries constraint by the diffeomorphism invariance, i.e. (\ref{fiamma}), all the other entries are just mass terms.

The two spin $2^-$ dof's are generically mixed.
The mixing can be eliminated by assuming 
$$
b^{TQ}_1=-3b^{TT}_1\qquad \mathrm{and}\qquad m^{TQ}_1=-2m^{TT}_1-m^{TT}_2\ .
$$
To avoid ghosts we must assume that 
$4b^{QQ}_1>3b^{TT}_1$ and $b^{TT}_1>0$.
In particular this condition can be satisfied 
by both dofs.

In order to propagate only the hook-antisymmetric component
$2^-_2$ 
, discussed in Section \ref{hookantisymtwominus}, we have to set
$$
b^{QQ}_1=3/4b^{TT}_1
$$
and then we must assume $b^{TT}_1>0$.
The mass squared term is proportional to $(2m^{TT}_1+m^{TT}_2)$.
In such a theory, the kinetic term for the state $2^-$ in the Lagrangian involves terms $\nabla T\nabla T$ , $\nabla Q\nabla Q$ and $\nabla T\nabla Q$.
\footnote
{This complication could be avoided by adopting another definition of hook (anti)symmetry.}

In order to propagate only the hook-symmetric component
$2^-_1$ 
, discussed in Section \ref{hooksymtwominus}, we have to set
$$
b^{TT}_1=0
$$
and assume $b^{QQ}_1>0$.
The mass squared term is proportional to $(6m^{TT}_1+3m^{TT}_2-8m^{QQ}_1+4m^{QQ}_2)$.
In such a theory, the kinetic term for the state $2^-$ in the Lagrangian involves only terms $\nabla Q\nabla Q$.

\subsection{Simple MAG with a \texorpdfstring{$3^-$}{3-minus} state}

We can remove all the terms proportional to $q^2$
in the coefficient matrices, except for those
that propagate the massless $3^-$ and $2^+$ d.o.f.'s.
This gives linear equations for the coefficients
that are solved by
\bea
&&
b^{TT}_i =  0 \quad\mathrm{for}\quad i=1,2,3,4,5,6,7,8,9
\nn
&&
b^{TQ}_i =0 \quad\mathrm{for}\quad i=1,2,3,4,5,6,7,8,9,10,11,12,13
\nn
&&
b^{QQ}_{2}  = 2 b^{QQ}_{1} \ ,  \ 
b^{QQ}_{3}  = -4 b^{QQ}_{1} \ , \ 
b^{QQ}_{4}  = - b^{QQ}_{1} \ ,  \ 
b^{QQ}_{5}  = -4 b^{QQ}_{1} \ , \ 
b^{QQ}_{6}  = -  b^{QQ}_{1} \ , \ 
\nn
&&
b^{QQ}_{7}  = -2 b^{QQ}_{1} \ , \ 
b^{QQ}_{8}  = -2   b^{QQ}_{1} \ , \ 
b^{QQ}_{9}  = -4 b^{QQ}_{1} \ , \ 
b^{QQ}_{10}  = 8   b^{QQ}_{1} \ , \ 
\nn
&&
b^{QQ}_{11}  = 
b^{QQ}_{12}  = 4 b^{QQ}_{1} \ , \ 
b^{QQ}_{13}  = 2 b^{QQ}_{1} \ , \ 
b^{QQ}_{15}  = 
b^{QQ}_{16} = 4 b^{QQ}_{14} \ , \
\eea
and further six relations for the $R\nabla T$ and $R\nabla Q$ that,
together withe Bianchi identities,
remove all the terms of this type.
This puts to zero the matrices $a(2^-)$, $a(1^+)$ and $a(0^-)$.
Further requiring that the ratio of 
the coefficients of $q^2$ in $a(3^-)$ and $a_{11}(0^+)$
be equal to $-9/2$,
as required by (\ref{Fronsdal2}),
fixes  $b^{QQ}_{14}=-1/2b^{QQ}_1$.
Then, the remaining coefficient matrices are
\begin{subequations}
\begin{align}
& a(3^-)= 12 b^{QQ}_1 (-q^2)\ ,
\\
&a(2^+) = (-q^2) \,\,\,
\left(
\begin{matrix}
0&0&0&0\\
0&0&0&0\\
0&0&0&0\\
0&0&0&\frac{m^R}{4}\\
\end{matrix}
\right)\,,
\\
&a(1^-) = (-q^2) \,\,\,
\left(
\begin{matrix}
-48b^{QQ}_1&0&0&0&0&0&0\\
0&0&0&0&0&0&0\\
0&0&0&0&0&0&0\\
0&0&0&0&0&0&0\\
0&0&0&0&0&0&0\\
0&0&0&0&0&0&0\\
0&0&0&0&0&0&0\\
\end{matrix}
\right)\,,
\\
& a(0^+) = (-q^2)\,\,\,
\left(
\begin{matrix}
-54b^{QQ}_1&0&0&-18b^{QQ}_1&0&0\\
0&0&0&0&0&0\\
0&0&0&0&0&0\\
-18b^{QQ}_1&0&0&-6b^{QQ}_1&0&0\\
0&0&0&0&-\frac12 m^R  &0\\
0&0&0&0&0&0\\
\end{matrix}
\right) \,.
\end{align}
\end{subequations}
Note that the first $4\times4$ block in $a(0^+)$
has rank one.
All the degrees of freedom are pure gauge, except
for the desired $2^+$ and $3^-$.
In such a theory, the kinetic term for the state $3^-$ in the Lagrangian involves only terms $\nabla Q\nabla Q$. 
We have chosen $b^{RR}_1=b^{RR}_2=b^{RR}_3=0$,
so the graviton propagator is as in GR.

Some comments are in order at this point.
The subject of higher spin theories is a thorny one.
Normally it is approached in a bottom-up fashion, starting from
a free theory in flat space and then trying to construct interactions.
In the process, one encounters numerous difficulties.
Here we have started from a ready-made nonlinear theory (MAG)
and tried to arrange its parameters so that at linearized level
it reproduces the known free spin-3 Lagrangian.
With our choice of coefficients,
the $\phi\phi$ part of the linearized action (\ref{linE}) is
\bea
S^{(2)} &=& -2b^{QQ}_1\int \frac{d^4q}{(2\pi)^4} \bigg[
q^2\Big(\frac14 Q_{\alpha\beta\gamma}Q^{\alpha\beta\gamma}
+\frac12 Q_{\alpha\beta\gamma}Q^{\beta\alpha\gamma}
\nonumber\\
&&
- \trQ12_\alpha\trQ12^\alpha
- \trQ12_\alpha\trQ23^\alpha
-\frac14 \trQ23_\alpha\trQ23^\alpha\Big)
\nonumber\\
&&
-\frac14 \divQ1_{\alpha\beta}\divQ1^{\alpha\beta}
- \divQ1_{\alpha\beta}\divQ2^{\alpha\beta}
-\divQ2_{(\alpha\beta)}\divQ2^{(\alpha\beta)}
\nonumber\\
&&
-\frac18 \trdivQ1_{\alpha\beta}\trdivQ1^{\alpha\beta}
-\frac12 \trdivQ1_{\alpha\beta}\trdivQ2^{\alpha\beta}
-\frac12 \trdivQ2_{\alpha\beta}\trdivQ2^{\alpha\beta}
\nonumber\\
&&
+ \divdivQ12_{\alpha}\trQ23^{\alpha}
+ \frac12\divdivQ23_{\alpha}\trQ23^{\alpha}
+ 2\divdivQ12_{\alpha}\trQ12^{\alpha}
+ \divdivQ23_{\alpha}\trQ12^{\alpha}
\bigg]\ ,
\nonumber
\eea
where now $\divQ1_{\alpha\beta}=iq^\lambda Q_{\lambda\alpha\beta}$ etc..
The standard description of the spin-3 particle is by a totally symmetric 3-tensor.
Thus in the formula above we replace $Q_{\alpha\beta\gamma}$ by 
$S_{\alpha\beta\gamma}=Q_{(\alpha\beta\gamma)}$, and set $b^{QQ}_1=1/3$
to obtain
\bea
S^{(2)} &=& \frac{1}{2}\int \frac{d^4q}{(2\pi)^4} \bigg[
-q^2S_{\alpha\beta\gamma}S^{\alpha\beta\gamma}
+3q^2\mathrm{tr}_{(12)}S_\alpha \mathrm{tr}_{(12)}S^\alpha
\nonumber\\
&&
+3\mathrm{div}_{(1)}S_{\alpha\beta}\mathrm{div}_{(1)}S^{\alpha\beta}
+\frac32 (\mathrm{trdiv}_{(1)}S)^2
-6\mathrm{div}_{(12)}S_\alpha \mathrm{tr}_{(12)}S^\alpha
\bigg] \,.
\eea
This is indeed the Fronsdal Lagrangian that correctly describes
a free massless spin-3 particle \cite{Fronsdal:1978rb}.
However, this is only a very limited success.
The ``higher spin symmetry''
$\delta S_{\alpha\beta\gamma}=\partial_{(\alpha}\Lambda_{\beta\gamma)}$,
that is a necessary invariance of a higher spin theory,
is only an accidental symmetry here.
More details on these issues,
the relation of this approach to earlier attempts
to embed higher-spin theory in MAG \cite{Baekler:2006vw,Marzo:2021esg}
and a discussion of the massive case will be given elsewhere.

\section{Conclusions}

Leaving aside the cosmological term,
and the possibility that distortion may contain a massless state,
the dynamics of MAGs at very low energies
(by which we mean energies below all the masses
that are present in the theory)
is dominated by the 12 dimension-two terms.
These comprise the Palatini term and terms quadratic in
distortion (or equivalently in torsion and nonmetricity).
In this regime the theory behaves like simple Palatini theory:
the equations of motion generically
imply that the connection has to be equal to the LC connection.
Thus, unless the distortion contains some massless state,
at sufficiently low energy the EFT of MAG becomes indistinguishable
form the EFT of metric theory of gravity.
If the masses of the distortion
(or equivalently of torsion and nonmetricity)
are much lower that the Planck mass, that we assume
to be the UV cutoff for this EFT,
there will be a regime where distortion could propagate.
For this one has to consider also the dimension-four terms,
of which, in a general MAG, there are over 900.

Already listing bases of independent terms requires
considerable work.
We have restricted our attention mainly to the terms 
of dimension 4 that
are quadratic in $(R,T,Q)$ (in the Einstein form)
or $(F,T,Q)$ (in the Cartan form).
These are the only terms that contribute to the propagator
in flat space.
We found that there are 47 independent invariants,
that have to be picked among 53 invariants in the Einstein form
of the theory and 99 invariants in the Cartan form.
Listing the independent terms in the Lagrangian implies a choice of basis and we have given two examples of such bases,
one containing all terms quadratic in $(T,Q)$, plus more,
and one containing all terms quadratic in $F$, plus more.

Even understanding the free propagator in a theory with so many
parameters is a very complex task.
One of our main results was the calculation of the matrices
of kinetic coefficient for the general MAG,
both in Einstein and Cartan form,
which are given in Appendix \ref{sec:app.coefficients}.
These coefficients enter in the description of the 
kinetic term for the individual spin/parity degrees of freedom
and are essential to understand the propagating
degrees of freedom.
In particular, it is important to understand the subspace of MAGs
that do not contain ghosts and tachyons.

We have then advocated a constructive method to arrive at
ghost- and tachyon-free Lagrangians.
We have listed in Appendix \ref{sec:app.propagators}
the standard forms of the propagators of particles of spin up to three
(including spin 2 with odd parity, which we could not find in
the literature and constructed with the aid of the spin projectors).
We then imposed constraints on the matrices of kinetic coefficient
requiring that they reproduce the propagators of the desired states. 
We have applied this method to construct ghost- and tachyon-free MAGs
with extra particles of spin/parity $2^-$ and $3^-$.
Much more work is possible in this direction to explore
the landscape of ghost- and tachyon-free MAGs.
We repeat that this construction is not unique and
only ensures consistency at the level of the free theory.
We have said nothing about the inclusion of terms
cubic and quartic in distortion.
In particular, it will be interesting to look with more care
at the spin-3 case where the well-known difficulties of 
higher spin theories are expected to appear.

The other main line of research that will have to be pursued in the future is the calculation of quantum properties of MAGs.
In particular, the possibility of a UV consistent quantum MAG 
will transcend the domain of EFT and rest
either on nonstandard perturbative arguments,
or on non-perturbative effects,
of the type that are already under consideration in 4DG,
and that are likely to invalidate the results of the
tree-level analysis.

\bigskip

\leftline{\bf Acknowledgements}
\noindent
We thank E. Sezgin, P. Novichkov, K. Falls for discussion
and E. Mielke, F. Hehl and J. Donoghue for useful correspondence.
This work would not have been possible without
extensive use of the free software packages
{\tt xAct}, {\tt xTensor}, {\tt xPert}, {\tt xTras}.

\vskip2cm
\goodbreak

\break

\begin{appendix}

\section{Linear relations}
\label{sec:app.lin}

\subsection{General MAG in Einstein form: Relation between couplings for \texorpdfstring{$\phi$}{phi} and \texorpdfstring{$TQ$}{TQ} variables}
\label{sec:app.lin.1}

Concerning the dimension-two terms,
the relation between the couplings in the Einstein form
and those in the Cartan form has already been given in 
(\ref{mTQtophi}).
Given the Lagrangian in the form (\ref{genPalE}),
it can be rewritten in the form (\ref{genPalETQ}),
where the couplings are related as follows:
\bea
m^{TT}_1&=&1/4(3 m^{\phi\phi}_1- 3 m^{\phi\phi}_2 + m^{\phi\phi}_3 + m^{\phi\phi}_4- m^{\phi\phi}_5 )\ ,
\nonumber\\
m^{TT}_2&=&1/2 (m^{\phi\phi}_1 - m^{\phi\phi}_2 + m^{\phi\phi}_3 + m^{\phi\phi}_4 - m^{\phi\phi}_5  )\ ,
\nonumber\\
m^{TT}_3&=&m^{\phi\phi}_6 + m^{\phi\phi}_7 - m^{\phi\phi}_9  \ ,
\nonumber\\ 
m^{QQ}_1&=&1/4(3 m^{\phi\phi}_1 - m^{\phi\phi}_2 + 3 m^{\phi\phi}_3 - m^{\phi\phi}_4 - m^{\phi\phi}_5  )\ ,
\nonumber\\ 
m^{QQ}_2&=&1/2(-m^{\phi\phi}_1 + m^{\phi\phi}_2 - m^{\phi\phi}_3 + m^{\phi\phi}_4 + m^{\phi\phi}_5  )\ ,
\nonumber\\ 
m^{QQ}_3&=&1/4(m^{\phi\phi}_6 + m^{\phi\phi}_7 + m^{\phi\phi}_8 - m^{\phi\phi}_9+m^{\phi\phi}_{10} - m^{\phi\phi}_{11}  )\ ,
\nonumber\\ 
m^{QQ}_4&=&m^{\phi\phi}_7\ ,
\nonumber\\ 
m^{QQ}_5&=&1/2(-2 m^{\phi\phi}_7+m^{\phi\phi}_9+m^{\phi\phi}_{11} )\ ,
\nonumber\\
m^{TQ}_1&=&-2 m^{\phi\phi}_1 + 2 m^{\phi\phi}_2 - 2 m^{\phi\phi}_3 + m^{\phi\phi}_5  \ ,
\nonumber\\ 
m^{TQ}_2&=&1/2(2 m^{\phi\phi}_6 + 2 m^{\phi\phi}_7 - 2 m^{\phi\phi}_9
+m^{\phi\phi}_{10}-m^{\phi\phi}_{11} )\ ,
\nonumber\\ 
m^{TQ}_3&=&-2 m^{\phi\phi}_7 + m^{\phi\phi}_9  \ .
\label{mphitoTQ}
\eea

Similarly the dimension-four terms are related as follows:
\footnote{We are choosing the basis \eqref{eingenbasis} and basis \eqref{eingenTQbasis}}:

\bea
b^{RT}_{3}&=& b^{R\phi}_{8}- b^{R\phi}_{9}\ , \qquad\qquad
b^{RT}_{5}= b^{R\phi}_{11}- b^{R\phi}_{12}\ , \nn 
b^{RQ}_{4}&=& 1/2( b^{R\phi}_{7}- b^{R\phi}_{8}+ b^{R\phi}_{9} )\ , 
\qquad 
b^{RQ}_{5}=  b^{R\phi}_{8}\ , \nn 
b^{RQ}_{6}&=& 1/2 (  b^{R\phi}_{10}-  b^{R\phi}_{11}+  b^{R\phi}_{12})\ ,
\qquad 
b^{RQ}_{7}=  b^{R\phi}_{11} \ ,
\eea

\bea
b^{TT}_1&=&1/4 (3 b^{\phi\phi}_1 - 3 b^{\phi\phi}_2 + b^{\phi\phi}_3 + b^{\phi\phi}_4 - b^{\phi\phi}_5)
\ , \nn
b^{TT}_2&=&1/2 (b^{\phi\phi}_1 - b^{\phi\phi}_2 + b^{\phi\phi}_3 + b^{\phi\phi}_4 - b^{\phi\phi}_5)
\ , \nn
b^{TT}_3&=& b^{\phi\phi}_6 + b^{\phi\phi}_7 - b^{\phi\phi}_9
\ , \nn
b^{TT}_4&=&1/2 (b^{\phi\phi}_{12} - b^{\phi\phi}_{13} + b^{\phi\phi}_{14} + b^{\phi\phi}_{15} + b^{\phi\phi}_{16} + b^{\phi\phi}_{17} - b^{\phi\phi}_{22} - b^{\phi\phi}_{23})
\ , \nn
b^{TT}_5&=& 
 1/2 (-b^{\phi\phi}_{12} + b^{\phi\phi}_{13} + b^{\phi\phi}_{14} + b^{\phi\phi}_{15} + b^{\phi\phi}_{16} + b^{\phi\phi}_{17} - b^{\phi\phi}_{22} - b^{\phi\phi}_{23})
\ , \nn
b^{TT}_6&=& 
 1/4 (b^{\phi\phi}_{12} - b^{\phi\phi}_{13} + b^{\phi\phi}_{14} - b^{\phi\phi}_{15} + b^{\phi\phi}_{16} - b^{\phi\phi}_{17} + b^{\phi\phi}_{18} - b^{\phi\phi}_{19} - b^{\phi\phi}_{20} + b^{\phi\phi}_{21} - b^{\phi\phi}_{22} + b^{\phi\phi}_{23})
\ , \nn
b^{TT}_7&=& 1/2 (2 b^{\phi\phi}_{12} - 2 b^{\phi\phi}_{13} + b^{\phi\phi}_{18} - b^{\phi\phi}_{19} - b^{\phi\phi}_{20} + b^{\phi\phi}_{21})
\ , \nn
b^{TT}_8&=&
 b^{\phi\phi}_{24} - b^{\phi\phi}_{25} - b^{\phi\phi}_{27} + b^{\phi\phi}_{28}
 \ , \nn
b^{TT}_9&=&b^{\phi\phi}_{34} + b^{\phi\phi}_{35} - b^{\phi\phi}_{38} \ ,
\eea
\bea
b^{QQ}_{1}&=& 1/4 (3 b^{\phi\phi}_1 - b^{\phi\phi}_2 + 3 b^{\phi\phi}_3 - b^{\phi\phi}_4 - b^{\phi\phi}_5)
\ , \nn
b^{QQ}_{2}&=&1/2 (-b^{\phi\phi}_1 + b^{\phi\phi}_2 - b^{\phi\phi}_3 + b^{\phi\phi}_4 + b^{\phi\phi}_5)
\ , \qquad\qquad
b^{QQ}_{3}=b^{\phi\phi}_7
\ , \nn
b^{QQ}_{4}&=& 1/4 (b^{\phi\phi}_6 + b^{\phi\phi}_7 + b^{\phi\phi}_8 - b^{\phi\phi}_9
+b^{\phi\phi}_{10}-b^{\phi\phi}_{11})
\ , \quad\quad
b^{QQ}_5= 1/2 ( - 2 b^{\phi\phi}_7 + b^{\phi\phi}_9+b^{\phi\phi}_{11})
\ , \nn
b^{QQ}_6&=& 1/4 (b^{\phi\phi}_{12} + b^{\phi\phi}_{13} + b^{\phi\phi}_{14} + b^{\phi\phi}_{15} + b^{\phi\phi}_{16} + b^{\phi\phi}_{17} - b^{\phi\phi}_{18} - b^{\phi\phi}_{19} + b^{\phi\phi}_{20} + b^{\phi\phi}_{21} - b^{\phi\phi}_{22} - b^{\phi\phi}_{23})
\ , \nn
b^{QQ}_7&=& 1/2 (b^{\phi\phi}_{12} - b^{\phi\phi}_{13} + b^{\phi\phi}_{14} + b^{\phi\phi}_{15} + b^{\phi\phi}_{16} - b^{\phi\phi}_{17} - b^{\phi\phi}_{20} + b^{\phi\phi}_{21})
\ , \nn
b^{QQ}_8&=&  1/2 (-b^{\phi\phi}_{12} + b^{\phi\phi}_{13} + b^{\phi\phi}_{14} + b^{\phi\phi}_{15} - b^{\phi\phi}_{16} + b^{\phi\phi}_{17} + b^{\phi\phi}_{20} - b^{\phi\phi}_{21})
\ , \nn
b^{QQ}_9&=&  
 1/2 (-2 b^{\phi\phi}_{14} - 2 b^{\phi\phi}_{15} + b^{\phi\phi}_{18} + b^{\phi\phi}_{19} + b^{\phi\phi}_{22} + b^{\phi\phi}_{23})
 \ , \nn
b^{QQ}_{10}&=&  b^{\phi\phi}_{28}
\ , \qquad
b^{QQ}_{11}= 1/2 (b^{\phi\phi}_{27} - b^{\phi\phi}_{28} + b^{\phi\phi}_{29})
\ , \quad
b^{QQ}_{12}= 1/2 (b^{\phi\phi}_{25} - b^{\phi\phi}_{28} + b^{\phi\phi}_{31})
\ , \nn
b^{QQ}_{13}&=&  1/4 (b^{\phi\phi}_{24} - b^{\phi\phi}_{25} + b^{\phi\phi}_{26} - b^{\phi\phi}_{27} + b^{\phi\phi}_{28} - b^{\phi\phi}_{29} + b^{\phi\phi}_{30} - b^{\phi\phi}_{31} + b^{\phi\phi}_{32})
\ , \nn
b^{QQ}_{14}&=& 1/4 (b^{\phi\phi}_{33} + b^{\phi\phi}_{34} + b^{\phi\phi}_{35} - b^{\phi\phi}_{36} + b^{\phi\phi}_{37} - b^{\phi\phi}_{38})
\ , \nn
b^{QQ}_{15}&=& b^{\phi\phi}_{34}
\ , \qquad\quad
b^{QQ}_{16}= 1/2 (-2 b^{\phi\phi}_{34} + b^{\phi\phi}_{36} + b^{\phi\phi}_{38}) \ ,
\eea
\bea
b^{TQ}_1&=& -2 b^{\phi\phi}_1 + 2 b^{\phi\phi}_2 - 2 b^{\phi\phi}_3 + b^{\phi\phi}_5
\ , \qquad
b^{TQ}_2= -2 b^{\phi\phi}_7 + b^{\phi\phi}_9
\ , \nn
b^{TQ}_3&=& b^{\phi\phi}_6 + b^{\phi\phi}_7 - b^{\phi\phi}_9
+ (b^{\phi\phi}_{10}-b^{\phi\phi}_{11})/2
\ , \nn
b^{TQ}_4&=& 1/2 (-2 b^{\phi\phi}_{14} - 2 b^{\phi\phi}_{15} - 2 b^{\phi\phi}_{16} - 2 b^{\phi\phi}_{17} + b^{\phi\phi}_{18} + b^{\phi\phi}_{19} - b^{\phi\phi}_{20} - b^{\phi\phi}_{21} + 
2 b^{\phi\phi}_{22} + 2 b^{\phi\phi}_{23})
\ , \nn
b^{TQ}_5&=& 1/2 (-2 b^{\phi\phi}_{12} + 2 b^{\phi\phi}_{13} - 2 b^{\phi\phi}_{16} + 2 b^{\phi\phi}_{17} - b^{\phi\phi}_{18} + b^{\phi\phi}_{19} + 2 b^{\phi\phi}_{20} - 2 b^{\phi\phi}_{21} + 
    b^{\phi\phi}_{22} - b^{\phi\phi}_{23})
    \ , \nn
b^{TQ}_6&=&  
 1/2 (-2 b^{\phi\phi}_{12} + 2 b^{\phi\phi}_{13} + 2 b^{\phi\phi}_{14} + 2 b^{\phi\phi}_{15} + b^{\phi\phi}_{20} - b^{\phi\phi}_{21} - b^{\phi\phi}_{22} - b^{\phi\phi}_{23})
 \ , \nn
b^{TQ}_7&=&  
 1/2 (2 b^{\phi\phi}_{12} - 2 b^{\phi\phi}_{13} + 2 b^{\phi\phi}_{14} + 2 b^{\phi\phi}_{15} - b^{\phi\phi}_{20} + b^{\phi\phi}_{21} - b^{\phi\phi}_{22} - b^{\phi\phi}_{23})
 \ , \nn
b^{TQ}_8&=& b^{\phi\phi}_{25} - b^{\phi\phi}_{28}
\ , \qquad
b^{TQ}_9= 1/2 (b^{\phi\phi}_{24} - b^{\phi\phi}_{25} + b^{\phi\phi}_{26} - b^{\phi\phi}_{27} + b^{\phi\phi}_{28} - b^{\phi\phi}_{29})
\ , \nn
b^{TQ}_{10}&=& b^{\phi\phi}_{27} - b^{\phi\phi}_{28}
 \ , \qquad
b^{TQ}_{11}=  
 1/2 (b^{\phi\phi}_{24} - b^{\phi\phi}_{25} - b^{\phi\phi}_{27} + b^{\phi\phi}_{28} + b^{\phi\phi}_{30} - b^{\phi\phi}_{31})
 \ , \nn
b^{TQ}_{12}&=&  
 1/2 (-2 b^{\phi\phi}_{34} - 2 b^{\phi\phi}_{35} + b^{\phi\phi}_{36} - b^{\phi\phi}_{37} + 2b^{\phi\phi}_{38})
\ , \qquad
b^{TQ}_{13}=2 b^{\phi\phi}_{34} - b^{\phi\phi}_{38} \ .
\eea

\subsection{General MAG in Cartan form: additional relations}
\label{sec:app.lin.2}

In addition to the relations coming from the Bianchi identities,
there are many more that can be obtained using the same procedure
as in the case of symmetric MAG (Section \ref{SMAGcartan}).
As in that section, these relations hold up to interaction terms.
Eliminating $R^2$ from the $F^2$ terms we find
the following ten relations:
\bea
L^{FF}_1 +L^{FF}_2&=&L^{QQ}_1-L^{QQ}_6\ ,\nonumber\\
L^{FF}_1-L^{FF}_3&=&3/2L^{QQ}_1-L^{QQ}_2-3/2L^{QQ}_6-L^{QQ}_7
+2L^{QQ}_9-4L^{TQ}_1+4L^{TQ}_4-4L^{TQ}_7
\nonumber\\
&&+3/2L^{TT}_1+L^{TT}_2-3L^{TT}_4-L^{TT}_5+1/2L^{TT}_6-2L^{TT}_7\ ,
\nonumber\\
L^{FF}_4+L^{FF}_5&=&-1/2L^{QQ}_1+L^{QQ}_2+1/2L^{QQ}_6+L^{QQ}_7
-2L^{QQ}_9+L^{TQ}_1-L^{TQ}_4+L^{TQ}_7\ ,\nonumber\\
L^{FF}_4 -L^{FF}_6&=&-L^{QQ}_1+L^{QQ}_2+L^{QQ}_6+L^{QQ}_7-2L^{QQ}_9
+2L^{TQ}_1-2L^{TQ}_4+2L^{TQ}_7\ ,
\nonumber\\
L^{FF}_7 -L^{FF}_8&=&1/4(L^{QQ}_4-L^{QQ}_{14})
+L^{TQ}_3-L^{TQ}_9+L^{TQ}_{12} 
+L^{TT}_3+1/2L^{TT}_6-2L^{TT}_8-L^{TT}_9\ ,
\nonumber
\eea
\bea
L^{FF}_{10}-L^{FF}_9&=&
-L^{QQ}_3 -1/4 L^{QQ}_4 +L^{QQ}_5 -L^{QQ}_7 +L^{QQ}_8 
+2L^{QQ}_{10}-L^{QQ}_{11}-2L^{QQ}_{12}+L^{QQ}_{13}  \nonumber\\&&
+ 1/4 L^{QQ}_{14}+L^{QQ}_{15} -L^{QQ}_{16}
+2L^{TQ}_2 -L^{TQ}_3 +2L^{TQ}_5 -2L^{TQ}_8 +L^{TQ}_9\nonumber\\&&
-2L^{TQ}_{10}+2L^{TQ}_{11}-L^{TQ}_{12} +2L^{TQ}_{13}
-L^{TT}_3 - 1/2L^{TT}_6 +2L^{TT}_8 +L^{TT}_9\ ,
\nonumber\\
L^{FF}_{12}-L^{FF}_{11}&=& 1/4L^{QQ}_4 - 1/2 L^{QQ}_5+1/2L^{QQ}_{11} -1/2 L^{QQ}_{13} 
- 1/4L^{QQ}_{14} +1/2 L^{QQ}_{16}\nonumber\\&&
 -L^{TQ}_2 
+L^{TQ}_3 -L^{TQ}_5+L^{TQ}_8 -L^{TQ}_9 + L^{TQ}_{10} -L^{TQ}_{11} +L^{TQ}_{12} -L^{TQ}_{13}\nonumber\\&& 
+L^{TT}_3 +1/2 L^{TT}_6 -2L^{TT}_8 -L^{TT}_9\ ,
\nonumber\\
L^{FF}_{13}&=&1/2(L^{QQ}_4-L^{QQ}_{14})\ ,\nonumber\\
L^{FF}_{14}&=&-1/4(L^{QQ}_4-L^{QQ}_{14})-1/2(L^{TQ}_3-L^{TQ}_9+L^{TQ}_{12})
\ ,
\nonumber\\
L^{FF}_{15}&=&1/4L^{QQ}_4-1/2L^{QQ}_5+1/2L^{QQ}_{11}-1/2L^{QQ}_{13}
-1/4L^{QQ}_{14}+1/2L^{QQ}_{16} \ ,\nonumber\\&&
+1/2(L^{TQ}_3-L^{TQ}_9+L^{TQ}_{12})\ .
\label{relFF}
\eea
Eliminating $R\nabla T$ from the $FDT$ terms we find
the following 15 relations:
\bea
L^{FT}_1 +L^{FT}_2&=&L^{TQ}_1-L^{TQ}_4\ ,
\nonumber\\
L^{FT}_1 -L^{FT}_4&=&1/2L^{TT}_1-L^{TT}_4
+1/2L^{TT}_6-L^{TT}_7-L^{TQ}_7\ ,
\nonumber\\
L^{FT}_3 -L^{FT}_5&=&1/2L^{TT}_1+L^{TT}_2-L^{TT}_4
-L^{TT}_5-1/2L^{TT}_6\nonumber\\ &&
-L^{TQ}_1+L^{TQ}_4+L^{TQ}_5-L^{TQ}_6-L^{TQ}_7
\ ,
\nonumber\\
L^{FT}_1 +L^{FT}_6&=&1/2L^{TT}_1-L^{TT}_4+1/2L^{TT}_6-L^{TT}_7
-L^{TQ}_5+L^{TQ}_6-L^{TQ}_7\ ,
\nonumber\\
L^{FT}_3 +L^{FT}_7&=&1/2L^{TT}_1+L^{TT}_2-L^{TT}_4-L^{TT}_5
-1/2L^{TT}_6\nonumber\\&&
-L^{TQ}_1+L^{TQ}_4+L^{TQ}_5-L^{TQ}_6+L^{TQ}_7
\ ,
\nonumber\\
L^{FT}_8 -L^{FT}_9&=&-L^{TT}_3+L^{TT}_8+L^{TT}_9
-1/2L^{TQ}_3-1/2L^{TQ}_{12}\ ,
\nonumber\\
L^{FT}_8+L^{FT}_{10}&=&-L^{TQ}_2+L^{TQ}_{10}\ ,
\nonumber\\
L^{FT}_8+L^{FT}_{11}&=&-L^{TT}_3+L^{TT}_8+L^{TT}_9
- 1/2L^{TQ}_3+L^{TQ}_{11}-1/2L^{TQ}_{12}+L^{TQ}_{13}
\ ,
\nonumber\\
L^{FT}_{13} -L^{FT}_{14}&=&L^{TT}_7-L^{TT}_8-1/2L^{TQ}_9\ ,\nonumber\\
L^{FT}_{13} +L^{FT}_{15}&=&L^{TQ}_6-L^{TQ}_8\ ,
\nonumber\\
L^{FT}_{13} +L^{FT}_{16}&=&L^{TT}_7-L^{TT}_8
+L^{TQ}_7-1/2L^{TQ}_9\ ,
\nonumber\\
L^{FT}_{17}&=&1/2L^{TT}_6-L^{TT}_8-1/2L^{TQ}_9\ ,
\nonumber\\
L^{FT}_{18}&=&-1/2L^{TT}_6+L^{TT}_8
+L^{TQ}_5-L^{TQ}_8+1/2L^{TQ}_9\ ,\nonumber\\
L^{FT}_{19}&=&1/2L^{TQ}_9\ 
,\nonumber\\
L^{FT}_{20}&=&L^{TQ}_9\ .
\label{relFT}
\eea
Eliminating $R\nabla Q$ from the $FDQ$ terms we find
the following 17 relations:
\bea
L^{FQ}_1&=&1/2(L^{QQ}_1-L^{QQ}_6)\ ,
\nonumber\\
L^{FQ}_2 +L^{FQ}_3&=&-L^{QQ}_2+L^{QQ}_9\ ,
\nonumber\\
L^{FQ}_2 +L^{FQ}_4&=&1/2L^{QQ}_1-L^{QQ}_2-1/2L^{QQ}_6-L^{QQ}_7
+2L^{QQ}_9-L^{TQ}_1+L^{TQ}_4-L^{TQ}_7\ ,
\nonumber\\
L^{FQ}_2 -L^{FQ}_5&=&1/2L^{QQ}_1-L^{QQ}_2+1/2L^{QQ}_6+L^{QQ}_9
-L^{TQ}_1+L^{TQ}_4-L^{TQ}_7\ ,
\nonumber\\
L^{FQ}_6 -L^{FQ}_7&=&-1/2L^{QQ}_5+1/2L^{QQ}_{16} 
-L^{TQ}_2+L^{TQ}_8-L^{TQ}_{13}\ ,
\nonumber\\
L^{FQ}_6+L^{FQ}_{10}&=&-L^{QQ}_3+L^{QQ}_{10}\ ,
\nonumber\\
L^{FQ}_6+L^{FQ}_{11}&=&-1/2L^{QQ}_5+L^{QQ}_{12}-L^{QQ}_{15} 
+1/2L^{QQ}_{16}-L^{TQ}_2+L^{TQ}_8-L^{TQ}_{13}\ ,
\nonumber\\
L^{FQ}_8 -L^{FQ}_9&=&-1/2L^{QQ}_4+1/2L^{QQ}_{14} 
-L^{TQ}_3+L^{TQ}_9-L^{TQ}_{12}\ ,
\nonumber\\
L^{FQ}_8+L^{FQ}_{12}&=&-L^{QQ}_5+L^{QQ}_{11}\ ,
\nonumber\\
L^{FQ}_8+L^{FQ}_{13}&=&-1/2L^{QQ}_4+L^{QQ}_{13}+1/2L^{QQ}_{14} 
-L^{QQ}_{16}-L^{TQ}_3+L^{TQ}_9-L^{TQ}_{12}\ ,
\nonumber\\
L^{FQ}_{14}&=&1/2(L^{QQ}_5-L^{QQ}_{16})\ ,
\nonumber\\
L^{FQ}_{15}&=&1/2(L^{QQ}_4-L^{QQ}_{14})\ ,\nn
L^{FQ}_{16} +L^{FQ}_{17}&=&L^{QQ}_9-L^{QQ}_{10}\ ,
\nonumber\\
L^{FQ}_{18} -L^{FQ}_{19}&=&-1/2L^{QQ}_{11}+1/2L^{QQ}_{13} 
+L^{TQ}_5-L^{TQ}_{10}+L^{TQ}_{11}\ ,
\nonumber\\
L^{FQ}_{18} +L^{FQ}_{20}&=&L^{QQ}_7-L^{QQ}_{10}\ ,
\nonumber\\
L^{FQ}_{18} +L^{FQ}_{21}&=&L^{QQ}_8-1/2L^{QQ}_{11}-L^{QQ}_{12} 
+1/2L^{QQ}_{13}+L^{TQ}_5-L^{TQ}_{10}+L^{TQ}_{11}\ ,
\nonumber\\
L^{FQ}_{22}&=&1/2(L^{QQ}_{11}-L^{QQ}_{13}) \ .
\label{relFQ}
\eea
Up to now we have collected 42 independent relations.
When taken together with the ones that come from the Bianchi
identities, they form a set of 68 relations,
of which only 50 turn out to be independent.

As in the case of symmetric MAG, there exist additional independent
relations involving simultaneously $FF$ and $FT$ or $FQ$.
Without writing them all out, let us pick just two:
\bea
L^{FF}_{10}+L^{FF}_{12}-L^{FQ}_6+L^{FQ}_{18}&=&
1/2 L^{QQ}_5+L^{QQ}_8-1/2L^{QQ}_{11}+2L^{QQ}_{12}
\nonumber\\&&+1/2L^{QQ}_{13}+L^{QQ}_{15}-1/2 L^{QQ}_{16}
\nonumber\\&&
+L^{TQ}_2+L^{TQ}_5-L^{TQ}_8-L^{TQ}_{10}+L^{TQ}_{11}+L^{TQ}_{13}\ ,
\eea
and
\footnote{Unlike the previous relations,
when one inserts $F=R+\ldots$ in this one,
the $R$ terms do not completely cancel but rather
form expression proportional to the Bianchi identities of $R$.}
\bea
L^{FF}_4+L^{FF}_5-L^{FF}_8-L^{FF}_{12}+L^{FF}_{14}+L^{FF}_{15}
&&\nonumber\\
+L^{FQ}_2-L^{FQ}_{11}+L^{FQ}_{14}+L^{FQ}_{16}-L^{FQ}_{24}
&=&L^{TQ}_5-L^{TQ}_{10}+L^{TQ}_{11}\ .
\eea

\break

\subsection{Antisymmetric MAG in Cartan form: the leftover terms}
\label{sec:app.lin.3}

In the end of Section \ref{AMAGC} we give two bases for 
the dimension-four terms of the type.
Here we give the formulas for the remaining invariants
as linear combinations of the basis elements.

Using the first basis \eqref{carantisymbasisonetone}
\bea
L^{FF}_3&=&L^{FF}_1-3/2L^{TT}_1-L^{TT}_2+3L^{TT}_4+L^{TT}_5
-1/2L^{TT}_6+2L^{TT}_7 \ , 
\nonumber\\
L^{FF}_4&=&1/2 (L^{FF}_1 - L^{TT}_1) +L^{TT}_4 \ , 
\nonumber\\
L^{FF}_8&=&L^{FF}_7 -L^{TT}_3-1/2 L^{TT}_6 +2L^{TT}_8+L^{TT}_9  \ , 
\nonumber\\
L^{FT}_{1}&=&L^{FT}_{13}+1/2L^{TT}_1-L^{TT}_4-L^{TT}_7+L^{TT}_8 \ ,
\nonumber\\
L^{FT}_3&=&2L^{FT}_{13}+1/2L^{TT}_1+L^{TT}_2 -L^{TT}_4 -L^{TT}_5 
- 1/2L^{TT}_6 -2L^{FF}_7+2L^{FF}_8\ , 
\nonumber\\
L^{FT}_4&=&L^{FT}_{13} -1/2L^{TT}_6 +L^{TT}_8 \ , 
\nonumber\\
L^{FT}_5&=&2 L^{FT}_{13} - 2L^{TT}_7 + 2 L^{TT}_8 \ , 
\nonumber\\
L^{FT}_8&=& -1/2L^{FT}_{21}-L^{TT}_3 + L^{TT}_8+L^{TT}_9 \ ,
\nonumber\\
L^{FT}_9&=&-1/2L^{FT}_{21} \ , 
\nonumber\\
L^{FT}_{14}&=&L^{FT}_{13} - L^{TT}_7 +L^{TT}_8 \ ,
\nonumber\\
L^{FT}_{17}&=&1/2L^{TT}_6 - L^{TT}_8 \ . 
\label{AMAGCbasis1}
\eea
Using the second basis \eqref{carantisymbasisFF}
\bea
L^{FT}_{3}  &=&  - L^{FF}_{3} +2  L^{FF}_{4} -2  L^{FF}_{7} +2  L^{FF}_{8} -2  L^{FT}_8 +2  L^{FT}_9 +2  L^{FT}_{13}  \ , \nn 
L^{FT}_{4}  &=& - L^{FF}_{7} + L^{FF}_{8} - L^{FT}_8 + L^{FT}_9 + L^{FT}_{13}  \ , \nn 
L^{FT}_{5}  &=&  - L^{FF}_{1} +2  L^{FF}_{4} +2  L^{FT}_{1}  \ , \nn 
L^{FT}_{14}  &=&  1/2 (- L^{FF}_{1} +2  L^{FF}_{4} +2  L^{FT}_{1} ) \ , \nn 
L^{FT}_{17}  &=&   L^{FF}_{7} - L^{FF}_{8} + L^{FT}_8 - L^{FT}_9  \ , \nn 
L^{FT}_{21}  &=& - 2  L^{FT}_9  \ , \nn
L^{TT}_{4}  &=&  1/2 (- L^{FF}_{1} +2  L^{FF}_{4} + L^{TT}_{1} ) \ , \nn 
L^{TT}_{6}  &=&  - L^{FF}_{1} +2  L^{FF}_{3} -2  L^{FF}_{4} +4  L^{FF}_{7} -4  L^{FF}_{8} +4  L^{FT}_{1} +4  L^{FT}_8 -4  L^{FT}_9 -4  L^{FT}_{13} +2  L^{TT}_{2} -2  L^{TT}_{5}  \ , \nn 
L^{TT}_{7}  &=&   L^{FF}_{3} -2  L^{FF}_{4} + L^{FF}_{7} - L^{FF}_{8} + L^{FT}_{1} + L^{FT}_8 - L^{FT}_9 - L^{FT}_{13} + L^{TT}_{2} - L^{TT}_{5}  \ , \nn 
L^{TT}_{8}  &=&  -1/2 L^{FF}_{1} + L^{FF}_{3} - L^{FF}_{4} + L^{FF}_{7} - L^{FF}_{8} +2 L^{FT}_{1} + L^{FT}_8 - L^{FT}_9 -2  L^{FT}_{13} + L^{TT}_{2} - L^{TT}_{5}  \ , \nn 
L^{TT}_{9}  &=&  1/2 L^{FF}_{1} - L^{FF}_{3} + L^{FF}_{4} - L^{FF}_{7} + L^{FF}_{8} -2  L^{FT}_{1} +2  L^{FT}_{13} - L^{TT}_{2} + L^{TT}_{3} + L^{TT}_{5} \ .
\label{AMAGCbasis2}
\eea

\subsection{Symmetric MAG in Cartan form: the leftover terms}
\label{sec:app.lin.4}

In the end of Section \ref{SMAGcartan} we give two bases for the dimension-four 
terms of the type.
Here we give the formulas for the remaining invariants
as linear combinations of the basis elements.

Using the first basis \eqref{carsymbasisonetone}
\bea
L^{FF}_2&=& -L^{FF}_1+L^{QQ}_1-L^{QQ}_6\ , \nn 
L^{FF}_3&=& 1/2
 (2 L^{FF}_1-3 L^{QQ}_1+2 L^{QQ}_2+3 L^{QQ}_6+2 L^{QQ}_7-4
L^{QQ}_9)\ , \nn 
L^{FF}_8&=& 1/4 (4 L^{FF}_7-L^{QQ}_4+L^{QQ}_{14})\ , \nn 
L^{FF}_9&=& L^{FF}_7-2 L^{FQ}_{18}+L^{FQ}_{24}+L^{QQ}_3-L^{QQ}_5+L^{QQ}_7-2
L^{QQ}_{10}+L^{QQ}_{12}-L^{QQ}_{15}+L^{QQ}_{16}\ , \nn
L^{FF}_{10}&=& 1/4 (4 L^{FF}_7-8 L^{FQ}_{18}+4 L^{FQ}_{24}-L^{QQ}_4+4 L^{QQ}_8-4 L^{QQ}_{11}-4 L^{QQ}_{12}+4L^{QQ}_{13}+L^{QQ}_{14})\ , \nn 
L^{FF}_{11}&=& 1/2 (-2 L^{FF}_7+2
L^{FQ}_{18}-L^{FQ}_{24}+L^{QQ}_5-L^{QQ}_{12}+L^{QQ}_{15}-L^{QQ}_{16})\ , \nn 
L^{FF}_{12}&=& 1/4 (-4
L^{FF}_7+4 L^{FQ}_{18}-2 L^{FQ}_{24}+L^{QQ}_4+2 L^{QQ}_{11}-2 L^{QQ}_{12}-2 L^{QQ}_{13}-L^{QQ}_{14}+2
L^{QQ}_{15})\ , \nn 
L^{FQ}_1&=&
1/2 (L^{QQ}_1-L^{QQ}_6)\ , \nn 
L^{FQ}_2&=&
-L^{FQ}_{16}+L^{FQ}_{18}+1/2L^{QQ}_1-L^{QQ}_2-1/2L^{QQ}_6-L^{QQ}_7+2
L^{QQ}_9\ , \nn 
L^{FQ}_3&=&  1/2(2 L^{FQ}_{16}-2
L^{FQ}_{18}-L^{QQ}_1+L^{QQ}_6+2 L^{QQ}_7-2 L^{QQ}_9)\ , \nn 
L^{FQ}_4&=& L^{FQ}_{16}-L^{FQ}_{18}\ , \nn 
L^{FQ}_5&=& -L^{FQ}_{16}+L^{FQ}_{18}-L^{QQ}_7+L^{QQ}_9\ , \nn 
L^{FQ}_6&=& 1/2
 (L^{FQ}_{24}-L^{QQ}_5+L^{QQ}_{12}-L^{QQ}_{15}+L^{QQ}_{16})\ , \nn 
L^{FQ}_7&=& 1/2
(L^{FQ}_{24}+L^{QQ}_{12}-L^{QQ}_{15})\ , \nn 
L^{FQ}_8&=& 1/2
(L^{FQ}_{23}-L^{QQ}_4+L^{QQ}_{13}+L^{QQ}_{14}-L^{QQ}_{16})\ , \nn
L^{FQ}_9&=& 1/2
(L^{FQ}_{23}+L^{QQ}_{13}-L^{QQ}_{16}) \ , \nn
L^{FQ}_{10}&=& 1/2 (-L^{FQ}_{24}-2 L^{QQ}_3+L^{QQ}_5+2
L^{QQ}_{10}-L^{QQ}_{12}+L^{QQ}_{15}-L^{QQ}_{16})\ , 
\nonumber
\eea
\bea
L^{FQ}_{11}&=& 1/2
(-L^{FQ}_{24}+L^{QQ}_{12}-L^{QQ}_{15})\ , \nn 
L^{FQ}_{12}&=& 1/2 (-L^{FQ}_{23}+L^{QQ}_4-2 L^{QQ}_5+2
L^{QQ}_{11}-L^{QQ}_{13}-L^{QQ}_{14}+L^{QQ}_{16})\ , \nn 
L^{FQ}_{13}&=& 1/2
(-L^{FQ}_{23}+L^{QQ}_{13}-L^{QQ}_{16})\ , \nn 
L^{FQ}_{14}&=& 1/2
(L^{QQ}_5-L^{QQ}_{16})\ , \nn 
L^{FQ}_{15}&=& 1/2 (L^{QQ}_4-L^{QQ}_{14})\ , \nn 
L^{FQ}_{17}&=&
-L^{FQ}_{16}+L^{QQ}_9-L^{QQ}_{10}\ , \nn 
L^{FQ}_{19}&=& 1/2(2
L^{FQ}_{18}+L^{QQ}_{11}-L^{QQ}_{13}) \ , \nn 
L^{FQ}_{20}&=& -L^{FQ}_{18}+L^{QQ}_7-L^{QQ}_{10}\ , \nn 
L^{FQ}_{21}&=& 1/2 (-2
L^{FQ}_{18}+2 L^{QQ}_8-L^{QQ}_{11}-2 L^{QQ}_{12}+L^{QQ}_{13})\ , \nn 
L^{FQ}_{22}&=& 1/2
(L^{QQ}_{11}-L^{QQ}_{13})\ .
\label{SMAGCbasis1}
\eea
Using the second basis \eqref{carsymbasisFF}
\bea
L^{FQ}_{1} &=& 1/2 (  L^{FF}_{1} +  L^{FF}_{2} ) \ , \nn  
L^{FQ}_{2} &=& -  L^{FF}_{2} -  L^{FF}_{3} -  L^{FQ}_{16} +  L^{FQ}_{18}  \ , \nn  
L^{FQ}_{3} &=& 1/2 (-  L^{FF}_{1}-  L^{FF}_{2} +2   L^{FF}_{9} +2
L^{FF}_{11}  +2   L^{FQ}_{10} -2   L^{FQ}_{17} ) \ , \nn  
L^{FQ}_{4} &=&
L^{FQ}_{16} -  L^{FQ}_{18}  \ , \nn  
L^{FQ}_{5} &=& -  L^{FF}_{9}  -  L^{FF}_{11} -  L^{FQ}_{10} +  L^{FQ}_{17}  \ , \nn  
L^{FQ}_{6} &=&
-  L^{FF}_{7} -  L^{FF}_{11} +  L^{FQ}_{18}  \ , \nn  
L^{FQ}_{7} &=&-  L^{FF}_{7} -  L^{FF}_{11}+  L^{FQ}_{14} +  L^{FQ}_{18}  \ , \nn  
L^{FQ}_{8} &=& -2 L^{FF}_{7} +2   L^{FF}_{8}  + 2L^{FF}_{11} -2   L^{FF}_{12}  +  L^{FQ}_{12} +  L^{FQ}_{23}  \ , \nn  
L^{FQ}_{9} &=& 2   L^{FF}_{11} -2
L^{FF}_{12} +  L^{FQ}_{12} +  L^{FQ}_{23}  \ , \nn  
L^{FQ}_{13} &=& 2   L^{FF}_{11} -2   L^{FF}_{12} +  L^{FQ}_{12}  \ , \nn  
L^{FQ}_{15} &=& 2
(  L^{FF}_{7} -  L^{FF}_{8} ) \ , \nn  
L^{FQ}_{19} &=&-  L^{FF}_{7} +  L^{FF}_{8}
-  L^{FF}_{11} +  L^{FF}_{12}  +  L^{FQ}_{14} +  L^{FQ}_{18}  \ , \nn 
L^{FQ}_{20} &=&  L^{FF}_{9}+
L^{FF}_{11}  +  L^{FQ}_{10}  \ , \nn  
L^{FQ}_{21} &=&
L^{FF}_{10} +  L^{FF}_{12} +  L^{FQ}_{11}  \ , \nn  
L^{FQ}_{22} &=&-  L^{FF}_{7} +  L^{FF}_{8}
-  L^{FF}_{11} +  L^{FF}_{12}  +  L^{FQ}_{14}  \ , \nn  
L^{FQ}_{24} &=&-  L^{FF}_{7}
-  L^{FF}_{11}  -  L^{FQ}_{11} +  L^{FQ}_{14} +  L^{FQ}_{18}  \ ,
\nonumber
\eea
\bea
L^{QQ}_{2} &=& 1/2 L^{FF}_{1}  + 3/2
L^{FF}_{2} +  L^{FF}_{3} -  L^{FF}_{9} -  L^{FF}_{11}-  L^{FQ}_{10} +2   L^{FQ}_{16} +2
L^{FQ}_{17} -  L^{FQ}_{18} +  L^{QQ}_{10}  \ , \nn  
L^{QQ}_{3} &=&  L^{FF}_{7}+
L^{FF}_{11}  -  L^{FQ}_{10} -  L^{FQ}_{18} +  L^{QQ}_{10}  \ , \nn  
L^{QQ}_{4} &=& 4   L^{FF}_{7} -4
L^{FF}_{8} +  L^{QQ}_{14}  \ , \nn  
L^{QQ}_{5} &=& 2   L^{FF}_{7} -2   L^{FF}_{8}-2   L^{FF}_{11} +2   L^{FF}_{12}  -2
L^{FQ}_{12} -  L^{FQ}_{23} +  L^{QQ}_{11}  \ , \nn  
L^{QQ}_{6} &=& -  L^{FF}_{1} -  L^{FF}_{2} +  L^{QQ}_{1}  \ , \nn  
L^{QQ}_{7} &=&  L^{FF}_{9} +
L^{FF}_{11} +  L^{FQ}_{10} +  L^{FQ}_{18} +  L^{QQ}_{10}  \ , \nn  
L^{QQ}_{8} &=& -  L^{FF}_{7} +  L^{FF}_{8} + L^{FF}_{10} -  L^{FF}_{11} +2
L^{FF}_{12}  +  L^{FQ}_{11} +  L^{FQ}_{14} +  L^{FQ}_{18} +  L^{QQ}_{12}  \ , \nn  
L^{QQ}_{9} &=& L^{FQ}_{16} +  L^{FQ}_{17} +  L^{QQ}_{10}  \ , \nn
L^{QQ}_{13} &=& 2   L^{FF}_{7} -2
L^{FF}_{8} + 2   L^{FF}_{11} -2   L^{FF}_{12}  -2   L^{FQ}_{14} +  L^{QQ}_{11}  \ , \nn  
L^{QQ}_{15} &=&  L^{FF}_{7}+
L^{FF}_{11}  -  L^{FQ}_{11} -  L^{FQ}_{14} -  L^{FQ}_{18} +  L^{QQ}_{12}  \ , \nn 
L^{QQ}_{16} &=& 2   L^{FF}_{7} -2   L^{FF}_{8}-2
L^{FF}_{11} +2   L^{FF}_{12}  -2   L^{FQ}_{12} -2
L^{FQ}_{14} -  L^{FQ}_{23} +  L^{QQ}_{11}  \ . 
\label{SMAGCbasis2}
\eea

\break

\subsection{General MAG in Cartan form: the leftover terms}
\label{sec:app.lin.5}

We give the formulas mentioned in the end of Section \ref{GMAGcartan}.

Using the first basis \eqref{cargenbasisonetone}
\bea
L^{FF}_{2}  \!\!&=&\!\!  - L^{FF}_{1} + L^{QQ}_{1} - L^{QQ}_{6}  \ , \nn  
L^{FF}_{3}  \!\!&=&\!\!   L^{FF}_{1} -3/2  L^{QQ}_{1}+ L^{QQ}_{2} +3/2  L^{QQ}_{6} + L^{QQ}_{7} -2  L^{QQ}_{9} -3/2  L^{TT}_{1}- L^{TT}_{2} +3  L^{TT}_{4} + L^{TT}_{5}  \nn
&&  -1/2 L^{TT}_{6}+2  L^{TT}_{7} +4  L^{TQ}_{1} -4  L^{TQ}_{4} +4  L^{TQ}_{7}    \ , \nn  
L^{FF}_{4}  \!\!&=&\!\! 1/2 L^{FF}_{1} - L^{QQ}_{1} + L^{QQ}_{2} + L^{QQ}_{6} + L^{QQ}_{7} -2  L^{QQ}_{9}-1/2 L^{TT}_{1}+ L^{TT}_{4}   +2  L^{TQ}_{1} -2  L^{TQ}_{4} +2  L^{TQ}_{7}   \ , \nn  
L^{FF}_{5}  \!\!&=&\!\! -1/2 L^{FF}_{1} +1/2 L^{QQ}_{1} -1/2 L^{QQ}_{6}  +1/2 L^{TT}_{1} - L^{TT}_{4} - L^{TQ}_{1} + L^{TQ}_{4} - L^{TQ}_{7}  \ , \nn  
L^{FF}_{6}  \!\!&=&\!\! 1/2 L^{FF}_{1} -1/2 L^{TT}_{1} + L^{TT}_{4}  \ , \nn  
L^{FF}_{8}  \!\!&=&\!\!  L^{FF}_{7} - 1/4 L^{QQ}_{4}+1/4 L^{QQ}_{14}  - L^{TT}_{3} -1/2 L^{TT}_{6}+2  L^{TT}_{8} + L^{TT}_{9}   - L^{TQ}_{3} + L^{TQ}_{9}- L^{TQ}_{12}   \ , \nn  
L^{FF}_{9}  \!\!&=&\!\!   L^{FF}_{7} -2  L^{FQ}_{18} + L^{FQ}_{24}  + L^{QQ}_{3} - L^{QQ}_{5} + L^{QQ}_{7}-2  L^{QQ}_{10} + L^{QQ}_{12} - L^{QQ}_{15} + L^{QQ}_{16} \nn
&&-2  L^{TQ}_{2} +2  L^{TQ}_{8}  -2  L^{TQ}_{13}  \ ,   \nn
L^{FF}_{10}  \!\!&=&\!\!   L^{FF}_{7} -2  L^{FQ}_{18} + L^{FQ}_{24} -1/4 L^{QQ}_{4} + L^{QQ}_{8}- L^{QQ}_{11} - L^{QQ}_{12} + L^{QQ}_{13} +1/4 L^{QQ}_{14}  \nn
&&- L^{TQ}_{3} +2  L^{TQ}_{5} + L^{TQ}_{9}-2  L^{TQ}_{10} +2 L^{TQ}_{11} - L^{TQ}_{12}  - L^{TT}_{3} -1/2 L^{TT}_{6} +2  L^{TT}_{8} + L^{TT}_{9}   \ , \nn  
L^{FF}_{11}  \!\!&=&\!\! - L^{FF}_{7} + L^{FQ}_{18}-1/2 L^{FQ}_{24}+1/2L^{QQ}_{5}-1/2 L^{QQ}_{12} +1/2 L^{QQ}_{15}-1/2 L^{QQ}_{16}    + L^{TQ}_{2} - L^{TQ}_{8}   + L^{TQ}_{13} \ , \nn  
L^{FF}_{12}  \!\!&=&\!\! - L^{FF}_{7} + L^{FQ}_{18} -1/2 L^{FQ}_{24}+ 1/4 L^{QQ}_{4}+1/2 L^{QQ}_{11} -1/2 L^{QQ}_{12} -1/2 L^{QQ}_{13} -1/4 L^{QQ}_{14}+ 1/2 L^{QQ}_{15}  \nn
&&+ L^{TT}_{3} +1/2 L^{TT}_{6} -2  L^{TT}_{8} - L^{TT}_{9} 
+ L^{TQ}_{3} - L^{TQ}_{5} - L^{TQ}_{9}+ L^{TQ}_{10} - L^{TQ}_{11} + L^{TQ}_{12}  \ , \nn  
L^{FF}_{13}  \!\!&=&\!\! 1/2 ( L^{QQ}_{4} - L^{QQ}_{14} )  \ , \nn  
L^{FF}_{14}  \!\!&=&\!\! 1/4( - L^{QQ}_{4}+L^{QQ}_{14} -2 L^{TQ}_{3} + 2 L^{TQ}_{9} -2 L^{TQ}_{12}  )  \ , \nn  
L^{FF}_{15}  \!\!&=&\!\! 1/4(  L^{QQ}_{4}-2 L^{QQ}_{5}+2 L^{QQ}_{11} -2 L^{QQ}_{13}- L^{QQ}_{14}+ 2 L^{QQ}_{16}   +2 L^{TQ}_{3}   -2 L^{TQ}_{9}  + 2 L^{TQ}_{12})  \ , 
\label{GMAGCbasis1}
\eea

\bea 
L^{FT}_{1}  \!\!&=&\!\!  L^{FT}_{13}  + 1/2 L^{TT}_{1} - L^{TT}_{4} - L^{TT}_{7} + L^{TT}_{8} - L^{TQ}_{7} +1/2 L^{TQ}_{9}  \ , \nn  
L^{FT}_{2}  \!\!&=&\!\! - L^{FT}_{13} -1/2 L^{TT}_{1} + L^{TT}_{4} + L^{TT}_{7} - L^{TT}_{8}+ L^{TQ}_{1} - L^{TQ}_{4} + L^{TQ}_{7}  -1/2 L^{TQ}_{9}   \ , \nn  
L^{FT}_{3}  \!\!&=&\!\!  2  L^{FT}_{13}  +1/2 L^{TT}_{1} + L^{TT}_{2} - L^{TT}_{4} - L^{TT}_{5} -1/2 L^{TT}_{6}-2  L^{TT}_{7} +2L^{TT}_{8}   \nn 
&&- L^{TQ}_{1} + L^{TQ}_{4} + L^{TQ}_{5} - L^{TQ}_{6} - L^{TQ}_{7} + L^{TQ}_{9} \ , \nn  
L^{FT}_{4}  \!\!&=&\!\!   L^{FT}_{13} -1/2 L^{TT}_{6}  + L^{TT}_{8} +1/2 L^{TQ}_{9}  \ , \nn 
L^{FT}_{5}  \!\!&=&\!\!  2  L^{FT}_{13}  -2  L^{TT}_{7} +2  L^{TT}_{8} + L^{TQ}_{9} \ , \nn  
L^{FT}_{6}  \!\!&=&\!\! - L^{FT}_{13}  +1/2 L^{TT}_{6} - L^{TT}_{8} - L^{TQ}_{5} + L^{TQ}_{6}-1/2 L^{TQ}_{9}   \ , \nn  
L^{FT}_{7}  \!\!&=&\!\!  -2  L^{FT}_{13}  +2 L^{TT}_{7} -2  L^{TT}_{8} +2  L^{TQ}_{7} - L^{TQ}_{9} \ , \nn  
L^{FT}_{8}  \!\!&=&\!\! 1/2( - L^{FT}_{21}   - 2L^{TT}_{3} +2L^{TT}_{8} + 2L^{TT}_{9} - L^{TQ}_{3}+ L^{TQ}_{11}- L^{TQ}_{12} + L^{TQ}_{13}  ) \ , \nn  
L^{FT}_{9}  \!\!&=&\!\! 1/2( - L^{FT}_{21} + L^{TQ}_{11} + L^{TQ}_{13} ) \ , \nn
L^{FT}_{10}  \!\!&=&\!\! 1/2 L^{FT}_{21} + L^{TT}_{3} - L^{TT}_{8} - L^{TT}_{9}- L^{TQ}_{2} +1/2 L^{TQ}_{3}  + L^{TQ}_{10} -1/2 L^{TQ}_{11} +1/2 L^{TQ}_{12} -1/2 L^{TQ}_{13}   \ , \nonumber
\eea
\bea
L^{FT}_{11}  \!\!&=&\!\!  1/2( L^{FT}_{21} + L^{TQ}_{11} + L^{TQ}_{13} )  \ , \nn  
L^{FT}_{12}  \!\!&=&\!\! 1/2 ( L^{TQ}_{3}+ L^{TQ}_{12} )  \ , \nn  
L^{FT}_{14}  \!\!&=&\!\!  L^{FT}_{13} - L^{TT}_{7} + L^{TT}_{8}  +1/2 L^{TQ}_{9}  \ , \nn  
L^{FT}_{15}  \!\!&=&\!\!  - L^{FT}_{13} + L^{TQ}_{6} - L^{TQ}_{8}  \ , \nn  
L^{FT}_{16}  \!\!&=&\!\! - L^{FT}_{13}  + L^{TT}_{7} - L^{TT}_{8} + L^{TQ}_{7}-1/2 L^{TQ}_{9}   \ , \nn
L^{FT}_{17}  \!\!&=&\!\!  1/2 L^{TT}_{6} - L^{TT}_{8}  -1/2 L^{TQ}_{9} \ , \nn  
L^{FT}_{18}  \!\!&=&\!\!   -1/2 L^{TT}_{6}+L^{TT}_{8} +L^{TQ}_{5} - L^{TQ}_{8} +1/2 L^{TQ}_{9}   \ , \nn  
L^{FT}_{19}  \!\!&=&\!\! 1/2 L^{TQ}_{9}   \ , \nn  
L^{FT}_{20}  \!\!&=&\!\!   L^{TQ}_{9} \ , 
\label{GMAGCbasis1bis}
\eea

\bea
L^{FQ}_{1}  \!\!&=&\!\!  1/2 ( L^{QQ}_{1} - L^{QQ}_{6} )  \ , \nn  
L^{FQ}_{2}  \!\!&=&\!\! - L^{FQ}_{16} + L^{FQ}_{18} +1/2 L^{QQ}_{1} - L^{QQ}_{2} -1/2 L^{QQ}_{6} - L^{QQ}_{7} +2  L^{QQ}_{9} - L^{TQ}_{1}+ L^{TQ}_{4}  - L^{TQ}_{7}    \ , \nn  
L^{FQ}_{3}  \!\!&=&\!\!  L^{FQ}_{16} - L^{FQ}_{18} -1/2 L^{QQ}_{1} +1/2 L^{QQ}_{6} + L^{QQ}_{7} - L^{QQ}_{9} + L^{TQ}_{1} - L^{TQ}_{4} + L^{TQ}_{7}   \ , \nn  
L^{FQ}_{4}  \!\!&=&\!\!   L^{FQ}_{16} - L^{FQ}_{18}  \ , \nn  
L^{FQ}_{5}  \!\!&=&\!\!  - L^{FQ}_{16} + L^{FQ}_{18} - L^{QQ}_{7} + L^{QQ}_{9}  \ , \nn  
L^{FQ}_{6}  \!\!&=&\!\! 1/2(  L^{FQ}_{24} - L^{QQ}_{5}+ L^{QQ}_{12}  - L^{QQ}_{15}+ L^{QQ}_{16} -2 L^{TQ}_{2} +2 L^{TQ}_{8} - 2L^{TQ}_{13}   ) \ , \nn  
L^{FQ}_{7}  \!\!&=&\!\! 1/2 ( L^{FQ}_{24} + L^{QQ}_{12} - L^{QQ}_{15}  ) \ , \nn  
L^{FQ}_{8}  \!\!&=&\!\! 1/2(  L^{FQ}_{23} - L^{QQ}_{4} + L^{QQ}_{13} + L^{QQ}_{14} - L^{QQ}_{16}  -2 L^{TQ}_{3} +2 L^{TQ}_{9} - 2L^{TQ}_{12} ) \ , \nn  
L^{FQ}_{9}  \!\!&=&\!\! 1/2(  L^{FQ}_{23} + L^{QQ}_{13} - L^{QQ}_{16} )  \ , \nn
L^{FQ}_{10}  \!\!&=&\!\! -1/2 L^{FQ}_{24}   - L^{QQ}_{3} +1/2 L^{QQ}_{5} + L^{QQ}_{10} -1/2 L^{QQ}_{12} +1/2 L^{QQ}_{15}-1/2 L^{QQ}_{16} + L^{TQ}_{2} - L^{TQ}_{8} + L^{TQ}_{13}  \ , \nn  
L^{FQ}_{11}  \!\!&=&\!\!  1/2(- L^{FQ}_{24} + L^{QQ}_{12} - L^{QQ}_{15} )  \ , \nn  
L^{FQ}_{12}  \!\!&=&\!\! -1/2 L^{FQ}_{23} +1/2 L^{QQ}_{4} - L^{QQ}_{5} + L^{QQ}_{11} -1/2 L^{QQ}_{13} -1/2 L^{QQ}_{14} +1/2 L^{QQ}_{16}  + L^{TQ}_{3} - L^{TQ}_{9} + L^{TQ}_{12}  \ ,\nn
L^{FQ}_{13}  \!\!&=&\!\!  1/2 (- L^{FQ}_{23} + L^{QQ}_{13} - L^{QQ}_{16} )  \ , \nn
L^{FQ}_{14}  \!\!&=&\!\! 1/2( L^{QQ}_{5} - L^{QQ}_{16} )  \ , \nonumber
\nonumber
\eea
\bea   
L^{FQ}_{15}  \!\!&=&\!\! 1/2 ( L^{QQ}_{4} - L^{QQ}_{14} )  \ , \nn  
L^{FQ}_{17}  \!\!&=&\!\!  - L^{FQ}_{16} + L^{QQ}_{9} - L^{QQ}_{10}  \ , \nn  
L^{FQ}_{19}  \!\!&=&\!\!  L^{FQ}_{18} +1/2 L^{QQ}_{11} -1/2 L^{QQ}_{13}- L^{TQ}_{5}+ L^{TQ}_{10} - L^{TQ}_{11}     \ , \nn  
L^{FQ}_{20}  \!\!&=&\!\!  - L^{FQ}_{18} + L^{QQ}_{7} - L^{QQ}_{10}  \ , \nn  
L^{FQ}_{21}  \!\!&=&\!\! - L^{FQ}_{18} + L^{QQ}_{8}-1/2 L^{QQ}_{11}- L^{QQ}_{12} +1/2 L^{QQ}_{13} + L^{TQ}_{5}  - L^{TQ}_{10} + L^{TQ}_{11}    \ , \nn  
L^{FQ}_{22}  \!\!&=&\!\! 1/2 (L^{QQ}_{11} - L^{QQ}_{13} )  \ .
\label{GMAGCbasis1bisbis}
\eea

Using the second basis \eqref{cargenbasisFF}
\bea
L^{FT}_{2} \!\!&=&\!\!  1/2 L^{FF}_{2} - L^{FF}_{5} - L^{FT}_{14}  \ , \nn  
L^{FT}_{3} \!\!&=&\!\! -L^{FF}_{3}+L^{FF}_{4}-L^{FF}_{5}-L^{FF}_{7}+L^{FF}_{8} -L^{FF}_{14}-L^{FT}_{8}+L^{FT}_{9} +L^{FT}_{13}-L^{FT}_{15}+L^{FT}_{18} \ , \nn  
L^{FT}_{4} \!\!&=&\!\!  -L^{FF}_{7}+L^{FF}_{8} -L^{FF}_{14}-L^{FT}_{8}+L^{FT}_{9} +L^{FT}_{13} \ , \nn  
L^{FT}_{5} \!\!&=&\!\!  2  L^{FT}_{14}  \ , \nn  
L^{FT}_{6} \!\!&=&\!\!  L^{FT}_{15} -  L^{FT}_{18}  \ , \nn  
L^{FT}_{7} \!\!&=&\!\!   L^{FF}_{1} -2  L^{FF}_{6} -2  L^{FT}_{1}  \ , \nn  
L^{FT}_{10} \!\!&=&\!\!  -L^{FF}_{11}+L^{FF}_{12}-L^{FF}_{15}+L^{FT}_{9} +L^{FT}_{18}+L^{FT}_{21} \ , \nn  
L^{FT}_{11} \!\!&=&\!\!  L^{FT}_{9} + L^{FT}_{21}  \ , \nn 
L^{FT}_{16} \!\!&=&\!\!  1/2 L^{FF}_{1} - L^{FF}_{6} - L^{FT}_{1}  \ , \nn  
L^{FT}_{17} \!\!&=&\!\!  L^{FF}_{7}-L^{FF}_{8}+L^{FF}_{14}+L^{FT}_{8}-L^{FT}_{9}  \ , \nn  
L^{FT}_{19} \!\!&=&\!\! 1/2 L^{FF}_{13} + L^{FF}_{14} + L^{FT}_{12}  \ , \nn  
L^{FT}_{20} \!\!&=&\!\!  L^{FF}_{13} +2  L^{FF}_{14} +2  L^{FT}_{12}  \ , 
\label{GMAGCbasis2}
\eea

\bea
L^{FQ}_{1} \!\!&=&\!\!  1/2\left( L^{FF}_{1} + L^{FF}_{2} \right)  \ , \nn  
L^{FQ}_{2} \!\!&=&\!\!  - L^{FF}_{4} - L^{FF}_{5} - L^{FQ}_{16} + L^{FQ}_{18}  \ , \nn  
L^{FQ}_{3} \!\!&=&\!\!  - L^{FF}_{5} - L^{FF}_{6} + L^{FF}_{9}+L^{FF}_{11} + L^{FQ}_{10} - L^{FQ}_{17}  \ , \nn  
L^{FQ}_{4} \!\!&=&\!\!   L^{FQ}_{16} - L^{FQ}_{18}  \ , \nn  
L^{FQ}_{5} \!\!&=&\!\! - L^{FF}_{9}- L^{FF}_{11}  - L^{FQ}_{10} + L^{FQ}_{17}  \ , \nn  
L^{FQ}_{6} \!\!&=&\!\!  - L^{FF}_{7} - L^{FF}_{11} + L^{FQ}_{18}  \ , \nn  
L^{FQ}_{7} \!\!&=&\!\! - L^{FF}_{8}- L^{FF}_{12}  + L^{FQ}_{19}  \ , \nn  
L^{FQ}_{8} \!\!&=&\!\!  2  L^{FF}_{14} -2  L^{FF}_{15} + L^{FQ}_{12} + L^{FQ}_{23}  \ , \nn  
L^{FQ}_{9} \!\!&=&\!\! -2L^{FF}_{15} + L^{FQ}_{12} + L^{FQ}_{23}  \ , \nn  
L^{FQ}_{13} \!\!&=&\!\!  -2L^{FF}_{15}+ L^{FQ}_{12}   \ , \nn  
L^{FQ}_{15} \!\!&=&\!\!   L^{FF}_{13}  \ , \nn 
L^{FQ}_{20} \!\!&=&\!\! L^{FF}_{9}+L^{FF}_{11}  + L^{FQ}_{10}  \ , \nn  
L^{FQ}_{21} \!\!&=&\!\!   L^{FF}_{10} + L^{FF}_{12} + L^{FQ}_{11}  \ , \nn  
L^{FQ}_{22} \!\!&=&\!\! L^{FF}_{14} + L^{FF}_{15} + L^{FQ}_{14}  \ , \nn  
L^{FQ}_{24} \!\!&=&\!\!  - L^{FF}_{8}- L^{FF}_{12}  - L^{FQ}_{11} + L^{FQ}_{19}  \ ,  
\label{GMAGCbasis2bis}
\eea

\bea
L^{TT}_{4} \!\!&=&\!\!    -1/2 L^{FF}_{1} + L^{FF}_{6} +1/2 L^{TT}_{1}  \ , \nn  
L^{TT}_{6}  \!\!&=&\!\!   - L^{FF}_{2} +2  L^{FF}_{3} -2
L^{FF}_{4} +4  L^{FF}_{5} +4  L^{FF}_{7} -4  L^{FF}_{8} +4  L^{FF}_{14}+4  L^{FT}_{8} -4 L^{FT}_{9} -4  L^{FT}_{13} \nn &&+4  L^{FT}_{14} +2  L^{TT}_{2} -2  L^{TT}_{5}  \ , \nn  
L^{TT}_{7}  \!\!&=&\!\!   -1/2 L^{FF}_{2} + L^{FF}_{3} - L^{FF}_{4} +2
L^{FF}_{5} + L^{FF}_{7} - L^{FF}_{8} +L^{FF}_{14} + L^{FT}_{8} - L^{FT}_{9}- L^{FT}_{13} + L^{FT}_{14} \nn && + L^{TT}_{2} - L^{TT}_{5}  \ , \nn  
L^{TT}_{8}  \!\!&=&\!\!      -1/2 L^{FF}_{2} + L^{FF}_{3} - L^{FF}_{4} +2
L^{FF}_{5} + L^{FF}_{7} - L^{FF}_{8} -1/2 L^{FF}_{13}+ L^{FT}_{8} - L^{FT}_{9} - L^{FT}_{12} \nn
&&-2  L^{FT}_{13} +2 L^{FT}_{14} + L^{TT}_{2} - L^{TT}_{5}  \ , \nn  
L^{TT}_{9}  \!\!&=&\!\!  1/2 L^{FF}_{2} - L^{FF}_{3} + L^{FF}_{4} -2
L^{FF}_{5} - L^{FF}_{7} + L^{FF}_{8} +1/2 L^{FF}_{13}+2  L^{FT}_{12} +2  L^{FT}_{13} -2
L^{FT}_{14} \nn &&- L^{TT}_{2} + L^{TT}_{3} + L^{TT}_{5} \ , \nn
L^{QQ}_{2} \!\!&=&\!\!   L^{FF}_{4} +2  L^{FF}_{5} + L^{FF}_{6} - L^{FF}_{9} - L^{FF}_{11} - L^{FQ}_{10} +2  L^{FQ}_{16} +2L^{FQ}_{17} - L^{FQ}_{18} + L^{QQ}_{10}  \ , \nn  
L^{QQ}_{3} \!\!&=&\!\!    L^{FF}_{7}+L^{FF}_{11}  - L^{FQ}_{10} - L^{FQ}_{18} + L^{QQ}_{10}  \ , \nn  
L^{QQ}_{4} \!\!&=&\!\!  2 L^{FF}_{13} + L^{QQ}_{14}  \ , \nn  
L^{QQ}_{5} \!\!&=&\!\!  -2  L^{FF}_{14} +2  L^{FF}_{15} -2  L^{FQ}_{12} - L^{FQ}_{23} + L^{QQ}_{11}  \ , \nn 
L^{QQ}_{6} \!\!&=&\!\! - L^{FF}_{1} - L^{FF}_{2} + L^{QQ}_{1}  \ , \nn  
L^{QQ}_{7} \!\!&=&\!\!    L^{FF}_{9} +L^{FF}_{11} + L^{FQ}_{10} + L^{FQ}_{18} + L^{QQ}_{10}  \ , \nn  
L^{QQ}_{8} \!\!&=&\!\! L^{FF}_{10} + L^{FF}_{12} + L^{FQ}_{11} + L^{FQ}_{19} + L^{QQ}_{12}  \ , \nn  
L^{QQ}_{9} \!\!&=&\!\!   L^{FQ}_{16} + L^{FQ}_{17} + L^{QQ}_{10}  \ , \nn  
L^{QQ}_{13} \!\!&=&\!\!  -2  L^{FF}_{14} -2  L^{FF}_{15} -2  L^{FQ}_{14} + L^{QQ}_{11}  \ , \nn  
L^{QQ}_{15} \!\!&=&\!\!  L^{FF}_{8}+L^{FF}_{12}  - L^{FQ}_{11} - L^{FQ}_{19} + L^{QQ}_{12}  \ , \nn  
L^{QQ}_{16} \!\!&=&\!\!  -2  L^{FF}_{14} +2  L^{FF}_{15} -2  L^{FQ}_{12} -2
L^{FQ}_{14} - L^{FQ}_{23} + L^{QQ}_{11}  \ , \nn  
L^{TQ}_{2} \!\!&=&\!\!    L^{FF}_{11} -  L^{FF}_{12} +  L^{FF}_{15} -  L^{FT}_{8} -  L^{FT}_{9}-  L^{FT}_{18} -  L^{FT}_{21}  +  L^{TQ}_{10} \ , \nn  
L^{TQ}_{3} \!\!&=&\!\!  2 L^{FT}_{12} - L^{TQ}_{12}  \ , \nn  
L^{TQ}_{4} \!\!&=&\!\!    -1/2 L^{FF}_{2} + L^{FF}_{5} - L^{FT}_{1} + L^{FT}_{14} + L^{TQ}_{1}  \ , \nn 
L^{TQ}_{5} \!\!&=&\!\! L^{FF}_{14} + L^{FF}_{15} + L^{FQ}_{14} + L^{FQ}_{18} - L^{FQ}_{19} + L^{TQ}_{10} - L^{TQ}_{11}  \ , \nn  
L^{TQ}_{6} \!\!&=&\!\!    -  L^{FF}_{7} +  L^{FF}_{8} +L^{FF}_{15}+ L^{FQ}_{14} + L^{FQ}_{18} - L^{FQ}_{19}  -  L^{FT}_{8} +  L^{FT}_{9}+  L^{FT}_{13} +  L^{FT}_{15} - L^{FT}_{18}  \nn
&& + L^{TQ}_{10} - L^{TQ}_{11}  \ , \nn
L^{TQ}_{7} \!\!&=&\!\!    1/2 L^{FF}_{1} - L^{FF}_{6} - L^{FT}_{1} + L^{FT}_{14}  \ , \nn  
L^{TQ}_{8} \!\!&=&\!\!   -  L^{FF}_{7} +  L^{FF}_{8}+ L^{FF}_{15} + L^{FQ}_{14} + L^{FQ}_{18} - L^{FQ}_{19}  -  L^{FT}_{8} +     L^{FT}_{9} -  L^{FT}_{18}+  L^{TQ}_{10} -  L^{TQ}_{11}   \ , \nn  
L^{TQ}_{9} \!\!&=&\!\! L^{FF}_{13} +2  L^{FF}_{14} +2  L^{FT}_{12}  \ , \nn 
L^{TQ}_{13} \!\!&=&\!\! 2  L^{FT}_{9} + L^{FT}_{21} - L^{TQ}_{11}  \ .
\label{GMAGCbasis2bisbis}
\eea

\subsection{The map}
\label{sec:app.lin.6}

Here we report the linear map between the coefficients of the general MAG Lagrangian 
in the Cartan form and in the Einstein form.
In order not to rely on a particular basis, we give the general relation between
the linearly dependent terms, namely the map from the 99 $c$-type coefficients
to the 53 $b$-type coefficients.
In order to derive the map between coefficients in fixed bases,
one has to remove from the r.h.s. all the $c$-coefficients
that are not part of the Cartan basis
and from the l.h.s. all the $b$-coefficients
that are not part of the Einstein basis
(this happens only in the $b^{RT}$ and $b^{RQ}$ sectors).

\bea
b^{RR}_1&=&c^{FF}_{1}-c^{FF}_{2}+c^{FF}_{3}
+(c^{FF}_{4}-c^{FF}_{5}+c^{FF}_{6})/2\ ,
\nn
b^{RR}_2&=&c^{FF}_{7}+c^{FF}_{8}+c^{FF}_{9}+c^{FF}_{10}-c^{FF}_{11}-c^{FF}_{12}\ ,
\nn
b^{RR}_3&=&c^{FF}_{16}\ ,
\label{themapRR}
\eea

\bea
b^{RT}_{1} +2 b^{RT}_{2} + b^{RT}_{3} &=&
8 c^{FF}_{1}-8 c^{FF}_{2} + 8 c^{FF}_{3} + 4 c^{FF}_{4} - 4 c^{FF}_{5}
+ 4 c^{FF}_{6} + 2 c^{FF}_{7} + 2 c^{FF}_{8} + 2 c^{FF}_{9} + 2 c^{FF}_{10}\nn
&&- 2 c^{FF}_{11} - 2 c^{FF}_{12}
+ c^{FT}_{1} - c^{FT}_{2} + 2 c^{FT}_{3} + c^{FT}_{4} + 2 c^{FT}_{5} \nn
&&- c^{FT}_{6} - 2 c^{FT}_{7} + c^{FT}_{13} + c^{FT}_{14} - c^{FT}_{15} - c^{FT}_{16} \,,\nn
-1/2 b^{RT}_{4} + b^{RT}_{5} &=&
c^{FF}_{7} + c^{FF}_{8} + c^{FF}_{9} + c^{FF}_{10} - c^{FF}_{11}
- c^{FF}_{12} + 4 c^{FF}_{16} \nn
&& -1/2 ( c^{FT}_{8} + c^{FT}_{9} )
+1/2 ( c^{FT}_{10} + c^{FT}_{11} ) + c^{FT}_{21} \,,\nn
b^{RQ}_{1} + b^{RQ}_{4} &=&
- 4 c^{FF}_{1} + 4 c^{FF}_{2} - 4 c^{FF}_{3} - 2 c^{FF}_{4}+2 c^{FF}_{5}
- 2 c^{FF}_{6} - c^{FF}_{7} - c^{FF}_{8} - c^{FF}_{9} - c^{FF}_{10} \nn
&&+ c^{FF}_{11} + c^{FF}_{12} - c^{FQ}_{2} + c^{FQ}_{3} + c^{FQ}_{4} - c^{FQ}_{5}
+ c^{FQ}_{16} - c^{FQ}_{17}\,,\nn
- b^{RQ}_{1} + b^{RQ}_{5} &=&  
4 c^{FF}_{1} - 4 c^{FF}_{2} + 4 c^{FF}_{3} + 2 c^{FF}_{4} - 2 c^{FF}_{5}
+ 2 c^{FF}_{6} + 2 c^{FF}_{7} + 2 c^{FF}_{8} - c^{FF}_{11} - c^{FF}_{12} \nn
&&+ c^{FQ}_{2} -  c^{FQ}_{3}-  c^{FQ}_{4}+ c^{FQ}_{5} + c^{FQ}_{18} + c^{FQ}_{19} - c^{FQ}_{20} - c^{FQ}_{21}  \,,\nn
b^{RQ}_{3} + 2b^{RQ}_{6} &=&
- c^{FF}_{7} - c^{FF}_{8} - c^{FF}_{9} - c^{FF}_{10} + c^{FF}_{11} + c^{FF}_{12} - 4 c^{FF}_{16} \nn
&&+ c^{FQ}_{8} +  c^{FQ}_{9} - c^{FQ}_{12} - c^{FQ}_{13} + 2 c^{FQ}_{23} \,,\nn
b^{RQ}_{2} + 2b^{RQ}_{7} &=&
2c^{FF}_{9}+ 2 c^{FF}_{10} - c^{FF}_{11} - c^{FF}_{12} + 4 c^{FF}_{16} + c^{FQ}_{6} + c^{FQ}_{7} - c^{FQ}_{10} \!- c^{FQ}_{11} \!+ 2c^{FQ}_{24} \,,
\label{themapRTQ}
\eea

\bea
b^{TT}_1&=&c^{TT}_1+(6 c^{FF}_1 - 6 c^{FF}_2 + c^{FF}_4 - c^{FF}_5 + c^{FF}_6
+2c^{FT}_{1}-2c^{FT}_2+2c^{FT}_3)/4\ , 
\nn
b^{TT}_2&=&c^{TT}_2+(2 c^{FF}_1 - 2 c^{FF}_2 + c^{FF}_4 - c^{FF}_5 + c^{FF}_6
+2c^{FT}_3)/2\ , 
\nn
b^{TT}_3&=&c^{TT}_3 + c^{FF}_7 + c^{FF}_9-c^{FF}_{11}-c^{FT}_{8}+c^{FT}_{10}\ , 
\nn
b^{TT}_4&=&c^{TT}_4+(-2 c^{FF}_1  + 2 c^{FF}_2 + 4 c^{FF}_3 + c^{FF}_4 - c^{FF}_5 +
c^{FF}_6 + c^{FF}_7 + c^{FF}_8 + c^{FF}_9+ c^{FF}_{10}
\nn
&&
- c^{FF}_{11}\! - c^{FF}_{12}\!-c^{FT}_1\!+c^{FT}_2\!+c^{FT}_4\!+2c^{FT}_5\!-c^{FT}_6\!-2c^{FT}_7\!
+ c^{FT}_{13}\! + c^{FT}_{14}\! - c^{FT}_{15}\!-c^{FT}_{16})/2\ , 
\nn
b^{TT}_5&=& c^{TT}_5+(2 c^{FF}_1-2 c^{FF}_2+4c^{FF}_3 + c^{FF}_4 - c^{FF}_5 + c^{FF}_6 + c^{FF}_7 + c^{FF}_8 + c^{FF}_9+ c^{FF}_{10} - c^{FF}_{11} 
\nn && - c^{FF}_{12}
+c^{FT}_1-c^{FT}_2+c^{FT}_4+2c^{FT}_5-c^{FT}_6-2c^{FT}_7
+c^{FT}_{13}+c^{FT}_{14}-c^{FT}_{15}-c^{FT}_{16})/2\ , 
\nn
b^{TT}_6&=&c^{TT}_6+(-2c^{FF}_1+2c^{FF}_2-4 c^{FF}_3-c^{FF}_4+c^{FF}_5-c^{FF}_6+c^{FF}_7 - c^{FF}_8+c^{FF}_9-c^{FF}_{10}
\nn && 
-c^{FF}_{11}+c^{FF}_{12}-2c^{FT}_3-2c^{FT}_4+2c^{FT}_6+2c^{FT}_{17}-2 c^{FT}_{18})/4 \ , 
\nn
b^{TT}_7&=&c^{TT}_7-2 c^{FF}_1 + 2 c^{FF}_2 - c^{FF}_4 + c^{FF}_5 - c^{FF}_6
\nn
&&+(-c^{FT}_1\!+c^{FT}_2\!-2c^{FT}_3\!+c^{FT}_4\!-2c^{FT}_5\!-c^{FT}_6\!+2c^{FT}_7\!
+c^{FT}_{13}\!-c^{FT}_{14}\!-c^{FT}_{15}\!+c^{FT}_{16})/2\ , 
\nn
b^{TT}_8&=&c^{TT}_8-2c^{FF}_7 -2c^{FF}_9+2c^{FF}_{11}
+c^{FT}_{8}-c^{FT}_{10}-c^{FT}_{13}+c^{FT}_{15}-c^{FT}_{17}+c^{FT}_{18}\ , 
\nn
b^{TT}_9&=&c^{TT}_9+c^{FF}_8+c^{FF}_{10} - c^{FF}_{12} + 4 c^{FF}_{16}
-c^{FT}_{9}+c^{FT}_{11}+2c^{FT}_{21} \ ,
\label{themapTT}
\eea

\bea
b^{QQ}_1&=&   c^{QQ}_1 +(6 c^{FF}_1 - 2 c^{FF}_2 - c^{FF}_4 - c^{FF}_5 + 3 c^{FF}_6)/4
+(c^{FQ}_1 + c^{FQ}_2 - c^{FQ}_3)/2\ , 
\nn
b^{QQ}_2&=& c^{QQ}_2+(-2 c^{FF}_1 + 2 c^{FF}_2 + c^{FF}_4 + c^{FF}_5 - c^{FF}_6)/2
-c^{FQ}_2 \ , 
\nn
b^{QQ}_3&=& c^{QQ}_3+c^{FF}_9-c^{FQ}_{10} \ , 
\nn
b^{QQ}_4&=& c^{QQ}_4+(c^{FF}_7 + c^{FF}_9-c^{FF}_{11} + 2 c^{FF}_{13} - c^{FF}_{14} + c^{FF}_{15})/4+(-c^{FQ}_8+c^{FQ}_{12}+c^{FQ}_{15})/2\ , 
\nn
b^{QQ}_5&=& c^{QQ}_5+(- 2 c^{FF}_9+c^{FF}_{11} - c^{FF}_{15}
-c^{FQ}_6+c^{FQ}_{10}- 2 c^{FQ}_{12} + c^{FQ}_{14})/2 \ , 
\nn
b^{QQ}_6&=& c^{QQ}_6+(-2 c^{FF}_1- 2 c^{FF}_2 + 4 c^{FF}_3 + 3 c^{FF}_4 - 
c^{FF}_5 - c^{FF}_6 + c^{FF}_7 + c^{FF}_8 + c^{FF}_9+c^{FF}_{10}
\nn && 
\qquad- c^{FF}_{11}- c^{FF}_{12})/4+(-c^{FQ}_1-c^{FQ}_4+c^{FQ}_5-c^{FQ}_{16}+c^{FQ}_{17})/2 \ , 
\nn
b^{QQ}_7&=& c^{QQ}_7+(-2c^{FF}_1+2c^{FF}_2+c^{FF}_4+c^{FF}_5
-c^{FF}_6+c^{FF}_7 + c^{FF}_8 +c^{FF}_9- c^{FF}_{10}
\nn&&
\qquad
-c^{FQ}_2+c^{FQ}_3-c^{FQ}_4-c^{FQ}_5+c^{FQ}_{18}+c^{FQ}_{19}+c^{FQ}_{20}-c^{FQ}_{21})/2 \ , 
\nn
b^{QQ}_8&=& c^{QQ}_8+(2 c^{FF}_1 - 2 c^{FF}_2 + 2 c^{FF}_3 + c^{FF}_4 - c^{FF}_5 + c^{FF}_6 + c^{FF}_7 +c^{FF}_8 - c^{FF}_9+ c^{FF}_{10}
\nn&&
\qquad
+c^{FQ}_2-c^{FQ}_3-c^{FQ}_4+c^{FQ}_5+c^{FQ}_{18}+c^{FQ}_{19}-c^{FQ}_{20}+c^{FQ}_{21})/2 \ , 
\nn
b^{QQ}_9&=& c^{QQ}_9- 2c^{FF}_3-2 c^{FF}_4-c^{FF}_7-c^{FF}_8
+(c^{FF}_{11} + c^{FF}_{12})/2
\nn&&
\qquad 
+(c^{FQ}_2+c^{FQ}_3+3c^{FQ}_4-c^{FQ}_5+2c^{FQ}_{16}-c^{FQ}_{18}-c^{FQ}_{19}+c^{FQ}_{20}+c^{FQ}_{21})/2
\ , \nn
b^{QQ}_{10}&=&c^{QQ}_{10}-2c^{FF}_9 +c^{FQ}_{10}-c^{FQ}_{17}-c^{FQ}_{20} \ , 
\nn
b^{QQ}_{11}&=&c^{QQ}_{11}+(2 c^{FF}_9-c^{FF}_{11} + c^{FF}_{15}
+2 c^{FQ}_{12} - c^{FQ}_{16} + c^{FQ}_{17} - c^{FQ}_{18}+c^{FQ}_{20}+c^{FQ}_{22})/2 \ , 
\nn
b^{QQ}_{12}&=&c^{QQ}_{12}+(2c^{FF}_9-2c^{FF}_{10}-c^{FF}_{11}-c^{FF}_{12}
+ c^{FQ}_6 + c^{FQ}_7-c^{FQ}_{10} + c^{FQ}_{11}-2 c^{FQ}_{21})/2
 \ , 
 \nn
b^{QQ}_{13}&=&c^{QQ}_{13}+(- c^{FF}_7 - c^{FF}_8 - c^{FF}_9+c^{FF}_{10} + c^{FF}_{11} - c^{FF}_{15}
\nn&&
\qquad
+ c^{FQ}_8 + c^{FQ}_9-c^{FQ}_{12} + c^{FQ}_{13} - c^{FQ}_{19} + c^{FQ}_{21}-c^{FQ}_{22})/2 \ , 
\nn
b^{QQ}_{14}&=&c^{QQ}_{14} +(c^{FF}_8+c^{FF}_{10} - c^{FF}_{12} - 2 c^{FF}_{13} + c^{FF}_{14} - c^{FF}_{15})/4+c^{FF}_{16}
\nn&&
\qquad
+(-c^{FQ}_9+c^{FQ}_{13}-c^{FQ}_{15}-2c^{FQ}_{23})/2\ , 
\nn
b^{QQ}_{15}&=& c^{QQ}_{15}+c^{FF}_{10} + c^{FF}_{16} -c^{FQ}_{11}+c^{FQ}_{24}\ , 
\label{themapQQ}
\\
b^{QQ}_{16}&=& c^{QQ}_{16}+(-2 c^{FF}_{10} + c^{FF}_{12} + c^{FF}_{15} - 4 c^{FF}_{16}
-c^{FQ}_7+c^{FQ}_{11} -2c^{FQ}_{13}- c^{FQ}_{14} + 2c^{FQ}_{23}-2c^{FQ}_{24})/2  \ ,
\nonumber
\eea

\bea
b^{TQ}_1&=& c^{TQ}_1-4 c^{FF}_1 + 4 c^{FF}_2 + c^{FF}_5 - 2 c^{FF}_6
+c^{FT}_2-c^{FT}_3-c^{FQ}_2+c^{FQ}_3 \ , 
\nn
b^{TQ}_2&=& c^{TQ}_2-2c^{FF}_9+c^{FF}_{11}-c^{FT}_{10}-c^{FQ}_6+c^{FQ}_{10} \ , 
\nn
b^{TQ}_3&=& c^{TQ}_3\!+c^{FF}_7\!+c^{FF}_9\!-c^{FF}_{11}\!-c^{FF}_{14}/2 +c^{FF}_{15}/2
+(-c^{FT}_{8}+c^{FT}_{10}+c^{FT}_{12})/2-c^{FQ}_8+c^{FQ}_{12}\ , 
\nn
b^{TQ}_4&=& c^{TQ}_4-4 c^{FF}_3 - 2 c^{FF}_4 + c^{FF}_5 - c^{FF}_7 - c^{FF}_8 - c^{FF}_9-c^{FF}_{10} + c^{FF}_{11} + c^{FF}_{12}  
\nn
&&+(-c^{FT}_1-c^{FT}_2-c^{FT}_4-2 c^{FT}_5+c^{FT}_6+2c^{FT}_7-c^{FT}_{13}-c^{FT}_{14}+c^{FT}_{15}+c^{FT}_{16})/2
\nn&&
+c^{FQ}_4-c^{FQ}_5+c^{FQ}_{16}-c^{FQ}_{17}\ , 
\nn
b^{TQ}_5&=& c^{TQ}_5+2 c^{FF}_1  - 2 c^{FF}_2 + 
 2 c^{FF}_3 + c^{FF}_4 - c^{FF}_5 + c^{FF}_6 - c^{FF}_9
 + c^{FF}_{10} +(c^{FF}_{11}-c^{FF}_{12})/2 
 \nn &&
 +c^{FT}_3-c^{FT}_6+c^{FT}_{18}
 +(c^{FQ}_2-c^{FQ}_3-c^{FQ}_4+c^{FQ}_5+c^{FQ}_{18}-c^{FQ}_{19}-c^{FQ}_{20}+c^{FQ}_{21})/2 \ , 
 \nn
b^{TQ}_6&=& c^{TQ}_6+2 c^{FF}_1 - 2 c^{FF}_2 + 
 2 c^{FF}_3 + c^{FF}_4 - c^{FF}_5 + c^{FF}_6 + c^{FF}_7 + c^{FF}_8
 -(c^{FF}_{11}+c^{FF}_{12})/2  
\nn
&&+(c^{FT}_1-c^{FT}_2+c^{FT}_4+2c^{FT}_5+c^{FT}_6-2c^{FT}_7+c^{FT}_{13}
+c^{FT}_{14}+c^{FT}_{15}-c^{FT}_{16}
\nn&&
+c^{FQ}_2-c^{FQ}_3-c^{FQ}_4+c^{FQ}_5+c^{FQ}_{18}+c^{FQ}_{19}-c^{FQ}_{20}-c^{FQ}_{21})/2\ , 
\nonumber
\eea
\bea
b^{TQ}_7&=&c^{TQ}_7 -2 c^{FF}_1+ 2 c^{FF}_2 + 
 2 c^{FF}_3 + c^{FF}_4 - c^{FF}_6 + c^{FF}_7 + c^{FF}_8
 - (c^{FF}_{11} +c^{FF}_{12})/2 
 \nn &&+(-c^{FT}_1+c^{FT}_2+c^{FT}_4+2c^{FT}_5-c^{FT}_6+2c^{FT}_7+c^{FT}_{13}+c^{FT}_{14}-c^{FT}_{15}+c^{FT}_{16})/2
 \nn&&
 +(-c^{FQ}_2+c^{FQ}_3-c^{FQ}_4+c^{FQ}_5+c^{FQ}_{18}+c^{FQ}_{19}-c^{FQ}_{20}-c^{FQ}_{21})/2\ , 
 \nn
b^{TQ}_8&=&c^{TQ}_8+ 2 c^{FF}_9-c^{FF}_{11}-c^{FT}_{15}-c^{FT}_{18}+c^{FQ}_6-c^{FQ}_{10} \ , 
\nn
b^{TQ}_9&=& c^{TQ}_9+(- 2 c^{FF}_7 - 2 c^{FF}_9+2 c^{FF}_{11} + c^{FF}_{14} - c^{FF}_{15}
\nn&&\qquad
-c^{FT}_{13}+c^{FT}_{15}-c^{FT}_{17}+c^{FT}_{18}  + c^{FT}_{19} + 2 c^{FT}_{20}
+2c^{FQ}_8-2c^{FQ}_{12} )/2 \ , 
\nn
b^{TQ}_{10}&=& c^{TQ}_{10}+2c^{FF}_9-c^{FF}_{11}
+c^{FT}_{10}-c^{FQ}_{16}+c^{FQ}_{17}-c^{FQ}_{18}+c^{FQ}_{20}\ , 
\nn
b^{TQ}_{11}&=&c^{TQ}_{11} - c^{FF}_7 - c^{FF}_8 - c^{FF}_9+c^{FF}_{10} + c^{FF}_{11}
+(c^{FT}_{8}+c^{FT}_{9}-c^{FT}_{10} + c^{FT}_{11})/2-c^{FQ}_{19}+c^{FQ}_{21} \ , 
\nn
b^{TQ}_{12}&=&c^{TQ}_{12}+(- 2 c^{FF}_8-2 c^{FF}_{10} + 2 c^{FF}_{12} - c^{FF}_{14} + c^{FF}_{15} - 8 c^{FF}_{16}
\nn&&
\qquad
+c^{FT}_{9}-c^{FT}_{11}+ c^{FT}_{12} - 2 c^{FT}_{21}
+2c^{FQ}_9-2c^{FQ}_{13}+4c^{FQ}_{23})/2 \ , 
\nn
b^{TQ}_{13}&=& c^{TQ}_{13}+2 c^{FF}_{10} - c^{FF}_{12} + 4 c^{FF}_{16}
 +c^{FT}_{11}+c^{FT}_{21}+c^{FQ}_7-c^{FQ}_{11}+2c^{FQ}_{24} \ .
\label{themapTQ}
\eea

\section{Conditions for vector invariance}
\label{sec:app.projective}

Here we report the conditions on the coefficients
that derive from imposing invariance of the action
under the vector transformations discussed in Section \ref{sec:vectr}.

\begin{fleqn}[0pt]
The conditions for projective invariance $\delta_1$ (\ref{pt1}) are
\begin{equation*}
\begin{aligned}
& 2 m^{TT}_{1} + m^{TT}_{2} + 3 m^{TT}_{3} + m^{TQ}_{1} - 4 m^{TQ}_{2} -  m^{TQ}_{3} = 0\,, \\ & 6 m^{TT}_{1} + 3 m^{TT}_{2} + 9 m^{TT}_{3} - 2 m^{TQ}_{1} - 6 m^{TQ}_{3} - 16 m^{QQ}_{1} + 4 m^{QQ}_{2} - 64 m^{QQ}_{3} + 4 m^{QQ}_{4} = 0\,, \\ & m^{TQ}_{1} + 3 m^{TQ}_{3} - 4 m^{QQ}_{2} - 4 m^{QQ}_{4} - 8 m^{QQ}_{5} =0\,, 
\deveq
& b^{RT}_{1} + 2 b^{RT}_{2} -  b^{RT}_{3} - 3 b^{RT}_{4} + 2 b^{RQ}_{1} + 2 b^{RQ}_{2} + 8 b^{RQ}_{3} + 2 b^{RQ}_{5} = 0\,, \\ & b^{RT}_{3} + 3 b^{RT}_{5} + 2 b^{RQ}_{4} + 8 b^{RQ}_{6} + 2 b^{RQ}_{7} = 0\,, \\ & 2 b^{TT}_{1} - 3 b^{TT}_{2} - 9 b^{TT}_{3} -  b^{TT}_{8} + 8 b^{TQ}_{3} + 2 b^{TQ}_{11} = 0\,, 
\deveq
& 8 b^{TT}_{2} + 24 b^{TT}_{3} + 3 b^{TT}_{8} + 2 b^{TQ}_{1} - 2 b^{TQ}_{2} - 24 b^{TQ}_{3} - 6 b^{TQ}_{11} = 0\,, \\ & 2 b^{TT}_{2} + 8 b^{TT}_{3} + b^{TT}_{8} - 8 b^{TQ}_{3} - 2 b^{TQ}_{11} + 2 b^{QQ}_{2} + 2 b^{QQ}_{3} = 0\,, \\ & 4 b^{TT}_{2} + 13 b^{TT}_{3} + 2 b^{TT}_{8} - 10 b^{TQ}_{3} - 4 b^{TQ}_{11} - 4 b^{QQ}_{1} - 16 b^{QQ}_{4} - 4 b^{QQ}_{5} = 0\,, 
\deveq
& 8 b^{TT}_{3} -  b^{TT}_{4} -  b^{TT}_{5} + 4 b^{TT}_{8} + 8 b^{TQ}_{2} - 8 b^{TQ}_{3} - 8 b^{TQ}_{9} - 2 b^{TQ}_{11} - 16 b^{QQ}_{5} - 4 b^{QQ}_{12} = 0\,, \\ & 2 b^{TT}_{3} + b^{TT}_{8} + 2 b^{TQ}_{2} - 2 b^{TQ}_{3} - 2 b^{TQ}_{9} - 2 b^{TQ}_{11} - 4 b^{QQ}_{5} + 4 b^{QQ}_{13} = 0\,, 
\deveq
& b^{TT}_{4} + 2 b^{TT}_{6} -  b^{TT}_{7} = 0\,, \\ & 3 b^{TT}_{4} -  b^{TT}_{5} - 6 b^{TT}_{7} - 4 b^{TQ}_{5} = 0\,, \\ & b^{TT}_{4} + 5 b^{TT}_{5} - 2 b^{TT}_{7} - 4 b^{TQ}_{6} = 0\,, \\ & 5 b^{TT}_{4} + b^{TT}_{5} + 2 b^{TT}_{7} - 4 b^{TQ}_{7} = 0\,, \\ & b^{TT}_{4} + b^{TT}_{5} - 3 b^{TT}_{8} + 2 b^{TQ}_{8} + 8 b^{TQ}_{9} = 0\,, 
\deveq
& 2 b^{TT}_{4} + 2 b^{TT}_{5} - 9 b^{TT}_{9} + 2 b^{TQ}_{4} - 8 b^{TQ}_{12} = 0\,, \\ & b^{TT}_{4} + b^{TT}_{5} + 2 b^{TT}_{7} - 4 b^{QQ}_{7} = 0\,, \\ & b^{TT}_{4} + b^{TT}_{5} - 2 b^{TT}_{7} - 4 b^{QQ}_{8} = 0\,, \\ & 3 b^{TT}_{4} + 3 b^{TT}_{5} - 9 b^{TT}_{9} - 8 b^{TQ}_{12} + 4 b^{QQ}_{9} = 0\,, 
\deveq
& b^{TT}_{4} + b^{TT}_{5} - 3 b^{TT}_{8} + 8 b^{TQ}_{9} - 4 b^{QQ}_{10} = 0\,, \\ & b^{TT}_{4} + b^{TT}_{5} - 13 b^{TT}_{9} - 16 b^{TQ}_{12} - 4 b^{QQ}_{6} - 16 b^{QQ}_{14} = 0\,, \\ & b^{TT}_{8} + 2 b^{TQ}_{10} = 0\,, \\ & 15 b^{TT}_{9} + 16 b^{TQ}_{12} + 2 b^{TQ}_{13} = 0\,, \\ & 4 b^{TT}_{9} + 4 b^{TQ}_{12} + b^{QQ}_{15} = 0\,, \\ & 17 b^{TT}_{9} + 18 b^{TQ}_{12} - 4 b^{QQ}_{16} = 0\,, \\ & b^{TQ}_{9} + 2 b^{QQ}_{11} = 0\,.
\end{aligned}
\end{equation*}

In the special case of the action that contains only the $F^2$ terms, 
the conditions reduce to those that had already been discussed in \cite{Percacci:2020ddy}.
The conditions for invariance under $\delta_2$ (\ref{pt2}) are
\begin{equation*}
\begin{aligned}
& m^{TQ}_{1} + 2 m^{TQ}_{2} + 5 m^{TQ}_{3} = 0\,, \\ & 10 m^{QQ}_{1} + 3 m^{QQ}_{2} - 4 m^{QQ}_{3} + 25 m^{QQ}_{4} = 0\,, \\ & 2 m^{QQ}_{1} + m^{QQ}_{2} + 5 m^{QQ}_{4} + m^{QQ}_{5} =0\,, \\
& b^{RQ}_{1} - 5 b^{RQ}_{2} - 2 b^{RQ}_{3} - 2 b^{RQ}_{4} - 3 b^{RQ}_{5} - 4 b^{RQ}_{6} - 10 b^{RQ}_{7} = 0\,, \\ & b^{TT}_{8} - 2 b^{TQ}_{11} = 0\,, 
\deveq
& b^{TT}_{8} - 2 b^{TQ}_{6} + b^{TQ}_{10} - 2 b^{TQ}_{12} - 5 b^{TQ}_{13} = 0\,, \\ & b^{TQ}_{1} + 5 b^{TQ}_{2} + 2 b^{TQ}_{3} + b^{TQ}_{10} = 0\,, \\ & b^{TQ}_{4} -  b^{TQ}_{5} + b^{TQ}_{6} = 0\,, \\ & 2 b^{TQ}_{4} + 2 b^{TQ}_{6} + 5 b^{TQ}_{8} + 2 b^{TQ}_{9} = 0\,, 
\deveq
& b^{TQ}_{4} + b^{TQ}_{6} + 2 b^{QQ}_{6} -  b^{QQ}_{8} = 0\,, \\ & b^{TQ}_{4} + b^{TQ}_{6} - 2 b^{QQ}_{8} -  b^{QQ}_{9} = 0\,, \\ & 2 b^{TQ}_{4} + 2 b^{TQ}_{6} + 5 b^{TQ}_{8} + 4 b^{QQ}_{13} = 0\,, \\ & b^{TQ}_{6} -  b^{TQ}_{7} = 0\,, \\ & b^{TQ}_{8} - 2 b^{QQ}_{12} = 0\,, \\ & 5 b^{TQ}_{8} + 6 b^{QQ}_{7} + 6 b^{QQ}_{8} + 5 b^{QQ}_{10} - 4 b^{QQ}_{14} + 25 b^{QQ}_{15} = 0\,, 
\deveq
& b^{TQ}_{8} + 2 b^{QQ}_{7} + 2 b^{QQ}_{8} + b^{QQ}_{10} + 10 b^{QQ}_{15} + 2 b^{QQ}_{16} = 0\,, \\ & 4 b^{QQ}_{1} + 2 b^{QQ}_{2} + 10 b^{QQ}_{3} + 2 b^{QQ}_{5} + b^{QQ}_{10} = 0\,, \\ & 10 b^{QQ}_{1} + 3 b^{QQ}_{2} + 25 b^{QQ}_{3} - 4 b^{QQ}_{4} + b^{QQ}_{7} -  b^{QQ}_{8} + 5 b^{QQ}_{10} = 0\,, \\ & 2 b^{QQ}_{7} - 2 b^{QQ}_{8} + 5 b^{QQ}_{10} + 2 b^{QQ}_{11} = 0 \,.
\end{aligned}
\end{equation*}

The conditions on the $m$-coefficients agree
with those of \cite{Percacci:2020ddy}).

The conditions for invariance under $\delta_3$ (\ref{pt3}) are
\begin{equation*}
\begin{aligned}
& 4 m^{TT}_{1} + 2 m^{TT}_{2} + 6 m^{TT}_{3} + m^{TQ}_{1} + 2 m^{TQ}_{2} + 5 m^{TQ}_{3} = 0\,, \\ 
& 6 m^{TT}_{1} + 3 m^{TT}_{2} + 9 m^{TT}_{3} + 5 m^{TQ}_{1} + 15 m^{TQ}_{3} + 10 m^{QQ}_{1} + 3 m^{QQ}_{2} - 4 m^{QQ}_{3} + 25 m^{QQ}_{4} = 0\,, \\ & m^{TQ}_{1} + 3 m^{TQ}_{3} + 4 m^{QQ}_{1} + 2 m^{QQ}_{2} + 10 m^{QQ}_{4} + 2 m^{QQ}_{5}=0\,, \\
& b^{RT}_{1} + 2 b^{RT}_{2} + b^{RT}_{3} - 3 b^{RT}_{4} + 6 b^{RT}_{5} + b^{RQ}_{1} - 5 b^{RQ}_{2} - 2 b^{RQ}_{3} - 2 b^{RQ}_{4} - 3 b^{RQ}_{5} - 4 b^{RQ}_{6} - 10 b^{RQ}_{7} = 0\,, 
\deveq
& 4 b^{TT}_{1} + 2 b^{TT}_{2} + 6 b^{TT}_{3} + b^{TT}_{8} + b^{TQ}_{1} + 5 b^{TQ}_{2} + 2 b^{TQ}_{3} + b^{TQ}_{10} = 0\,, \\ 
& 6 b^{TT}_{1} + 3 b^{TT}_{2} + 9 b^{TT}_{3} -  b^{TT}_{4} + b^{TT}_{5} + 4 b^{TT}_{6} + 3 b^{TT}_{8} + 5 b^{TQ}_{1} + 15 b^{TQ}_{2} + 2 b^{TQ}_{5} + 5 b^{TQ}_{8} \\ & \quad\quad + 3 b^{TQ}_{10} + 10 b^{QQ}_{1} + 3 b^{QQ}_{2} + 25 b^{QQ}_{3} - 4 b^{QQ}_{4} + b^{QQ}_{7} -  b^{QQ}_{8} + 5 b^{QQ}_{10} = 0\,, 
\deveq
& 2 b^{TT}_{4} - 2 b^{TT}_{5} + 2 b^{TT}_{7} + b^{TQ}_{6} -  b^{TQ}_{7} = 0\,, \\ 
& 2 b^{TT}_{4} + 2 b^{TT}_{5} -  b^{TT}_{8} + 6 b^{TT}_{9} -  b^{TQ}_{6} -  b^{TQ}_{7} + b^{TQ}_{10} + 2 b^{TQ}_{11} - 2 b^{TQ}_{12} - 5 b^{TQ}_{13} = 0\,, 
\deveq
& 2 b^{TT}_{4} - 2 b^{TT}_{5} - 8 b^{TT}_{6} - 4 b^{TQ}_{5} + b^{TQ}_{6} + b^{TQ}_{7} + 4 b^{QQ}_{8} + 2 b^{QQ}_{9} = 0\,, \\
& 2 b^{TT}_{4} - 4 b^{TT}_{6} + b^{TQ}_{4} - 3 b^{TQ}_{5} + b^{TQ}_{7} - 3 b^{TQ}_{10} - 2 b^{QQ}_{7} + 2 b^{QQ}_{8} - 5 b^{QQ}_{10} - 2 b^{QQ}_{11} = 0\,, \\
& 2 b^{TT}_{4} - 2 b^{TT}_{5} - 8 b^{TT}_{6} - 2 b^{TQ}_{5} + b^{TQ}_{6} -  b^{TQ}_{7} - 6 b^{TQ}_{11} - 10 b^{QQ}_{12} - 4 b^{QQ}_{13} = 0\,, 
\deveq
& 4 b^{TT}_{4} + 4 b^{TT}_{5} + 9 b^{TT}_{9} + b^{TQ}_{6} + b^{TQ}_{7} + 5 b^{TQ}_{8} - 6 b^{TQ}_{12} - 6 b^{QQ}_{7} - 6 b^{QQ}_{8} - 5 b^{QQ}_{10} - 10 b^{QQ}_{12} \\ & \quad\quad + 4 b^{QQ}_{14} - 25 b^{QQ}_{15} = 0\,, \\ 
& 6 b^{TT}_{4} + 6 b^{TT}_{5} - 3 b^{TT}_{8} + 18 b^{TT}_{9} + 2 b^{TQ}_{6} + 2 b^{TQ}_{7} + 5 b^{TQ}_{8} + 3 b^{TQ}_{10} + 6 b^{TQ}_{11} - 6 b^{TQ}_{12} - 10 b^{QQ}_{7} \\ & \quad\quad - 10 b^{QQ}_{8} - 5 b^{QQ}_{10} - 10 b^{QQ}_{12} - 50 b^{QQ}_{15} - 10 b^{QQ}_{16} = 0\,, 
\deveq
& 2 b^{TT}_{5} + 4 b^{TT}_{6} + 3 b^{TT}_{8} + b^{TQ}_{4} + b^{TQ}_{5} + b^{TQ}_{7} + 5 b^{TQ}_{8} + 2 b^{TQ}_{9} = 0\,, \\ & b^{TT}_{5} + 2 b^{TT}_{6} + b^{TQ}_{4} + b^{TQ}_{5} + 2 b^{QQ}_{6} -  b^{QQ}_{8} = 0\,, \\ 
& b^{TQ}_{1} + 3 b^{TQ}_{2} + b^{TQ}_{8} + 4 b^{QQ}_{1} + 2 b^{QQ}_{2} + 10 b^{QQ}_{3} + 2 b^{QQ}_{5} + b^{QQ}_{10} = 0 \,.
\end{aligned}
\end{equation*}
Again, the conditions on the $m$-coefficients agree
with those of \cite{Percacci:2020ddy}).
\end{fleqn}

\section{Projectors for vectors and rank-two tensors}
\label{sec:pro12sa}

For a vector
\be
L_{\mu \nu} = \frac{q_\mu q_\nu}{q^2} \,,
\qquad
T_{\mu \nu} = \eta_{\mu \nu}-\frac{q_\mu q_\nu}{q^2} \ .
\ee
For symmetric $2$-rank tensors
\bea
P_s\left(2^+\right){}_{\mu \nu}{}^{\rho \sigma} \!&=&\! T_{(\mu}^{(\rho} T_{\nu)}^{\rho)} 
- \frac{1}{d-1}T_{\mu \nu} T^{\rho \sigma}\,,
\\
P_s\left(1^-\right){}_{\mu \nu}{}^{\rho \sigma} \!&=&\! 2 T_{(\mu}^{(\rho} L_{\nu)}^{\rho)}\,,
\\
P_s\left(0^+,ss\right){}_{\mu \nu}{}^{\rho \sigma} \!&=&\! \frac{1}{d-1}T_{\mu \nu} T^{\rho \sigma}\,,
\\
P_s\left(0^+,ww\right){}_{\mu \nu}{}^{\rho \sigma} \!&=&\! L_{\mu \nu} L^{\rho \sigma}\,,
\\
P_s\left(0^+,sw\right){}_{\mu \nu}{}^{\rho \sigma} \!&=&\! \frac{1}{\sqrt{d-1}}T_{\mu \nu} L^{\rho \sigma} \,,
\\
P_s\left(0^+,ws\right){}_{\mu \nu}{}^{\rho \sigma} \!&=&\! \frac{1}{\sqrt{d-1}}L_{\mu \nu} T^{\rho \sigma} \,,
\eea
such that
\be
P_s\left(2^+\right){}_{\mu \nu}{}^{\rho \sigma} + P_s\left(1^-\right){}_{\mu \nu}{}^{\rho \sigma} + 
P_s\left(0^+,ss\right){}_{\mu \nu}{}^{\rho \sigma} + P_s\left(0^+,ww\right){}_{\mu \nu}{}^{\rho \sigma} =
\delta_{(\mu}^{(\rho} \delta_{\nu)}^{\sigma)} \,.
\ee
Diagonal terms of projectors between symmetric $2$-rank tensors and vectors
\bea
P_{sv}\left(1^-\right){}_{\mu\nu}{}^{\rho} \!&=&\! \frac{\sqrt{2}}{|q|}q_{(\mu} T_{\nu)}^{\rho} \,,
\\
P_{sv}\left(0^+,sv\right){}_{\mu \nu}{}^{\rho } \!&=&\! \frac{1}{\sqrt{d-1}}\frac{q^\rho}{|q|}  T_{\mu\nu} \,,
\\
P_{sv}\left(0^+,wv\right){}_{\mu \nu}{}^{\rho} \!&=&\! \frac{q^\rho}{|q|} L_{\mu\nu} \,,
\eea

For antisymmetric $2$-rank tensors
\bea
P_a\left(1^+\right){}_{\mu \nu}{}^{\rho \sigma} \!&=&\! T_{[\mu}^{[\rho} T_{\nu]}^{\rho]} \,,
\\
P_a\left(1^-\right){}_{\mu \nu}{}^{\rho \sigma} \!&=&\! 2 T_{[\mu}^{[\rho} L_{\nu]}^{\rho]}\,,
\eea
such that
\be
P_a\left(1^+\right){}_{\mu \nu}{}^{\rho \sigma} + P_a\left(1^-\right){}_{\mu \nu}{}^{\rho \sigma}  =
\delta_{[\mu}^{[\rho} \delta_{\nu]}^{\sigma]}\,.
\ee
Diagonal terms of projectors between antisymmetric $2$-rank tensors and symmetric $2$-rank tensors
\bea
P_{sa}\left(1^+\right){}_{\mu \nu}{}^{\rho \sigma} \!&=&\! 2 T_{(\mu}^{[\rho} L_{\nu)}^{\rho]}\,,
\\
P_{as}\left(1^+\right){}_{\mu \nu}{}^{\rho \sigma} \!&=&\! 2 T_{[\mu}^{(\rho} L_{\nu]}^{\rho)}\,.
\eea
Diagonal terms of projectors between antisymmetric $2$-rank tensors and vectors
\be
P_{av}\left(1^-\right){}_{\mu\nu}{}^{\rho} = \frac{\sqrt{2}}{|q|}q_{[\mu} T_{\nu]}^{\rho} \,.
\ee

\section{The standard propagators for fields of
spin \texorpdfstring{$0,1,2,3$}{1,2,3,4}}
\label{sec:app.propagators}

The information about the particle content of MAG,
when linearized around flat space,
is contained the coefficient matrices.
However, these do not give the final form of the propagator.
For this one has to use the explicit form of the spin projectors.
These contain $q^{-2}$, so in general the propagator
could contain terms up to $q^{-8}$.
In a theory without ghosts, all the terms containing
higher inverse powers of $q^2$ must cancel out, and the terms
of order $q^{-2}$ must appear in special combinations,
dictated by the spin and parity.
It is therefore useful to have a list of the standard forms
of the propagators for each spin/parity.
This is the content of the present appendix.
While much of this material is standard,
we could not find in the literature
a discussion of the spin $2^-$ case.

Let us remark that the parity states discussed here
are not obtained by just attributing an intrinsic parity
(a quantum-mechanical phase) to the wave function.
States with same spin but different parities are different geometric objects.
The standard way of describing a spin-$s$ object, with integer $s$, 
is by means of totally symmetric rank-$s$ tensor.
Such a state has parity $(-1)^s$,
so for spin 0 and 2 it has even parity while for spin 1 and 3
it has odd parity.
In order to describe the other states one needs:
a totally antisymmetric 3-tensor for $0^-$,
an antisymmetric 2-tensor for $1^+$,
a 3-tensor that is antisymmetric in two indices for $2^-$.

We also emphasize that the material in this section
is not directly related to MAGs or even to gravity
but rather belongs to the general subject of
(special) relativistic wave equations.

\subsection{\texorpdfstring{Spin/parity $0^+$}{0-plus}}
Given the Lagrangian
\be
{\cal L} = -\frac12 \partial_\mu \phi \partial^\mu \phi -\frac12 m^2 \phi^2 +J \phi\ ,
\ee
the Fourier transformed field equation takes the form
\be
( q^2+m^2 ) \phi = J\ .
\ee
It follows that the saturated propagator is
\be
\Pi = \frac{1}{2}\int \frac{d^4q}{(2\pi)^4}\ J(-q) \frac{1}{q^2+m^2} J(q)\ .
\ee
%

\subsection{Spin/parity \texorpdfstring{$0^-$}{0-minus}}
A degree of freedom $0^-$ is carried by a totally antisymmetric rank three tensor. Let's consider the following Lagrangian
\bea
{\cal L} \!&=&\!-\frac{1}{8} H_{\mu\nu\rho\sigma} H^{\mu\nu\rho\sigma} -\frac{1}{2} m^2 B_{\mu\nu\rho} B^{\mu \nu\rho} 
+J^{\mu\nu\rho} B_{\mu\nu\rho}\ , \nn
H_{\mu\nu\rho \sigma} \!&=&\! 4 \partial_{[\mu} B_{\nu \rho \sigma]} \,,
\eea
where $B$ is totally antisymmetric.
The Fourier transform of the Lagrangian is
\bea
{\cal L} \!&=&\! 
-\frac{1}{2}B^{\mu\nu\rho} 
\left[\left(q^2+ m^2\right) P\left(0^-\right)
+m^2 P_{33}\left(1^+\right) \right]{}_{\mu \nu \rho}{}^{\alpha\beta\gamma} B_{\alpha\beta\gamma} +J^{\mu\nu\rho} B_{\mu\nu\rho} \,,
\eea
and the Fourier transformed field equation takes the form
\be
\left[\left(q^2+ m^2\right) P\left(0^-\right)
+m^2 P_{33}\left(1^+\right) \right] \cdot B= J\ .
\ee
We see that in the absence of sources
the $1^+$ component vanishes on shell.
The saturated propagator is
\be
\Pi = \frac{1}{2}\int \frac{d^4q}{(2\pi)^4}\ J(-q) \cdot\left[ \frac{1}{q^2+m^2} P\left(0^-\right) + \frac{1}{m^2} P_{33}\left(1^+\right) \right] \cdot J(q)\ .
\ee
Substituting the expressions for $P\left(0^-\right)$ and $P_{33}\left(1^+\right)$ gives
\bea
\Pi \!&=&\! \frac{1}{2}\int \frac{d^4q}{(2\pi)^4}\ J^{\mu\nu\rho}(-q)\,\frac{P(0^-,m^2){}_{\mu\nu\rho}{}^{\alpha\beta\gamma}}{q^2+m^2} \,J_{\alpha\beta\gamma}(q) \ ,
\eea
where
\be
P(0^-,m^2){}_{\mu\nu\rho}{}^{\alpha\beta\gamma} 
\equiv P\left(0^-\right){}_{\mu\nu\rho}{}^{\alpha\beta\gamma}  \Big|_{q^2 \to -m^2}
=
\left( \delta_{[\mu}^{[\alpha} \delta_{\nu}^{\beta} \delta_{\rho]}^{\gamma]}
+3 \frac{q_{[\mu} q^{[\alpha} \delta_{\nu}^{\beta}\delta_{\rho]}^{\gamma]}}{m^2} \right)\ .
\ee

Turning to massless case, the saturated propagator is then given by
\be
\Pi = \frac{1}{2} \int \frac{d^4q}{(2\pi)^4}\, J (-q)\cdot \frac{1}{q^2}\, P\left(0^-\right) \cdot J(q)\ ,
\ee
and the source obeys the constraint
\be
P_{33}\left(1^+\right)\cdot J=0\ .
\ee
Using this constraint for the saturated propagator, one readily finds that
\bea
\Pi  \!&=&\! \frac{1}{2}\int \frac{d^4q}{(2\pi)^4}\, J (-q)\cdot \frac{1}{q^2}\,  P\left(0^-\right) \cdot J(q) 
\nn
\!&=&\! \frac{1}{2} \int \frac{d^4q}{(2\pi)^4}\, J^{\mu\nu\rho} (-q) \frac{1}{q^2}\, \eta_{\mu\alpha}\eta_{\nu\beta}\eta_{\rho\gamma}\, J^{\alpha\beta\gamma}(q)\,.
\eea
In the massless case the projector $P(1^+)_{33}$ is absent.
This is related to the fact that the Lagrangian is 
then invariant under the gauge transformation 
\bea
&& B_{\mu\nu\rho} \to B_{\mu\nu\rho} + \partial_{[\mu} \Omega_{\nu\rho]} \,,
\eea
where $\Omega_{\nu\rho} = -\Omega_{\rho \nu}$ and $\partial_\nu \Omega^{\nu\rho}=0$. 
This makes the $1^+$ components gauge degrees of freedom.

\subsection{Spin/parity \texorpdfstring{$1^-$}{1-minus}}
Given the Proca Lagrangian
\be
{\cal L} = -\frac14 F_{\mu\nu} F^{\mu\nu} -\frac12 m^2 A_\mu A^\mu +J^\mu A_\mu\ ,
\ee
the Fourier transformed field equation takes the form
\be
\left[ ( q^2+m^2 ) T +m^2 L \right] \cdot A= J\ .
\ee
It follows that the saturated propagator is
\be
\Pi = \frac{1}{2}\int \frac{d^4q}{(2\pi)^4}\ J(-q) \cdot\left[ \frac{1}{q^2+m^2} T + \frac{1}{m^2} L \right] \cdot J(q)\ .
\ee
Substituting the expressions for $T$ and $L$ gives
\bea
\Pi \!&=&\! \frac{1}{2}\int \frac{d^4q}{(2\pi)^4}\ J^\mu(-q)\,\frac{P\left(1^-,m^2 \right)_{\mu\nu}}{q^2+m^2} \,J^\nu(q) \ ,
\eea
where
\be
P\left(1^-,m^2 \right){}_{\mu\nu} \equiv T{}_{\mu\nu}\Big|_{q^2 \to -m^2}
= \left(\eta_{\mu\nu} + \frac{q_\mu q_\nu}{m^2} \right)\ .
\label{Ponemass}
\ee

Turning to the massless case, the saturated propagator is then given by
\be
\Pi = \frac{1}{2} \int \frac{d^4q}{(2\pi)^4}\, J (-q)\cdot \frac{1}{q^2}\,T \cdot J(q)\ ,
\ee
and the source obeys the constraint
\be
q\cdot J=0\ .
\ee
Using this constraint for the saturated propagator, one readily finds that
\bea
\Pi  \!&=&\! \frac{1}{2}\int \frac{d^4q}{(2\pi)^4}\, J (-q)\cdot \frac{1}{q^2}\,  T \cdot J(q) 
\nn
\!&=&\! \frac{1}{2} \int \frac{d^4q}{(2\pi)^4}\, J^{\mu} (-q) \frac{1}{q^2}\, \eta_{\mu\nu}\, J^{\nu}(q)\,.
\eea
The source constraint above, in the massless case is related to invariance under the gauge transformation $A_\mu \to A_\mu +\partial_\mu \phi $. 

\subsection{Spin/parity \texorpdfstring{$1^+$}{1-plus}}
Since a degree of freedom $1^+$ is contained in an 
antisymmetric rank-2 tensor $B_{\mu\nu}$, let's consider the following Lagrangian
\bea
{\cal L} \!&=&\!-\frac{1}{6} H_{\mu\nu\rho} H^{\mu\nu\rho} -\frac12 m^2 B_{\mu\nu} B^{\mu \nu} 
+J^{\mu\nu} B_{\mu\nu}\ , \nn
H_{\mu\nu\rho} \!&=&\! 3 \partial_{[\mu} B_{\nu \rho]} \ .
\eea
Using the spin projectors of Appendix~\ref{sec:pro12sa},
the Fourier transform of the Lagrangian is
\bea
{\cal L} \!&=&\! 
-\frac{1}{2}B^{\mu\nu} 
\left[\left(q^2+ m^2\right) P_a\left(1^+\right)
+m^2 P_a\left(1^-\right) \right]{}_{\mu \nu}{}^{\rho \sigma} B_{\rho \sigma} +J^{\mu\nu} B_{\mu\nu}
\,,
\eea
the Fourier transformed field equation takes the form
\be
\left[ ( q^2+m^2 ) P_a\left(1^+\right) +m^2 P_a\left(1^-\right) \right] \cdot B= J\ .
\ee
We see that in the absence of sources the $1^-$
component vanishes on shell.
The saturated propagator is
\be
\Pi = \frac{1}{2}\int \frac{d^4q}{(2\pi)^4}\ J(-q) \cdot\left[ \frac{1}{q^2+m^2} P_a\left(1^+\right) + \frac{1}{m^2} P_a\left(1^-\right) \right] \cdot J(q)\ .
\ee
Substituting the expressions for $P_a\left(1^+\right)$ and $P_a\left(1^-\right)$ gives
\bea
\Pi \!&=&\! \frac{1}{2}\int \frac{d^4q}{(2\pi)^4}\ J^{\mu\nu}(-q)\,\frac{P\left(1^+,m^2\right){}_{\mu\nu}{}^{\rho\sigma}}{q^2+m^2} \,J_{\rho\sigma}(q) \ ,
\eea
where
\be
P\left(1^+,m^2\right){}_{\mu\nu}{}^{\rho\sigma} 
\equiv P_a(1^+){}_{\mu\nu}{}^{\rho\sigma}\Big|_{q^2 \to -m^2}
=
\left( \delta_{[\mu}^{[\rho} \delta_{\nu]}^{\sigma]} 
+ 2\frac{\delta_{[\mu}^{[\rho}q_{\nu]} q^{\sigma]}}{m^2} \right)\ .
\ee

In the massless case, $P_a(1^-)$ drops out and 
the $1^-$ components become pure gauge.
This is related to invariance of the Lagrangian 
under the transformation
\be
B_{\rho\sigma} \to B_{\rho\sigma} + \partial_{[\rho} \xi_{\sigma]} \,.
\ee
In this case the propagator becomes
\be
\Pi = \frac{1}{2} \int \frac{d^4q}{(2\pi)^4}\, J (-q)\cdot \frac{1}{q^2}\, P_{a}\left(1^+\right) \cdot J(q)\ ,
\ee
where the source obeys the constraint
\be
P_{a}\left(1^-\right)\cdot J=0\ .
\ee
Using this constraint, one readily finds that
\bea
\Pi  \!&=&\! \frac{1}{2}\int \frac{d^4q}{(2\pi)^4}\, J (-q)\cdot \frac{1}{q^2}\,  \, P_{a}\left(1^+\right) \cdot J(q) 
\nn
\!&=&\! \frac{1}{2} \int \frac{d^4q}{(2\pi)^4}\, J^{\mu\nu} (-q) \frac{1}{q^2}\, \eta_{\mu\rho}\eta_{\nu\sigma}\, J^{\rho\sigma}(q)\,.
\label{1+satprop}
\eea
A generic 2nd rank antisymmetric tensor in four dimension has $6$ independent components.
Since $\xi$ can be transverse or longitudinal, we can remove $2+3$ components and so there is only one physical degrees of freedom \cite{Kalb:2009fj}.

\subsection{Spin/parity \texorpdfstring{$2^+$}{2-plus}} 
\label{2plus}
The massive Fierz-Pauli Lagrangian with a source is
\be
{\cal L} = -\frac12 h^{\mu\nu} \Box h_{\mu\nu} + \frac12 h\Box h +(\partial_\mu h^{\mu\nu}) \partial_\nu h - (\partial_\mu h^{\mu\nu}) \partial^\rho h_{\nu\rho} + \frac12 m^2 (h_{\mu\nu} h^{\mu\nu} -h^2) + J^{\mu\nu} h_{\mu\nu}\ .
\ee
Using the spin projectors written in Appendix~\ref{sec:pro12sa},
the Fourier transform of the Lagrangian is
\bea
\cL&=&
\frac{1}{2}(q^2+m^2) h^{\mu\nu} 
\left[ P_s(2^+)-2 P_s(0^+,ss)
\right] {}_{\mu \nu}{}^{\rho \sigma} h_{\rho \sigma}
\nn
&&\qquad
+\frac{1}{2}m^2 h^{\mu\nu} \left[ P_s(1^-)-{\sqrt 3} P_s(0^+,sw) -{\sqrt 3} P_s(0^+,ws) \right]{}_{\mu \nu}{}^{\rho \sigma} h_{\rho \sigma} 
+J^{\mu\nu} h_{\mu\nu} \,,
\nonumber
\eea
the resulting Fourier transformed field equation is 
\be
 \left[(q^2\!+\!m^2)\left(P_s(2^+)\!-\!2P_s(0^+,ss) \right)
\! +\!m^2 \left(P_s(1^-)\!-\!{\sqrt 3} (P_s(0^+,sw) \!+\! P_s(0^+,ws)) \right) \right] \cdot h = -J \ .
\ee
Thus the saturated propagator is
\bea
\Pi \!&=&\! -\frac{1}{2}\int \frac{d^4q}{(2\pi)^4}\, J (-q) \cdot \frac{1}{q^2+m^2} \Big[  P_s(2^+)
+ \frac{q^2+m^2}{m^2} P_s(1^-) + \frac{2(q^2+m^2)^2}{3m^4}  P_s(0^+,ww)
\nn
&&  -\frac{\left(q^2+m^2\right)}{ \sqrt{3}\,m^2} P_s(0^+,sw) -\frac{\left(q^2+m^2\right)}{ \sqrt{3}\,m^2}  P_s(0^+,ws) \Big] \cdot J(q) \,.
\eea
Substituting the expressions for the spin projection operators gives 
\be
\Pi = - \frac{1}{2}\int \frac{d^4q}{(2\pi)^4}\, J^{\mu\nu} (-q)  \frac{1}{q^2+m^2} P(2^+,m^2){}_{\mu\nu}{}^{\rho\sigma}  J_{\rho\sigma}(q)\ ,
\ee
where
\bea
P(2^+,m^2){}_{\mu\nu}{}^{\rho\sigma} &\equiv& P_s(2^+){}_{\mu\nu}{}^{\rho\sigma}\Big|_{q^2 \to -m^2}
\\
\!&=&\! P(1^-,m^2)_{(\mu}{}^{(\rho} P(1^-,m^2)_{\nu)}{}^{\sigma)}
-\frac13 P(1^-,m^2)_{\mu\nu} P(1^-,m^2)^{\rho\sigma} \,,
\nonumber
\eea
where we used definition (\ref{Ponemass}).

Turning to the massless case, the saturated propagator is then given by
\be
\Pi = -\frac{1}{2} \int \frac{d^4q}{(2\pi)^4}\, J (-q)\cdot \frac{1}{q^2}\left( P_s(2^+) - \frac12 P_s(0^+,ss)\right) \cdot J(q)\ ,
\label{m2b}
\ee
and the sources obey the constraints
\be
P_s(1^-)\cdot J=0\ ,\qquad P_s(0^+,ww)\cdot J=0\ ,\qquad P_s(0^+,sw)\cdot J=0\ .
\ee
It is possible to show that these constraints are equivalent to $q^\mu J_{\mu\nu}(q)=0$. 
Using this constraint for the saturated propagator, one readily finds that
\bea
\Pi  \!&=&\! -\frac{1}{2}\int \frac{d^4q}{(2\pi)^4}\, J (-q)\cdot \frac{1}{q^2}\, \left(P_s(2^+) -\frac12 P_s(0^+,ss) \right)\cdot J(q) 
\nn
\!&=&\! -\frac{1}{2} \int \frac{d^4q}{(2\pi)^4}\, J^{\mu\nu} (-q) \frac{1}{q^2}\left( \eta_{\mu\rho}\eta_{\nu\sigma}-\frac12 \eta_{\mu\nu}\eta_{\rho\sigma}\right) J^{\rho\sigma}(q)\,.
\label{m2c}
\eea
Note that the absence of $P_s(1^-)$ and $P_s(0^+,ww)$ is connected to the invariance under diffeomorphism
\be
h_{\mu\nu} \to h_{\mu \nu} + \partial_{(\mu} \xi_{\nu)} \,.
\ee
Note also that the ratio between coefficients of $P_s(2^+)$ and $P_s(0^+,ss)$ is $-2$ : 
this particular choice is the only one that ensures locality of the Lagrangian.
This means that we cannot have a local theory where only $P_s(2^+)$ appears.
To show this, one can consider the following EOM
\be
 q^2\left\{P_s(2^+)+c_1\,P_s(0^+,ss) +c_2\, P_s(1^-) +c_3\,P_s(0^+,ww) \right\} \cdot h = -J \,.
\ee
Invariance under diffeomorphism implies $c_2=c_3 =0$. 
Then we can expand the remaining spin projectors 
\bea
&&q^2\,h_{\mu\nu} +\frac{(c_1-1)}{3}\left(\eta_{\mu\nu}q^2-q_\mu q_\nu \right) h
- q_\mu q_\lambda h_\nu^\lambda - q_\nu q_\lambda h_\mu^\lambda
- \frac{(c_1-1)}{3} \eta_{\mu\nu}\,q_\rho q_\sigma h^{\rho\sigma}
\nn
&&\qquad+\frac{(c_1+2)}{3} \frac{q_\mu q_\nu}{q^2} q_\rho q_\sigma h^{\rho\sigma}
= -J_{\mu\nu} \,.
\eea
Now it is clear that only for $c_1=-2$ we have a local EOM.

The presence of $P_s(0^+,ss)$ does not imply that in such a theory there are propagating states with spin $0^+$.\\
Starting from a generic symmetric tensor we have $10$ independent components, then diffeomorphism invariance reduces them to $1+5$, where the $1$ is coming from $0^+$ and $5$ from $2^+$ .\\
Residual gauge consists on transverse $\partial_{(\mu} \xi_{\nu)}$ 
\bea
P_s(2^+)_{\mu\nu}{}^{\rho\sigma} \, q_{(\rho} \xi_{\sigma)} \!&=&\! 
q_{(\mu} \xi_{\nu)} -\frac{2}{3} T_{\mu\nu} q_\rho \xi^\rho \,,
\nn
P_s(0^+,ss)_{\mu\nu}{}^{\rho\sigma} \, q_{(\rho} \xi_{\sigma)} \!&=&\! 
\frac{2}{3} T_{\mu\nu} q_\rho \xi^\rho \,,
\nonumber
\eea
and it can remove $1+3$ components.
So only $2$ degrees of freedom survive and they correspond to the two polarizations of $2^+$ spin states.

\subsubsection{Changing the mass term}
Let's insert a parameter $b$ inside the mass term:
$\frac12 m^2 (h_{\mu\nu} h^{\mu\nu} - b \,h^2)$.
Now the Fourier transform of the Lagrangian is
\bea
{\cal L} &=& 
\frac{1}{2}(q^2+m^2) h \cdot P_s(2^+) \cdot h
-\left(q^2+\frac{(3b-1)}{2}m^2 \right) h \cdot 
P_s(0^+,ss) \cdot h
\nn
&&
+\frac{1}{2}m^2 h \cdot \left[ P_s(1^-) +(1-b)\, P_s(0^+,ww)
-b\,{\sqrt 3} (P_s(0^+,sw) +P_s(0^+,ws)) \right]\cdot h
+J\cdot h \,.
\nonumber
\eea
Thus the saturated propagator is
\bea
\Pi \!&=&\! -\frac{1}{2}\int \frac{d^4q}{(2\pi)^4}\, J (-q) \cdot \frac{1}{q^2+m^2} \Big\{  P_s(2^+)
+ \frac{q^2+m^2}{m^2} P_s(1^-) 
\nn
&&+ \frac{(q^2+ m^2)}{2(1-b)\,q^2 +(7b-3b^2-1)\,m^2} \left[
\frac{2(q^2+ \frac{(3b-1)}{2} m^2)}{m^2}  P_s(0^+,ww)
- (1-b)  P_s(0^+,ss) \right]
\nn
&&  -\frac{b \sqrt{3}\left(q^2+m^2\right)}{2(1-b)\,q^2 +(7b-3b^2-1)\,m^2} 
\left[ P_s(0^+,sw) +  P_s(0^+,ws) \right]
 \Big\} \cdot J(q) \,.
\nonumber
\eea
Substituting the expressions for the spin projection operators gives 
\be
\Pi = - \frac{1}{2}\int \frac{d^4q}{(2\pi)^4}\, J (-q)\cdot  \left[ 
\frac{1}{q^2+m^2} P(2^+,m^2) 
-\frac{1}{6} \frac{1}{q^2 +\left(\frac{7b-3b^2-1}{2(1-b)}\right)\,m^2} P(0^+,m^2) 
\right]\cdot J(q)\ ,
\ee
where
\bea
P(0^+,m^2) &\equiv& 
P(0^+,ss)
+3 \left(\frac{4q^4 + 2(3b-1)m^2\,q^2+(6b+1)m^4 }{m^4}\right) P(0^+,ww)
\nn
&&
-\sqrt{3} \left(\frac{2q^2+(3b-1)\,m^2}{m^2} \right) \left[ P(0^+,sw) + P(0^+,ws) \right]
\,.
\nonumber
\eea
We see that in general the theory propagates also a scalar ghost.
Sending $b \to 1$, in the second term the mass diverges and the ghost decouples.

\subsection{Spin/parity \texorpdfstring{$2^-$}{2-minus}}
\label{2minus}
We see from Table \ref{t1} that a degree of freedom $2^-$ is contained either in a hook anti-symmetric or a hook symmetric 3rd rank tensor.
However, such tensor contains more than just the $2^-$, so we have to impose some constraints.
For the massless case we impose gauge invariances to kill the undesired degrees of freedom. This, plus locality, uniquely pins down the action.
Then, to get the massive case we add a mass term to the massless case and we impose the following requirement:
\be
P(2^-,m^2) \equiv 
P_{11}(2^-)\Big|_{q^2 \to -m^2}
\ee
for the symmetric case, and the same with $P(2^-)_{22}$
in the antisymmetric case.
By construction, the massless case is achieved by setting to zero the mass parameter.

\subsubsection{Using a hook-antisymmetric tensor}
\label{hookantisymtwominus}

We deal with the massless case first.
Form Table \ref{t1} the general Lagrangian compatible 
with hook-antisymmetry is
\bea
{\cal L}\!&=&\! -\frac{1}{2} B \cdot q^2\left[P_{22}\left(2^-\right)
+c_1 \,P_{33}\left(1^-\right)  
+ c_2\,P_{22}(1^+) +c_3\,P_{33}(2^+)+c_4\,P_{33}(0^+) 
 \right.
\nn
&&\left. 
+c_5 \,P_{66}(1^-)
+ c_6 \, \left( P_{36}(1^-) + P_{63}(1^-) \right)
\right]\cdot B + J\cdot B \,.
\nonumber
\eea
Then we require the following invariances
\bea
B_{\nu \rho \sigma} &\to& B_{\nu \rho \sigma} +2 \partial_\nu \Omega_{\rho \sigma}
-\partial_\rho \Omega_{ \sigma\nu}-\partial_\sigma \Omega_{\nu \rho}\,
\implies\, c_2=0 \,,
\\
B_{\nu \rho \sigma} &\to& B_{\nu \rho \sigma} + \partial_\sigma \psi_{\nu\rho} 
- \partial_\rho \psi_{\nu\sigma}\,
\implies\, c_3=c_4=0 \,,
\\
B_{\nu \rho \sigma} &\to& B_{\nu \rho \sigma} + \partial_\nu \partial_\rho \xi_\sigma
- \partial_\nu \partial_\sigma \xi_\rho \,
\implies\, c_5=c_6=0 \,,
\eea
where $\Omega$ is antisymmetric and transverse,
$\psi$ is symmetric and transverse, $\xi$ is transverse.
These transformations are constructed starting from the 
properties of the field they are designed to kill.
We can expand the remaining spin projectors to obtain
\bea
{\cal L}\!&=&\!-\frac{1}{3} q^2 B_{\nu \rho \sigma} B^{\nu \rho \sigma}
-\frac{1}{3} q^2 B_{\nu \rho \sigma}  B^{\rho \nu \sigma}
-\frac{(c_1-1)}{2} q^2 B_\alpha{}^\alpha{}_\nu B_\beta{}^\beta{}^\nu
+\frac{1}{3} q_\mu q_\nu B^{\mu}{}_{\rho\sigma}  B^{\nu\rho\sigma}
\nn
&&+\frac{2}{3} q_\mu q_\nu B^{\mu}{}_{\rho\sigma}  B^{\rho\nu\sigma}
+\frac{2}{3} q_\mu q_\nu B_\rho{}^\mu{}_\sigma  B^{\rho\nu\sigma}
+\frac{1}{3} q_\mu q_\nu B_\rho{}^\mu{}_\sigma  B^{\sigma\nu\rho}
+\frac{(c_1-1)}{2} q_\mu q_\nu B^{\alpha}{}_\alpha{}^\mu  B^{\beta}{}_\beta{}^\nu
\nn
&&
-(c_1-1) q_\mu q_\nu B^{\mu\nu}{}_\rho  B^{\alpha}{}_\alpha{}^\rho
-\frac{(c_1+1)}{2q^2} q_\mu q_\nu q_\rho q_\sigma B^{\mu\nu\lambda}B^{\rho\sigma}{}_\lambda
+J^{\nu\rho\sigma}B_{\nu \rho \sigma} \,.
\eea
It is clear that only for $c_1=-1$ we have locality.
Therefore, the massless Lagrangian is
\bea
&&{\cal L} = -\frac{1}{3}\partial_\mu B_{\nu \rho \sigma} \partial^\mu B^{\nu \rho \sigma}
-\frac{1}{3}\partial_\mu B_{\nu \rho \sigma} \partial^\mu B^{\rho \nu \sigma}
+ \partial_\mu B^\alpha{}_\alpha{}_\nu \partial^\mu B_\beta{}^\beta{}^\nu
+\frac{1}{3} \partial_\mu B^{\mu}{}_{\rho\sigma} \partial_\nu B^{\nu\rho\sigma}
\nn
&&
+\frac{2}{3} \partial_\mu B^{\mu}{}_{\rho\sigma} \partial_\nu B^{\rho\nu\sigma}
+\frac{2}{3} \partial_\mu B_\rho{}^\mu{}_\sigma \partial_\nu B^{\rho\nu\sigma}
+\frac{1}{3} \partial_\mu B_\rho{}^\mu{}_\sigma \partial_\nu B^{\sigma\nu\rho}
- \partial_\mu B^{\alpha}{}_\alpha{}^\mu \partial_\nu B^{\beta}{}_\beta{}^\nu
\nn
&&
+2 \partial_\mu B^{\mu\nu}{}_\rho \partial_\nu B^{\alpha}{}_\alpha{}^\rho
+J^{\mu\nu\rho} B_{\mu\nu\rho}\, ,
\eea
and its Fourier transform can be written simply
\bea
{\cal L} \!&=&\! 
-\frac{1}{2}B \cdot 
\,q^2\left[P_{22}\left(2^-\right)
-P_{33}\left(1^-\right) \right] \cdot B 
+J \cdot B\,.
\eea

Now let's consider the massive case.
We add to the massless case a generic mass term
\bea \label{eq:2minusmassgen}
{\cal L}\!&=&\! -\frac{1}{2} B \cdot q^2\left\{ P_{22}\left(2^-\right) 
-\,P_{33}\left(1^-\right)  \right\} \cdot B+ J\cdot B 
\\
&&
-\frac{m^2}{2} B \cdot \left\{ P_{22}\left(2^-\right) 
+ a_1\,P_{33}\left(1^-\right)  
+ a_2\,P_{22}(1^+) +a_3\,P_{33}(2^+)+a_4\,P_{33}(0^+) +a_5 \,P_{66}(1^-)
 \right.
\nonumber\\
&&\left. 
\quad+ a_6 \, \left( P_{36}(1^-) + P_{63}(1^-) \right)
\right\} \cdot B \,.
\nonumber
\eea
Then, the saturated propagator is 
\bea
&&\Pi = \frac{1}{2}\int \frac{d^4q}{(2\pi)^4}\, J (-q) \cdot \frac{1}{q^2+m^2} \Big[  P_{22}(2^-)
\!+\! \frac{q^2+m^2}{m^2} \left( \frac{1}{a_3} P_{33}\left(2^+\right)
\!+\!\frac{1}{a_4} P_{33}\left(0^+\right) 
\!+\! \frac{1}{a_2} P_{22}(1^+)\right)
\nn
&&  + \frac{q^2+m^2}{a_5\,(q^2-a_1\,m^2)+a_6^2\,m^2} \left(\! \frac{q^2-a_1\,m^2}{m^2}\,P_{66}(1^-)\!-\!a_5 P_{33}(1^-)
\!+\! a_6\left(P_{36}(1^-) \!+\!  P_{63}(1^-) \right) \right) \Big] \cdot J(q) \,.
\nonumber
\eea
At this point we define the term inside the square bracket as $P_{ha}(2^-,m^2)$ and we require 
\be
P_{ha}(2^-,m^2) \equiv 
P_{22}(2^-)\Big|_{q^2 \to -m^2} \,,
\ee
obtaining
\begin{align}
 & a_1=-1\,,\,\,\,\,a_2 = 1\,,\,\,\,\,  a_3 = 1\,,\,\,\,\, a_4 = -2\,,\,\,\,\, a_5 = 0\,,\,\,\,\, a_6 = -\sqrt{2}\,.
\end{align}

Therefore, the massive Lagrangian for spin $2^-$ is
\bea
{\cal L} \!&=&\! -\frac{1}{3}\partial_\mu B_{\nu \rho \sigma} \partial^\mu B^{\nu \rho \sigma}
\!-\!\frac{1}{3}\partial_\mu B_{\nu \rho \sigma} \partial^\mu B^{\rho \nu \sigma}
\!+\! \partial_\mu B^\alpha{}_\alpha{}_\nu \partial^\mu B_\beta{}^\beta{}^\nu
\!+\!\frac{1}{3} \partial_\mu B^{\mu}{}_{\rho\sigma} \partial_\nu B^{\nu\rho\sigma}
\!+\!\frac{2}{3} \partial_\mu B^{\mu}{}_{\rho\sigma} \partial_\nu B^{\rho\nu\sigma}
\nn
&&
+\frac{2}{3} \partial_\mu B_\rho{}^\mu{}_\sigma \partial_\nu B^{\rho\nu\sigma}
+\frac{1}{3} \partial_\mu B_\rho{}^\mu{}_\sigma \partial_\nu B^{\sigma\nu\rho}
- \partial_\mu B^{\alpha}{}_\alpha{}^\mu \partial_\nu B^{\beta}{}_\beta{}^\nu
+2 \partial_\mu B^{\mu\nu}{}_\rho \partial_\nu B^{\alpha}{}_\alpha{}^\rho
\nn
&&
-\frac{1}{3}m^2 \left( B_{\nu\rho\sigma}B^{\nu\rho\sigma} +B_{\nu\rho\sigma}B^{\rho\nu\sigma}
- 3 B^{\alpha}{}_{\alpha\nu} B^{\beta}{}_\beta{}^\nu \right)
+J^{\mu\nu\rho} B_{\mu\nu\rho}\, ,
\eea
and its Fourier transform reads simply
\bea
{\cal L} \!&=&\! 
-\frac{1}{2}B \cdot 
\,(q^2 +m^2)
\left[  P_{22}\left(2^-\right)
-P_{33}\left(1^-\right) \right] \cdot B +J \cdot B
\\
&& -\frac{1}{2}
m^2 B \cdot \left[  P_{22}\left(1^+\right) + P_{33}\left(2^+\right)
-2 P_{33}\left(0^+\right)
-\sqrt{2}\left(P_{36}\left(1^-\right) + P_{63}\left(1^-\right)\right)\right] \cdot B 
\,,\nonumber
\eea
and the saturated propagator is
\be
\Pi =  \frac{1}{2}\int \frac{d^4q}{(2\pi)^4}\, J (-q)\cdot  \frac{1}{q^2+m^2} P_{ha}(2^-,m^2) \cdot  J(q)\ ,
\ee
where
\begin{align}
P_{ha}(2^-,m^2) \equiv 
P_{22}(2^-)\Big|_{q^2 \to -m^2} &= P_{22}(2^-)
+ \frac{q^2+m^2}{m^2} \left( P_{22}\left(1^+\right) + P_{33}\left(2^+\right)
-\frac{1}{2} P_{33}\left(0^+\right) \right) 
\nn
& + \frac{(q^2+m^2)^2}{2m^4}  P_{66}(1^-)
-\frac{\left(q^2+m^2\right)}{ \sqrt{2}\,m^2} \left( P_{36}(1^-) +  P_{63}(1^-) \right) \,.
\end{align}

The presence of $P_{33}(1^-)$ does not imply that in this theory there are propagating states with spin $1^-$.
A generic $(ha)$ 3rd rank tensor has $20$ independent components,
and the previous invariances reduce them to 8, of which 3 come from $1^-$ and 5 from $2^-$ .
The residual gauge invariance consists on transverse 
\bea
P_{22}(2^-)_{\alpha\beta\gamma}{}^{\nu\rho\sigma} \,( 2 q_\nu \Omega_{\rho \sigma}
-q_\rho \Omega_{ \sigma\nu}-q_\sigma \Omega_{\nu \rho} ) \!&=&\! 
2 q_\alpha \Omega_{\beta\gamma}
-q_\beta \Omega_{ \gamma\alpha}-q_\gamma \Omega_{\alpha \beta} \,,
\nn
P_{22}(2^-)_{\alpha\beta\gamma}{}^{\nu\rho\sigma}\, ( q_\sigma \psi_{\nu\rho} 
- q_\rho \psi_{\nu\sigma} ) \!&=&\! q_\gamma \psi_{\alpha\beta} 
- q_\beta \psi_{\alpha\gamma} \,,
\nn
P_{22}(2^-)_{\alpha\beta\gamma}{}^{\nu\rho\sigma} \,( q_\nu q_\rho \xi_\sigma
- q_\nu q_\sigma \xi_\rho ) \!&=&\! 
q_\alpha q_\beta \xi_\gamma -q_\alpha q_\gamma \xi_\beta
-q^2\,T_{\alpha[\beta}  \xi_{\gamma]} \,,
\nn
P_{33}(1^-)_{\alpha\beta\gamma}{}^{\nu\rho\sigma} \,( 2 q_\nu \Omega_{\rho \sigma}
-q_\rho \Omega_{ \sigma\nu}-q_\sigma \Omega_{\nu \rho} ) \!&=&\! 0 \,,
\nn
P_{33}(1^-)_{\alpha\beta\gamma}{}^{\nu\rho\sigma} \, ( q_\sigma \psi_{\nu\rho} 
- q_\rho \psi_{\nu\sigma} ) \!&=&\! 0 \,,
\nn
P_{33}(1^-)_{\alpha\beta\gamma}{}^{\nu\rho\sigma} \, ( q_\nu q_\rho \xi_\sigma
- q_\nu q_\sigma \xi_\rho ) \!&=&\! q^2\,T_{\alpha[\beta}  \xi_{\gamma]} \,,
\nonumber
\eea
and it removes $1+1+1+3$ components.
So only $2$ degrees of freedom survive, corresponding to the two polarizations of $2^-$ spin states.

\subsubsection{Using a hook symmetric tensor}
\label{hooksymtwominus}
For the massless case, we write a general combination of
hook-symmetric projectors:
\bea
{\cal L}\!&=&\! -\frac{1}{2} B \cdot q^2
\left\{ P_{11}\left(2^-\right)
+c_1 \,P_{22}\left(1^-\right) + c_2\,P_{11}(1^+)
+c_3\,P_{22}(2^+)+c_4\,P_{22}(0^+) 
 +c_5 \,P_{55}\left(1^-\right)
\right.
\nn
&&\left. 
 +c_6 \left( P_{25}\left(1^-\right) + P_{52}\left(1^-\right) \right)
\right\} \cdot B + J\cdot B \,.
\nonumber
\eea
Then we require the following invariances
\bea
B_{\nu \rho \sigma} &\to& B_{\nu \rho \sigma} +2 \partial_\nu \psi_{\rho \sigma}
-\partial_\rho \psi_{ \sigma\nu}-\partial_\sigma \psi_{\nu \rho}\,\implies\,
c_3=c_4=0\,,
\\
B_{\nu \rho \sigma} &\to& B_{\nu \rho \sigma} + \partial_\sigma \Omega_{\nu\rho} 
+ \partial_\rho \Omega_{\nu\sigma}\,\implies\,
c_2=0\,,
\\
B_{\nu \rho \sigma} &\to& B_{\nu \rho \sigma} + \partial_\nu \partial_\rho \xi_\sigma
+ \partial_\nu \partial_\sigma \xi_\rho \,\implies\,
c_5=c_6=0\,,
\eea
where $\Omega$ is antisymmetric and transverse,
$\psi$ is symmetric and transverse, 
$\xi$ is transverse.
As before, note that these transformations are constructed starting from the symmetry properties of the field and are designed to kill the desired degrees of freedom.
Expand the remaining spin projectors we get
\bea
{\cal L}\!&=&\!\frac{1-c_1}{6} q^2 B_{\nu}{}^\alpha{}_\alpha  B^{\nu\beta}{}_\beta 
-\frac{1}{3}q^2 B_{\nu \rho \sigma}  B^{\nu \rho \sigma}
+\frac{1}{3}q^2 B_{\nu \rho \sigma} B^{\rho \nu \sigma}
- \frac{1-c_1}{3}q^2 B^\alpha{}_\alpha{}_\nu  B{}^\nu{}^\beta{}_\beta
\\
&&
+\frac{1-c_1}{6} q^2 B^\alpha{}_\alpha{}_\nu  B{}^\beta{}_\beta{}^\nu
+\frac{1}{3} q_\mu q_\nu B^{\mu}{}_{\rho\sigma} B^{\nu\rho\sigma}
-\frac{1-c_1}{6} q_\mu q_\nu B{}^\mu{}^\alpha{}_\alpha B{}^\nu{}^\beta{}_\beta
-\frac{2}{3} q_\mu q_\nu B^{\mu}{}_{\rho\sigma} B^{\rho\nu\sigma}
\nn
&&
+\frac{1-c_1}{3} q_\mu q_\nu B^{\mu\nu \rho} B_\rho{}^\alpha{}_\alpha
-\frac{1-c_1}{3} q_\mu q_\nu B^{\rho\mu\nu } B_\rho{}^\alpha{}_\alpha
+\frac{2}{3}q_\mu q_\nu B_{\rho}{}^\mu{}_{\sigma} B^{\rho\nu\sigma}
-\frac{1}{3}q_\mu q_\nu B_{\rho}{}^\mu{}_{\sigma} B^{\sigma\nu\rho}
\nn
&&
+(1-c_1)\left(\frac{1}{3}q_\mu q_\nu B^\mu{}^\alpha{}_\alpha B^\beta{}_\beta{}^\nu
-\frac{1}{6}q_\mu q_\nu B^\alpha{}_\alpha{}^\mu B^\beta{}_\beta{}^\nu
-\frac{1}{3}q_\mu q_\nu B^{\mu\nu\rho} B^\beta{}_\beta{}_\rho
+\frac{1}{3}q_\mu q_\nu B^{\rho\mu\nu}  B^\beta{}_\beta{}_\rho \right)
\nn
&&
+(c_1+1)\left(-\frac{1}{6q^2} q_\mu q_\nu q_\rho q_\sigma B^{\mu\nu\lambda}B^{\rho\sigma}{}_\lambda
+\frac{1}{3q^2} q_\mu q_\nu q_\rho q_\sigma B^{\mu\nu\lambda}B{}_\lambda{}^{\rho\sigma}
-\frac{1}{6q^2} q_\mu q_\nu q_\rho q_\sigma B^{\lambda\mu\nu}B{}_\lambda{}^{\rho\sigma}
\right)
\nn
&&
+J^{\nu\rho\sigma}B_{\nu \rho \sigma} \,.
\nonumber
\eea
Now it is clear that only for $c_1=-1$ we have locality.

Therefore, the massless Lagrangian for spin $2^-$ is
\bea
{\cal L} \!&=&\! 
\frac{1}{3} \partial_\mu B_{\nu}{}^\alpha{}_\alpha \partial^\mu B^{\nu\beta}{}_\beta 
-\frac{1}{3}\partial_\mu B_{\nu \rho \sigma} \partial^\mu B^{\nu \rho \sigma}
+\frac{1}{3}\partial_\mu B_{\nu \rho \sigma} \partial^\mu B^{\rho \nu \sigma}
- \frac{2}{3}\partial_\mu B^\alpha{}_\alpha{}_\nu \partial^\mu B{}^\nu{}^\beta{}_\beta
\\
&&
+\frac{1}{3} \partial_\mu B^\alpha{}_\alpha{}_\nu \partial^\mu B{}^\beta{}_\beta{}^\nu
+\frac{1}{3} \partial_\mu B^{\mu}{}_{\rho\sigma} \partial_\nu B^{\nu\rho\sigma}
-\frac{1}{3} \partial_\mu B{}_\mu{}^\alpha{}_\alpha \partial^\nu B{}^\nu{}^\beta{}_\beta
-\frac{2}{3} \partial_\mu B^{\mu}{}_{\rho\sigma} \partial_\nu B^{\rho\nu\sigma}
\nn
&&
+\frac{2}{3} \partial_\mu B^{\mu\nu \rho} \partial_\nu B_\rho{}^\alpha{}_\alpha
-\frac{2}{3} \partial_\mu B^{\rho\mu\nu } \partial_\nu B_\rho{}^\alpha{}_\alpha
+\frac{2}{3}\partial_\mu B_{\rho}{}^\mu{}_{\sigma} \partial_\nu B^{\rho\nu\sigma}
-\frac{1}{3}\partial_\mu B_{\rho}{}^\mu{}_{\sigma} \partial_\nu B^{\sigma\nu\rho}
\nn
&&
+\frac{2}{3}\partial_\mu B^\mu{}^\alpha{}_\alpha \partial_\nu B^\beta{}_\beta{}^\nu
-\frac{1}{3}\partial_\mu B^\alpha{}_\alpha{}^\mu \partial_\nu B^\beta{}_\beta{}^\nu
-\frac{2}{3}\partial_\mu B^{\mu\nu\rho} \partial_\nu B^\beta{}_\beta{}_\rho
+\frac{2}{3}\partial_\mu B^{\rho\mu\nu} \partial_\nu B^\beta{}_\beta{}_\rho
\nn
&&
+J^{\mu\nu\rho} B_{\mu\nu\rho}\, ,
\nonumber
\eea
and its Fourier transform is simply
\bea
{\cal L} \!&=&\! 
-\frac{1}{2}B \cdot 
\,q^2\left[  P_{11}\left(2^-\right)
-P_{22}\left(1^-\right)\right] \cdot B 
 +J \cdot B \,.
\eea

Now we come to the massive case.
We add to the massless Lagrangian a generic mass term
\bea
{\cal L} \!&=&\! 
-\frac{1}{2}B \cdot 
\,q^2\left[  P_{11}\left(2^-\right)
-P_{22}\left(1^-\right)\right] \cdot B 
\\
&& -\frac{m^2}{2} B \cdot
\left\{ P_{11}\left(2^-\right)
+a_1 \,P_{22}\left(1^-\right) + a_2\,P_{11}(1^+)
+a_3\,P_{22}(2^+)+c_4\,P_{22}(0^+) 
 +a_5 \,P_{55}\left(1^-\right)
\right.
\nn
&&\left. 
+a_6 \left( P_{25}\left(1^-\right) + P_{52}\left(1^-\right) \right)
\right\} \cdot B + J\cdot B \,.
\nonumber
\eea
We calculate the saturated propagator and we define $P_{hs}(2^-,m^2)$ 
\bea
\Pi \!&=&\! \frac{1}{2}\int \frac{d^4q}{(2\pi)^4}\, J (-q) \cdot \frac{1}{q^2+m^2} P_{hs}(2^-,m^2) \cdot J(q)\,.
\eea
Then we require 
\be
P_{hs}(2^-,m^2) =
P_{11}(2^-)\Big|_{q^2 \to -m^2} \,,
\ee
obtaining
\begin{align}
 & a_1=-1\,,\,\,\,\,a_2 = 1\,,\,\,\,\,  a_3 = 1\,,\,\,\,\, a_4 = -2\,,\,\,\,\, a_5 = 0\,,\,\,\,\, a_6 = -\sqrt{2}\,.
\end{align}

Therefore, the massive Lagrangian for spin $2^-$ is
\bea
{\cal L} \!&=&\! 
\frac{1}{3} \partial_\mu B_{\nu}{}^\alpha{}_\alpha \partial^\mu B^{\nu\beta}{}_\beta 
-\frac{1}{3}\partial_\mu B_{\nu \rho \sigma} \partial^\mu B^{\nu \rho \sigma}
+\frac{1}{3}\partial_\mu B_{\nu \rho \sigma} \partial^\mu B^{\rho \nu \sigma}
- \frac{2}{3}\partial_\mu B^\alpha{}_\alpha{}_\nu \partial^\mu B{}^\nu{}^\beta{}_\beta
\\
&&
+\frac{1}{3} \partial_\mu B^\alpha{}_\alpha{}_\nu \partial^\mu B{}^\beta{}_\beta{}^\nu
+\frac{1}{3} \partial_\mu B^{\mu}{}_{\rho\sigma} \partial_\nu B^{\nu\rho\sigma}
-\frac{1}{3} \partial_\mu B{}_\mu{}^\alpha{}_\alpha \partial^\nu B{}^\nu{}^\beta{}_\beta
-\frac{2}{3} \partial_\mu B^{\mu}{}_{\rho\sigma} \partial_\nu B^{\rho\nu\sigma}
\nn
&&
+\frac{2}{3} \partial_\mu B^{\mu\nu \rho} \partial_\nu B_\rho{}^\alpha{}_\alpha
-\frac{2}{3} \partial_\mu B^{\rho\mu\nu } \partial_\nu B_\rho{}^\alpha{}_\alpha
+\frac{2}{3}\partial_\mu B_{\rho}{}^\mu{}_{\sigma} \partial_\nu B^{\rho\nu\sigma}
-\frac{1}{3}\partial_\mu B_{\rho}{}^\mu{}_{\sigma} \partial_\nu B^{\sigma\nu\rho}
\nn
&&
+\frac{2}{3}\partial_\mu B^\mu{}^\alpha{}_\alpha \partial_\nu B^\beta{}_\beta{}^\nu
-\frac{1}{3}\partial_\mu B^\alpha{}_\alpha{}^\mu \partial_\nu B^\beta{}_\beta{}^\nu
-\frac{2}{3}\partial_\mu B^{\mu\nu\rho} \partial_\nu B^\beta{}_\beta{}_\rho
+\frac{2}{3}\partial_\mu B^{\rho\mu\nu} \partial_\nu B^\beta{}_\beta{}_\rho
\nn
&&
+\frac{1}{3}m^2 \left( B_\nu{}^{\alpha}{}_{\alpha} B{}^\nu{}^{\beta}{}_\beta -B_{\nu\rho\sigma}B^{\nu\rho\sigma} + B_{\nu\rho\sigma}B^{\rho\nu\sigma}
- 2 B^{\alpha}{}_{\alpha \nu} B{}^\nu{}^{\beta}{}_\beta
+  B^{\alpha}{}_{\alpha \nu} B^{\beta}{}_\beta{}^\nu \right)
+J^{\mu\nu\rho} B_{\mu\nu\rho}\, .
\nonumber
\eea
The Fourier transform of the Lagrangian can be written
\bea
{\cal L} \!&=&\! 
-\frac{1}{2}B \cdot 
\,(q^2+m^2)\left[  P_{11}\left(2^-\right)
-P_{22}\left(1^-\right)\right] \cdot B +J \cdot B
\\
&& -\frac{1}{2}
m^2 B \cdot \left[  P_{11}\left(1^+\right) + P_{22}\left(2^+\right)
-2 P_{22}\left(0^+\right)
-\sqrt{2}P_{25}\left(1^-\right) -\sqrt{2} P_{52}\left(1^-\right)\right] \cdot B  \,,
\nonumber
\eea
and the saturated propagator is
\be
\Pi =  \frac{1}{2}\int \frac{d^4q}{(2\pi)^4}\, J (-q)\cdot  \frac{1}{q^2+m^2} P_{hs}(2^-,m^2) \cdot  J(q)\ ,
\ee
where
\begin{align}
P_{hs}(2^-,m^2) \equiv 
P_{11}(2^-)\Big|_{q^2 \to -m^2} &= P_{11}(2^-)
+ \frac{q^2+m^2}{m^2} \left( P_{11}\left(1^+\right) + P_{22}\left(2^+\right)
-\frac{1}{2} P_{22}\left(0^+\right) \right) 
\nn
& + \frac{(q^2+m^2)^2}{2m^4}  P_{55}(1^-)
-\frac{\left(q^2+m^2\right)}{ \sqrt{2}\,m^2} \left( P_{25}(1^-) +  P_{52}(1^-) \right) \,.
\end{align}
%

\subsection{Spin/parity \texorpdfstring{$3^-$}{3-minus}} 

A quadratic Lagrangian that describes a single massive spin 3 field is due to Singh and Hagen \cite{Singh:1974qz}, 
and while it can be expressed in different ways by using different set of auxiliary fields, on-shell they are all equivalent. Let us consider the formulation that requires the minimal number of auxiliary field, namely a single real scalar field, which in four dimensional spacetime is given by 
\bea
{\cal L} \!&=&\! -\frac12 \Big[ (-\Phi^{\mu\nu\rho} \Box \Phi_{\mu\nu\rho}  +3 \Phi^\mu \Box \Phi_\mu + 6 \left( \partial_\rho \Phi^{\mu\nu\rho} \right)\left( \partial_\mu \Phi_\nu\right)  - 3 \left(\partial_\rho \Phi^{\mu\nu\rho} \right) \left( \partial^\sigma \Phi_{\mu\nu\sigma}\right)-\frac32 \left(\partial_\mu \Phi^\mu\right)^2\, \Big]
\nn
&& +\phi(-\Box+ 4 m^2) \phi + m \Phi^\mu \partial_\mu \phi -\frac12 m^2 (\Phi_{\mu\nu\rho}\Phi^{\mu\nu\rho} -3 \Phi_\mu \Phi^\mu ) + J\cdot \Phi + j\cdot \phi \ ,
\label{ms3L}
\eea
where $\Phi_{\mu\nu\rho}= \Phi_{(\mu\nu\rho)}$ and $\Phi_\mu \equiv \Phi_{\mu\nu\rho} \eta^{\nu\rho}$. The first line describes the massless spin 3 field. 

We can first solve for $\phi$ through its EOM and back substitute the result into the action. This produces the term $- \frac14 m^2(\partial_\mu\Phi^\mu) (-\Box+4m^2)^{-1}(\partial_\nu\Phi^\nu)$. Substituting this back into the action, Fourier transforming and expressing the result in terms of the spin projection operators gives \cite{Mendonca:2019gco}
\bea
S \!&=&\!-\frac12 \int \frac{d^4q}{(2\pi)^4}\, \Big( \Phi \cdot \Big[(q^2+m^2)P(3^-) +m^2 P_{11}(2^+) -4(q^2+m^2) P_{11}(1^-) -{\sqrt 5}m^2 P_{14}(1^-)
\nn
&&-{\sqrt 5}m^2 P_{41}(1^-) - \left(\frac92 q^2+ 2m^2   -\frac{m^2 q^2}{2(q^2+4m^2)}\right) P_{11}(0^+)
\nn
&& - \left(\frac12 q^2+ 2m^2   -\frac{ m^2 q^2}{2(q^2+4m^2)}\right)P_{44}(0^+)
-3 \left( (\frac{q^2}{2}+m^2)-\frac{m^2q^2}{6(q^2+4m^2)}\right) P_{14}(0^+)
\nn
&& -3 \left( (\frac{q^2}{2}+m^2)-\frac{m^2q^2}{6(q^2+4m^2)}\right) P_{41}(0^+) \Big] \cdot \Phi + J\cdot \Phi\Big)\ .
\eea
This gives the saturated propagator
\bea
\Pi \!&=&\! \int \frac{d^4q}{(2\pi)^4}\, J \cdot \Big[ \frac{1}{q^2+m^2} P(3^-) + \frac{1}{m^2} P_{11}(2^+) +\frac{4}{5m^4} (q^2+m^2)  P_{44}(1^-)
\nn
&& -\frac{1}{{\sqrt 5} m^2} P(1^-)_{14} -\frac{1}{{\sqrt 5} m^2} P_{41}(1^-) +\frac{1}{40m^6} \left( (q^2)^2 +7 m^2 q^2+16m^4\right) P_{11}(0^+)
\nn
&& +\frac{1}{40m^6} \left(9 (q^2)^2 +39 m^2 q^2+16 m^4\right) P_{22}(0^+)
\nn
&& -\frac{1}{40m^6} (q^2+3m^2)(3q^2+8 m^2)\left( P_{14}(0^+)+P_{41}(0^+) \right) \Big] \cdot J\ .
\eea
Substituting the expressions for the spin projection operators gives (see, for example, \cite{Huang:2004we})
\bea
\Pi \!&=&\! \int \frac{d^4q}{(2\pi)^4}\, J \cdot \frac{P(3^-,m^2)}{q^2+m^2} \cdot J
\label{ms3}
\eea
where
\bea
P(3^-,m^2)&&^{\mu\nu\rho}{}_{\lambda\tau\sigma} 
\equiv P(3^-)^{\mu\nu\rho}{}_{\lambda\tau\sigma} \Big|_{q^2 \to -m^2}
\\
\!&=&\! P(1^-,m^2)_{\mu}{}^{\lambda}P(1^-,m^2)_{\nu}{}^{\tau}P(1^-,m^2)_{\rho}{}^{\sigma} -\frac35 P(1^-,m^2)_{\mu\nu} P(1^-,m^2)^{\lambda\tau} P(1^-,m^2)_{\rho}{}^{\sigma} \ ,
\nonumber
\eea
where we used definition (\ref{Ponemass}).

The Lagrangian for the massless spin 3 field is obtained bu setting $m=0$ and $\phi=0$ in \eqref{ms3L}. Then, Fourier transformed action in terms of spin projection operators is 
\bea
S \,\,&=&\,\,  \frac12 \int \frac{d^4q}{(2\pi)^4} \left( q^2 \Phi\cdot
\left[-P(3^-)+4P_{11}(1^-)+\frac92 P_{11}(0^+)+\frac32 P_{41}(0^+)+\frac32 P_{14}(0^+)
\right.\right.\nn
&& \left.\left.\vspace{2cm}+\frac12 P_{44}(0^+) \right]\cdot \Phi 
+ J\cdot \Phi\right) \ .
\label{Fronsdal2}
\eea
It readily follows that this implies the following source constraints as a result of the spin 3 gauge invariance:
\be
P_{11}(2^+)\cdot J=0\ ,\quad P_{44}(1^-) \cdot J=0\ ,\qquad P_{14}(1^-)\cdot J=0\ , \quad \left[ P_{11}(0^+)-3P_{44}(0^+)\right]\cdot J= 0\ .
\ee
Using a gauge in which $P_{11}(0^+)\cdot \Phi=0$ and $P_{41}(0^+) \cdot \Phi=0$, the saturated propagator takes the form
\be
\Pi = \int \frac{d^4q}{(2\pi)^4}\,  \frac{1}{q^2}\, J\cdot \Big[ -P(3^-,q) +\frac14 P_{11}(1^-,q) + 2 P_{44}(0^+,q) \Big] \cdot J \,.
\ee 
Substituting the expressions for the spin projectors and assuming that the source constraints are equivalent to the condition $q^\mu J_{\mu\nu\rho} -{\rm trace}=0$, one finds
\bea
\Pi \!&=&\!\! \int \frac{d^4q}{(2\pi)^4}\,  \frac{1}{q^2}\, J \cdot \Big[ -P(3^-,\eta) +\frac14 P_{11}(1^-,\eta) \Big] \cdot J
\nn
\!&=&\! \int \frac{d^4q}{(2\pi)^4}\,  \frac{1}{q^2}\, J^{\mu\nu\rho}(-q) \Big[ -\eta_{\mu\lambda} \eta_{\nu\tau} \eta_{\rho\sigma} +\frac34 \eta_{\mu\nu}\eta_{\lambda\tau} \eta_{\rho^\sigma} \Big] J^{\lambda\tau\sigma}(q)\ .
\eea
This is in agreement with the formula (4.12) in \cite{Ponomarev:2016jqk}. 
The assumption on the source condition follows directly from the known spin 3 symmetry of the action given by $\delta \Phi_{\mu\nu\rho}=\partial_{(\mu} \xi_{\nu\rho)}$, where $\xi_{\mu\nu}$ is traceless.

\section{Matrix Coefficients}
\label{sec:app.coefficients}

\subsection{Einstein formulation}

\small 

\begin{fleqn}[0pt]
\begin{equation*}
a(3^{-}) = 4 (- q^2) ( b^{QQ}_{1} + b^{QQ}_{2}) - 4 ( m^{QQ}_{1} + m^{QQ}_{2})
\end{equation*}

\begin{equation*}
\begin{aligned}
a(2^{+})_{11} & = \frac{4}{3} (-q^{2}{}) (3 b^{QQ}_{1} + 3 b^{QQ}_{2} + b^{QQ}_{6} + b^{QQ}_{7} + b^{QQ}_{8} + b^{QQ}_{9}) - 4 ( m^{QQ}_{1} + m^{QQ}_{2}) \\
a(2^{+})_{12} & = \frac{1}{3 \sqrt{2}}(-q^{2}{}) (3 b^{TQ}_{4} + 3 b^{TQ}_{6} + 3 b^{TQ}_{7} + 8 b^{QQ}_{6} - 4 b^{QQ}_{7} - 4 b^{QQ}_{8} + 2 b^{QQ}_{9}) \\
a(2^{+})_{13} & = \frac{1}{\sqrt{6}}(-q^{2}{}) (b^{TQ}_{4} + b^{TQ}_{6} + b^{TQ}_{7}) 
\deveq
a(2^{+})_{14} & = \frac{i q^3}{2 \sqrt{3}} (b^{RQ}_{4} + b^{RQ}_{5})
\deveq
a(2^{+})_{22} & = \frac{1}{6} (-q^{2}{}) \left(18 b^{TT}_{1} + 9 b^{TT}_{2} + 9 b^{TT}_{4} + 9 b^{TT}_{5} + 18 b^{TQ}_{1} + 12 b^{TQ}_{4} - 6 b^{TQ}_{6} - 6 b^{TQ}_{7} + 24 b^{QQ}_{1} - 12 b^{QQ}_{2} \right.\\ & \left.+ 16 b^{QQ}_{6} + 4 b^{QQ}_{7} + 4 b^{QQ}_{8} - 8 b^{QQ}_{9} \right) + \frac{1}{2} \left( - 6 m^{TT}_{1} - 3 m^{TT}_{2} - 6 m^{TQ}_{1} - 8 m^{QQ}_{1} + 4 m^{QQ}_{2}\right)
\deveq
a(2^{+})_{23} & = \frac{1}{2 \sqrt{3}}(-q^{2}{}) (6 b^{TT}_{1} + 3 b^{TT}_{2} + 3 b^{TT}_{4} + 3 b^{TT}_{5} + 3 b^{TQ}_{1} + 2 b^{TQ}_{4} -  b^{TQ}_{6} -  b^{TQ}_{7}) \\
& - \frac{\sqrt{3}}{2} (2 m^{TT}_{1} + m^{TT}_{2} + m^{TQ}_{1}) 
\deveq
a(2^{+})_{24} & = \frac{i q^3}{4 \sqrt{6}} (3 b^{RT}_{1} + 6 b^{RT}_{2} + 3 b^{RT}_{3} + 6 b^{RQ}_{1} + 4 b^{RQ}_{4} - 2 b^{RQ}_{5}) 
\deveq
a(2^{+})_{33} & = \frac{1}{2} (-q^{2}{}) (2 b^{TT}_{1} + b^{TT}_{2} + b^{TT}_{4} + b^{TT}_{5}) - \frac{1}{2} \left(2 m^{TT}_{1} +  m^{TT}_{2}\right) 
\deveq
a(2^{+})_{34} & = \frac{i q^3}{4 \sqrt{2}} (b^{RT}_{1} + 2 b^{RT}_{2} + b^{RT}_{3}) 
\deveq
a(2^{+})_{44} & = \frac{1}{4} (-q^{2}{})  m^{R}-  q^4  \left( b^{RR}_{1} + \frac{1}{4} b^{RR}_{2}\right)
\end{aligned}
\end{equation*}

\begin{equation*}
\begin{aligned}
a(2^{-})_{11} & = (-q^{2}{}) (3 b^{TT}_{1} + \frac{3}{2} b^{TT}_{2} + 3 b^{TQ}_{1} + 4 b^{QQ}_{1} - 2 b^{QQ}_{2}) - 3 m^{TT}_{1} - \frac{3}{2} m^{TT}_{2} - 3 m^{TQ}_{1} - 4 m^{QQ}_{1} + 2 m^{QQ}_{2}
\deveq
a(2^{-})_{12} & = \frac{\sqrt{3}}{2}  (-q^{2}{}) (2 b^{TT}_{1} + b^{TT}_{2} + b^{TQ}_{1}) - \frac{\sqrt{3}}{2}  (2 m^{TT}_{1} +  m^{TT}_{2} +  m^{TQ}_{1}) 
\deveq
a(2^{-})_{22} & = \frac{1}{2} (-q^{2}{}) (2 b^{TT}_{1} + b^{TT}_{2}) - \frac{1}{2} (2 m^{TT}_{1} +  m^{TT}_{2}) 
\end{aligned}    
\end{equation*}

\begin{equation*}
\begin{aligned}
a(1^{+})_{11} & = \frac{1}{2} (-q^{2}{}) (6 b^{TT}_{1} + 3 b^{TT}_{2} + b^{TT}_{4} -  b^{TT}_{5} + 4 b^{TT}_{6} + 2 b^{TT}_{7} + 6 b^{TQ}_{1} + 4 b^{TQ}_{5} + 2 b^{TQ}_{6} - 2 b^{TQ}_{7} \\
& + 8 b^{QQ}_{1} - 4 b^{QQ}_{2} + 4 b^{QQ}_{7} - 4 b^{QQ}_{8}) - \frac{1}{2} (6 m^{TT}_{1} + 3 m^{TT}_{2} + 6 m^{TQ}_{1} + 8 m^{QQ}_{1} - 4 m^{QQ}_{2})
\deveq
a(1^{+})_{12} & = - \frac{1}{2 \sqrt{3}}(-q^{2}{}) (6 b^{TT}_{1} + 3 b^{TT}_{2} + b^{TT}_{4} -  b^{TT}_{5} + 4 b^{TT}_{6} + 2 b^{TT}_{7} + 3 b^{TQ}_{1} + 2 b^{TQ}_{5} + b^{TQ}_{6} -  b^{TQ}_{7}) \\
& + \frac{\sqrt{3}}{2}(2 m^{TT}_{1} + m^{TT}_{2} + m^{TQ}_{1}) \\
a(1^{+})_{13} & = \frac{1}{\sqrt{6}}(-q^{2}{}) (2 b^{TT}_{4} - 2 b^{TT}_{5} - 4 b^{TT}_{6} + b^{TT}_{7} - 2 b^{TQ}_{5} + 2 b^{TQ}_{6} - 2 b^{TQ}_{7})
\deveq
a(1^{+})_{22} & = \frac{1}{6} (-q^{2}{}) (6 b^{TT}_{1} + 3 b^{TT}_{2} + b^{TT}_{4} -  b^{TT}_{5} + 4 b^{TT}_{6} + 2 b^{TT}_{7}) - m^{TT}_{1} - \frac{1}{2} m^{TT}_{2} 
\deveq
a(1^{+})_{23} & = \frac{1}{3 \sqrt{2}}(-q^{2}{}) (- 2 b^{TT}_{4} + 2 b^{TT}_{5} + 4 b^{TT}_{6} - b^{TT}_{7}) 
\deveq
a(1^{+})_{33} & = \frac{4}{3} (-q^{2}{}) (3 b^{TT}_{1} - 3 b^{TT}_{2} + b^{TT}_{4} -  b^{TT}_{5} + b^{TT}_{6} -  b^{TT}_{7}) - 4 m^{TT}_{1} + 4 m^{TT}_{2}
\end{aligned}
\end{equation*}

\begin{equation*}
\begin{aligned}
a(1^{-})_{11} & = \frac{4}{3} (-q^{2}{}) (3 b^{QQ}_{1} + 3 b^{QQ}_{2} + 5 b^{QQ}_{3} +5 b^{QQ}_{4} + 5 b^{QQ}_{5}) - \frac{4}{3}(3 m^{QQ}_{1} + 3 m^{QQ}_{2} + 5 m^{QQ}_{3} + 5 m^{QQ}_{4} + 5 m^{QQ}_{5}) 
\deveq
a(1^{-})_{12} & = \frac{\sqrt{5}}{3}  (-q^{2}{}) (3 b^{TQ}_{2} + 3 b^{TQ}_{3} + 4 b^{QQ}_{3} - 8 b^{QQ}_{4} - 2 b^{QQ}_{5}) \\ & + \frac{\sqrt{5}}{3} (-3 m^{TQ}_{2} - 3 m^{TQ}_{3} + 8 m^{QQ}_{3} - 4 m^{QQ}_{4} + 2 m^{QQ}_{5}) 
\deveq
a(1^{-})_{13} & = \sqrt{5/3} (-q^{2}{}) (b^{TQ}_{2} + b^{TQ}_{3}) - \sqrt{5/3}  ( m^{TQ}_{2} +  m^{TQ}_{3})
\deveq
a(1^{-})_{14} & = \frac{2\sqrt{5}}{3}  (-q^{2}{}) (2 b^{QQ}_{3} + 2 b^{QQ}_{4} + 2 b^{QQ}_{5} + b^{QQ}_{10} + b^{QQ}_{11} + b^{QQ}_{12} + b^{QQ}_{13})  - \frac{2\sqrt{5}}{3}  (2 m^{QQ}_{3} + 2 m^{QQ}_{4} + 2 m^{QQ}_{5}) 
\deveq
a(1^{-})_{15} & = \frac{\sqrt{5}}{3 \sqrt{2}} (-q^{2}{}) (3 b^{TQ}_{2} + 3 b^{TQ}_{3} + 3 b^{TQ}_{8} + 3 b^{TQ}_{9} + 4 b^{QQ}_{3} - 8 b^{QQ}_{4} - 2 b^{QQ}_{5} + 2 b^{QQ}_{10} + 2 b^{QQ}_{11} \\ & - 4 b^{QQ}_{12} - 4 b^{QQ}_{13}) + \frac{\sqrt{5}}{3 \sqrt{2}} (-3 m^{TQ}_{2} - 3 m^{TQ}_{3} + 8 m^{QQ}_{3} - 4 m^{QQ}_{4} + 2 m^{QQ}_{5}) 
\deveq
a(1^{-})_{16} & = \sqrt{5/6} (-q^{2}{}) (b^{TQ}_{2} + b^{TQ}_{3} + b^{TQ}_{8} + b^{TQ}_{9}) - \sqrt{5/6} ( m^{TQ}_{2} +  m^{TQ}_{3}) \\
a(1^{-})_{17} & = 0 
\deveq
a(1^{-})_{22} & = (-q^{2}{}) (3 b^{TT}_{1} + \frac{3}{2} b^{TT}_{2} + 3 b^{TT}_{3} + 3 b^{TQ}_{1} + 2 b^{TQ}_{2} - 4 b^{TQ}_{3} + 4 b^{QQ}_{1} - 2 b^{QQ}_{2} + \frac{4}{3} b^{QQ}_{3} \\ & + \frac{16}{3} b^{QQ}_{4} -  \frac{8}{3} b^{QQ}_{5})  - 3 m^{TT}_{1} -  \frac{3}{2} m^{TT}_{2} - 3 m^{TT}_{3} - 3 m^{TQ}_{1} + 4 m^{TQ}_{2} - 2 m^{TQ}_{3} - 4 m^{QQ}_{1} \\ & + 2 m^{QQ}_{2} -  \frac{16}{3} m^{QQ}_{3} -  \frac{4}{3} m^{QQ}_{4} + \frac{8}{3} m^{QQ}_{5} 
\deveq
a(1^{-})_{23} & = \frac{1}{2 \sqrt{3}}(-q^{2}{}) (6 b^{TT}_{1} + 3 b^{TT}_{2} + 6 b^{TT}_{3} + 3 b^{TQ}_{1} + 2 b^{TQ}_{2} - 4 b^{TQ}_{3}) \\
& + \frac{1}{2 \sqrt{3}}(-6 m^{TT}_{1} - 3 m^{TT}_{2} - 6 m^{TT}_{3} - 3 m^{TQ}_{1} + 4 m^{TQ}_{2} - 2 m^{TQ}_{3}) 
\deveq
a(1^{-})_{24} & = \frac{1}{3} (-q^{2}{}) (3 b^{TQ}_{2} + 3 b^{TQ}_{3} + 3 b^{TQ}_{10} + 3 b^{TQ}_{11} + 4 b^{QQ}_{3} - 8 b^{QQ}_{4} - 2 b^{QQ}_{5} + 2 b^{QQ}_{10} - 4 b^{QQ}_{11} \\ & + 2 b^{QQ}_{12} - 4 b^{QQ}_{13}) + \frac{1}{3} ( - 3 m^{TQ}_{2} - 3 m^{TQ}_{3} + 8 m^{QQ}_{3} - 4 m^{QQ}_{4} + 2 m^{QQ}_{5}) 
\deveq
a(1^{-})_{25} & = \frac{1}{6 \sqrt{2}}(-q^{2}{}) (18 b^{TT}_{3} + 9 b^{TT}_{8} + 12 b^{TQ}_{2} - 24 b^{TQ}_{3} + 6 b^{TQ}_{8} - 12 b^{TQ}_{9} + 6 b^{TQ}_{10} \\ & - 12 b^{TQ}_{11} + 8 b^{QQ}_{3} + 32 b^{QQ}_{4} - 16 b^{QQ}_{5} + 4 b^{QQ}_{10} - 8 b^{QQ}_{11} - 8 b^{QQ}_{12} + 16 b^{QQ}_{13}) \\ & + \frac{ 1}{3 \sqrt{2}}(-9 m^{TT}_{3} + 12 m^{TQ}_{2} - 6 m^{TQ}_{3} - 16 m^{QQ}_{3} - 4 m^{QQ}_{4} + 8 m^{QQ}_{5}) 
\deveq
a(1^{-})_{26} & = \frac{1}{2 \sqrt{6}}(-q^{2}{}) (6 b^{TT}_{3} + 3 b^{TT}_{8} + 2 b^{TQ}_{2} - 4 b^{TQ}_{3} + 2 b^{TQ}_{8} - 4 b^{TQ}_{9}) + \frac{1}{2 \sqrt{6}}(-6 m^{TT}_{3} + 4 m^{TQ}_{2} - 2 m^{TQ}_{3}) 
\deveq
a(1^{-})_{27} & = 0
\deveq
a(1^{-})_{33} & = (-q^{2}{}) (b^{TT}_{1} + \frac{1}{2} b^{TT}_{2} + b^{TT}_{3}) -  m^{TT}_{1} -  \frac{1}{2} m^{TT}_{2} -  m^{TT}_{3} 
\deveq
a(1^{-})_{34} & = \frac{1}{\sqrt{3}}(-q^{2}{}) (b^{TQ}_{2} + b^{TQ}_{3} + b^{TQ}_{10} + b^{TQ}_{11}) + \frac{1}{\sqrt{3}}(- m^{TQ}_{2} -  m^{TQ}_{3}) 
\deveq
a(1^{-})_{35} & = \frac{1}{2 \sqrt{6}}(-q^{2}{}) (6 b^{TT}_{3} + 3 b^{TT}_{8} + 2 b^{TQ}_{2} - 4 b^{TQ}_{3} + 2 b^{TQ}_{10} - 4 b^{TQ}_{11}) + \frac{1}{2 \sqrt{6}}(-6 m^{TT}_{3} + 4 m^{TQ}_{2} - 2 m^{TQ}_{3}) 
\deveq
a(1^{-})_{36} & = \frac{1}{2 \sqrt{2}}(-q^{2}{}) (2 b^{TT}_{3} + b^{TT}_{8}) -  \frac{1}{\sqrt{2}}m^{TT}_{3} 
\deveq
a(1^{-})_{37} & = 0 
\deveq
a(1^{-})_{44} & = \frac{4}{3} (-q^{2}{}) (3 b^{QQ}_{1} + 3 b^{QQ}_{2} + b^{QQ}_{3} + b^{QQ}_{4} + b^{QQ}_{5} + 2 b^{QQ}_{6} + 2 b^{QQ}_{7} + 2 b^{QQ}_{8} + 2 b^{QQ}_{9}  + b^{QQ}_{10} \\ & + b^{QQ}_{11} + b^{QQ}_{12} + b^{QQ}_{13}) - \frac{4}{3} \left( 3 m^{QQ}_{1} + 3 m^{QQ}_{2} + m^{QQ}_{3} + m^{QQ}_{4} + m^{QQ}_{5} \right) 
\deveq
a(1^{-})_{45} & = \frac{1}{3 \sqrt{2}}(-q^{2}{}) (3 b^{TQ}_{2} + 3 b^{TQ}_{3} + 3 b^{TQ}_{4} + 3 b^{TQ}_{6} + 3 b^{TQ}_{7} + 3 b^{TQ}_{8} + 3 b^{TQ}_{9} + 3 b^{TQ}_{10} + 3 b^{TQ}_{11} + 4 b^{QQ}_{3} \\ & - 8 b^{QQ}_{4} - 2 b^{QQ}_{5}  + 8 b^{QQ}_{6} - 4 b^{QQ}_{7} - 4 b^{QQ}_{8} + 2 b^{QQ}_{9} + 4 b^{QQ}_{10} - 2 b^{QQ}_{11} - 2 b^{QQ}_{12} - 8 b^{QQ}_{13}) \\ & + \frac{1}{3 \sqrt{2}}(-3 m^{TQ}_{2} - 3 m^{TQ}_{3} + 8 m^{QQ}_{3} - 4 m^{QQ}_{4} + 2 m^{QQ}_{5}) 
\deveq
a(1^{-})_{46} & = \frac{1}{\sqrt{6}}(-q^{2}{}) (b^{TQ}_{2} + b^{TQ}_{3} + b^{TQ}_{4} + b^{TQ}_{6} + b^{TQ}_{7} + b^{TQ}_{8} + b^{TQ}_{9} + b^{TQ}_{10} + b^{TQ}_{11}) - \frac{1}{\sqrt{6}}(m^{TQ}_{2} +  m^{TQ}_{3}) 
\deveq
a(1^{-})_{47} & = 0
\deveq
a(1^{-})_{55} & = \frac{1}{6} (-q^{2}{}) (18 b^{TT}_{1} + 9 b^{TT}_{2} + 9 b^{TT}_{3} + 9 b^{TT}_{4} + 18 b^{TT}_{6} + 9 b^{TT}_{7} + 9 b^{TT}_{8} + 18 b^{TQ}_{1} + 6 b^{TQ}_{2} - 12 b^{TQ}_{3} \\ & + 6 b^{TQ}_{4} + 18 b^{TQ}_{5} + 6 b^{TQ}_{6} - 12 b^{TQ}_{7} + 6 b^{TQ}_{8} - 12 b^{TQ}_{9} + 6 b^{TQ}_{10} - 12 b^{TQ}_{11} + 24 b^{QQ}_{1} - 12 b^{QQ}_{2} + 4 b^{QQ}_{3} \\ & + 16 b^{QQ}_{4} - 8 b^{QQ}_{5} + 8 b^{QQ}_{6} + 20 b^{QQ}_{7} - 16 b^{QQ}_{8} - 4 b^{QQ}_{9} + 4 b^{QQ}_{10} - 8 b^{QQ}_{11} - 8 b^{QQ}_{12} + 16 b^{QQ}_{13}) \\ & + \frac{1}{6} (-18 m^{TT}_{1} - 9 m^{TT}_{2} - 9 m^{TT}_{3} - 18 m^{TQ}_{1} + 12 m^{TQ}_{2} - 6 m^{TQ}_{3} \\ & - 24 m^{QQ}_{1} + 12 m^{QQ}_{2} - 16 m^{QQ}_{3} - 4 m^{QQ}_{4} + 8 m^{QQ}_{5}) 
\deveq
a(1^{-})_{56} & = \frac{1}{2 \sqrt{3}}(-q^{2}{}) (6 b^{TT}_{1} + 3 b^{TT}_{2} + 3 b^{TT}_{3} + 3 b^{TT}_{4} + 6 b^{TT}_{6} + 3 b^{TT}_{7} + 3 b^{TT}_{8} + 3 b^{TQ}_{1} + b^{TQ}_{2} - 2 b^{TQ}_{3} \\ & + b^{TQ}_{4} + 3 b^{TQ}_{5} + b^{TQ}_{6} - 2 b^{TQ}_{7} + b^{TQ}_{8} - 2 b^{TQ}_{9} + b^{TQ}_{10} - 2 b^{TQ}_{11}) \\ & + \frac{1}{2 \sqrt{3}}(-6 m^{TT}_{1} - 3 m^{TT}_{2} - 3 m^{TT}_{3} - 3 m^{TQ}_{1} + 2 m^{TQ}_{2} -  m^{TQ}_{3}) 
\deveq
a(1^{-})_{57} & = 0 
\deveq
a(1^{-})_{66} & = \frac{1}{2} (-q^{2}{}) (2 b^{TT}_{1} + b^{TT}_{2} + b^{TT}_{3} + b^{TT}_{4} + 2 b^{TT}_{6} + b^{TT}_{7} + b^{TT}_{8}) - \frac{1}{2} (2 m^{TT}_{1} + m^{TT}_{2} + m^{TT}_{3}) 
\deveq
a(1^{-})_{67} & = 0 
\deveq
a(1^{-})_{77} & = 0 
\end{aligned}
\end{equation*}

\begin{equation*}
\begin{aligned}
a(0^{+})_{11} & = \frac{4}{3} (-q^{2}{}) (3 b^{QQ}_{1} + 3 b^{QQ}_{2} + 3 b^{QQ}_{3} + 3 b^{QQ}_{4} + 3 b^{QQ}_{5} + b^{QQ}_{6} + b^{QQ}_{7} + b^{QQ}_{8} + b^{QQ}_{9} + 3 b^{QQ}_{14} \\ & + 3 b^{QQ}_{15} + 3 b^{QQ}_{16}) - 4 ( m^{QQ}_{1} + m^{QQ}_{2} + m^{QQ}_{3} + m^{QQ}_{4} + m^{QQ}_{5} )
\deveq
a(0^{+})_{12} & = \frac{1}{3 \sqrt{2}}(-q^{2}{}) (-9 b^{TQ}_{2} - 9 b^{TQ}_{3} + 3 b^{TQ}_{4} + 3 b^{TQ}_{6} + 3 b^{TQ}_{7} + 9 b^{TQ}_{12} + 9 b^{TQ}_{13} - 12 b^{QQ}_{3} + 24 b^{QQ}_{4} \\ & + 6 b^{QQ}_{5} + 8 b^{QQ}_{6} - 4 b^{QQ}_{7} - 4 b^{QQ}_{8} + 2 b^{QQ}_{9} + 24 b^{QQ}_{14} - 12 b^{QQ}_{15} + 6 b^{QQ}_{16}) \\ & + \frac{1}{3 \sqrt{2}}(9 m^{TQ}_{2} + 9 m^{TQ}_{3} - 24 m^{QQ}_{3} + 12 m^{QQ}_{4} - 6 m^{QQ}_{5}) 
\deveq
a(0^{+})_{13} & = \frac{1}{\sqrt{6}}(-q^{2}{}) (-3 b^{TQ}_{2} - 3 b^{TQ}_{3} + b^{TQ}_{4} + b^{TQ}_{6} + b^{TQ}_{7} + 3 b^{TQ}_{12} + 3 b^{TQ}_{13}) + \sqrt{\frac{3}{2}}( m^{TQ}_{2} +  m^{TQ}_{3}) \\
a(0^{+})_{14} & = 2 (-q^{2}{}) (2 b^{QQ}_{3} + 2 b^{QQ}_{4} + 2 b^{QQ}_{5} + b^{QQ}_{10} + b^{QQ}_{11} + b^{QQ}_{12} + b^{QQ}_{13} + 2 b^{QQ}_{14} + 2 b^{QQ}_{15} + 2 b^{QQ}_{16}) \\ & - 4 (m^{QQ}_{3} + m^{QQ}_{4} + m^{QQ}_{5}) 
\deveq
a(0^{+})_{15} & = \frac{i q^3}{2 \sqrt{3}}(3 b^{RQ}_{2} + 3 b^{RQ}_{3} + b^{RQ}_{4} + b^{RQ}_{5} + 6 b^{RQ}_{6} + 6 b^{RQ}_{7}) \\
a(0^{+})_{16} & = 0
\deveq
a(0^{+})_{22} & = (-q^{2}{}) (3 b^{TT}_{1} + \frac{3}{2} b^{TT}_{2} + \frac{9}{2} b^{TT}_{3} + \frac{3}{2} b^{TT}_{4} + \frac{3}{2} b^{TT}_{5} + \frac{9}{2} b^{TT}_{9} + 3 b^{TQ}_{1} + 3 b^{TQ}_{2} - 6 b^{TQ}_{3} + 2 b^{TQ}_{4} \\ & - b^{TQ}_{6} -  b^{TQ}_{7}  + 6 b^{TQ}_{12} - 3 b^{TQ}_{13} + 4 b^{QQ}_{1} - 2 b^{QQ}_{2} + 2 b^{QQ}_{3} + 8 b^{QQ}_{4} - 4 b^{QQ}_{5} + \frac{8}{3} b^{QQ}_{6} + \frac{2}{3} b^{QQ}_{7} \\ & + \frac{2}{3} b^{QQ}_{8} -  \frac{4}{3} b^{QQ}_{9} + 8 b^{QQ}_{14} + 2 b^{QQ}_{15} - 4 b^{QQ}_{16}) - 3 m^{TT}_{1} -  \frac{3}{2} m^{TT}_{2} -  \frac{9}{2} m^{TT}_{3} - 3 m^{TQ}_{1} + 6 m^{TQ}_{2} \\ & - 3 m^{TQ}_{3} - 4 m^{QQ}_{1} + 2 m^{QQ}_{2} - 8 m^{QQ}_{3} - 2 m^{QQ}_{4} + 4 m^{QQ}_{5} 
\deveq
a(0^{+})_{23} & = \frac{1}{2 \sqrt{3}}(-q^{2}{}) (6 b^{TT}_{1} + 3 b^{TT}_{2} + 9 b^{TT}_{3} + 3 b^{TT}_{4} + 3 b^{TT}_{5} + 9 b^{TT}_{9} + 3 b^{TQ}_{1} + 3 b^{TQ}_{2} - 6 b^{TQ}_{3} + 2 b^{TQ}_{4} \\ & -  b^{TQ}_{6} -  b^{TQ}_{7}  + 6 b^{TQ}_{12} - 3 b^{TQ}_{13}) + \frac{1}{2 \sqrt{3}}(-6 m^{TT}_{1} - 3 m^{TT}_{2} - 9 m^{TT}_{3} - 3 m^{TQ}_{1} + 6 m^{TQ}_{2} - 3 m^{TQ}_{3}) 
\deveq
a(0^{+})_{24} & = \frac{1}{\sqrt{2}}(-q^{2}{}) (-3 b^{TQ}_{2} - 3 b^{TQ}_{3} - 3 b^{TQ}_{10} - 3 b^{TQ}_{11} + 3 b^{TQ}_{12} + 3 b^{TQ}_{13} - 4 b^{QQ}_{3} + 8 b^{QQ}_{4} + 2 b^{QQ}_{5} \\ & - 2 b^{QQ}_{10} + 4 b^{QQ}_{11} - 2 b^{QQ}_{12} + 4 b^{QQ}_{13} + 8 b^{QQ}_{14} - 4 b^{QQ}_{15} + 2 b^{QQ}_{16}) \\ & + \frac{1}{\sqrt{2}}(3 m^{TQ}_{2} + 3 m^{TQ}_{3} - 8 m^{QQ}_{3} + 4 m^{QQ}_{4} - 2 m^{QQ}_{5}) 
\deveq
a(0^{+})_{25} & = \frac{iq^3}{4 \sqrt{6}} (3 b^{RT}_{1} + 6 b^{RT}_{2} + 3 b^{RT}_{3} - 9 b^{RT}_{4} + 18 b^{RT}_{5} \\ & + 6 b^{RQ}_{1} - 6 b^{RQ}_{2} + 12 b^{RQ}_{3} + 4 b^{RQ}_{4} - 2 b^{RQ}_{5} + 24 b^{RQ}_{6} - 12 b^{RQ}_{7}) 
\deveq
a(0^{+})_{26} & = 0
\deveq
a(0^{+})_{33} & = \frac{1}{2} (-q^{2}{}) (2 b^{TT}_{1} + b^{TT}_{2} + 3 b^{TT}_{3} + b^{TT}_{4} + b^{TT}_{5} + 3 b^{TT}_{9}) - \frac{1}{2} (2 m^{TT}_{1} + m^{TT}_{2} + 3 m^{TT}_{3}) 
\deveq
a(0^{+})_{34} & = \sqrt{\frac{3}{2}} (-q^{2}{}) (- b^{TQ}_{2} -  b^{TQ}_{3} -  b^{TQ}_{10} -  b^{TQ}_{11} + b^{TQ}_{12} + b^{TQ}_{13}) + \sqrt{\frac{3}{2}} (m^{TQ}_{2} + m^{TQ}_{3}) 
\deveq
a(0^{+})_{35} & = \frac{i q^3}{4 \sqrt{2}} (b^{RT}_{1} + 2 b^{RT}_{2} + b^{RT}_{3} - 3 b^{RT}_{4} + 6 b^{RT}_{5}) 
\deveq
a(0^{+})_{36} & = 0 
\deveq
a(0^{+})_{44} & = 4 (-q^{2}{}) (b^{QQ}_{1} + b^{QQ}_{2} + b^{QQ}_{3} + b^{QQ}_{4} + b^{QQ}_{5} + b^{QQ}_{6} + b^{QQ}_{7} + b^{QQ}_{8} + b^{QQ}_{9} + b^{QQ}_{10} + b^{QQ}_{11} \\ & + b^{QQ}_{12} + b^{QQ}_{13} + b^{QQ}_{14} + b^{QQ}_{15} + b^{QQ}_{16}) - 4 ( m^{QQ}_{1} + m^{QQ}_{2} + m^{QQ}_{3} + m^{QQ}_{4} + m^{QQ}_{5}) 
\deveq
a(0^{+})_{45} & = \frac{\sqrt{3}}{2} i  q^3 (b^{RQ}_{2} + b^{RQ}_{3} + b^{RQ}_{4} + b^{RQ}_{5} + 2 b^{RQ}_{6} + 2 b^{RQ}_{7}) 
\deveq
a(0^{+})_{46} & = 0 
\deveq
a(0^{+})_{55} & = -  q^4 (b^{RR}_{1} + b^{RR}_{2} + 3 b^{RR}_{3}) - \frac{1}{2} m^{R} (-q^{2}{}) 
\deveq
a(0^{+})_{56} & = 0 
\deveq
a(0^{+})_{66} & = 0 
\end{aligned}
\end{equation*}

\begin{equation*}
a(0^{-}) = 4 (-q^{2}{}) (b^{TT}_{1} -  b^{TT}_{2}) + 4 (- m^{TT}_{1} + m^{TT}_{2})
\end{equation*}
\end{fleqn}

\normalsize

\subsection{Cartan formulation}
\small 

\begin{fleqn}[0pt]
\begin{equation*}
\begin{aligned}
a(3^{-}) & = (-q^{2}{}) (2 c^{FF}_{1} + 2 c^{FF}_{2} + c^{FF}_{4} + c^{FF}_{5} + c^{FF}_{6} + 2 c^{FQ}_{1} - 2 c^{FQ}_{2} - 2 c^{FQ}_{3} + 4 c^{QQ}_{1} + 4 c^{QQ}_{2}) \\ & - a^{R} - 4 a^{QQ}_{1} - 4 a^{QQ}_{2}
\end{aligned}
\end{equation*}

\begin{equation*}
\begin{aligned}
a(2^{+})_{11} & = \frac{1}{3} (-q^{2}{}) (4 c^{FF}_{1} + 4 c^{FF}_{2} + 2 c^{FF}_{4} + 2 c^{FF}_{5} + 2 c^{FF}_{6} + c^{FF}_{7} + c^{FF}_{8} + c^{FF}_{9} + c^{FF}_{10} + c^{FF}_{11} \\ & + c^{FF}_{12} + 4 c^{FQ}_{1} - 4 c^{FQ}_{2} - 4 c^{FQ}_{3} + 2 c^{FQ}_{16} + 2 c^{FQ}_{17} + 2 c^{FQ}_{18} + 2 c^{FQ}_{19} + 2 c^{FQ}_{20} + 2 c^{FQ}_{21} \\ & + 12 c^{QQ}_{1} + 12 c^{QQ}_{2} + 4 c^{QQ}_{6} + 4 c^{QQ}_{7} + 4 c^{QQ}_{8} + 4 c^{QQ}_{9}) - a^{R} - 4 a^{QQ}_{1} - 4 a^{QQ}_{2}
\deveq
a(2^{+})_{12} & = \frac{1}{6 \sqrt{2}}(-q^{2}{}) ( - 8 c^{FF}_{1} - 8 c^{FF}_{2} - 4 c^{FF}_{4} - 4 c^{FF}_{5} - 4 c^{FF}_{6} - 2 c^{FF}_{7} - 2 c^{FF}_{8} - 2 c^{FF}_{9} - 2 c^{FF}_{10} \\ & - 2 c^{FF}_{11} - 2 c^{FF}_{12} - 3 c^{FT}_{1} - 3 c^{FT}_{2} + 3 c^{FT}_{4} + 6 c^{FT}_{5} + 3 c^{FT}_{6} + 6 c^{FT}_{7} + 3 c^{FT}_{13} + 3 c^{FT}_{14} + 3 c^{FT}_{15} \\ & + 3 c^{FT}_{16} - 8 c^{FQ}_{1} + 2 c^{FQ}_{2}  + 2 c^{FQ}_{3} + 6 c^{FQ}_{4} + 6 c^{FQ}_{5} + 2 c^{FQ}_{16} + 2 c^{FQ}_{17} - 4 c^{FQ}_{18} - 4 c^{FQ}_{19} \\ & - 4 c^{FQ}_{20} - 4 c^{FQ}_{21} + 6 c^{TQ}_{4} + 6 c^{TQ}_{6} + 6 c^{TQ}_{7} + 16 c^{QQ}_{6} - 8 c^{QQ}_{7} - 8 c^{QQ}_{8} + 4 c^{QQ}_{9})
\deveq
a(2^{+})_{13} & = \frac{1}{2 \sqrt{6}}(-q^{2}{}) (2 c^{FF}_{7} + 2 c^{FF}_{8} - 2 c^{FF}_{9} - 2 c^{FF}_{10} -  c^{FT}_{1} -  c^{FT}_{2} + c^{FT}_{4} + 2 c^{FT}_{5} + c^{FT}_{6} + 2 c^{FT}_{7} + c^{FT}_{13} \\ & + c^{FT}_{14} + c^{FT}_{15} + c^{FT}_{16} + 2 c^{FQ}_{16} - 2 c^{FQ}_{17} + 2 c^{FQ}_{18} + 2 c^{FQ}_{19} - 2 c^{FQ}_{20} - 2 c^{FQ}_{21} + 2 c^{TQ}_{4} + 2 c^{TQ}_{6} + 2 c^{TQ}_{7})
\deveq
a(2^{+})_{14} & = \frac{i q^3}{2 \sqrt{3}} \left(- c^{FQ}_{2} -  c^{FQ}_{3} + c^{FQ}_{4} + c^{FQ}_{5} + c^{FQ}_{16} + c^{FQ}_{17} + 4 c^{QQ}_{1} + 4 c^{QQ}_{2} + 4 c^{QQ}_{6} + 2 c^{QQ}_{9} \right) \\& + \frac{i q}{2 \sqrt{3}} (a^{R} + 4 a^{QQ}_{1} + 4 a^{QQ}_{2})
\deveq
a(2^{+})_{22} & = \frac{1}{6} (-q^{2}{}) (4 c^{FF}_{1} + 4 c^{FF}_{2} + 2 c^{FF}_{4} + 2 c^{FF}_{5} + 2 c^{FF}_{6} + c^{FF}_{7} + c^{FF}_{8} + c^{FF}_{9} + c^{FF}_{10} + c^{FF}_{11} \\ & + c^{FF}_{12} + 3 c^{FT}_{1} + 3 c^{FT}_{2} - 3 c^{FT}_{4} - 6 c^{FT}_{5} - 3 c^{FT}_{6} - 6 c^{FT}_{7} - 3 c^{FT}_{13} - 3 c^{FT}_{14} - 3 c^{FT}_{15} - 3 c^{FT}_{16} \\ & + 4 c^{FQ}_{1} + 2 c^{FQ}_{2} + 2 c^{FQ}_{3} - 6 c^{FQ}_{4} - 6 c^{FQ}_{5} - 4 c^{FQ}_{16} - 4 c^{FQ}_{17} + 2 c^{FQ}_{18} + 2 c^{FQ}_{19} + 2 c^{FQ}_{20} + 2 c^{FQ}_{21} \\ & + 18 c^{TT}_{1} + 9 c^{TT}_{2} + 9 c^{TT}_{4} + 9 c^{TT}_{5} + 18 c^{TQ}_{1} + 12 c^{TQ}_{4} - 6 c^{TQ}_{6} - 6 c^{TQ}_{7} + 24 c^{QQ}_{1} - 12 c^{QQ}_{2} \\ & + 16 c^{QQ}_{6} + 4 c^{QQ}_{7} + 4 c^{QQ}_{8} - 8 c^{QQ}_{9}) + \frac{1}{2} ( a^{R} - 6 a^{TT}_{1} - 3 a^{TT}_{2} - 6 a^{TQ}_{1} - 8 a^{QQ}_{1} + 4 a^{QQ}_{2} )
\deveq
a(2^{+})_{23} & = \frac{1}{2 \sqrt{3}}(-q^{2}{}) (- c^{FF}_{7} -  c^{FF}_{8} + c^{FF}_{9} + c^{FF}_{10} + 2 c^{FT}_{1} -  c^{FT}_{2} + 3 c^{FT}_{3} + c^{FT}_{4} + 2 c^{FT}_{5} \\ & - 2 c^{FT}_{6} - 4 c^{FT}_{7} + c^{FT}_{13} + c^{FT}_{14} - 2 c^{FT}_{15} - 2 c^{FT}_{16} - 3 c^{FQ}_{2} + 3 c^{FQ}_{3} + 3 c^{FQ}_{4} - 3 c^{FQ}_{5} \\ & + 2 c^{FQ}_{16} - 2 c^{FQ}_{17} -  c^{FQ}_{18} -  c^{FQ}_{19} + c^{FQ}_{20} + c^{FQ}_{21} + 6 c^{TT}_{1} + 3 c^{TT}_{2} + 3 c^{TT}_{4} + 3 c^{TT}_{5} + 3 c^{TQ}_{1} \\ & + 2 c^{TQ}_{4} -  c^{TQ}_{6} -  c^{TQ}_{7}) - \frac{ \sqrt{3}}{2}(2 a^{TT}_{1} + a^{TT}_{2} + a^{TQ}_{1})
\deveq
a(2^{+})_{24} & = \frac{i q^3}{2 \sqrt{6}} (c^{FQ}_{2} + c^{FQ}_{3} - c^{FQ}_{4} - c^{FQ}_{5} - c^{FQ}_{16} - c^{FQ}_{17} + 3 c^{TQ}_{1} + 3 c^{TQ}_{4} \\ & + 8 c^{QQ}_{1} - 4 c^{QQ}_{2} + 8 c^{QQ}_{6} - 2 c^{QQ}_{9}) + \frac{i q}{2 \sqrt{6}} ( - a^{R} + 3 a^{TQ}_{1} + 8 a^{QQ}_{1} - 4 a^{QQ}_{2})
\deveq
a(2^{+})_{33} & = \frac{1}{2} (-q^{2}{}) (4 c^{FF}_{1} - 4 c^{FF}_{2} + 4 c^{FF}_{3} + 2 c^{FF}_{4} - 2 c^{FF}_{5} + 2 c^{FF}_{6} + c^{FF}_{7} + c^{FF}_{8} + c^{FF}_{9} \\ & + c^{FF}_{10} -  c^{FF}_{11} -  c^{FF}_{12} + c^{FT}_{1} -  c^{FT}_{2} + 2 c^{FT}_{3} + c^{FT}_{4} + 2 c^{FT}_{5} -  c^{FT}_{6} - 2 c^{FT}_{7} + c^{FT}_{13} \\ & + c^{FT}_{14} -  c^{FT}_{15} -  c^{FT}_{16}  + 2 c^{TT}_{1} + c^{TT}_{2} + c^{TT}_{4} + c^{TT}_{5}) + \frac{1}{2} (a^{R} - 2 a^{TT}_{1} -  a^{TT}_{2})
\deveq
a(2^{+})_{34} & = \frac{i q^3}{2 \sqrt{2}} (- c^{FQ}_{2} + c^{FQ}_{3} + c^{FQ}_{4} -  c^{FQ}_{5} + c^{FQ}_{16} -  c^{FQ}_{17} + c^{TQ}_{1} + c^{TQ}_{4}) + \frac{i}{2 \sqrt{2}}q (a^{R} + a^{TQ}_{1}) \\
a(2^{+})_{44} & = - q^4 (c^{QQ}_{1} + c^{QQ}_{6}) + (-q^{2}{}) a^{QQ}_{1}
\end{aligned}
\end{equation*}

\begin{equation*}
\begin{aligned}
a(2^{-})_{11} & = \frac{1}{2} (-q^{2}{}) (4 c^{FF}_{1} + 4 c^{FF}_{2} -  c^{FF}_{4} -  c^{FF}_{5} -  c^{FF}_{6} + 3 c^{FT}_{1} + 3 c^{FT}_{2}  + 4 c^{FQ}_{1} + 2 c^{FQ}_{2} + 2 c^{FQ}_{3} + 6 c^{TT}_{1} \\ & + 3 c^{TT}_{2} + 6 c^{TQ}_{1} + 8 c^{QQ}_{1} - 4 c^{QQ}_{2}) + \frac{1}{2} ( a^{R} - 6 a^{TT}_{1} - 3 a^{TT}_{2} - 6 a^{TQ}_{1} - 8 a^{QQ}_{1} + 4 a^{QQ}_{2} )
\deveq
a(2^{-})_{12} & = \frac{\sqrt{3}}{2}  (-q^{2}{}) (c^{FF}_{4} -  c^{FF}_{6} + c^{FT}_{1} + c^{FT}_{3} - c^{FQ}_{2} + c^{FQ}_{3} + 2 c^{TT}_{1} + c^{TT}_{2} + c^{TQ}_{1}) \\ & + \frac{\sqrt{3}}{2}  (-2 a^{TT}_{1} -  a^{TT}_{2} -  a^{TQ}_{1})
\deveq
a(2^{-})_{22} & = \frac{1}{2} (-q^{2}{}) (4 c^{FF}_{1} - 4 c^{FF}_{2} + c^{FF}_{4} -  c^{FF}_{5} + c^{FF}_{6} + c^{FT}_{1} -  c^{FT}_{2} + 2 c^{FT}_{3} + 2 c^{TT}_{1} + c^{TT}_{2}) \\ & + \frac{1}{2} (a^{R} - 2 a^{TT}_{1} -  a^{TT}_{2})
\end{aligned}    
\end{equation*}

\begin{equation*}
\begin{aligned}
a(1^{+})_{11} & = \frac{1}{2} (-q^{2}{}) (4 c^{FF}_{1} + 4 c^{FF}_{2} + c^{FF}_{7} -  c^{FF}_{8} + c^{FF}_{9} -  c^{FF}_{10} + c^{FF}_{11} - c^{FF}_{12} + 3 c^{FT}_{1} + 3 c^{FT}_{2} - c^{FT}_{4} \\ & - 2 c^{FT}_{5} -  c^{FT}_{6} - 2 c^{FT}_{7} + c^{FT}_{13} - c^{FT}_{14} + c^{FT}_{15} -  c^{FT}_{16} + 2 c^{FT}_{17} + 2 c^{FT}_{18}  + 4 c^{FQ}_{1} + 2 c^{FQ}_{2} + 2 c^{FQ}_{3} \\ & - 2 c^{FQ}_{4} - 2 c^{FQ}_{5} + 2 c^{FQ}_{18} - 2 c^{FQ}_{19} + 2 c^{FQ}_{20} - 2 c^{FQ}_{21} + 6 c^{TT}_{1} + 3 c^{TT}_{2} + c^{TT}_{4} - c^{TT}_{5} + 4 c^{TT}_{6} \\ & + 2 c^{TT}_{7} + 6 c^{TQ}_{1} + 4 c^{TQ}_{5} + 2 c^{TQ}_{6} - 2 c^{TQ}_{7} + 8 c^{QQ}_{1} - 4 c^{QQ}_{2} + 4 c^{QQ}_{7} - 4 c^{QQ}_{8}) \\ & + \frac{1}{2} (a^{R} - 6 a^{TT}_{1} - 3 a^{TT}_{2} - 6 a^{TQ}_{1} - 8 a^{QQ}_{1} + 4 a^{QQ}_{2}) 
\deveq
a(1^{+})_{12} & = \frac{1}{2 \sqrt{3}}(-q^{2}{}) ( - 2 c^{FF}_{4} + 2 c^{FF}_{6} - c^{FF}_{7} + c^{FF}_{8} + c^{FF}_{9} - c^{FF}_{10} - 2 c^{FT}_{1} - c^{FT}_{2} - c^{FT}_{3} +  c^{FT}_{4} \\ & + 2 c^{FT}_{5} - c^{FT}_{13} + c^{FT}_{14} - 2 c^{FT}_{17} + c^{FQ}_{2} - c^{FQ}_{3} + c^{FQ}_{4} - c^{FQ}_{5} - c^{FQ}_{18} + c^{FQ}_{19} + c^{FQ}_{20} - c^{FQ}_{21} \\ & - 6 c^{TT}_{1} - 3 c^{TT}_{2} - c^{TT}_{4} + c^{TT}_{5} - 4 c^{TT}_{6} - 2 c^{TT}_{7} - 3 c^{TQ}_{1} - 2 c^{TQ}_{5} - c^{TQ}_{6} + c^{TQ}_{7}) \\ & + \frac{\sqrt{3}}{2}(2 a^{TT}_{1} + a^{TT}_{2} + a^{TQ}_{1}) 
\deveq
a(1^{+})_{13} & = \frac{1}{2 \sqrt{6}}(-q^{2}{}) (- 4 c^{FF}_{4} + 4 c^{FF}_{6} - 2 c^{FF}_{7} + 2 c^{FF}_{8} + 2 c^{FF}_{9} - 2 c^{FF}_{10} - c^{FT}_{1} +  c^{FT}_{2} - 2 c^{FT}_{3} \\ & + 5 c^{FT}_{4} - 2 c^{FT}_{5} + 3 c^{FT}_{6} - 6 c^{FT}_{7} +  c^{FT}_{13} - c^{FT}_{14} + 3 c^{FT}_{15} - 3 c^{FT}_{16} - 4 c^{FT}_{17} + 2 c^{FQ}_{2} - 2 c^{FQ}_{3} \\ & + 2 c^{FQ}_{4} - 2 c^{FQ}_{5} - 2 c^{FQ}_{18} + 2 c^{FQ}_{19} + 2 c^{FQ}_{20} - 2 c^{FQ}_{21} + 4 c^{TT}_{4} - 4 c^{TT}_{5} - 8 c^{TT}_{6} + 2 c^{TT}_{7} \\ & - 4 c^{TQ}_{5} + 4 c^{TQ}_{6} - 4 c^{TQ}_{7}) 
\deveq
a(1^{+})_{22} & = \frac{1}{6} (-q^{2}{}) (4 c^{FF}_{1} - 4 c^{FF}_{2} - 4 c^{FF}_{3} + c^{FF}_{7} -  c^{FF}_{8} + c^{FF}_{9} -  c^{FF}_{10} -  c^{FF}_{11} + c^{FF}_{12} + c^{FT}_{1} \\ & - c^{FT}_{2} + 2 c^{FT}_{3} -  c^{FT}_{4} - 2 c^{FT}_{5} + c^{FT}_{6} + 2 c^{FT}_{7} + c^{FT}_{13} -  c^{FT}_{14} -  c^{FT}_{15} + c^{FT}_{16} + 2 c^{FT}_{17} - 2 c^{FT}_{18} \\ & + 6 c^{TT}_{1} + 3 c^{TT}_{2} + c^{TT}_{4} -  c^{TT}_{5} + 4 c^{TT}_{6} + 2 c^{TT}_{7}) + \frac{1}{2} ( a^{R} - 2 a^{TT}_{1} - a^{TT}_{2} )
\deveq
a(1^{+})_{23} & = \frac{1}{6 \sqrt{2}}(-q^{2}{}) (8 c^{FF}_{1} - 8 c^{FF}_{2} - 8 c^{FF}_{3} + 2 c^{FF}_{7} - 2 c^{FF}_{8} + 2 c^{FF}_{9} - 2 c^{FF}_{10} - 2 c^{FF}_{11} + 2 c^{FF}_{12} \\ & + 5 c^{FT}_{1} - 5 c^{FT}_{2} - 2 c^{FT}_{3} - 5 c^{FT}_{4} + 2 c^{FT}_{5} + 5 c^{FT}_{6} - 2 c^{FT}_{7} -  c^{FT}_{13} + c^{FT}_{14} + c^{FT}_{15} -  c^{FT}_{16} \\ & + 4 c^{FT}_{17} - 4 c^{FT}_{18} - 4 c^{TT}_{4} + 4 c^{TT}_{5} + 8 c^{TT}_{6} - 2 c^{TT}_{7}) 
\deveq
a(1^{+})_{33} & = \frac{1}{3} (-q^{2}{}) (4 c^{FF}_{1} - 4 c^{FF}_{2} - 4 c^{FF}_{3} + c^{FF}_{7} -  c^{FF}_{8} + c^{FF}_{9} -  c^{FF}_{10} -  c^{FF}_{11} + c^{FF}_{12} + 4 c^{FT}_{1} \\ & - 4 c^{FT}_{2} - 4 c^{FT}_{3} - 4 c^{FT}_{4} + 4 c^{FT}_{5} + 4 c^{FT}_{6} - 4 c^{FT}_{7} - 2 c^{FT}_{13} + 2 c^{FT}_{14} + 2 c^{FT}_{15} - 2 c^{FT}_{16} + 2 c^{FT}_{17} \\ & - 2 c^{FT}_{18} + 12 c^{TT}_{1} - 12 c^{TT}_{2} + 4 c^{TT}_{4} - 4 c^{TT}_{5} + 4 c^{TT}_{6} - 4 c^{TT}_{7}) - a^{R} - 4 a^{TT}_{1} + 4 a^{TT}_{2} 
\end{aligned}
\end{equation*}

\begin{equation*}
\begin{aligned}
a(1^{-})_{11} & = \frac{1}{3} (-q^{2}{}) (6 c^{FF}_{1} + 6 c^{FF}_{2} + 3 c^{FF}_{4} + 3 c^{FF}_{5} + 3 c^{FF}_{6} + 5 c^{FF}_{7} + 5 c^{FF}_{9} + 5 c^{FF}_{11} + 10 c^{FF}_{13} \\ & - 5 c^{FF}_{14} - 5 c^{FF}_{15} + 6 c^{FQ}_{1} - 6 c^{FQ}_{2} - 6 c^{FQ}_{3} - 10 c^{FQ}_{6} - 10 c^{FQ}_{8} - 10 c^{FQ}_{10} - 10 c^{FQ}_{12} \\ & + 10 c^{FQ}_{14} + 10 c^{FQ}_{15} + 12 c^{QQ}_{1} + 12 c^{QQ}_{2} + 20 c^{QQ}_{3} + 20 c^{QQ}_{4} + 20 c^{QQ}_{5}) \\ & + \frac{1}{3} (2 a^{R} - 12 a^{QQ}_{1} - 12 a^{QQ}_{2} - 20 a^{QQ}_{3} - 20 a^{QQ}_{4} - 20 a^{QQ}_{5}) 
\deveq
a(1^{-})_{12} & = \frac{1}{6} \sqrt{5} (-q^{2}{}) (2 c^{FF}_{7} + 2 c^{FF}_{9} + 2 c^{FF}_{11} - 8 c^{FF}_{13} + c^{FF}_{14} + c^{FF}_{15} - 3 c^{FT}_{8} - 3 c^{FT}_{10} + 3 c^{FT}_{12} \\ & - 4 c^{FQ}_{6} + 2 c^{FQ}_{8} - 4 c^{FQ}_{10} + 2 c^{FQ}_{12} - 2 c^{FQ}_{14} - 8 c^{FQ}_{15} + 6 c^{TQ}_{2} + 6 c^{TQ}_{3} + 8 c^{QQ}_{3} - 16 c^{QQ}_{4} \\ & - 4 c^{QQ}_{5}) + \frac{1}{6} \sqrt{5} (2 a^{R} - 6 a^{TQ}_{2} - 6 a^{TQ}_{3} + 16 a^{QQ}_{3} - 8 a^{QQ}_{4} + 4 a^{QQ}_{5}) 
\deveq
a(1^{-})_{13} & = - \frac{\sqrt{5}}{2 \sqrt{3}} (-q^{2}{}) (-2 c^{FF}_{7} + 2 c^{FF}_{9} + c^{FF}_{14} -  c^{FF}_{15} + c^{FT}_{8} + c^{FT}_{10} -  c^{FT}_{12} + 2 c^{FQ}_{6} \\ & + 2 c^{FQ}_{8} - 2 c^{FQ}_{10} - 2 c^{FQ}_{12} - 2 c^{TQ}_{2} - 2 c^{TQ}_{3}) -  \sqrt{5/3} (a^{TQ}_{2} + a^{TQ}_{3}) 
\deveq
a(1^{-})_{14} & = \frac{1}{3} \sqrt{5} (-q^{2}{}) (- c^{FF}_{8} -  c^{FF}_{10} -  c^{FF}_{12} + 2 c^{FF}_{13} -  c^{FF}_{14} -  c^{FF}_{15} -  c^{FQ}_{6} + c^{FQ}_{7} - c^{FQ}_{8} + c^{FQ}_{9} - c^{FQ}_{10} \\ & + c^{FQ}_{11} -  c^{FQ}_{12} + c^{FQ}_{13} + 2 c^{FQ}_{14} + 2 c^{FQ}_{15} -  c^{FQ}_{16} -  c^{FQ}_{17} -  c^{FQ}_{18} -  c^{FQ}_{19} -  c^{FQ}_{20} -  c^{FQ}_{21} + 4 c^{QQ}_{3} \\ & + 4 c^{QQ}_{4} + 4 c^{QQ}_{5} + 2 c^{QQ}_{10} + 2 c^{QQ}_{11} + 2 c^{QQ}_{12} + 2 c^{QQ}_{13}) + \frac{1}{3} \sqrt{5} (a^{R} - 4 a^{QQ}_{3} - 4 a^{QQ}_{4} - 4 a^{QQ}_{5}) 
\deveq
a(1^{-})_{15} & = \frac{\sqrt{5}}{6 \sqrt{2}} (-q^{2}{}) \left(4 c^{FF}_{8} + 4 c^{FF}_{10} + 4 c^{FF}_{12} - 8 c^{FF}_{13} + 4 c^{FF}_{14} + 4 c^{FF}_{15} - 3 c^{FT}_{8} - 3 c^{FT}_{10} + 3 c^{FT}_{12} - 3 c^{FT}_{13} \right.\\ & - 3 c^{FT}_{15} - 3 c^{FT}_{17} - 3 c^{FT}_{18} + 3 c^{FT}_{19} + 6 c^{FT}_{20} - 2 c^{FQ}_{6} - 4 c^{FQ}_{7} + 4 c^{FQ}_{8} - 4 c^{FQ}_{9} - 2 c^{FQ}_{10} - 4 c^{FQ}_{11} \\ & + 4 c^{FQ}_{12} - 4 c^{FQ}_{13}  - 2 c^{FQ}_{14} - 8 c^{FQ}_{15} - 2 c^{FQ}_{16} - 2 c^{FQ}_{17} - 2 c^{FQ}_{18} + 4 c^{FQ}_{19} - 2 c^{FQ}_{20} + 4 c^{FQ}_{21} + 6 c^{FQ}_{22} \\ & \left.+ 6 c^{TQ}_{2} + 6 c^{TQ}_{3} + 6 c^{TQ}_{8} + 6 c^{TQ}_{9}  + 8 c^{QQ}_{3} - 16 c^{QQ}_{4} - 4 c^{QQ}_{5} + 4 c^{QQ}_{10} + 4 c^{QQ}_{11} - 8 c^{QQ}_{12} - 8 c^{QQ}_{13}\right) \\ & + \frac{\sqrt{10}}{6} (a^{R} - 3 a^{TQ}_{2} - 3 a^{TQ}_{3} + 8 a^{QQ}_{3} - 4 a^{QQ}_{4} + 2 a^{QQ}_{5}) 
\deveq
a(1^{-})_{16} & =  \frac{\sqrt{5}}{2 \sqrt{6}} (-q^{2}{}) ( - c^{FT}_{8} - c^{FT}_{10} + c^{FT}_{12} - c^{FT}_{13} - c^{FT}_{15} - c^{FT}_{17} - c^{FT}_{18} + c^{FT}_{19} + 2 c^{FT}_{20} + 2 c^{TQ}_{2} + 2 c^{TQ}_{3} \\ & + 2 c^{TQ}_{8} + 2 c^{TQ}_{9}) -  \sqrt{5/6} (a^{TQ}_{2} + a^{TQ}_{3}) 
\deveq
a(1^{-})_{17} & =  \frac{i q^3 }{2} \sqrt{\frac{5}{6}} \left( - c^{FQ}_{6} - c^{FQ}_{10} + c^{FQ}_{14} - c^{FQ}_{16} - c^{FQ}_{17} - c^{FQ}_{18} - c^{FQ}_{20} + c^{FQ}_{22} + 4 c^{QQ}_{3} + 2 c^{QQ}_{5} \right. \\ & \left. + 2 c^{QQ}_{10} + 2 c^{QQ}_{11} \right) + \frac{i q }{2} \sqrt{\frac{5}{6}} ( - a^{R} + 4 a^{QQ}_{4} + 2 a^{QQ}_{5}) 
\deveq
a(1^{-})_{22} & = \frac{1}{6} (-q^{2}{}) \left(12 c^{FF}_{1} + 12 c^{FF}_{2} - 3 c^{FF}_{4} - 3 c^{FF}_{5} - 3 c^{FF}_{6} + 2 c^{FF}_{7} + 2 c^{FF}_{9} + 2 c^{FF}_{11} + 16 c^{FF}_{13} + 4 c^{FF}_{14} \right.\\ & + 4 c^{FF}_{15} + 9 c^{FT}_{1} + 9 c^{FT}_{2} - 6 c^{FT}_{8} - 6 c^{FT}_{10} - 12 c^{FT}_{12} + 12 c^{FQ}_{1} + 6 c^{FQ}_{2} + 6 c^{FQ}_{3} - 4 c^{FQ}_{6} + 8 c^{FQ}_{8} \\ & - 4 c^{FQ}_{10} + 8 c^{FQ}_{12} - 8 c^{FQ}_{14} + 16 c^{FQ}_{15} + 18 c^{TT}_{1} + 9 c^{TT}_{2} + 18 c^{TT}_{3} + 18 c^{TQ}_{1} + 12 c^{TQ}_{2} - 24 c^{TQ}_{3} \\ & \left. + 24 c^{QQ}_{1} - 12 c^{QQ}_{2} + 8 c^{QQ}_{3} + 32 c^{QQ}_{4} - 16 c^{QQ}_{5} \right) + \frac{1}{6} \left(5 a^{R} - 18 a^{TT}_{1} - 9 a^{TT}_{2} - 18 a^{TT}_{3} \right. \\ & \left. - 18 a^{TQ}_{1} + 24 a^{TQ}_{2} - 12 a^{TQ}_{3} - 24 a^{QQ}_{1} + 12 a^{QQ}_{2} - 32 a^{QQ}_{3} - 8 a^{QQ}_{4} + 16 a^{QQ}_{5} \right) 
\deveq
a(1^{-})_{23} & = \frac{1}{2 \sqrt{3}}(-q^{2}{}) \left(3 c^{FF}_{4} - 3 c^{FF}_{6} + 2 c^{FF}_{7} - 2 c^{FF}_{9} + 2 c^{FF}_{14} - 2 c^{FF}_{15} + 3 c^{FT}_{1} + 3 c^{FT}_{3} - 4 c^{FT}_{8} + 2 c^{FT}_{10} \right. \\ & - 2 c^{FT}_{12} - 3 c^{FQ}_{2} + 3 c^{FQ}_{3} - 2 c^{FQ}_{6} + 4 c^{FQ}_{8} + 2 c^{FQ}_{10} - 4 c^{FQ}_{12} + 6 c^{TT}_{1} + 3 c^{TT}_{2} + 6 c^{TT}_{3} + 3 c^{TQ}_{1} \\ & \left.+ 2 c^{TQ}_{2} - 4 c^{TQ}_{3}\right) + \frac{1}{2 \sqrt{3}}(-6 a^{TT}_{1} - 3 a^{TT}_{2} - 6 a^{TT}_{3} - 3 a^{TQ}_{1} + 4 a^{TQ}_{2} - 2 a^{TQ}_{3}) 
\deveq
a(1^{-})_{24} & = \frac{1}{6} (-q^{2}{}) \left(-2 c^{FF}_{8} - 2 c^{FF}_{10} - 2 c^{FF}_{12} - 8 c^{FF}_{13} + c^{FF}_{14} + c^{FF}_{15} + 3 c^{FT}_{9} + 3 c^{FT}_{11} + 3 c^{FT}_{12} - 2 c^{FQ}_{6} \right.\\ &  + 2 c^{FQ}_{7} - 2 c^{FQ}_{8} - 4 c^{FQ}_{9} - 2 c^{FQ}_{10} + 2 c^{FQ}_{11} - 2 c^{FQ}_{12} - 4 c^{FQ}_{13} - 2 c^{FQ}_{14} - 8 c^{FQ}_{15} - 2 c^{FQ}_{16} - 2 c^{FQ}_{17} \\ & - 2 c^{FQ}_{18} - 2 c^{FQ}_{19} - 2 c^{FQ}_{20} - 2 c^{FQ}_{21} + 6 c^{TQ}_{2} + 6 c^{TQ}_{3} + 6 c^{TQ}_{10} + 6 c^{TQ}_{11} + 8 c^{QQ}_{3} - 16 c^{QQ}_{4} - 4 c^{QQ}_{5} \\ & \left.+ 4 c^{QQ}_{10} - 8 c^{QQ}_{11} + 4 c^{QQ}_{12} - 8 c^{QQ}_{13}\right) + \frac{1}{3} (a^{R} - 3 a^{TQ}_{2} - 3 a^{TQ}_{3} + 8 a^{QQ}_{3} - 4 a^{QQ}_{4} + 2 a^{QQ}_{5}) 
\deveq
a(1^{-})_{25} & = \frac{1}{6 \sqrt{2}}(-q^{2}{}) \left(4 c^{FF}_{8} + 4 c^{FF}_{10} + 4 c^{FF}_{12} + 16 c^{FF}_{13} - 2 c^{FF}_{14} - 2 c^{FF}_{15} - 3 c^{FT}_{8} - 6 c^{FT}_{9} - 3 c^{FT}_{10} - 6 c^{FT}_{11} \right.\\ & - 12 c^{FT}_{12}  - 3 c^{FT}_{13} - 3 c^{FT}_{15} - 3 c^{FT}_{17} - 3 c^{FT}_{18} - 6 c^{FT}_{19} - 12 c^{FT}_{20} - 2 c^{FQ}_{6} - 4 c^{FQ}_{7} + 4 c^{FQ}_{8} + 8 c^{FQ}_{9}  \\ & - 2 c^{FQ}_{10} - 4 c^{FQ}_{11} + 4 c^{FQ}_{12} + 8 c^{FQ}_{13} - 8 c^{FQ}_{14} + 16 c^{FQ}_{15} - 2 c^{FQ}_{16} - 2 c^{FQ}_{17} - 2 c^{FQ}_{18} + 4 c^{FQ}_{19} - 2 c^{FQ}_{20} \\ & + 4 c^{FQ}_{21} - 12 c^{FQ}_{22} + 18 c^{TT}_{3} + 9 c^{TT}_{8}+ 12 c^{TQ}_{2} - 24 c^{TQ}_{3} + 6 c^{TQ}_{8} - 12 c^{TQ}_{9} + 6 c^{TQ}_{10} \\ & \left.- 12 c^{TQ}_{11} + 8 c^{QQ}_{3} + 32 c^{QQ}_{4} - 16 c^{QQ}_{5} + 4 c^{QQ}_{10} - 8 c^{QQ}_{11} - 8 c^{QQ}_{12} + 16 c^{QQ}_{13} \right) \\ & + \frac{1}{3 \sqrt{2}}( a^{R} - 9 a^{TT}_{3} + 12 a^{TQ}_{2} - 6 a^{TQ}_{3} - 16 a^{QQ}_{3} - 4 a^{QQ}_{4} + 8 a^{QQ}_{5}) 
\deveq
a(1^{-})_{26} & = - \frac{1}{2 \sqrt{6}}(-q^{2}{}) \left(c^{FT}_{8} + c^{FT}_{10} + 2 c^{FT}_{12} + c^{FT}_{13} + c^{FT}_{15} + c^{FT}_{17} + c^{FT}_{18} + 2 c^{FT}_{19} + 4 c^{FT}_{20} - 6 c^{TT}_{3} \right.\\ & \left.- 3 c^{TT}_{8} - 2 c^{TQ}_{2} + 4 c^{TQ}_{3} - 2 c^{TQ}_{8} + 4 c^{TQ}_{9}\right) + \frac{1}{\sqrt{6}}(-3 a^{TT}_{3} + 2 a^{TQ}_{2} -  a^{TQ}_{3}) 
\deveq
a(1^{-})_{27} & = - \frac{i q^3}{2 \sqrt{6}} (c^{FQ}_{6} + c^{FQ}_{10} + 2 c^{FQ}_{14} + c^{FQ}_{16} + c^{FQ}_{17} + c^{FQ}_{18} + c^{FQ}_{20} + 2 c^{FQ}_{22} - 3 c^{TQ}_{2} - 3 c^{TQ}_{10} \\ & - 4 c^{QQ}_{3} + 4 c^{QQ}_{5} - 2 c^{QQ}_{10} + 4 c^{QQ}_{11}) + \frac{i q}{2 \sqrt{6}} ( - a^{R} + 3 a^{TQ}_{3} + 4 a^{QQ}_{4} - 4 a^{QQ}_{5}) 
\deveq
a(1^{-})_{33} & = \frac{1}{2} (-q^{2}{}) (4 c^{FF}_{1} - 4 c^{FF}_{2} + c^{FF}_{4} -  c^{FF}_{5} + c^{FF}_{6} + 2 c^{FF}_{7} + 2 c^{FF}_{9} - 2 c^{FF}_{11} + c^{FT}_{1} - c^{FT}_{2} \\ & + 2 c^{FT}_{3} - 2 c^{FT}_{8} + 2 c^{FT}_{10} + 2 c^{TT}_{1} + c^{TT}_{2} + 2 c^{TT}_{3}) - \frac{1}{2} ( a^{R} + 2 a^{TT}_{1} +  a^{TT}_{2} + 2 a^{TT}_{3}) 
\deveq
a(1^{-})_{34} & = \frac{1}{2 \sqrt{3}}(-q^{2}{}) (-2 c^{FF}_{8} + 2 c^{FF}_{10} -  c^{FF}_{14} + c^{FF}_{15} + c^{FT}_{9} + c^{FT}_{11} + c^{FT}_{12} - 2 c^{FQ}_{6} - 2 c^{FQ}_{8} + 2 c^{FQ}_{10} \\ & + 2 c^{FQ}_{12} - 2 c^{FQ}_{16} + 2 c^{FQ}_{17} - 2 c^{FQ}_{18} - 2 c^{FQ}_{19} + 2 c^{FQ}_{20} + 2 c^{FQ}_{21} + 2 c^{TQ}_{2} + 2 c^{TQ}_{3} \\ & + 2 c^{TQ}_{10} + 2 c^{TQ}_{11}) - \frac{1}{\sqrt{3}}( a^{TQ}_{2} + a^{TQ}_{3}) 
\deveq
a(1^{-})_{35} & = \frac{1}{2 \sqrt{6}}(-q^{2}{}) (4 c^{FF}_{8} - 4 c^{FF}_{10} + 2 c^{FF}_{14} - 2 c^{FF}_{15} - 3 c^{FT}_{8} - 2 c^{FT}_{9} + 3 c^{FT}_{10} - 2 c^{FT}_{11} - 2 c^{FT}_{12} - 3 c^{FT}_{13} \\ & + 3 c^{FT}_{15} - 3 c^{FT}_{17} + 3 c^{FT}_{18} - 2 c^{FQ}_{6} + 4 c^{FQ}_{8} + 2 c^{FQ}_{10} - 4 c^{FQ}_{12} - 2 c^{FQ}_{16} + 2 c^{FQ}_{17} - 2 c^{FQ}_{18} + 4 c^{FQ}_{19} + 2 c^{FQ}_{20} \\ & - 4 c^{FQ}_{21} + 6 c^{TT}_{3} + 3 c^{TT}_{8} + 2 c^{TQ}_{2} - 4 c^{TQ}_{3} + 2 c^{TQ}_{10} - 4 c^{TQ}_{11}) + \frac{1}{\sqrt{6}}(-3 a^{TT}_{3} + 2 a^{TQ}_{2} -  a^{TQ}_{3}) 
\deveq
a(1^{-})_{36} & = \frac{1}{2 \sqrt{2}}(-q^{2}{}) (- c^{FT}_{8} + c^{FT}_{10} -  c^{FT}_{13} + c^{FT}_{15} -  c^{FT}_{17} + c^{FT}_{18} + 2 c^{TT}_{3} + c^{TT}_{8}) - \frac{1}{\sqrt{2}}( a^{R} + a^{TT}_{3})
\deveq
a(1^{-})_{37} & = \frac{i q^3}{2 \sqrt{2}} (- c^{FQ}_{6} + c^{FQ}_{10} -  c^{FQ}_{16} + c^{FQ}_{17} -  c^{FQ}_{18} + c^{FQ}_{20} + c^{TQ}_{2} + c^{TQ}_{10}) + \frac{i q}{2 \sqrt{2}} (- a^{R} + a^{TQ}_{3})
\deveq
a(1^{-})_{44} & = \frac{1}{3} (-q^{2}{}) (2 c^{FF}_{1} + 2 c^{FF}_{2} + c^{FF}_{4} + c^{FF}_{5} + c^{FF}_{6} + c^{FF}_{7} + c^{FF}_{9} + c^{FF}_{11} + 2 c^{FF}_{13} -  c^{FF}_{14} -  c^{FF}_{15} \\ & + 2 c^{FQ}_{1} - 2 c^{FQ}_{2} - 2 c^{FQ}_{3} + 2 c^{FQ}_{7} + 2 c^{FQ}_{9} + 2 c^{FQ}_{11} + 2 c^{FQ}_{13} + 2 c^{FQ}_{14} + 2 c^{FQ}_{15} + 2 c^{FQ}_{16} \\ & + 2 c^{FQ}_{17} + 2 c^{FQ}_{18} + 2 c^{FQ}_{19} + 2 c^{FQ}_{20} + 2 c^{FQ}_{21} + 12 c^{QQ}_{1} + 12 c^{QQ}_{2} + 4 c^{QQ}_{3} + 4 c^{QQ}_{4} \\ & + 4 c^{QQ}_{5} + 8 c^{QQ}_{6} + 8 c^{QQ}_{7} + 8 c^{QQ}_{8} + 8 c^{QQ}_{9} + 4 c^{QQ}_{10} + 4 c^{QQ}_{11} + 4 c^{QQ}_{12} + 4 c^{QQ}_{13}) \\ & - \frac{2}{3} (a^{R} + 6 a^{QQ}_{1} + 6 a^{QQ}_{2} + 2 a^{QQ}_{3} + 2 a^{QQ}_{4} + 2 a^{QQ}_{5}) 
\deveq
a(1^{-})_{45} & = \frac{1}{6 \sqrt{2}}(-q^{2}{}) (-8 c^{FF}_{1} - 8 c^{FF}_{2} - 4 c^{FF}_{4} - 4 c^{FF}_{5} - 4 c^{FF}_{6} - 4 c^{FF}_{7} - 4 c^{FF}_{9} - 4 c^{FF}_{11} - 8 c^{FF}_{13} + 4 c^{FF}_{14} \\ & + 4 c^{FF}_{15} - 3 c^{FT}_{1} - 3 c^{FT}_{2} + 3 c^{FT}_{4} + 6 c^{FT}_{5} + 3 c^{FT}_{6} + 6 c^{FT}_{7} + 3 c^{FT}_{9} + 3 c^{FT}_{11} + 3 c^{FT}_{12} + 3 c^{FT}_{14} \\ & + 3 c^{FT}_{16} - 3 c^{FT}_{17} - 3 c^{FT}_{18} + 3 c^{FT}_{19} + 6 c^{FT}_{20} - 8 c^{FQ}_{1} + 2 c^{FQ}_{2} + 2 c^{FQ}_{3} + 6 c^{FQ}_{4} + 6 c^{FQ}_{5} - 2 c^{FQ}_{7} \\ &  - 8 c^{FQ}_{9} - 2 c^{FQ}_{11} - 8 c^{FQ}_{13} - 2 c^{FQ}_{14} - 8 c^{FQ}_{15} - 2 c^{FQ}_{16} - 2 c^{FQ}_{17} - 8 c^{FQ}_{18} - 2 c^{FQ}_{19} - 8 c^{FQ}_{20} - 2 c^{FQ}_{21} \\ & + 6 c^{FQ}_{22} + 6 c^{TQ}_{2} + 6 c^{TQ}_{3} + 6 c^{TQ}_{4} + 6 c^{TQ}_{6} + 6 c^{TQ}_{7} + 6 c^{TQ}_{8} + 6 c^{TQ}_{9} + 6 c^{TQ}_{10} + 6 c^{TQ}_{11} + 8 c^{QQ}_{3} \\ & - 16 c^{QQ}_{4} - 4 c^{QQ}_{5} + 16 c^{QQ}_{6} - 8 c^{QQ}_{7} - 8 c^{QQ}_{8}  + 4 c^{QQ}_{9} + 8 c^{QQ}_{10} - 4 c^{QQ}_{11} - 4 c^{QQ}_{12} - 16 c^{QQ}_{13}) \\ & + \frac{1}{3 \sqrt{2}}( a^{R} - 3 a^{TQ}_{2} - 3 a^{TQ}_{3} + 8 a^{QQ}_{3} - 4 a^{QQ}_{4} + 2 a^{QQ}_{5}) 
\deveq
a(1^{-})_{46} & = \frac{1}{2 \sqrt{6}}(-q^{2}{}) (- c^{FT}_{1} -  c^{FT}_{2} + c^{FT}_{4} + 2 c^{FT}_{5} + c^{FT}_{6} + 2 c^{FT}_{7} + c^{FT}_{9} + c^{FT}_{11} + c^{FT}_{12} + c^{FT}_{14} \\ & + c^{FT}_{16} -  c^{FT}_{17} -  c^{FT}_{18} + c^{FT}_{19} + 2 c^{FT}_{20} + 2 c^{TQ}_{2} + 2 c^{TQ}_{3} + 2 c^{TQ}_{4} + 2 c^{TQ}_{6} + 2 c^{TQ}_{7} + 2 c^{TQ}_{8} \\ & + 2 c^{TQ}_{9} + 2 c^{TQ}_{10} + 2 c^{TQ}_{11}) - \frac{1}{ \sqrt{6}}( a^{TQ}_{2} + a^{TQ}_{3}) 
\deveq
a(1^{-})_{47} & = \frac{i q^3}{2 \sqrt{6}} (- c^{FQ}_{2} -  c^{FQ}_{3} + c^{FQ}_{4} + c^{FQ}_{5} + c^{FQ}_{7} + c^{FQ}_{11} + c^{FQ}_{14} + c^{FQ}_{16} + c^{FQ}_{17} + c^{FQ}_{19} + c^{FQ}_{21} \\ & + c^{FQ}_{22} + 8 c^{QQ}_{1} + 8 c^{QQ}_{2} + 4 c^{QQ}_{3} + 2 c^{QQ}_{5} + 8 c^{QQ}_{6} + 4 c^{QQ}_{7} + 4 c^{QQ}_{8} + 6 c^{QQ}_{9} + 4 c^{QQ}_{10} \\ & + 2 c^{QQ}_{11} + 2 c^{QQ}_{12}) + \frac{i q}{2 \sqrt{6}} (a^{R} + 8 a^{QQ}_{1} + 8 a^{QQ}_{2} + 4 a^{QQ}_{4} + 2 a^{QQ}_{5}) 
\deveq
a(1^{-})_{55} & = \frac{1}{6} (-q^{2}{}) (8 c^{FF}_{1} + 8 c^{FF}_{2} + 4 c^{FF}_{4} + 4 c^{FF}_{5} + 4 c^{FF}_{6} + 4 c^{FF}_{7} + 4 c^{FF}_{9} + 4 c^{FF}_{11} + 8 c^{FF}_{13} - 4 c^{FF}_{14} - 4 c^{FF}_{15} \\ & + 6 c^{FT}_{1} + 6 c^{FT}_{2} - 6 c^{FT}_{4} - 12 c^{FT}_{5} - 6 c^{FT}_{6} - 12 c^{FT}_{7} - 6 c^{FT}_{9} - 6 c^{FT}_{11} - 6 c^{FT}_{12} - 6 c^{FT}_{14} - 6 c^{FT}_{16} + 6 c^{FT}_{17} \\ & + 6 c^{FT}_{18} - 6 c^{FT}_{19} - 12 c^{FT}_{20} + 8 c^{FQ}_{1} + 4 c^{FQ}_{2} + 4 c^{FQ}_{3} - 12 c^{FQ}_{4} - 12 c^{FQ}_{5} - 4 c^{FQ}_{7} + 8 c^{FQ}_{9} - 4 c^{FQ}_{11} \\ & + 8 c^{FQ}_{13} - 4 c^{FQ}_{14} + 8 c^{FQ}_{15} - 4 c^{FQ}_{16} - 4 c^{FQ}_{17} + 8 c^{FQ}_{18} - 4 c^{FQ}_{19} + 8 c^{FQ}_{20} - 4 c^{FQ}_{21} - 12 c^{FQ}_{22} + 18 c^{TT}_{1} \\ & + 9 c^{TT}_{2} + 9 c^{TT}_{3} + 9 c^{TT}_{4} + 18 c^{TT}_{6} + 9 c^{TT}_{7} + 9 c^{TT}_{8} + 18 c^{TQ}_{1} + 6 c^{TQ}_{2} - 12 c^{TQ}_{3} + 6 c^{TQ}_{4} + 18 c^{TQ}_{5} \\ & + 6 c^{TQ}_{6} - 12 c^{TQ}_{7} + 6 c^{TQ}_{8} - 12 c^{TQ}_{9} + 6 c^{TQ}_{10} - 12 c^{TQ}_{11} + 24 c^{QQ}_{1} - 12 c^{QQ}_{2} + 4 c^{QQ}_{3} + 16 c^{QQ}_{4} - 8 c^{QQ}_{5} \\ & + 8 c^{QQ}_{6} + 20 c^{QQ}_{7} - 16 c^{QQ}_{8} - 4 c^{QQ}_{9} + 4 c^{QQ}_{10} - 8 c^{QQ}_{11}  - 8 c^{QQ}_{12} + 16 c^{QQ}_{13}) + \frac{1}{6} (4 a^{R} - 18 a^{TT}_{1} - 9 a^{TT}_{2} \\ & - 9 a^{TT}_{3} - 18 a^{TQ}_{1} + 12 a^{TQ}_{2} - 6 a^{TQ}_{3} - 24 a^{QQ}_{1} + 12 a^{QQ}_{2} - 16 a^{QQ}_{3} - 4 a^{QQ}_{4} + 8 a^{QQ}_{5}) 
\deveq
a(1^{-})_{56} & = \frac{1}{2 \sqrt{3}}(-q^{2}{}) (c^{FT}_{1} + c^{FT}_{2} -  c^{FT}_{4} - 2 c^{FT}_{5} -  c^{FT}_{6} - 2 c^{FT}_{7} -  c^{FT}_{9} -  c^{FT}_{11} -  c^{FT}_{12} - c^{FT}_{14} - c^{FT}_{16} \\ & + c^{FT}_{17} + c^{FT}_{18} -  c^{FT}_{19} - 2 c^{FT}_{20} + 6 c^{TT}_{1} + 3 c^{TT}_{2} + 3 c^{TT}_{3} + 3 c^{TT}_{4} + 6 c^{TT}_{6} + 3 c^{TT}_{7} + 3 c^{TT}_{8} \\ & + 3 c^{TQ}_{1} + c^{TQ}_{2} - 2 c^{TQ}_{3} + c^{TQ}_{4} + 3 c^{TQ}_{5} + c^{TQ}_{6} - 2 c^{TQ}_{7} + c^{TQ}_{8} - 2 c^{TQ}_{9} + c^{TQ}_{10} - 2 c^{TQ}_{11}) \\ & + \frac{1}{2 \sqrt{3}}(-6 a^{TT}_{1} - 3 a^{TT}_{2} - 3 a^{TT}_{3} - 3 a^{TQ}_{1} + 2 a^{TQ}_{2} -  a^{TQ}_{3})
\deveq
a(1^{-})_{57} & = - \frac{i q^3}{4 \sqrt{3}} (-2 c^{FQ}_{2} - 2 c^{FQ}_{3} + 2 c^{FQ}_{4} + 2 c^{FQ}_{5} + 2 c^{FQ}_{7} + 2 c^{FQ}_{11} + 2 c^{FQ}_{14} + 2 c^{FQ}_{16} + 2 c^{FQ}_{17} \\ & + 2 c^{FQ}_{19} + 2 c^{FQ}_{21} + 2 c^{FQ}_{22} - 3 c^{TQ}_{1} - 3 c^{TQ}_{2} - 3 c^{TQ}_{4} - 3 c^{TQ}_{5} - 3 c^{TQ}_{6} - 3 c^{TQ}_{8} - 3 c^{TQ}_{10} \\ & - 8 c^{QQ}_{1} + 4 c^{QQ}_{2} - 4 c^{QQ}_{3} + 4 c^{QQ}_{5} - 8 c^{QQ}_{6} - 4 c^{QQ}_{7} + 8 c^{QQ}_{8} - 4 c^{QQ}_{10} + 4 c^{QQ}_{11} + 4 c^{QQ}_{12}) \\ & -  \frac{i}{4 \sqrt{3}}q (2 a^{R} - 3 a^{TQ}_{1} - 3 a^{TQ}_{3} - 8 a^{QQ}_{1} + 4 a^{QQ}_{2} - 4 a^{QQ}_{4} + 4 a^{QQ}_{5}) 
\deveq
a(1^{-})_{66} & = \frac{1}{2} (-q^{2}{}) (2 c^{TT}_{1} + c^{TT}_{2} + c^{TT}_{3} + c^{TT}_{4} + 2 c^{TT}_{6} + c^{TT}_{7} + c^{TT}_{8}) - \frac{1}{2} (2 a^{TT}_{1} + a^{TT}_{2} +  a^{TT}_{3}) 
\deveq
a(1^{-})_{67} & = \frac{i q^3}{4} (c^{TQ}_{1} + c^{TQ}_{2} + c^{TQ}_{4} + c^{TQ}_{5} + c^{TQ}_{6} + c^{TQ}_{8} + c^{TQ}_{10}) + \frac{i q}{4}(a^{TQ}_{1} + a^{TQ}_{3}) 
\deveq
a(1^{-})_{77} & = \frac{1}{2} q^4 (-2 c^{QQ}_{1} -  c^{QQ}_{2} -  c^{QQ}_{3} - 2 c^{QQ}_{6} -  c^{QQ}_{7} -  c^{QQ}_{9} -  c^{QQ}_{10}) + \frac{1}{2} (-q^{2}{}) (2 a^{QQ}_{1} + a^{QQ}_{2} + a^{QQ}_{4}) \\
\end{aligned}
\end{equation*}

\begin{equation*}
\begin{aligned}
a(0^{+})_{11} & = \frac{2}{3} (-q^{2}{}) (2 c^{FF}_{1} + 2 c^{FF}_{2} + c^{FF}_{4} + c^{FF}_{5} + c^{FF}_{6} + 2 c^{FF}_{7} + 2 c^{FF}_{8} + 2 c^{FF}_{9} + 2 c^{FF}_{10} + 2 c^{FF}_{11} + 2 c^{FF}_{12} \\ & + 2 c^{FQ}_{1} - 2 c^{FQ}_{2} - 2 c^{FQ}_{3} - 3 c^{FQ}_{6} - 3 c^{FQ}_{7} - 3 c^{FQ}_{8} - 3 c^{FQ}_{9} - 3 c^{FQ}_{10} - 3 c^{FQ}_{11} - 3 c^{FQ}_{12} - 3 c^{FQ}_{13} + c^{FQ}_{16} \\ & + c^{FQ}_{17} + c^{FQ}_{18} + c^{FQ}_{19} + c^{FQ}_{20} + c^{FQ}_{21} + 6 c^{QQ}_{1} + 6 c^{QQ}_{2} + 6 c^{QQ}_{3} + 6 c^{QQ}_{4} + 6 c^{QQ}_{5} + 2 c^{QQ}_{6} + 2 c^{QQ}_{7} \\ & + 2 c^{QQ}_{8} + 2 c^{QQ}_{9} + 6 c^{QQ}_{14} + 6 c^{QQ}_{15} + 6 c^{QQ}_{16}) - 4 ( a^{QQ}_{1} + a^{QQ}_{2} + a^{QQ}_{3} + a^{QQ}_{4} + a^{QQ}_{5}) 
\deveq
a(0^{+})_{12} & = \frac{1}{6 \sqrt{2}}(-q^{2}{}) (-8 c^{FF}_{1} - 8 c^{FF}_{2} - 4 c^{FF}_{4} - 4 c^{FF}_{5} - 4 c^{FF}_{6} - 8 c^{FF}_{7} - 8 c^{FF}_{8} - 8 c^{FF}_{9} - 8 c^{FF}_{10} - 8 c^{FF}_{11} \\ & - 8 c^{FF}_{12} - 3 c^{FT}_{1} - 3 c^{FT}_{2} + 3 c^{FT}_{4} + 6 c^{FT}_{5} + 3 c^{FT}_{6} + 6 c^{FT}_{7} + 9 c^{FT}_{8} + 9 c^{FT}_{9} + 9 c^{FT}_{10} + 9 c^{FT}_{11} \\ & + 3 c^{FT}_{13} + 3 c^{FT}_{14} + 3 c^{FT}_{15} + 3 c^{FT}_{16} - 8 c^{FQ}_{1} + 2 c^{FQ}_{2} + 2 c^{FQ}_{3} + 6 c^{FQ}_{4} + 6 c^{FQ}_{5} + 12 c^{FQ}_{6} + 12 c^{FQ}_{7} \\ & - 6 c^{FQ}_{8} - 6 c^{FQ}_{9} + 12 c^{FQ}_{10}  + 12 c^{FQ}_{11} - 6 c^{FQ}_{12} - 6 c^{FQ}_{13} + 2 c^{FQ}_{16} + 2 c^{FQ}_{17} - 4 c^{FQ}_{18} - 4 c^{FQ}_{19} \\ & - 4 c^{FQ}_{20} - 4 c^{FQ}_{21} - 18 c^{TQ}_{2} - 18 c^{TQ}_{3} + 6 c^{TQ}_{4} + 6 c^{TQ}_{6} + 6 c^{TQ}_{7} + 18 c^{TQ}_{12} + 18 c^{TQ}_{13} - 24 c^{QQ}_{3} \\ & + 48 c^{QQ}_{4} + 12 c^{QQ}_{5} + 16 c^{QQ}_{6} - 8 c^{QQ}_{7} - 8 c^{QQ}_{8} + 4 c^{QQ}_{9} + 48 c^{QQ}_{14} - 24 c^{QQ}_{15} + 12 c^{QQ}_{16}) \\ & + \frac{1}{ \sqrt{2}}(- a^{R} + 3 a^{TQ}_{2} + 3 a^{TQ}_{3} - 8 a^{QQ}_{3} + 4 a^{QQ}_{4} - 2 a^{QQ}_{5}) 
\deveq
a(0^{+})_{13} & = \frac{1}{2 \sqrt{6}}(-q^{2}{}) (-4 c^{FF}_{7} - 4 c^{FF}_{8} + 4 c^{FF}_{9} + 4 c^{FF}_{10} -  c^{FT}_{1} -  c^{FT}_{2} + c^{FT}_{4} + 2 c^{FT}_{5} + c^{FT}_{6} + 2 c^{FT}_{7} \\ & + 3 c^{FT}_{8} + 3 c^{FT}_{9} + 3 c^{FT}_{10} + 3 c^{FT}_{11} + c^{FT}_{13} + c^{FT}_{14} + c^{FT}_{15} + c^{FT}_{16} + 6 c^{FQ}_{6} + 6 c^{FQ}_{7} + 6 c^{FQ}_{8} + 6 c^{FQ}_{9} \\ & - 6 c^{FQ}_{10} - 6 c^{FQ}_{11} - 6 c^{FQ}_{12} - 6 c^{FQ}_{13} + 2 c^{FQ}_{16} - 2 c^{FQ}_{17} + 2 c^{FQ}_{18} + 2 c^{FQ}_{19} - 2 c^{FQ}_{20} - 2 c^{FQ}_{21} + 12 c^{FQ}_{23} \\ & + 12 c^{FQ}_{24} - 6 c^{TQ}_{2} - 6 c^{TQ}_{3} + 2 c^{TQ}_{4} + 2 c^{TQ}_{6} + 2 c^{TQ}_{7} + 6 c^{TQ}_{12} + 6 c^{TQ}_{13}) + \frac{\sqrt{6}}{2 }( a^{TQ}_{2} +  a^{TQ}_{3}) 
\deveq
a(0^{+})_{14} & = (-q^{2}{}) (- c^{FQ}_{6} -  c^{FQ}_{7} -  c^{FQ}_{8} -  c^{FQ}_{9} -  c^{FQ}_{10} -  c^{FQ}_{11} -  c^{FQ}_{12} -  c^{FQ}_{13} -  c^{FQ}_{16} -  c^{FQ}_{17} - c^{FQ}_{18} \\ & -  c^{FQ}_{19} -  c^{FQ}_{20} -  c^{FQ}_{21} + 4 c^{QQ}_{3} + 4 c^{QQ}_{4} + 4 c^{QQ}_{5} + 2 c^{QQ}_{10} + 2 c^{QQ}_{11} + 2 c^{QQ}_{12} + 2 c^{QQ}_{13} \\ & + 4 c^{QQ}_{14} + 4 c^{QQ}_{15} + 4 c^{QQ}_{16}) + a^{R} - 4 a^{QQ}_{3} - 4 a^{QQ}_{4} - 4 a^{QQ}_{5} 
\deveq
a(0^{+})_{15} & = \frac{i q^3}{2 \sqrt{3}} (- c^{FQ}_{2} -  c^{FQ}_{3} + c^{FQ}_{4} + c^{FQ}_{5} - 3 c^{FQ}_{8} - 3 c^{FQ}_{9} - 3 c^{FQ}_{12} - 3 c^{FQ}_{13} + c^{FQ}_{16} \\ & + c^{FQ}_{17} + 4 c^{QQ}_{1} + 4 c^{QQ}_{2} + 12 c^{QQ}_{4} + 6 c^{QQ}_{5} + 4 c^{QQ}_{6} + 2 c^{QQ}_{9} + 12 c^{QQ}_{14} + 6 c^{QQ}_{16}) \\ & + \frac{i q}{2 \sqrt{3}} (a^{R} + 4 a^{QQ}_{1} + 4 a^{QQ}_{2} + 12 a^{QQ}_{3} + 6 a^{QQ}_{5}) 
\deveq
a(0^{+})_{16} & = - \frac{i q^3}{2} (c^{FQ}_{6} + c^{FQ}_{7} + c^{FQ}_{8} + c^{FQ}_{9} + c^{FQ}_{10} + c^{FQ}_{11} + c^{FQ}_{12} + c^{FQ}_{13} + c^{FQ}_{16} + c^{FQ}_{17} \\ & + c^{FQ}_{18} + c^{FQ}_{19} + c^{FQ}_{20} + c^{FQ}_{21} - 4 c^{QQ}_{3} - 4 c^{QQ}_{4} - 4 c^{QQ}_{5} - 2 c^{QQ}_{10} - 2 c^{QQ}_{11} - 2 c^{QQ}_{12} \\ & - 2 c^{QQ}_{13} - 4 c^{QQ}_{14} - 4 c^{QQ}_{15} - 4 c^{QQ}_{16}) -  \frac{i q}{2} (a^{R} - 4 a^{QQ}_{3} - 4 a^{QQ}_{4} - 4 a^{QQ}_{5}) 
\deveq
a(0^{+})_{22} & = \frac{1}{6} (-q^{2}{}) (4 c^{FF}_{1} + 4 c^{FF}_{2} + 2 c^{FF}_{4} + 2 c^{FF}_{5} + 2 c^{FF}_{6} + 4 c^{FF}_{7} + 4 c^{FF}_{8} + 4 c^{FF}_{9} + 4 c^{FF}_{10} + 4 c^{FF}_{11} \\ & + 4 c^{FF}_{12} + 3 c^{FT}_{1} + 3 c^{FT}_{2} - 3 c^{FT}_{4} - 6 c^{FT}_{5} - 3 c^{FT}_{6} - 6 c^{FT}_{7} - 9 c^{FT}_{8} - 9 c^{FT}_{9} - 9 c^{FT}_{10} - 9 c^{FT}_{11} \\ & - 3 c^{FT}_{13} - 3 c^{FT}_{14} - 3 c^{FT}_{15} - 3 c^{FT}_{16} + 4 c^{FQ}_{1} + 2 c^{FQ}_{2} + 2 c^{FQ}_{3} - 6 c^{FQ}_{4} - 6 c^{FQ}_{5} - 6 c^{FQ}_{6} - 6 c^{FQ}_{7} \\ & + 12 c^{FQ}_{8} + 12 c^{FQ}_{9} - 6 c^{FQ}_{10} - 6 c^{FQ}_{11} + 12 c^{FQ}_{12} + 12 c^{FQ}_{13} - 4 c^{FQ}_{16} - 4 c^{FQ}_{17}+ 2 c^{FQ}_{18} + 2 c^{FQ}_{19} + 2 c^{FQ}_{20} \\ & + 2 c^{FQ}_{21} + 18 c^{TT}_{1} + 9 c^{TT}_{2} + 27 c^{TT}_{3} + 9 c^{TT}_{4} + 9 c^{TT}_{5} + 27 c^{TT}_{9} + 18 c^{TQ}_{1}  + 18 c^{TQ}_{2} - 36 c^{TQ}_{3} + 12 c^{TQ}_{4} \\ & - 6 c^{TQ}_{6} - 6 c^{TQ}_{7} + 36 c^{TQ}_{12} - 18 c^{TQ}_{13} + 24 c^{QQ}_{1} - 12 c^{QQ}_{2} + 12 c^{QQ}_{3} + 48 c^{QQ}_{4} - 24 c^{QQ}_{5} + 16 c^{QQ}_{6} \\ & + 4 c^{QQ}_{7} + 4 c^{QQ}_{8} - 8 c^{QQ}_{9} + 48 c^{QQ}_{14} + 12 c^{QQ}_{15} - 24 c^{QQ}_{16}) + a^{R} - 3 a^{TT}_{1} - \frac{3}{2} a^{TT}_{2} - \frac{9}{2} a^{TT}_{3} \\ & - 3 a^{TQ}_{1} + 6 a^{TQ}_{2} - 3 a^{TQ}_{3} - 4 a^{QQ}_{1} + 2 a^{QQ}_{2} - 8 a^{QQ}_{3} - 2 a^{QQ}_{4} + 4 a^{QQ}_{5}
\deveq
a(0^{+})_{23} & = \frac{1}{2 \sqrt{3}}(-q^{2}{}) (2 c^{FF}_{7} + 2 c^{FF}_{8} - 2 c^{FF}_{9} - 2 c^{FF}_{10} + 2 c^{FT}_{1} -  c^{FT}_{2} + 3 c^{FT}_{3} + c^{FT}_{4} + 2 c^{FT}_{5} - 2 c^{FT}_{6} \\ & - 4 c^{FT}_{7} - 6 c^{FT}_{8} - 6 c^{FT}_{9} + 3 c^{FT}_{10} + 3 c^{FT}_{11} + c^{FT}_{13} + c^{FT}_{14} - 2 c^{FT}_{15} - 2 c^{FT}_{16} + 9 c^{FT}_{21} - 3 c^{FQ}_{2} \\ & + 3 c^{FQ}_{3} + 3 c^{FQ}_{4} - 3 c^{FQ}_{5} - 3 c^{FQ}_{6} - 3 c^{FQ}_{7} + 6 c^{FQ}_{8} + 6 c^{FQ}_{9} + 3 c^{FQ}_{10} + 3 c^{FQ}_{11} - 6 c^{FQ}_{12} - 6 c^{FQ}_{13} \\ & + 2 c^{FQ}_{16} - 2 c^{FQ}_{17} -  c^{FQ}_{18} -  c^{FQ}_{19} + c^{FQ}_{20} + c^{FQ}_{21} + 12 c^{FQ}_{23} - 6 c^{FQ}_{24} + 6 c^{TT}_{1} + 3 c^{TT}_{2} + 9 c^{TT}_{3}\\ & + 3 c^{TT}_{4} + 3 c^{TT}_{5} + 9 c^{TT}_{9} + 3 c^{TQ}_{1} + 3 c^{TQ}_{2} - 6 c^{TQ}_{3} + 2 c^{TQ}_{4} -  c^{TQ}_{6} -  c^{TQ}_{7} + 6 c^{TQ}_{12} - 3 c^{TQ}_{13}) \\ & + \frac{ \sqrt{3}}{2}(-2 a^{TT}_{1} -  a^{TT}_{2} - 3 a^{TT}_{3} -  a^{TQ}_{1} + 2 a^{TQ}_{2} -  a^{TQ}_{3}) 
\deveq
a(0^{+})_{24} & = \frac{1}{\sqrt{2}}(-q^{2}{}) (c^{FQ}_{6} + c^{FQ}_{7} + c^{FQ}_{8} + c^{FQ}_{9} + c^{FQ}_{10} + c^{FQ}_{11} + c^{FQ}_{12} + c^{FQ}_{13} + c^{FQ}_{16} + c^{FQ}_{17} + c^{FQ}_{18} \\ & + c^{FQ}_{19} + c^{FQ}_{20} + c^{FQ}_{21} - 3 c^{TQ}_{2} - 3 c^{TQ}_{3} - 3 c^{TQ}_{10} - 3 c^{TQ}_{11} + 3 c^{TQ}_{12} + 3 c^{TQ}_{13} - 4 c^{QQ}_{3} \\ & + 8 c^{QQ}_{4} + 2 c^{QQ}_{5} - 2 c^{QQ}_{10} + 4 c^{QQ}_{11} - 2 c^{QQ}_{12} + 4 c^{QQ}_{13} + 8 c^{QQ}_{14} - 4 c^{QQ}_{15} + 2 c^{QQ}_{16}) \\ & + \frac{1}{\sqrt{2}}(- a^{R} + 3 a^{TQ}_{2} + 3 a^{TQ}_{3} - 8 a^{QQ}_{3} + 4 a^{QQ}_{4} - 2 a^{QQ}_{5})
\deveq
a(0^{+})_{25} & = \frac{i~q^3}{2 \sqrt{6}} (c^{FQ}_{2} + c^{FQ}_{3} -  c^{FQ}_{4} -  c^{FQ}_{5} + 3 c^{FQ}_{8} + 3 c^{FQ}_{9} + 3 c^{FQ}_{12} + 3 c^{FQ}_{13} -  c^{FQ}_{16} -  c^{FQ}_{17} + 3 c^{TQ}_{1} \\ & - 9 c^{TQ}_{3} + 3 c^{TQ}_{4} + 9 c^{TQ}_{12} + 8 c^{QQ}_{1} - 4 c^{QQ}_{2} + 24 c^{QQ}_{4} - 6 c^{QQ}_{5} + 8 c^{QQ}_{6} - 2 c^{QQ}_{9} + 24 c^{QQ}_{14} - 6 c^{QQ}_{16}) \\ & + \frac{i~q}{2 \sqrt{6}} (- a^{R} + 3 a^{TQ}_{1} - 9 a^{TQ}_{2} + 8 a^{QQ}_{1} - 4 a^{QQ}_{2} + 24 a^{QQ}_{3} - 6 a^{QQ}_{5}) 
\deveq
a(0^{+})_{26} & = \frac{i~q^3}{2 \sqrt{2}} (c^{FQ}_{6} + c^{FQ}_{7} + c^{FQ}_{8} + c^{FQ}_{9} + c^{FQ}_{10} + c^{FQ}_{11} + c^{FQ}_{12} + c^{FQ}_{13} + c^{FQ}_{16} + c^{FQ}_{17} + c^{FQ}_{18} \\ & + c^{FQ}_{19} + c^{FQ}_{20} + c^{FQ}_{21} - 3 c^{TQ}_{2} - 3 c^{TQ}_{3} - 3 c^{TQ}_{10} - 3 c^{TQ}_{11} + 3 c^{TQ}_{12} + 3 c^{TQ}_{13} - 4 c^{QQ}_{3} \\ & + 8 c^{QQ}_{4} + 2 c^{QQ}_{5}  - 2 c^{QQ}_{10} + 4 c^{QQ}_{11} - 2 c^{QQ}_{12} + 4 c^{QQ}_{13} + 8 c^{QQ}_{14} - 4 c^{QQ}_{15} + 2 c^{QQ}_{16}) \\ & + \frac{i~q}{2 \sqrt{2}} (a^{R} - 3 a^{TQ}_{2} - 3 a^{TQ}_{3} + 8 a^{QQ}_{3} - 4 a^{QQ}_{4} + 2 a^{QQ}_{5}) 
\deveq
a(0^{+})_{33} & = \frac{1}{2} (-q^{2}{}) (4 c^{FF}_{1} - 4 c^{FF}_{2} + 4 c^{FF}_{3} + 2 c^{FF}_{4} - 2 c^{FF}_{5} + 2 c^{FF}_{6} + 4 c^{FF}_{7} + 4 c^{FF}_{8} + 4 c^{FF}_{9} + 4 c^{FF}_{10} \\ & - 4 c^{FF}_{11} - 4 c^{FF}_{12} + 12 c^{FF}_{16} + c^{FT}_{1} -  c^{FT}_{2} + 2 c^{FT}_{3} + c^{FT}_{4} + 2 c^{FT}_{5} -  c^{FT}_{6} - 2 c^{FT}_{7} - 3 c^{FT}_{8} \\ & - 3 c^{FT}_{9} + 3 c^{FT}_{10} + 3 c^{FT}_{11} + c^{FT}_{13} + c^{FT}_{14} -  c^{FT}_{15} -  c^{FT}_{16} + 6 c^{FT}_{21} + 2 c^{TT}_{1} + c^{TT}_{2} + 3 c^{TT}_{3} \\ & + c^{TT}_{4} + c^{TT}_{5} + 3 c^{TT}_{9}) - \frac{1}{2} (2 a^{R} + 2 a^{TT}_{1} + a^{TT}_{2} + 3 a^{TT}_{3}) 
\deveq
a(0^{+})_{34} & = - \sqrt{3/2} q^{5}{} (c^{FQ}_{6} + c^{FQ}_{7} + c^{FQ}_{8} + c^{FQ}_{9} -  c^{FQ}_{10} -  c^{FQ}_{11} -  c^{FQ}_{12} -  c^{FQ}_{13} + c^{FQ}_{16} -  c^{FQ}_{17} + c^{FQ}_{18} + c^{FQ}_{19} \\ & - c^{FQ}_{20} -  c^{FQ}_{21} + 2 c^{FQ}_{23} + 2 c^{FQ}_{24} -  c^{TQ}_{2} -  c^{TQ}_{3} -  c^{TQ}_{10} -  c^{TQ}_{11} + c^{TQ}_{12} + c^{TQ}_{13}) + \sqrt{3/2} (a^{TQ}_{2} + a^{TQ}_{3})
\deveq
a(0^{+})_{35} & = - \frac{i q^3}{2 \sqrt{2}} (c^{FQ}_{2} -  c^{FQ}_{3} -  c^{FQ}_{4} + c^{FQ}_{5} - 3 c^{FQ}_{8} - 3 c^{FQ}_{9} + 3 c^{FQ}_{12} + 3 c^{FQ}_{13} -  c^{FQ}_{16} + c^{FQ}_{17} \\ & - 6 c^{FQ}_{23} - c^{TQ}_{1} + 3 c^{TQ}_{3} -  c^{TQ}_{4} - 3 c^{TQ}_{12}) -  \frac{i q}{2 \sqrt{2}} (2 a^{R} -  a^{TQ}_{1} + 3 a^{TQ}_{2}) 
\deveq
a(0^{+})_{36} & = - \frac{i q^3 }{2 } \sqrt{\frac{3}{2}}  (- c^{FQ}_{6} -  c^{FQ}_{7} -  c^{FQ}_{8} -  c^{FQ}_{9} + c^{FQ}_{10} + c^{FQ}_{11} + c^{FQ}_{12} + c^{FQ}_{13} -  c^{FQ}_{16} + c^{FQ}_{17} -  c^{FQ}_{18} - c^{FQ}_{19} \\ & + c^{FQ}_{20} + c^{FQ}_{21} - 2 c^{FQ}_{23} - 2 c^{FQ}_{24} + c^{TQ}_{2} + c^{TQ}_{3} + c^{TQ}_{10} + c^{TQ}_{11} -  c^{TQ}_{12} -  c^{TQ}_{13}) - \frac{i q}{2 } \sqrt{\frac{3}{2}} (a^{TQ}_{2} + a^{TQ}_{3}) 
\deveq
a(0^{+})_{44} & = 4 (-q^{2}{}) (c^{QQ}_{1} + c^{QQ}_{2} + c^{QQ}_{3} + c^{QQ}_{4} + c^{QQ}_{5} + c^{QQ}_{6} + c^{QQ}_{7} + c^{QQ}_{8} + c^{QQ}_{9} + c^{QQ}_{10} + c^{QQ}_{11} \\ & + c^{QQ}_{12} + c^{QQ}_{13} + c^{QQ}_{14} + c^{QQ}_{15} + c^{QQ}_{16}) - 4 (a^{QQ}_{1} + a^{QQ}_{2} + a^{QQ}_{3} + a^{QQ}_{4} + a^{QQ}_{5}) 
\deveq
a(0^{+})_{45} & = i q^3 \sqrt{3}  (2 c^{QQ}_{4} + c^{QQ}_{5} + c^{QQ}_{11} + c^{QQ}_{13} + 2 c^{QQ}_{14} + c^{QQ}_{16}) + i q \sqrt{3}  (2 a^{QQ}_{3} + a^{QQ}_{5}) 
\deveq
a(0^{+})_{46} & = 2i q^3 (c^{QQ}_{1} + c^{QQ}_{2} + c^{QQ}_{3} + c^{QQ}_{4} + c^{QQ}_{5} + c^{QQ}_{6} + c^{QQ}_{7} + c^{QQ}_{8} + c^{QQ}_{9} + c^{QQ}_{10} + c^{QQ}_{11} + c^{QQ}_{12} \\ & + c^{QQ}_{13} + c^{QQ}_{14} + c^{QQ}_{15} + c^{QQ}_{16}) + 2i q (a^{QQ}_{1} + a^{QQ}_{2} + a^{QQ}_{3} + a^{QQ}_{4} + a^{QQ}_{5}) 
\deveq
a(0^{+})_{55} & = q^4 (- c^{QQ}_{1} - 3 c^{QQ}_{4} -  c^{QQ}_{6} - 3 c^{QQ}_{14}) + (-q^{2}{}) (a^{QQ}_{1} + 3 a^{QQ}_{3}) \deveq
a(0^{+})_{56} & = \frac{\sqrt{3}}{2}  q^4 (-2 c^{QQ}_{4} -  c^{QQ}_{5} -  c^{QQ}_{11} -  c^{QQ}_{13} - 2 c^{QQ}_{14} -  c^{QQ}_{16}) + \frac{\sqrt{3}}{2}  (-q^{2}{}) (2 a^{QQ}_{3} + a^{QQ}_{5}) 
\deveq
a(0^{+})_{66} & = - q^4 ( c^{QQ}_{1} + c^{QQ}_{2} + c^{QQ}_{3} + c^{QQ}_{4} + c^{QQ}_{5} + c^{QQ}_{6} + c^{QQ}_{7} + c^{QQ}_{8} + c^{QQ}_{9} + c^{QQ}_{10} + c^{QQ}_{11} + c^{QQ}_{12} \\ & + c^{QQ}_{13} + c^{QQ}_{14} + c^{QQ}_{15} +  c^{QQ}_{16}) + (-q^{2}{}) (a^{QQ}_{1} + a^{QQ}_{2} + a^{QQ}_{3} + a^{QQ}_{4} + a^{QQ}_{5}) 
\end{aligned}
\end{equation*}

\begin{equation*}
a(0^{-}) = (-q^{2}{}) (2 c^{FF}_{1} - 2 c^{FF}_{2} -  c^{FF}_{4} + c^{FF}_{5} -  c^{FF}_{6} + 2 c^{FT}_{1} - 2 c^{FT}_{2} - 2 c^{FT}_{3} + 4 c^{TT}_{1} - 4 c^{TT}_{2}) - a^{R} - 4 a^{TT}_{1} + 4 a^{TT}_{2}
\end{equation*}

\end{fleqn}

\normalsize

\end{appendix}


\end{document}